\documentclass[]{aastex631}
\usepackage{amsmath}
\usepackage{graphics}
\usepackage[T1]{fontenc}
\pdfoutput=1
\usepackage{xcolor}
\shorttitle{The LSST era of supermassive black holes accretion-disk reverberation mapping}
\shortauthors{Kova{\v c}evi{\'c} et al.}
\graphicspath{{./}{figures/}}

\begin{document}

\title{The LSST era of supermassive black holes accretion-disk reverberation mapping }

\author[0000-0001-5139-1978]{Andjelka B. Kova{\v c}evi{\'c}}
\affiliation{University of Belgrade-Faculty of Mathematics, Department of astronomy,
Studentski trg 16
Belgrade, Serbia}

\author[0000-0002-1172-128X]{Viktor Radovi{\'c}}
\affiliation{University of Belgrade-Faculty of Mathematics, Department of astronomy,
Studentski trg 16
Belgrade, Serbia}
\affiliation{PIFI Research Fellow, Key Laboratory for Particle Astrophysics, Institute of High Energy Physics, Chinese Academy of Sciences,19B Yuquan Road, 100049 Beijing, China}

\author[0000-0002-1134-4015]{Dragana Ili{\'c}}
\affiliation{University of Belgrade-Faculty of Mathematics, Department of astronomy,
Studentski trg 16
Belgrade, Serbia}
\affiliation{Hamburger Sternwarte, Universitat Hamburg, Gojenbergsweg 112, 21029 Hamburg, Germany}

\author[0000-0003-2398-7664]{Luka {\v C}. Popovi{\'c}}
\affiliation{University of Belgrade-Faculty of Mathematics, Department of astronomy,
Studentski trg 16
Belgrade, Serbia}
\affiliation{Astronomical Observatory, Volgina 7, 11000 Belgrade, Serbia}

\author[0000-0002-9508-3667]{Roberto J.~Assef}
\affiliation{N\'ucleo de Astronom\'ia de la Facultad de Ingenier\'ia y Ciencias\\ Universidad Diego Portales, Av. Ej\'ercito Libertador 441, Santiago, Chile}

\author[0000-0003-0820-4692]{Paula  S\'anchez-S\'aez}
\affiliation{Directorate for Science, European Southern Observatory, Karl-Schwarzschild-Strasse 2, Garching bei M{\"u}nchen, Germany}
\affiliation{Millennium Institute of Astrophysics (MAS), Nuncio Monse\~nor S\'otero Sanz 100, Providencia, Santiago, Chile}

\author[0000-0002-7052-6900]{Robert Nikutta}
\affiliation{NSF's NOIRLab, 950 N. Cherry Ave., Tucson, AZ 85719, USA}

\author[0000-0003-1784-2784]{Claudia M. Raiteri}
\affiliation{INAF-Osservatorio Astrofisico di Torino, Via Osservatorio 20, I-10025 Pino Torinese, Italy}

\author[0000-0001-9163-0064]{Ilsang Yoon}
\affiliation{The National Radio Astronomy Observatory, 520 Edgemont Road Charlottesville, VA 22903, USA}

\author[0000-0002-0957-7151]{Yasaman Homayouni}
\affiliation{The Pennsylvania State University, 201 Old Main, University Park, PA 16802, USA}

\author[0000-0001-5841-9179]{Yan-Rong Li}
\affiliation{Key Laboratory for Particle Astrophysics, Institute of High Energy Physics, Chinese Academy of Sciences, 19B Yuquan Road, Beijing 100049, China}

\author[0000-0003-3287-5250]{Neven Caplar}
\affiliation{Astrophysical Department, Princeton University,
4 Ivy Lane, Princeton, NJ, USA
}

\author[0000-0001-5848-4333]{Bozena Czerny}
\affiliation{Center for Theoretical Physics, Polish Academy of Sciences, Al.Lotnikow 32/46, 02-668 Warsaw, Poland}

\author[0000-0002-5854-7426]{Swayamtrupta Panda}
\affiliation{Center for Theoretical Physics, Polish Academy of Sciences, Al.Lotnikow 32/46, 02-668 Warsaw, Poland}
\affiliation{Nicolaus Copernicus Astronomical Center, Polish Academy of Sciences, ul. Bartycka 18, 00-716 Warsaw, Poland}
\affiliation{Laborat\'orio Nacional de Astrof\'isica - MCTIC, R. dos Estados Unidos, 154 - Na\c{c}\~oes, Itajub\'a - MG, 37504-364, Brazil}

\author{Claudio Ricci}
\affiliation{N\'ucleo de Astronom\'ia de la Facultad de Ingenier\'ia, Universidad Diego
Portales, Av. Ej\'ercito Libertador 441, Santiago, Chile}
\affiliation{Kavli Institute for Astronomy and Astrophysics, Peking University, Beijing
100871, China}

\author{Isidora Jankov}
\affiliation{University of Belgrade-Faculty of Mathematics, Department of astronomy,
Studentski trg 16
Belgrade, Serbia}

\author{Hermine Landt}
\affiliation{Centre for Extragalactic Astronomy, Department of Physics, Durham University, South Road, Durham DH1 3LE, UK}

\author[0000-0002-4569-016X]{Christian Wolf}
\affiliation{Research School of Astronomy and Astrophysics, Australian National University, Weston Creek ACT 2611, Australia}
\affiliation{Centre for Gravitational Astrophysics, Australian National University, Canberra ACT
2600, Australia}

\author{Jelena Kova{\v c}evi{\'c}-Doj{\v c}inovi{\' c}}
\affiliation{Astronomical Observatory, Volgina 7, 11000 Belgrade, Serbia}

\author[0000-0002-8231-0963]{Ma{\v s}a Laki{\'c}evi{\'c}}
\affiliation{Astronomical Observatory, Volgina 7, 11000 Belgrade, Serbia}

\author[0000-0003-0880-8963]{Đorđe V.\,Savić}
\affiliation{Institut d’Astrophysique et de Géophysique, Université de Liège, Allée du 6 Août 19c, 4000 Liège, Belgium}
\affiliation{Astronomical Observatory, Volgina 7, 11000 Belgrade, Serbia}

\author{Oliver Vince}
\affiliation{Astronomical Observatory, Volgina 7, 11000 Belgrade, Serbia}

\author[0000-0001-7453-2016]{Sa{\v s}a Simi{\'c}}
\affiliation{Faculty of Science, University of Kragujevac, Radoja Domanovi\'ca 12,
34000 Kragujevac, Serbia}

\author{Iva {\v C}vorovi{\'c}-Hajdinjak}
\affiliation{University of Belgrade-Faculty of Mathematics, Department of astronomy,
Studentski trg 16
Belgrade, Serbia}
\author{Sladjana Mar{\v c}eta-Mandi{\'c}}
\affiliation{Astronomical Observatory, Volgina 7, 11000 Belgrade, Serbia}



\begin{abstract}


The Vera C. Rubin Observatory's Legacy Survey of Space and Time (LSST) will detect an unprecedentedly large sample of actively accreting supermassive black holes with typical accretion disk (AD) sizes of a few light days. This brings us to face challenges in the reverberation mapping (RM) measurement of AD sizes {in active galactic nuclei (AGNs)} using interband continuum delays. We examine the effect of LSST cadence strategies on AD RM using our metric \texttt{AGN\_TimeLagMetric}. It accounts for redshift, cadence, the magnitude limit, and magnitude corrections for dust extinction. Running our metric on different LSST cadence strategies, we produce an atlas of the performance estimations for LSST photometric RM {measurements}. We provide an upper limit on the estimated number of quasars for which the AD time lag can be computed within $0<z<7$ using the features of our metric. We forecast that the total counts of such objects will increase as the mean sampling rate of the survey decreases. {The AD time lag measurements are expected for  $>1000$ sources in each Deep Drilling field (DDF, ($10\,\,  \rm{deg}^{2}$)) in any filter, with the redshift distribution of these sources peaking at  $z \approx 1$.} We find the LSST observation strategies with a good cadence ($\lesssim 5$ days) and a long cumulative season ($\sim$ 9 yr), as proposed for LSST DDF, are favored for the AD size measurement. We create synthetic LSST light curves for the most suitable DDF cadences and determine RM time lags to demonstrate the impact of the best cadences based on the proposed metric.


\end{abstract}



\section{Introduction} \label{sec:intro}

 The first Event Horizon Telescope \citep[EHT,][]{2019ApJ...875L...4E} observations of M87 revealed a prominent  ring of $\sim 40\mu$as in diameter, compatible with a lensed photon {ring} encircling the shadow of a
supermassive black hole (SMBH).
This iconic image, in fact, portrays the accretion disk (AD) surrounding {the} SMBH.
As a result, we are now closer than ever to
{constructing} a definitive model of accretion disks.
{Inflowing} gas quickly accretes from {a few parsecs} down to {a few Schwarzschild radii of the SMBH} forming a ring-like structure (the AD)  that characterizes the phenomenon of {an active galactic nucleus (AGN)}.
There is only a short ($\sim$2 Myr) active phase\footnote{For luminous quasars, the activity period can be of order 10 Myrs \citep[e.g.,][]{2005ApJ...630..705H}.} for these objects, followed by prolonged periods of inactivity \citep[see][]{2021ApJ...917...53A} \footnote{{Continuous accretion is proposed} to
cope with the demands of growing black holes in the early universe \citep{2021MNRAS.501.4289Z}.}.
Besides theoretical investigations, 
a variety of activity cycles have been observed in radio and optical {active galactic nuclei (AGNs)}. The longest timescales of $10^{7}-10^{8}$ years have been traced using radio wavelengths, and short timescales of $10^{4}-10^{5}$ years (and down to years) have been estimated using {radio} wavelengths \citep[$\sim 10^{6}$ yr,][]{2021A&A...650A.170B}, {radio and optical wavelengths}
\citep[central engine already quiescent, but the remnant of the past activity is on scales of $\sim 10^{4}$yr,][review of multiple cycles]{2019ApJ...870...65I, 2017NatAs...1..596M},  {optical wavelengths} \citep[e.g., changing look AGN where the continuum changes  over a time-scale of years,][]{1985A&A...146L..11K, 2020A&A...638A..13I,2019MNRAS.483..558O, 2022arXiv220610056P}. \, {A recent review on the  AGN  active phase timescales is reported by \cite{2022A&A...662A..28M}.}

The most luminous AGNs, known as quasars \citep{1963ApJ...138...30M,1963Natur.197.1040S}, are thought to be  SMBH accreting matter at  high rates through  accretion disks that {are} both geometrically thin and optically thick \citep{1989ApJ...346...68S}.
In contrast, most AGNs in the local Universe\footnote{{We denote as local Universe the regions within 300\,Mpc $(z<0.1)$ from Earth} \citep{2016AstBu..71..155T}.}, such as the {one in} M87, {correspond to} SMBHs fed by hot plasma at significantly lower accretion rates than quasars \citep{1977ApJ...214..840I,1999MNRAS.303L...1B, 2014ARA&A..52..529Y}.


We already have a typical several-component model of AGN that includes the AD, broad line region (BLR), corona, molecular torus, narrow line region (NLR), jets, and captures the observed features of spectral energy distributions for AGN. Nevertheless, the precise geometry and behavior of these components are mostly unknown \citep[for a review, see][]{2017A&ARv..25....2P}. {Even though the entirety of the large scale structure has been identified, the features of the central engine are unclear due to its small dimensions, even for the nearest AGNs. As a result, additional information must be gathered indirectly, through the comparison of theoretical model predictions and observations.}
Recent developments in instrumentation have allowed  the BLR to be {spatially} resolved in the case of nearby active galaxies  \citep{2018Natur.563..657G, 2019ApJ...875L...4E}.
However,  at least in the near future, it will be difficult to resolve a significant sample of more distant AGNs. Fortunately, the SMBH AD may be examined using either microlensing studies of gravitationally lensed quasars to determine the accretion disk's half light radii at different wavelengths \citep{2012ApJ...756...52M, 2015ApJ...806..258M}, or reverberation mapping (RM) investigations of the SMBH's immediate surroundings \citep{1982ApJ...255..419B}. RM offers information on the structure and kinematics of the BLR and accretion disk \citep[see early works of][]{1973ApL....13..165C,1986ApJ...305..175G} in the time domain. According to the RM premises, the main driver of fluctuations in the BLR lines is accretion disk emission variability. Observing the time delays between disk and broad line emission can be used to calculate the BLR dimension {which in turn allows to estimate the SMBH mass}. More than 100 robust dynamical black hole mass {estimates} have been published in the literature, slowly building up to a statistically significant sample \citep{2015ApJ...806...22D,2000ApJ...533..631K,2019A&A...625A..62T,2019FrASS...6...75P,2004ApJ...613..682P}.

The \citet{1973A&A....24..337S} model {was} the first to provide a more comprehensive picture of an optically thick, geometrically thin accretion disk, {albeit in the context of} stellar-mass black holes. Soon, \citet{1979ApJ...229..318G} {proposed} magnetic reconnection in a corona above a \citet{1973A&A....24..337S} disk to explain the  hard X-ray  emission {observed in AGN}. The central corona {can} directly illuminate and heat the outer AGN disk \citep[e.g.,][]{2002apa..book.....F},  resulting in the "lamppost/reprocessing" paradigm\footnote{Electrons heated by a magnetic field transfer their energy to optical/UV photons and upscatter them into the X-band}.
According to this hypothesis, the fluctuating X-ray energy from the corona can be reprocessed and detected in the UV/optical light from the disk.

Also, \citet{2019ApJ...887...10S} and \citet{2019ApJ...876..155S} {proposed a relation between SMBH mass and central velocity dispersion that is dependent on morphological type, capable of more precisely estimating SMBH masses in other galaxies, and has implications for calibrating virial f-factors.}
Similarly to the case of the BLR,  the light
travel time across the AD can be inferred {through} RM time delays \citep[see e.g.,][]{1998ApJ...500..162C}.
In particular, in classic AD theory, effective temperature is coupled to black hole mass and accretion rate \citep{2007MNRAS.380..669C,2008ApJ...677..884L, 2010ApJ...712.1129M, 2016ApJ...821...56F,2018ApJ...866..115P}.
When compared to the outer and cooler AD regions, the reprocessing model demonstrates that radiation from the innermost part of the accretion disk, closer to the SMBH, has a peak of emission at shorter wavelengths, and its variability is detected \citep{2007MNRAS.380..669C, 2013ApJ...772....9C, 2015ApJ...806..129E}  with a mean delay scaling $\tau\sim \lambda ^{4/3}$ \citep{1998ApJ...500..162C, 2016MNRAS.456.1960S}.
An alternate AGN variability model proposed by \citet{2011ApJ...727L..24D} assumes large local temperature fluctuations in {the} AD as the origin of flux variability.
While it can explain the amplitude and time scales of observed variability \citep{2010ApJ...721.1014M}, it cannot explain the {wavelength dependence of the time lags} \citep{2016MNRAS.456.1960S}.
{\citet{2022arXiv220108943Y} model AGN variability as a Damped Harmonic Oscillator and offer more insight; the modeled short-term variability behaves differently in the vicinity of 2500A, which could be an imprint of a different accretion flow geometry rather than a homogeneous thin disk. }

 While thin ADs are the most robust explanation for  AGN luminosity and variability, multiple studies have found that the AD sizes depart  from the immediate prediction of the
model.
Estimated accretion disk sizes from RM campaigns are three times larger than the expected based on thin-disk theory \citep{2005ApJ...622..129S, 2014ApJ...788...48S, 2015ApJ...806..129E, 2016ApJ...821...56F, 2017ApJ...836..186J,2018ApJ...857...53C, 2018ApJ...862..123M}. Noticeably, microlensing studies of luminous lensed quasars have independently obtained comparable conclusions \citep{2006ApJ...648...67P, 2007ApJ...661...19P,2010ApJ...712.1129M,2013ApJ...769...53M,2016AN....337..356C}.

{Using data from the Neil Gehrels Swift Observatory} \citep[see][]{2019ApJ...870..123E}, \citet{2014ApJ...788...48S} and \citet{2014MNRAS.444.1469M} found that the UV {leads} the optical emission in NGC 2617 and NGC 5548, respectively.
Following these campaigns,  \citet{2017ApJ...840...41E} and \citet{2019ApJ...870..123E} observed {four} AGNs (NGC 5548, NGC 4151, NGC 4593, and Mrk
509) {and found} significantly correlated UV/optical fluctuations, and the same overall pattern of interband delays growing from UV to optical wavelengths.
While these lags are two times larger than expected {from classical theory}, the uncertainties on the predicted lags are large, so it is unclear whether this is {inconsistent with} the usual thin disk picture.
Using Pan-STARRS light curves, \citet{2017ApJ...836..186J} measured the continuum lags of $\sim 200$ quasars {(the mean lags between the \textit{g} band and the
\textit{r, i}, and \textit{z} bands for the whole sample are 1.1, 2.1, and 3.0 days)}, whereas \citet{2018ApJ...862..123M} used the Dark Energy Survey's (DES) photometric data  to measure the disk sizes of 15 quasars
{(e.g. lags $\leq 3.0$ days in \textit{r,i} band, and $\leq5.0$ days in \textit{z} band relative to \textit{g}-band)}.
\citet{2019ApJ...880..126H} recently measured the continuum lags of 95 quasars using photometric data from the Sloan Digital Sky Survey (SDSS) RM project. They found that average size of  measured disks is consistent with the  \citet{1973A&A....24..337S} analytic thin-disk model. {However, it was shown that there is a wide variety of disk sizes, both small and large, that are in excess of the measurement uncertainties.}

\citet{2020ApJS..246...16Y} assessed the time lags and accretion disk sizes of 22 quasars with {SMBH} masses ranging from $10^{7}$ to $10^{9} M_{\odot}$ with the data from
from the Dark Energy Survey (DES) standard-star fields and the supernova C fields.
They concluded that when the disk variability is taken into account,  most of the reported accretion disk diameters are compatible with the predictions of the classic thin-disk model.
\citet{2022arXiv220108533G} compiled a sample of 92 AGNs at $z<0.75$ with  \textit{gri}  photometric light curves from the archival data of the Zwicky Transient Facility and found that AD sizes are systematically larger than expected, although the ratio of the measured to theoretical disk sizes depend on using the emissivity -- or responsivity -- weighted disk radius. 

 According to a recent study conducted by \citet{2021ApJ...922..151K}, short CIV and Ly$\alpha$ delays in the spectra of Mrk 817 show that the accretion disk extends beyond the UV broad-line region. {Also, \citet{2022MNRAS.509.2637N}  suggested that the large bulk of observed continuum delays are due to varying diffuse BLR emission rather than {from disk emission} \citep[see also][]{2001ApJ...553..695K,2018MNRAS.481..533L}. }


We are about to enter an era in which an abundance of new data provides the opportunity to settle the disk size dispute and unravel diverse effects. The Vera C. Rubin Observatory opens a window of discovery in our dynamic universe through its Legacy Survey of Space and Time (LSST).
The LSST  project  {is intended to investigate dark energy and dark matter, the solar system inventory, the transient optical sky, and to map the Milky Way \citep[see][]{2019ApJ...873..111I}.  The distribution of revisit times (i.e., the cadence of
observations) is among the  main constraints on data and system properties placed
by the four main science themes \citep[see][]{2019ApJ...873..111I}}.
The {LSST community} has conducted a thorough  investigation of prospective cadences  \citep{2014SPIE.9150E..14C}.
Metrics developed by the LSST community provide a scientific  testbed for evaluating {a suite of} cadence simulations \citep[see details in][]{2022ApJS..258....1B} employing the  Metrics Analysis Framework \citep[MAF][]{2014SPIE.9149E..0BJ,Jon20} and The Operations Simulator  \citep[OpSim][]{2016SPIE.9911E..25R}. 
{Both codes are open source. The  \texttt{OpSim} generates databases containing 10-year pointing histories, complete with weather, seeing, and sky
brightness information.  It is more often now known  as the ‘Feature Based Scheduler’ or FBS,  and its fifth version is used for this analysis. The \, \texttt{MAF}  offers an application programming interface (API) for inferring image properties (e.g.,
seeing, proper motion) over various spatial scales, as well as tools for predicting the properties of photometric measurements for objects and light curves inserted in survey simulations 
 \citep[see details in][]{2022ApJS..258....1B}.}
{LSST} is expected {to tightly constrain} the radial dependence of the effective temperature in  accretion disks {through its \textit{u,g,r,i,z,y} } light curves with daily cadences {of} a large sample of AGNs  \citep[see details in AGN SC Roadmap][]{Shemmer18}. 
Assuming a $5\sigma$ depth of 24.6 mag in the \textit{r}-band, \citet{2021A&A...645A.103D} estimated a sky density of $6.2\times 10^6$ AGNs for the LSST main survey of $\sim 18000 \mathrm{deg}^{2}$. Specific metrics on the number of quasars that will be detected in the 10-year co-added \textit{i}-band
observations as a function of \texttt{OpSim}  run have been developed by \citet{Assef21}.


The LSST AGN and Transient and Variable Stars (TVS) Scientific Collaborations (SCs) devised a variety of metrics\footnote{The metrics have been presented at the LSST Project and Community Workshop 2021 in Survey Strategy III: Domain-Specific Cadence Optimization Discussions \url{project.lsst.org/meetings/rubin2021/content/survey-strategy-iii-domain-specific-cadence-optimization-discussions}. Joint session of AGN, TVS microlensing, and Strong Lensing scientific collaborations titles Together we stand! was held during The second SCOC science collaboration workshop \url{https://project.lsst.org/meetings/scoc-sc-workshop2/agenda}.}  to quantify the impact of LSST observing strategies on   different variability and photometric redshift measurements as well as quasar counts.
The Blazar Variability metric by \citet{2022ApJS..258....3R} evaluates the performance of LSST observing strategies for extracting information from data linked to the shortest time and spatial scales of blazars (less than 1 hr to years). Related to largest scales, \citet{Assef212, Assef21}
have developed two metrics to evaluate the 10-year co-added depths expected for each band in the context of
photometric redshifts for type-1 quasars.
 \citet{Richards21} have created a model-independent metric that quantifies the degree to which we can deduce the structure function of variable sources (e.g., AGNs) {observed by LSST)}. 
 The metric described in {the following sections} complements this suite by quantifying the impact of LSST observing strategies on the AD and BLR  photometric 
reverberation mapping, {covering time scales from a few  up to 400  light days. In addition, as  LSST  will be able to measure AGN dust sublimation radii which recovery critically depends on the cadence as well as the continuity of observations \citep[see][]{2014ApJ...784L...4H}, our metric could be applied on the corresponding time scales of several tens of days.  The caveat, however, is that if most accretion discs dominate the AGN 
emission in the NIR \citep[see e.g.,][]{10.1111/j.1365-2966.2011.18383.x}, the dust sublimation 
radius reverberation signal will be difficult to isolate.}

\citet{2021MNRAS.505.5012K} developed {a} metric {to assess the impact of} different LSST cadences {on the detection of} periodicity and time lags for {AGN}.
While  modeling specific object classes is useful for assessing the science implications, they  may not cover the entire discovery space.
When paired with survey design, metrics that provide quantized estimates rather than continuous probability distributions can be useful for survey design optimization \citep{2022ApJS..258...13B}.

 Based on the AGN SC roadmap premises as well as a need for  metrics, the goal of this work is to evaluate different LSST cadences for the science case of measuring AGN interband time delays, specifically {in the} Deep Drilling Fields (DDFs), and to identify  {suboptimal cadences}. {
 The LSST Deep Drilling Fields will have the deepest co-add depths of 
the survey \citep[see][]{2018arXiv181106542B}, and are expected to exceed those achieved by 8m class 
observatories in the same areas \citep[see][]{10.1093/pasj/psab122}.
  Roughly 10-20\% 
of the LSST observing time will be used in mini-surveys and in the six 
DDFs.  
Through a combination of depth and cadence, the DDFs are expected 
to provide the most complete AGN catalogs and their host galaxies
\citep{2018arXiv181106542B, 2022arXiv220606432Z}.
 }
 
 We used our metric \citep{2021MNRAS.505.5012K} simulation in the LSST \texttt{MAF}  to test various cadence strategies. For the best strategies revealed by our metric \texttt{MAF}  simulations, {we simulated AGN light curves using a damped random walk (DRW) model}, as the simplest Gaussian random process model that can  describe  the optical light curves of
quasars more-or-less reasonably well \citep[DRW,][]{2009ApJ...698..895K, 2010ApJ...708..927K,2013ApJ...765..106Z}. However, the following limitation of the DRW model should be kept in mind:
as \cite{2017A&A...597A.128K} identified, it is important that the time baseline of the light curve is considerably longer than the damping timescale.
Also,  \cite{2022arXiv220102762S} imply that even a 20-year baseline is unlikely to be long enough to constrain the damping timescale in some of the quasars in their sample. This could be due to long-term trends in the  light curve, or  that these processes are more complex than a basic DRW  with only one characteristic timescale \citep[see][]{2022arXiv220102762S}.
 We also used the AGN lamppost\footnote{{The lamppost is typically considered a point-like source hovering over the black hole spin axis and associated with} the X-ray corona.} AD model \citep{2007MNRAS.380..669C} to account for {wavelength dependence} and  replicate delayed LSST multiband light curves. Following that, we used \texttt{OpSim}  1.7-2.0 to produce data points for the delayed multiband light curves based on observational sequences of different cadences. We used a shifting method inspired by \texttt{PyCS} \citep{2013A&A...553A.120T} but with Partial Curve Mapping  \citep[PCM,][]{Jekel2019} to determine the similarity of shifted curves in order to calculate the time delay from the simulated observations.

The following is a general outline of the structure of the article.
{Section \ref{sec:expectation} answers the question of what lags are expected for different black hole masses.}
Section \ref{sec:model} {describes the assumed accretion disk model,  and how the synthetic
light curves, which are used to probe the  optimal cadences indicated by our metric, are constructed. }
Section  \ref{sec:estimator} {describes two tiers of estimators of cadence strategies. The first tier shows our metric run, which quantifies the optimal LSST cadences for AD time lag measurements using the LSST \texttt{MAF}. {In the} second, LSST multiband light curves are created under the assumption of generic quasar parameters and optimal LSST cadences, and their time delays are assessed with respect to the \texttt{u}-band.}
Section \ref{sec:results} presents the results, which  are interpreted and discussed in Section \ref{sec:discussion} in more detail. 
{Section \ref{sec:final}  concludes the article and presents further implications of this work.}



 


\section{Distribution in time lag  of LSST observable AGNs}\label{sec:expectation}

{For reverberation mapping, it is not only interesting to know  the efficiency of the survey observing strategies, but also what lags are expected for different black hole masses. For this purpose,} we use the quasar luminosity function \citep[QLF, ][]{2007ApJ...654..731H} to derive a probability density of AGNs per measured time lag. 
The probability
for measuring the time lags $p_{\tau}=p(M, L,z)$ {is} estimated from a
uniform sample of SMBH masses $M\in(10^{6},10^{10}) M_\odot$, {whose luminosities are inferred according to procedure described  in \citet{2021MNRAS.505.5012K}} and use the standard thin-disk model to calculate the expected mean time lags corresponding to LSST filters. {The probability  $p_{\tau}$ in our approach  equals to our metric (containing expected mean time lags with respect to \textit{u} band)  divided by total number of measurements. The distribution in time lag ($\tau$) of LSST observable AGNs is given by} :

\begin{equation}
   \frac{dN}{d\tau}={4\pi}\frac{d^{2}V}{d\tau dz}\int^{+\infty}_{\log L(z)} \frac{d^{2}\mathcal{N}}{d\log L dV}p_{\tau} d\log L dz 
\end{equation}

\noindent where 
$\frac{d^{2}\mathcal{N}}{d\log L dV}$ is a
double-power-law QLF with redshift-dependent slopes from
\citet[][see their Table 4, second column]{2020MNRAS.495.3252S} and
$\frac{dV}{dz}$ is
the co-moving volume per redshift (calculated with \texttt{CosmoloPy} \footnote{\url{http://roban.github.com/CosmoloPy/}}).

\begin{figure}
\includegraphics[width=0.48\textwidth]{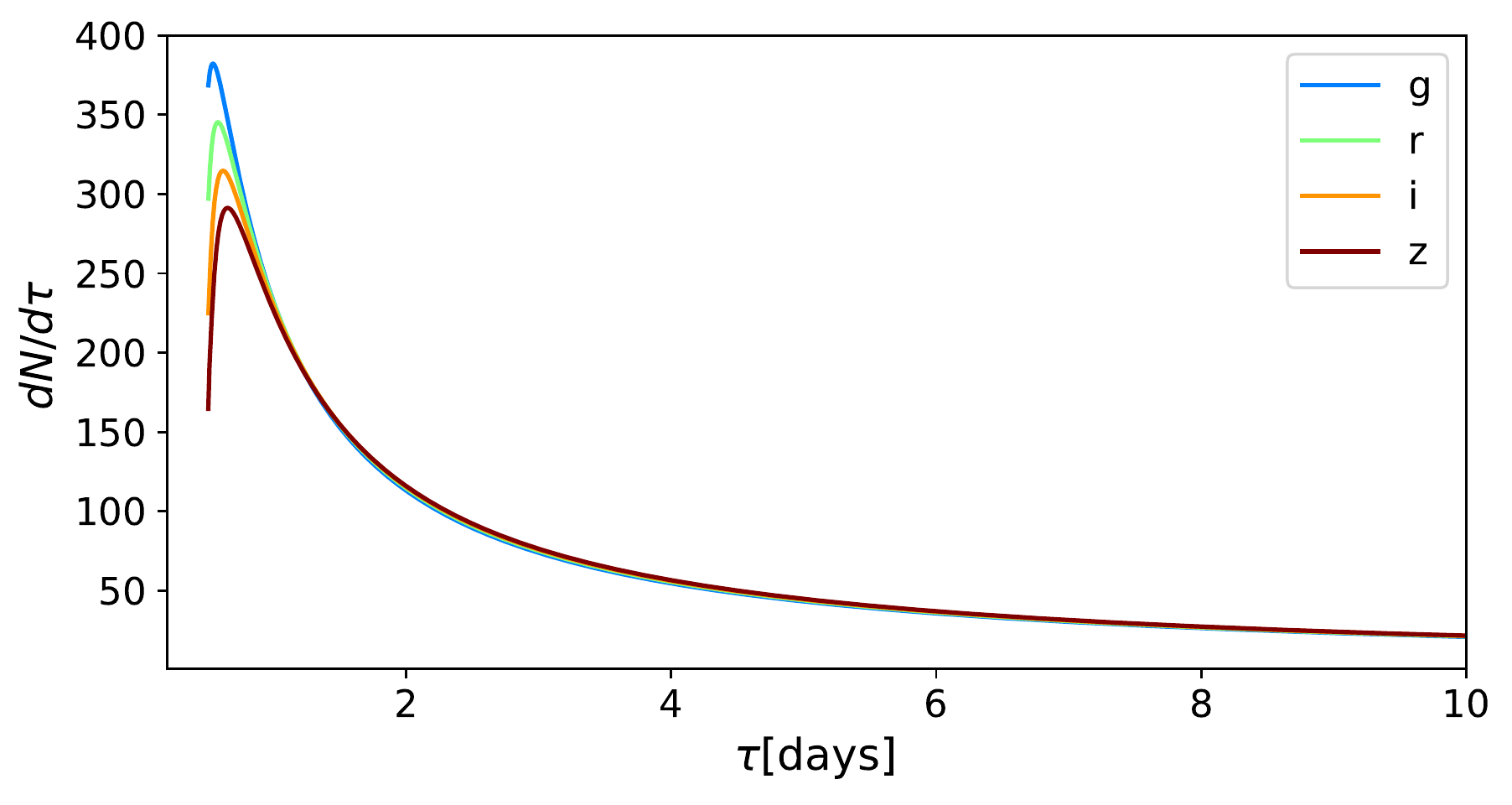}
\includegraphics[width=0.48\textwidth]{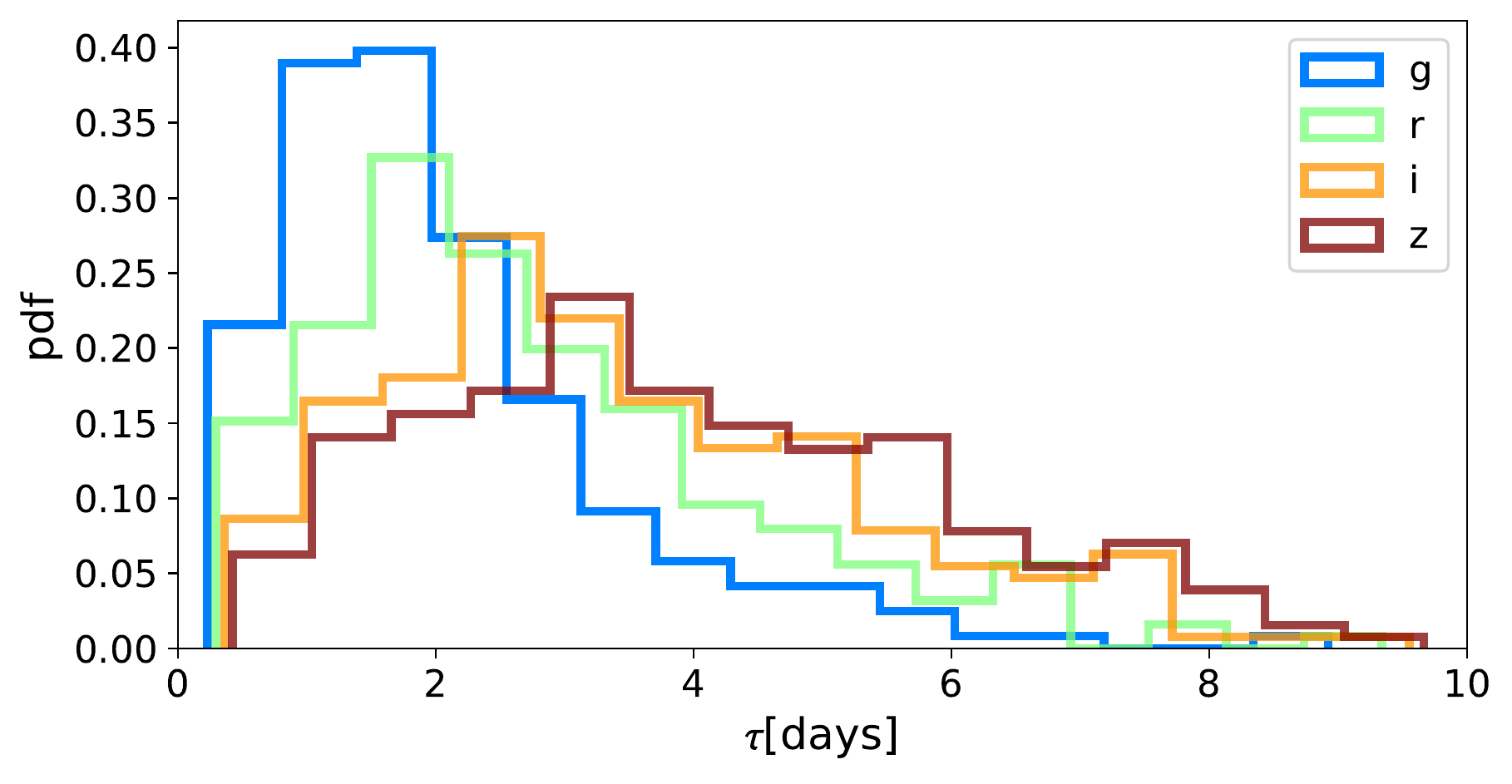}
\caption{Left: Distribution of  quasars per expected time lag  with respect to \textit{u}-band filter  based on ideal set of uniform distributed masses in the range $10^{6}-10^{10}M_{\odot}$. Right:  PDF of  expected time lag with respect to \textit{u}-band for the  208 quasars sample from \cite{2019ApJS..243...21L} and \cite{2011ApJS..194...45S} catalogues located in \texttt{COSMOS}.
\label{fig:pdfnum}}
\end{figure}

{Based on this, we summarize the expected} distribution in time lag  of LSST-observable AGNs in Figure \ref{fig:pdfnum}  (left). 
Besides this general distribution, as a forecast for the distribution of the observed time lags in LSST DDFs, we built a sample of 208 broad-line AGNs {located within \
{the} \texttt{COSMOS} field}, with estimated optical spectral parameters from the SDSS DR7 catalogs of \cite{2019ApJS..243...21L} and \cite{2011ApJS..194...45S}. If an object is identified in both catalogues, we used the data from \cite{2019ApJS..243...21L}  as it provides {the most up-to-date measurements of the AGN spectral parameters}. The resulting sample covers a large range of redshifts $(z = 0.02 - 3.82)$ and black hole masses $(\log(M/M_\odot) = 6.6 - 10.3)$.
As in the case of an ideal uniform set of SMBH masses, we  use the standard thin-disk model to calculate the expected time lags in LSST \textit{ugriz} bands. The right panel of Figure \ref{fig:pdfnum} shows the {probability density function (PDF)} of the predicted time lag.
Both distributions show that a majority of AGNs will have lags  less than 5 days.
We can expect that  higher observational cadences, such as \texttt{multi\_short}, may  increase the rate of detected time lags.
As the LSST will provide   higher-resolution images  and  a multi-visit-multi band data sets, our estimate is most likely a lower threshold.

\section{Synthetic light curves for probing LSST optimal cadences}\label{sec:model}
Here, we outline  the calculation of synthetic time-delayed AD light curves using a combination  of {a} theoretical AD transfer function (Section \ref{sec:explag}) and  {a} damped random walk (DRW, Section \ref{sec:driving}).

\subsection{Accretion disk model and expected time lag} \label{sec:explag}
\begin{figure}
\includegraphics[width=\textwidth]{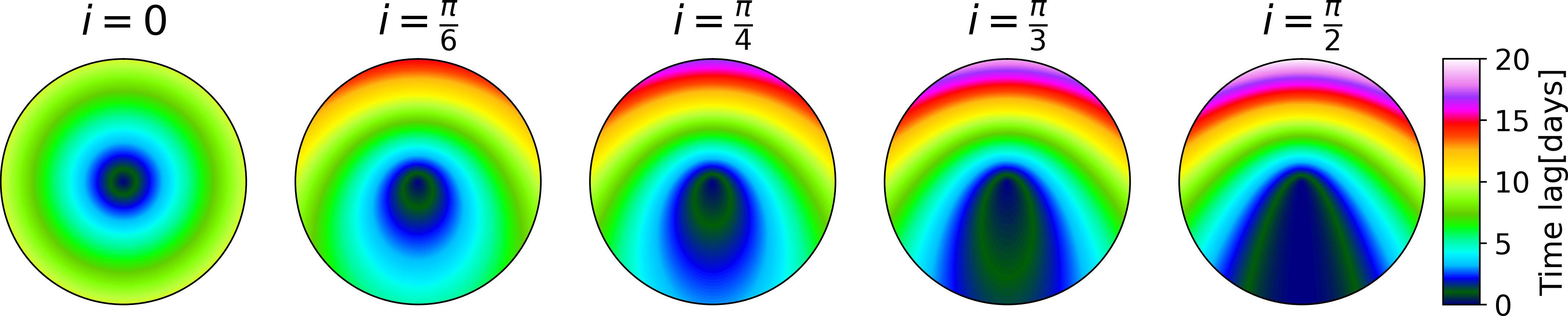}
\caption{Flat accretion disk time delay distributions  ($\tau (r,\theta, i)=r(1+\cos\theta \sin\ i)/c$), as seen by distant observer, for {various} inclinations {$i$ given on top of each plot}.  {The theoretical
observer position is  to the bottom at infinity.} {The colorbar (in units of  days) indicates time lag values}. Distributions are deformed from circular to elliptical and becomes parabolic for the extreme inclination $i=\pi/2$. {The observer will witness the re-emitted light arriving with delays ranging from 0 to $2r(1+\cos\theta \sin\ i)/c$, with a bulk average delay of $r(1+\cos\theta \sin\ i)/c$ for radius $r$ and azimuth $\theta$ of the reprocessing site, with $r \in [0,10]$ light days}. Our disk model reiterates the concept of isodelay surfaces as described in \citet{2006LNP...693...77P, 2016MNRAS.456.1960S}. }
 \label{fig:adlag}
\end{figure}

We assume a non-relativistic thin-disk model \citep{1973A&A....24..337S} that emits {black-body radiation}  and has a {compact variable} source (a "lamppost"), {whose photons hit the disk and  cause a local temperature increase; full details can be found in the literature  \citep{2007MNRAS.380..669C, 2015A&A...576A..73P, 2016MNRAS.456.1960S, 2017ApJ...835...65S}. Here, we {summarize} {the key points}.}
{We assume an optically
thick and geometrically thin accretion disk.}
The total observed {energy} flux $\Xi$ from the accretion disk is a sum of  the energy flux radiated due to a viscous heating process $\Xi_{V}$ from a surface unit  and
 irradiated flux $\Xi_{I}$ of the disk photo sphere,
 $\Xi=\Xi_{V}+\Xi_{I}$ \citep{2019MNRAS.490.3936P}.

If $R$ is the radial distance from the disk's innermost radius, $R_{0}=\frac{6GM}{c^{2}}$ {for a non-spinning SMBH}, $G$ is the gravitational constant, $M$ is the SMBH's mass, and $\dot{M}$ is the mass accretion rate, then

$$\Xi_{V}=\frac{3GM\dot{M}}{8\pi R^{3}}\Big(1-\sqrt{\frac{R_{0}}{R}}\Big).$$

Because the geometry  of the emitting source is unknown in the lamppost
model, a crude approximation is established by setting the source of
luminosity  $L^{\star}$ \footnote{\citet{2009MNRAS.399.1553V} estimated that about $5\%$ of total flux comes from X-ray source.}  at a height $h^{\star}$ along the SMBH's rotational axis. Then the irradiative  flux might be calculated using the disk's albedo $A$  as follows \citep{2019MNRAS.490.3936P}: 

$$\Xi_{I}=\frac{L^{\star}(1-A)}{4\pi R^{3}}h^{\star} \cos\theta$$

\noindent so that  $\theta$ is the incident  angle between the disk surface normal and radiation of the emitting source \citep{2013peag.book.....N}.
For viscous heating alone, when $R>>R_{0}$ the disk temperature
profile should follow
$$T(R)=\Big(\frac{3GM\dot{M}}{8\pi R^{3}\sigma_{\mathrm{SB}}}\Big)^{1/4}$$

\noindent {where $\sigma_{\mathrm{SB}}$ is the Stefan–Boltzmann constant.}
Similarly, in the case of irradiation of the disk from a height $h^{\star}$ above the disk, the total temperature profile is a superposition of the temperature
due to irradiation and the temperature due to viscous heating

$$T^{4}(R)=\Big(\frac{3GM\dot{M}}{8\pi R^{3}\sigma_{\mathrm{SB}}}\Big) +\frac{L^{\star}(1-A)}{4\pi R^{3}\sigma_{\mathrm{SB}}}h^{\star} \cos\theta .
$$
That is,  $$T(R)\propto \Big(\frac{L^{\star}h^{\star}}{R^{3} }\Big)^{1/4}$$
for $R>>h^{\star}$ {and  the viscous heating flux is much smaller than the irradiated flux} \citep{2019MNRAS.490.3936P}. Thus, in both cases, we expect that the
temperature profile follows the relation $T(R)\propto R^{-3/4}$
\citep{2007MNRAS.380..669C}.
{We assume that reprocessing has a minor effect on deviations from the viscous temperature and the $T(R)\propto R^{-3/4}$ temperature structure \citep[see][]{1998ApJ...500..162C,2016MNRAS.456.1960S}}:
\begin{equation}
T(R)=\Big(\frac{3GM\dot{M}}{8\pi\sigma_{\mathrm{SB}} R_{\lambda}^{3}}\Big)^{1/4}\Big(\frac{R_{\lambda}}{R}\Big)^{3/4}\sim \Big(\frac{3GM^{2}L}{8\pi\eta c^{2}L_{\rm Edd}\sigma_{\mathrm{SB}} R_{\lambda}^{3}} \Big)^{\frac{1}{4}}\Big(\frac{R_{\lambda}}{R}\Big)^{3/4}.
\label{eq:temp}
\end{equation}
Here, $L_{Edd}$ is the Eddington
luminosity ($L_{Edd}\approx 1.3\cdot10^{38} \frac{M}{M\odot} \mathrm{erg}\mathrm{s}^{-1}$), $\eta$ is the accretion efficiency.
If we
are observing at rest wavelength  $\lambda^{rest}$, {the radius $R_{\lambda}$ is usually approximated as the radius} where the disk temperature matches the photon wavelength 

$$kT=\frac{hc}{\lambda^{rest}};$$
\noindent here, $h$  and $k$ are  the Planck and Boltzmann constants, respectively.
Then the radius corresponding to the photon wavelength of LSST filters is 
\begin{equation}
R_{\lambda}\sim\Big(\frac{k\lambda}{hc}\Big)^{\frac{4}{3}}\Big(\frac{3GM\dot{M}}{8\pi\sigma_{\mathrm{SB}}}\Big)^{1/3}\\
\sim \Big(\frac{k\lambda}{hc}\Big)^{\frac{4}{3}}\Big(\frac{3GM^{2}L}{8\pi\eta c^{2}L_{\rm Edd}\sigma_{\mathrm{SB}}}\Big)^{\frac{1}{3}}.
\end{equation}



The disk transfer function $\psi(\tau,\lambda)$  describes
the response of the disk to changes at a given wavelength and time delay \citep{2007MNRAS.380..669C}:

\begin{equation}
\psi({\tau,\lambda})=\int^{R_{out}}_{R_{in}}\int^{2\pi}_{0} \frac{\partial{B}}{\partial{T}}\frac{\partial{T}}{\partial{L}} \frac{R dR d\theta \cos i}{D^2}
\delta({\tau-\frac{R}{c}(1+\sin i \cos{\theta})})
\end{equation}

\noindent where $L$ is the driving luminosity, $i$ is the inclination angle of the disk, $\theta$ is the azimuthal angle of emission patch, $D$ is  the distance of the object, and  ($R_{in}$, $R_{out}$) are inner and outer boundary of integration  \citep[see][and references therein]{2007MNRAS.380..669C}.  Also, we assume the accretion disk flux arises from the blackbody radiation described by Planck’s law 
$$B(T)=\frac{2hc}{\lambda^5}  \frac{1}{e^{\frac{hc}{\lambda k T}-1}}$$ with the temperature profile given by Eq.~\ref{eq:temp}.

If we assume the temperature variations
to be small, then the derivative is
$$\left|\frac{\partial{B}}{\partial{T}}\frac{\partial{T}}{\partial{L}} \right|\sim \frac{q\cdot x^{-1/4}e^{qx^{3/4}}}{4 (e^{qx^{3/4}}-1)^{2}},$$ 

\noindent where $x=\frac{R}{R_{\lambda}}$, $q=hc/\lambda k$. 

For our blackbody disk model, the time-delay
centroid $\langle \tau \rangle$ of the transfer function is\footnote{Taking into account that {the} integral $I(a)\sim \int \frac{s^{a}se^{s} \,ds}{(e^s-1)^{2}}$ can be evaluated numerically \citep{2007MNRAS.380..669C}, {and that the} centroid time delay can be approximated as follows:
$$ \langle \tau \rangle \sim\frac{R_{\lambda}}{c}\bigg(\frac{\lambda k}{hc}\bigg)^{4/3}\frac{I(8/3)}{I(4/3)}$$
where $I(8/3)=5.139, I(4/3)=2.895$.}

\begin{equation}
\langle \tau \rangle=\frac{\int \psi(\tau,\lambda) \tau d\tau}{\int \psi(\tau,\lambda)  d\tau}\sim
\frac{\int \psi(\tau,\lambda) \frac{R^2}{c} dR}{\int \psi(\tau,\lambda) R dR}.
\label{eq:meantime delay}
\end{equation}

{Figure \ref{fig:adlag} illustrates the differentiation of the time lag distributions within the disk with regard to different inclinations. Each region of distribution is labeled with the time delay it represents (so that we would observe relative to the continuum source). This is a well-known notion of isodelay surfaces, as described in \citet{2006LNP...693...77P} and \citet{2016MNRAS.456.1960S}. The line of sight to the observer is oriented towards the bottom of each disk plot, except in the left
panel, where it is exactly vertical to the plot plane.}

\begin{figure}[ht!]
\includegraphics[width=0.7\textwidth]{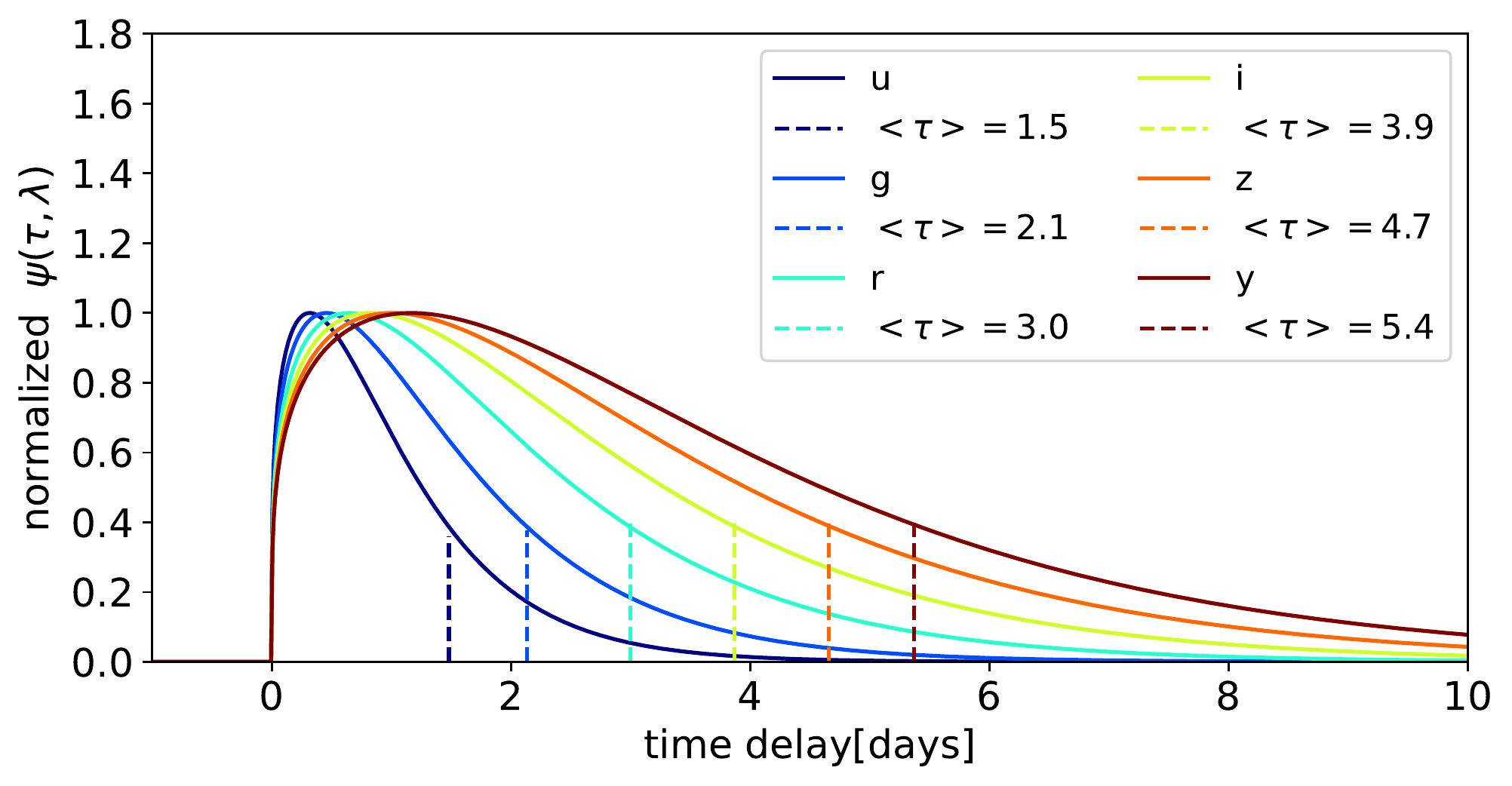}\\
\includegraphics[width=0.7\textwidth]{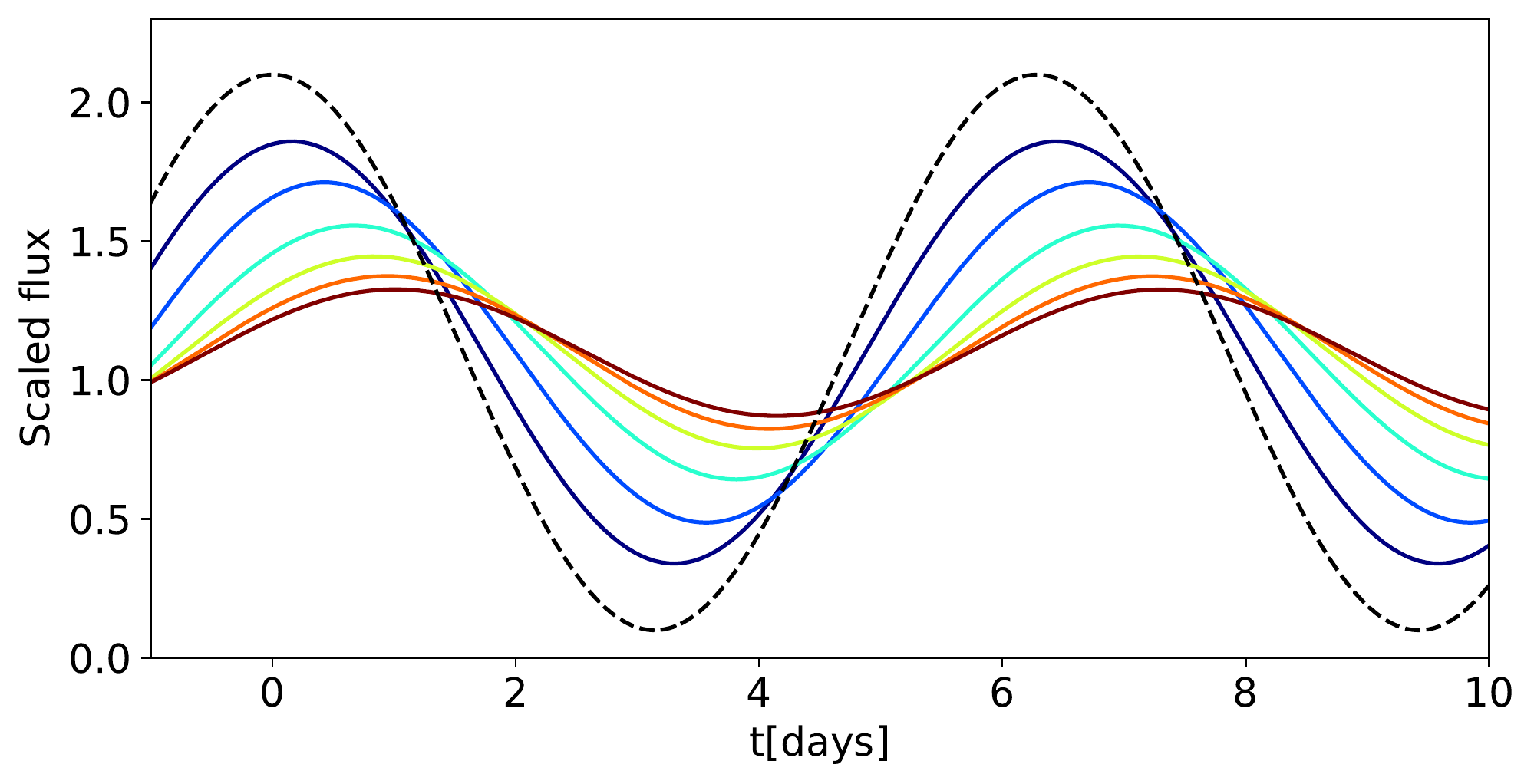}\\
\includegraphics[width=0.71\textwidth]{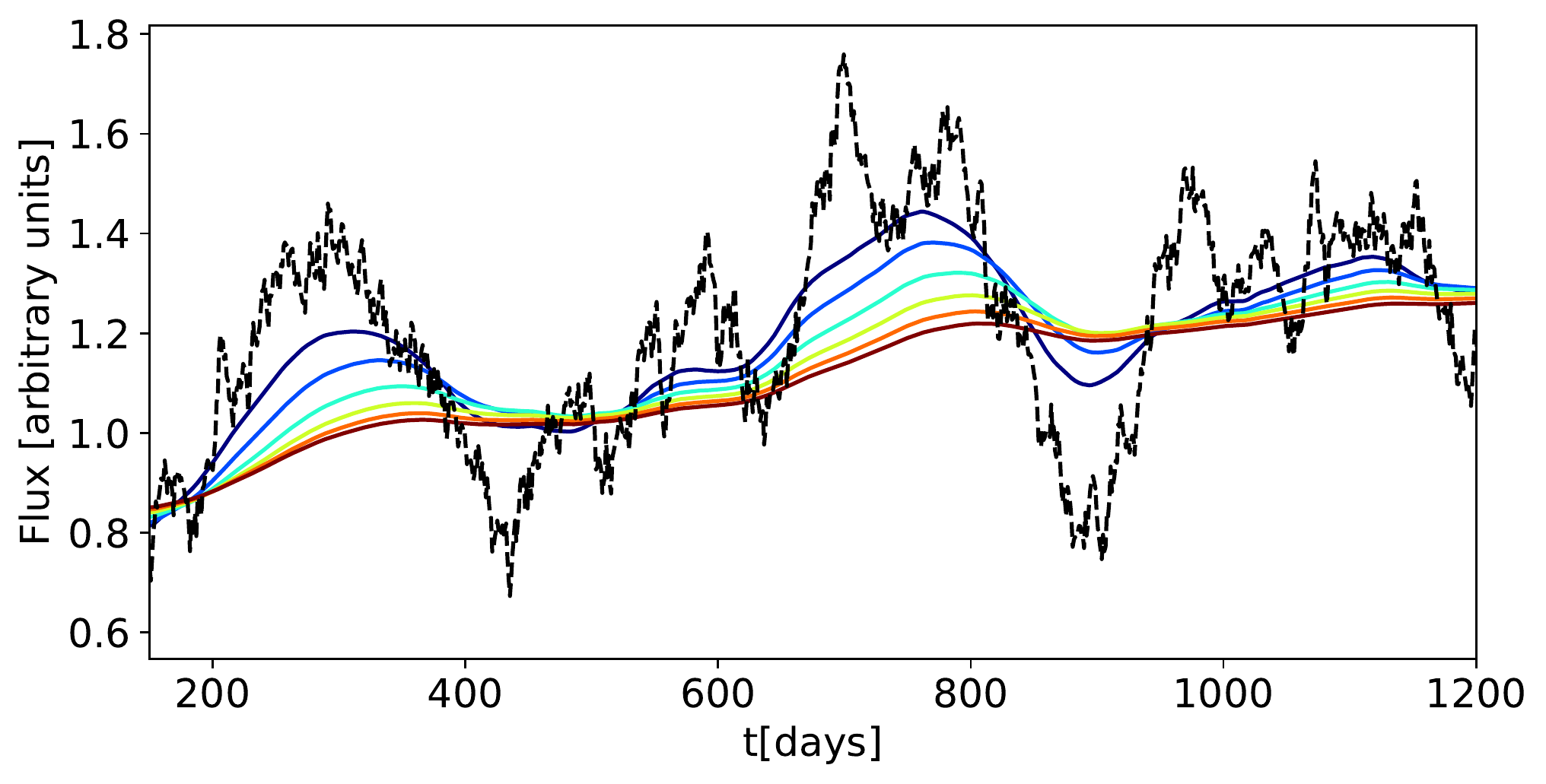}
\caption{Top: Numerical accretion disk transfer function for LSST \textit{ugrizy} filters {calculated for the following set of parameters:} SMBH mass $M=2\cdot10^{8}M_{\odot}$, $L/L_{\rm Edd}\sim 0.1$, $\eta=0.1$, disk inclination of  $45^{\circ}$ and azimuth angle of $0^{\circ}$. Dashed vertical lines mark the theoretical centroid delay (Equation \ref{eq:meantime delay}) for comparison with the peaks {for each transfer function (values given in upper right corner)}. Middle:  convolution (in colors)  of  transfer function and a sinusoid of 2.5 days period  (shown in dashed black).
Bottom: Synthetic LSST light curves  obtained by convolution of driving light curve which flux is scaled  (black line) and inclined AD transfer function {shown} in top panel. {Different bands are plotted with the same color coding in all plots.}} 
\label{fig:general}
\end{figure}

Also, we plot example response functions showing the LSST wavelength dependence in Figure \ref{fig:general}  {for the following set of parameters: SMBH mass $M=2\cdot10^{8}M_{\odot}$, $L/L_{\rm Edd}\sim 0.1$, $\eta=0.1$, disk inclination of  $45^{\circ}$ and azimuth angle of $0^{\circ}$}. As shown in Figure \ref{fig:general}, the mean delay, indicated by the vertical lines, and the width of the response functions increase with wavelength. These effects occur due to the inverse scaling between temperature and radius \citep{2016MNRAS.456.1960S}.

{In addition}, a larger inclination of the disk (see Figure \ref{fig:adlag}) makes the response function more skewed \citep{2016MNRAS.456.1960S}. The peaks will move toward shorter lags and develop a tail towards large lags. This  occurs because larger inclinations increase the lag on the far side and shorten the lag of the near side of the disk \citep{2016MNRAS.456.1960S} relative to  smaller inclination (see Figure \ref{fig:adlag}). 
A simple sinusoidal with a period of 2.5 days was constructed to demonstrate how its peak positions and amplitudes change while convolving with the time-delay transfer function of a disk with an inclination of $\pi/4$ and for different LSST filters (see Figure \ref{fig:general}).

\begin{figure}
    \centering
    \includegraphics[width=0.5\textwidth]{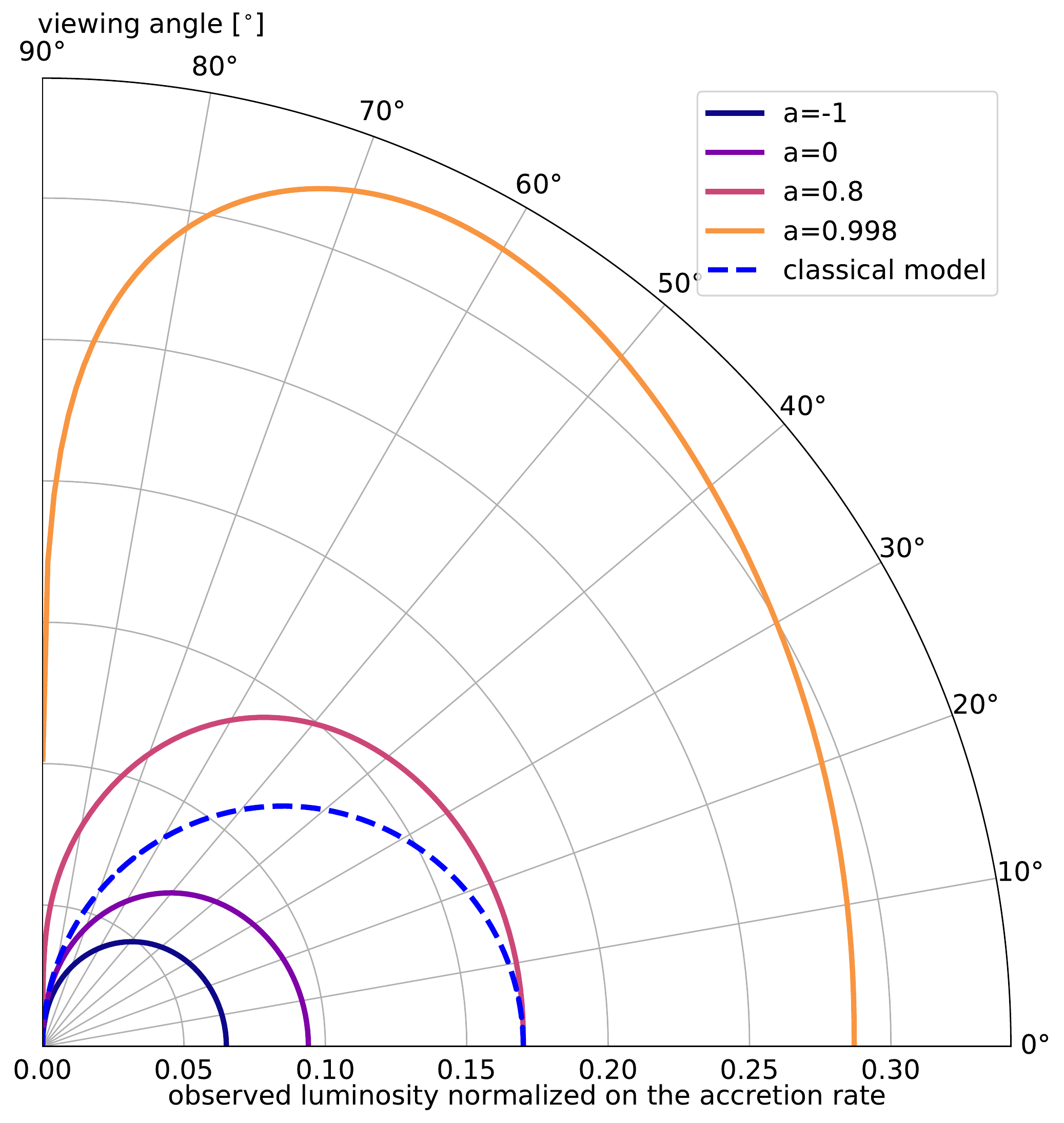}
    \caption{Schematic comparison of  emission patterns across viewing angles for different AD models as given by the phenomenological approximation of \citet[][see their Figure 2]{2018A&A...612A..59C}. The classical nonrelativistic  model (dashed blue line) is compared
with relativistic models with $a=-1$ (dark blue curve), $a=0$ (dark purple curve), $a\sim 0.8$ (red curve) and $ a=0.998$ (orange curve). The radial axis is the observed efficiency $\eta_{\rm{o}}=
\frac{L^{\rm{o}}}{\dot{M}c^{2}}$, i.e. given by the observed luminosity normalized on the accretion rate (which is the same for each model). The plot plane is the accretion disk plane. 
The main axis of all models is in direction of $0^{\circ}$, assuming azimuthal symmetry with respect to this direction. A distant observer is at the left of the figure. 
}
    \label{fig:polar}
\end{figure}

{Figure \ref{fig:polar} demonstrates potential effects of a non-relativistic AD setup by  comparing the  observed luminosity normalized to the accretion rate for classical and Kerr-metric black-body disk emission as given by the \citet[][see their Figure 2]{2018A&A...612A..59C} as a function of SMBH spin $a$ and viewing angle $\theta$.  It is easy to notice that for $a < 0.8$, the  relativistic angular
pattern of radiation  is very similar to the classical behavior. The classical model, when observed at $\theta = 0$ is well approximated by the relativistic model with $a \sim 0.8$, but at larger viewing angles, the relativistic model is brighter because of the strong relativistic effects.
When spin $a\rightarrow 1$, the radiation pattern is very steep,  even for the largest viewing
angles. This is due to the combination of different relativistic effects (Doppler beaming, gravitational redshift, and light-bending)
along with the black hole spin: the trajectories of the energetic
photons coming from the innermost region of the disk (close to the event horizon) are bent in all directions
and the intensity of radiation is almost the same at all viewing angles.}

{To compare  time lags in curved and flat spacetime,  we calculate their ratio. Time lags in curved space are obtained by combining analytical approximation of time delay for Kerr black hole in flat spacetime \citep[e.g.,][]{1995MNRAS.272..585C, 2018MNRAS.475.4027M} and approximation for light bending  \citep{2006MNRAS.373..836P}}:
\begin{equation}
c\Delta t=r y\cdot\Bigg(1+\frac{uy}{8}+\frac{uy^{2}}{8}(\frac{1}{3}-\frac{u}{14})\Bigg)
\label{eq:reltl}
\end{equation}
\noindent where  $r$ and 
$\theta$ are radius and azimuth of reprocessing site,  $u=\frac{2GM}{c^{2} r}$, and  $y=1+\cos(\theta)\sin(i)$ for a given inclination $i$. Note that $c\Delta t=ry$ corresponds to the time delay one would expect in flat space time (see Figure \ref{fig:polar}).

\begin{figure}
\centering
\includegraphics[width=0.5\textwidth]{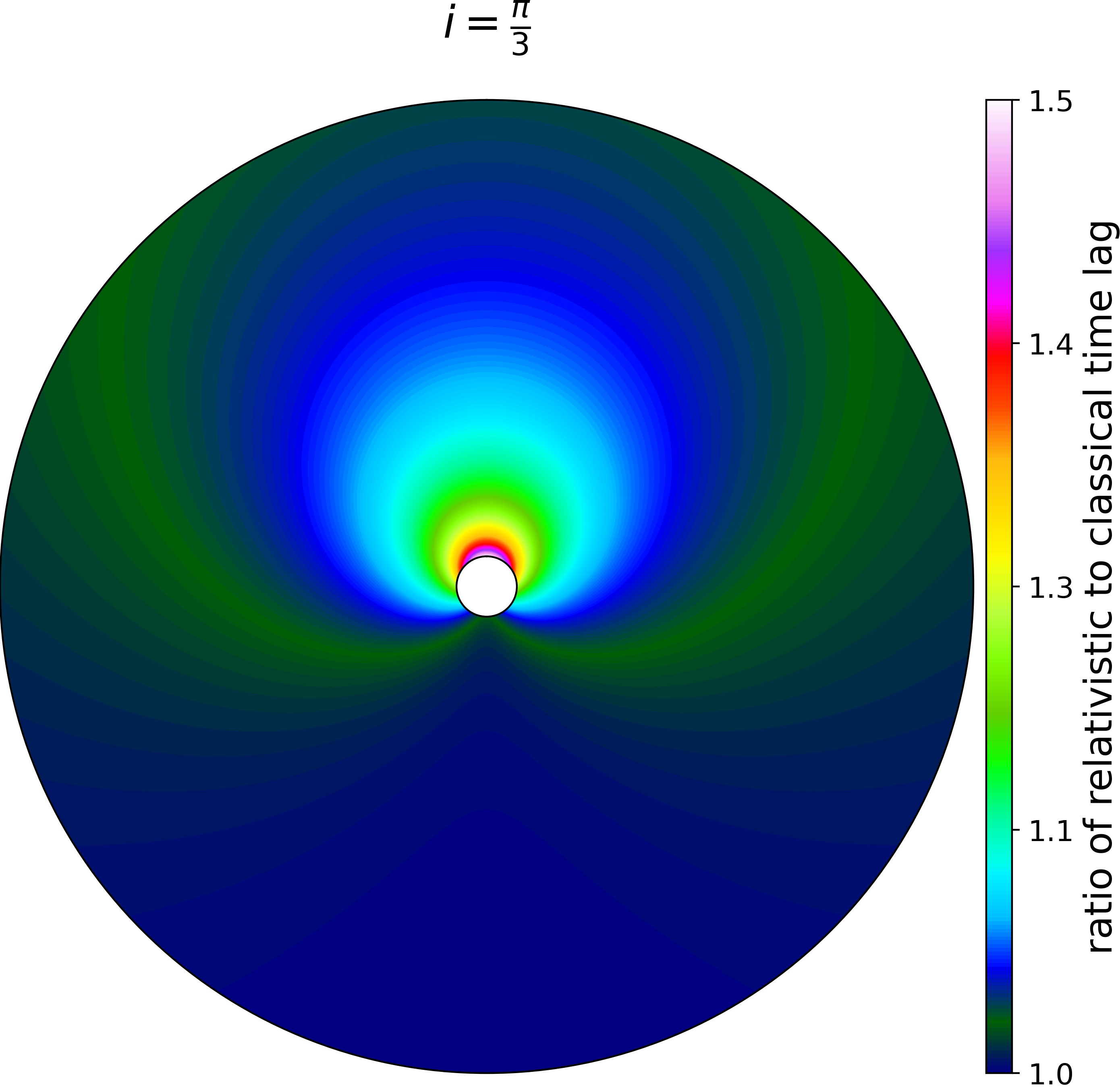}
\caption{Distribution of ratio of the time delay calculated from a relativistic and a classical spacetime approximation, when $a\rightarrow 1$. The inclination of the observer is set to $i=\frac{\pi}{3}$, the region extent to $10 r_{\rm g}$, and the SMBH mass is $10^{8}M_{\odot}$.  The dark green color depicts a ratio value of  $\sim 1.075$. The small white circle is a 'plunging region' inside the marginally stable orbit of $r_{\rm ms}=1.24 r_{\rm g}$. The color bar shows the ratio between relativistic and classical time lags. The ratio is highest where the light bending effects are largest, i.e., near the SMBH. As the distance increases,  the ratio approaches 1, because spacetime is almost flat.
  }
 \label{fig:difflag}
\end{figure}

{Figure \ref{fig:difflag}  depicts a schematic view of the 2D distribution of the ratio between the relativistic time delay (Equation \ref{eq:reltl}) and its classical equivalent. The ratio values  are  sharply increasing with decrease of distance to  the center of mass. This is due to light bending which is largest there. }
{For a non-rotating black hole ($a\rightarrow0$), the marginally stable radius is $r_{\rm ms}=6\frac{GM}{c^{2}}=6r_{\rm g}$, whilst  for a maximally rotating black hole ($a\rightarrow1$) it takes a value of  $r_{\rm ms}=1 \frac{GM}{c^{2}}= r_{\rm g}$. As theoretical predictions state that the maximal spin
for a rotating black hole can not exceed 0.998,  then the marginally stable radius of fast rotating black hole is $r_{ms} = 1.24 r_{\rm g}$ \citep{1974ApJ...191..507T}.}
{
Figure \ref{fig:difflag} corresponds to fast rotating $a\rightarrow1$ SMBH, so that the small white inner circle with
radius of the marginally stable orbit $r_{\rm ms}=1.24 r_{\rm g}$ assumes that no source of light can orbit around SMBH.}
{
As for non-rotating $a\rightarrow0$ SMBH, at $6 r_{\rm g}$ and larger distances there will be no effects (dark green and dark blue color). The upper limit of the navy blue region is at $6 r_{\rm g}$. 
Therefore, for a fast-rotating SMBH relativistic corrections should be
taken into account, because  $r_{\rm ms}$ is closer to the event horizon.} {Even though it looks like an extreme possibility, there are  well-known cases like MCG -6-30-15
and NGC 4051, whose X-ray emission could be explained with light-bending effects \citep[e.g.,][]{2010ApJ...709..278D,2012ApJ...757..137C,  2014MNRAS.439L..26K}. Also, the published EHT observations of M87 and Sgr A$^{\star}$ are consistent with the Kerr paradigm, indicating that fast spinning black holes might be ubiquitously  case \citep{2019ApJ...875L...4E,eht22}. }


  \subsection{Simulating driving and multiband light curves}\label{sec:driving}
  We provide here a {summary of our }  light curve modeling. A full
description of synthetic light curve modeling, and {of tests using these} synthetic light curves,
are presented by \citet{2021MNRAS.505.5012K}.

 The time variability of AGN and quasar continuum emission can be approximated using a damped random walk \citep[DRW,][]{2009ApJ...698..895K, 2010ApJ...708..927K,2013ApJ...765..106Z}. A DRW is defined by a stochastic differential equation with a broken power law power spectrum and a slope that steepens from 0 to 2 above a break time-scale. The break time-scale in AGNs measured with the SDSS is up to 100 days. {It} correlates favorably with black hole mass and wavelength, and is a weak function of brightness \citep{2010ApJ...721.1014M, 2021Sci...373..789B}.
However, light curves of AGNs observed with Kepler and ground-based surveys show
steeper power spectral density (PSD) than the DRW model at timescales up to $\sim 1$ month \citep{2011ApJ...743L..12M, 2018ApJ...857..141S,2017ApJ...834..111C,2018ApJ...864...87S}.
 {
\citet{2017A&A...597A.128K} demonstrated that light curves with insufficient time length cannot consistently constrain the damped variability timescale, and advised that light curves time length should be at least tenfold that of the true damped variability time scale.
Based on 1384 variable AGNs from the QUEST-La Silla AGN variability survey, \citet{2018ApJ...864...87S} showed that  the DRW variability amplitude is influenced by the length of the light curve, and that the DRW model can represent the variability of 74\% of AGNs in their specified sample. On the other hand, the work of \citet{2019ApJ...877...23L}  suggests that only 35\% of light curves in their RM AGN sample of 113 objects can be derived with accurate DRW model parameters. This proportion is much lower than that found by  \citet{2018ApJ...864...87S}, which could be related to the RM AGN shorter light-curve length (one year) in \citet{2019ApJ...877...23L} than  for the sample of  \citet{2018ApJ...864...87S}.
In this work,
the probed damped variability time scales are {short}, {up to}  200 days with respect to LSST's 10 year time baseline}, so that  we can
assume that DRW is an approximate working description of quasar
variability for  optical light curves \citep{2021ApJ...907...96S}.
Although  {the} damping mechanism is unknown yet, \citet{2009ApJ...698..895K} show that the damping timescale might be connected to the thermal timescale of the accretion disk, assuming that 
its thermal fluctuations are responsible for the AGN variability.
Because of this, we concentrate our analysis on the DRW model, based on its implementation in \citet{2011ApJ...735...80Z,2013ApJ...765..106Z} {and} \citet{2021ApJ...907...96S}.

{\citet{2022arXiv220110565N} investigated the effect of AD temperature fluctuations on AGN light curves, revealing that the light curves closely reflect the expectations for a lamppost model, albeit with slower variability time-scales of the in-going waves. This also means that longer time-scale variability signals will increasingly diverge from conventional models, since the smoothing of slower-moving waves reduces significantly as their period or spatial wavelength grows. LSST will operate {much longer than} most  RM campaigns, providing the first well-characterized multi-wavelength light curves to investigate what happens on long time scales.}

We first generate a uniform time grid $\Delta t_{i}=1$ day on which to evaluate the driving light curves $F_{\rm DRW}$. This is convenient for evaluating the
convolution integral.
{As the AD transfer functions asymmetry increases with an increase in the wavelengths of the filters, light curves features are still evident in the \textit{u}-band for the shortest wavelength, but they are blurred out for the \textit{y}-band (see Figure \ref{fig:general} and Section \ref{sec:discussion})}.
Since the \textit{y}-band light curve distortion is the greatest, determining the time lag requires additional care (due to large possible errors). 
Thus, for the sake of simplicity,  we continue by generating quintets of  multi-band light curves based on the LSST \textit{ugriz} bands using a transfer function as described in Section \ref{sec:explag}.

The multiband continuum light curves ($\mathcal{F}(t,\lambda)$) are then obtained from the driver ($F_{\rm DRW}$) by convolution with the {asymmetric} response function  $\psi_{\lambda}(\tau)$ (see an example in Figure \ref{fig:general}):

\begin{equation}
\mathcal{F}(t,\lambda)=\sum^{m}_{i=0} \psi(\tau_{i},\lambda) F_{\rm DRW}(t-\tau_i)\Delta \tau.
\label{eq:convol}
\end{equation}

{We assume   the following set of parameters  {for the transfer function}: SMBH mass $M=2\cdot10^{8}M_{\odot}$, $L/L_{\rm Edd}\sim 0.1$, $\eta=0.1$, disk inclination of  $45^{\circ}$ and azimuth angle of $0^{\circ}$. For the driving source emission we assume  a characteristic restframe timescale $\tau_{\rm DRW}\sim 200$ days and a structure function at infinity
$\sigma_{\rm DRW}=2.5$ (flux unit), which describes the long-term variability of
the light curve. These DRW parameters are typical of the quasars
in the DES sample \citep{2020ApJS..246...16Y}. In general, \citet{2010ApJ...721.1014M} found that both parameters follow a power law relationship with AGN luminosity, measured wavelength, and black hole mass \citep[see][for a compact representation of these relations]{2015MNRAS.453.1701K}. }

 A realization of the driving source emission  for given parameters is displayed in the bottom panel of Figure \ref{fig:general}  in black.  Using the  transfer functions for LSST bands, we then generated the multiband light curves\footnote{In a work by  \citet{Jankov21}, we determined which LSST filters contain continuum at different redshifts (see e.g. their Figure 1)} as  shown  in the bottom panel of Figure \ref{fig:general}.
In practice, light curves where the lags  are not clearly visible are likely to have delays that are {shorter} than the cadence of the light curve.

\section{Estimator of cadence strategies}\label{sec:estimator}
In this section, we characterize distinct LSST observing strategies. 
Section \ref{sec:metric}  describes our metric implemented in \texttt{MAF}  that quantifies  {the impact of cadences} for {LSST RM time lag measurements}.
Additional calculations of time lag measurement of simulated LSST continuum light curves using  {the best cadences identified by} our metric in two DDFs are outlined in Section \ref{sec:bayes}.

\subsection{Metric simulation in \texttt{MAF} }\label{sec:metric}

The Metrics Analysis Framework (\texttt{MAF} )\footnote{\url{https://rubin-sim.lsst.io/rs_maf/index.html}} is a software package designed to enable writing code to evaluate  {the performance of different survey strategies.}
Our metric is integrated into \texttt{MAF}   through the module \texttt{AGN\_TimeLagMetric}\footnote{\url{https://rubin-sim.lsst.io/rs_maf/metricList.html}}.

The metric includes the quasar selection based on {the limiting magnitude of the observation}\footnote{The 5 $\sigma$ depth in the \texttt{OpSim}   is the limiting magnitude of the visit -- meaning,
objects fainter than this limit will not be visible. However, objects
brighter than that limit are definitely still visible, and at higher SNR
than 5. } in the $r$  filter. 
The magnitudes were corrected for the Galactic dust-extinction.
 The sky regions with dust extinction  $E(B-V)> 0.2$ are excluded because this amount of dust extinction is problematic for extragalactic science for two reasons: it reduces the effective coadded {$5\sigma$} depth\footnote{Unit is magnitude, see \url{https://confluence.lsstcorp.org/display/SIM/Summary+Table+Column+Descriptions}.}, and the total amount and wavelength dependence of dust extinction are not always well characterized, making it difficult to calibrate the effect on the background galaxies \citep[see][]{Jon20}.

{
The depth of LSST single visits is assumed to be similar to that of the COSMOS field {observations} by the VLT Survey Telescope  \citep{2021A&A...645A.103D}. \citet{2021A&A...645A.103D} {found} that 91\% of {AGNs} have $r > 21$ mag, putting them {beyond} the average depth of most existing variability studies. 
Thus we impose in our \texttt{MAF}  simulations a {cut for $5\sigma$ depth in the {r}  band $> 21.0$ {mag}}. \footnote{Otherwise, adding a cut for $5\sigma$ depth in the {r}  band $< 21.0$ mag means that the (vast majority) of
visits with depth which is $> 21.0$ {mag} would not be counted, yet could still detect
objects which are as bright as 21.0 {mag}.}
This enables us to investigate how well AGN time lag measurements work for fainter objects.}
{
Our later analysis reveals that the most suitable DDF cadence  strategies in the \textit{r}-band have mean $5\sigma$ depths {of} 21.5  (see discussion in Section \ref{sec:discussion}, and Figures \ref{fig:figcomparison1}-\ref{fig:figcomparison3}). {Application of our metric} with concurrent limitations in all filters could eliminate the vast majority of visits for {time lags up to 5 days} due to the Nyquist criterion. {Otherwise}, for larger time delays  (order of tens of days) simultaneous $5\sigma$ cuts {can be used in more than one band}. {Thus}, we relaxed the limitation in bands other than \textit{r} in order to have a broader pool of feasible cadences in the other bands. When we compared the mean  $5\sigma$ depths of  \textit{r} and other  bands within the DDFs, we {found} that they are  {within the range of 18 and 21.5 mag in DDFs} ( Figures \ref{fig:figcomparison1}-\ref{fig:figcomparison3}). This means that objects as bright as 21 mag will still be seen in those bands for these overlapping cadence strategies. }

When analyzing observations of variable objects, two major issues arise. First, because of weather or other technical constraints, sampling times can be perturbed. Second, 
objects  can have a wide range of  variability characteristics and behaviors. As a result, it is important  {to establish} a threshold {to search for} variability {on} the time scales of interest \citep{1999A&AS..135....1E}.
By assuming that there is no correlation 
between  flux variation $F_\mathrm{var}$ \citep[as defined in e.g.,][]{1998ApJ...509..163O}, the measured time delay $\tau_\mathrm{obs}$ and  its uncertainty $\sigma_{\tau_\mathrm{obs}}$, \citet{2021MNRAS.505.5012K}  developed the metric
 $\phi_{\tau}$ :
\begin{equation}
\log \phi_{\tau}=\log\frac{\sigma_{\tau_\mathrm{obs}}}{\tau_\mathrm{obs}}\propto A+C_{1}\frac{F_\mathrm{var}}{\sigma}+C_{2}\frac{ \tau_\mathrm{obs}}{(1+z)\Delta t^{c}},
\label{eq:89}
\end{equation}
\noindent where the  time sampling (or cadence) of the light curve is $\Delta t^{c}$ and the mean photometric error of light curve points is  $\sigma$.
If the observed variation is
intrinsic to the quasars at cosmological distances, one would expect a time dilation effect.
{The time delay is reduced to the rest frame, taking into account the cosmological redshift z in Equation \ref{eq:89} by a factor $(1+z)^{-1}$.}

As the expected {size of the AD in the UV/optical  } is on the  order of $\lesssim 10$ light days {where the Nyquist criterion becomes important, we modify our metric}\footnote{The other terms could be kept  less than an arbitrary constant, e.g., assuming limits of $F_{\mathrm{var}}\sim 0.4$ and $\sigma\sim 0.005$mag. }
as follows:
\begin{equation}
\log \phi_{\tau}\propto \frac{ \tau_\mathrm{obs}}{(1+z)\Delta t^{c}}.
\label{eq:891}
\end{equation}
{Note} that the fraction term on the right hand side of Equation \ref{eq:891} {represents} a sampling rate.
There are different opinions in the literature when it comes to the sampling rate  ($\Delta t^{c}$) for irregularly sampled data. Some authors \citep[e.g.,][]{1986ApJ...302..757H,1992nrca.book.....P} associate the Nyquist frequency  with $1/2 \langle t_{sam} \rangle$, where $\langle t_{sam} \rangle$  is the averaged sample rate, or with
$1/(2s)$, where $s$ is the smallest time sampling in the sample \citep{1987AJ.....93..968R, 1982ApJ...263..835S}.

For irregular sampling, \citet{1999A&AS..135....1E}  defined the Nyquist frequency  as $f_{Ny}=1/(2p)>1/(2s)$, where  $p$ is a greatest common divisor (gcd) for all $t_{i}-t_{1}$
of  time of observation $t_i$.
{Since} gcd calculation is very time consuming, we will use {the definition based on} the smallest cadence in the sample \citep{1987AJ.....93..968R, 1982ApJ...263..835S}.

It is critical that the light curves are sufficiently well sampled {for} RM. Recalling Nyquist's Theorem, the sampling rate of the time lag measurement must be  $$\frac{ \tau_\mathrm{obs}}{(1+z)\Delta t^{c}} > 2.2$$ in order for the time lag to be evaluated \citep{2011A&A...535A..73H}. 
As a result, our metric evaluates whether the sampling rate for a required time lag measurement is greater than the  Nyquist threshold\footnote{Due to data loss to weather or technical difficulties in reality, practical sampling might require a stiffer criterion, known
as the engineer's Nyquist \citep[see e.g.,][]{Srini98}:
$$\frac{ \tau_\mathrm{obs}}{(1+z)\Delta t^{c}} > 2.5.$$}.


\subsection{Recovering time lags}\label{sec:bayes}

For simulated quintets of AGN light curves in {the} LSST \textit{ugriz} bands (Section \ref{sec:driving}),  we
estimate the time delay relative to \textit{u}-band and compare it to the {input} value. We chose the \textit{u} band as the reference band since its has the shortest wavelength. 
Moreover, as the cadence of \texttt{u}-band observations is a critical issue in
the LSST observing strategy optimization \citep{2018arXiv181106542B}, we  aim  to  probe  if the \texttt{u}-band  can  be  used  to evaluate AD time lags in
the DDFs.

In addition to the standard cross correlation methodologies for determining time lags \citep{1986ApJ...305..175G,1987ApJS...65....1G, 1997ASSL..218..163A}, other methods have been developed {as those built on the assumption of a DRW process} \citep[e.g.,][]{2016ApJ...819..122Z},  the DRW process combined with expansion of  the transfer function into a 
series of Gaussian functions \citep{2016ApJ...831..206L},  {and} shifting techniques \citep{2013A&A...553A.120T}, to deal with  complex  light curves \citep{2020A&A...636A..52C}.
For example, \citet{2021arXiv210512420K, 2021MNRAS.505.5012K} adapted 
the functionality of {the} widely used z-transformed discrete correlation function technique \citep{1997ASSL..218..163A} to take into account general Gaussian {processes in the} modeling of {the} light curves.
We employed our shifting method to {estimate}  time delays from the mock data, which {is} inspired by {the} \texttt{PyCS} {toolbox}  developed by \citet{2013A&A...553A.120T}.
One of the \texttt{PyCS} {implementation}s {is optimized} to find the best set of time delays that minimizes the variability
of the difference between Gaussian-process regressions on {the} light curves.   This is motivated {by} the standard problem that there is a different number of {light curve data points}. 
An alternative objective function is needed to measure the
similarity between the curves.
As  multiband light curves are mainly deformed due to {the} asymmetric transfer function, our  basic
idea of the optimizer is to  shift the data iteratively in time
and magnitude with respect to a reference band, and  apply the Partial Curve Mapping \footnote{\url{https://github.com/cjekel/similarity_measures}}  \citep[PCM,][]{Jekel2019} method to determine the similarity between curves and  {estimate} a time delay through minimization of the PCM.

For measuring the time lag $\tau$ between  two noisy light curves $f_{1}(t), f_{2}(t)$ we fit the relation $f_{1}(t) = a \cdot f_{2}(t-\tau
)$ where $a$ is a scaling coefficient.
Denoting the LSST light curves as $X= \left\{x_{j}(t_i)|i=1,..,N, j\in(u,g,r,i,z)\right\}$, the scaling coefficients and the lags as
$a_k$ and $\tau_{k}, k\in{g,r,i,z}$, respectively, and finally the variances of the
noise as $v_k$, we can 
fit the quartet of multiband light curves $\mathbf{x}=x_{g},x_{r},x_{i},x_{z}$ with one predictor time series in the shortest wavelength $x_u$.

{A prior  estimate of the time delay range of light curves can be obtained through preprocessing with less precise algorithms
\citep[see e.g.,][]{2013A&A...553A.120T}, available prior knowledge  \citep[see an example of R-L relation use in][]{2022MNRAS.509.4008P}, or expected ranges obtained from assumed  AD model  $L=\rm {math.floor}(<\tau>_{k}-<\tau_{u}>),   k=\textit{g,r,i,z}$ from AD model: $[L,L+1]$ .}
{Considering the}  time lags that will be probed, we {adopt the latter strategy}, such that the uniform priors (in days) are: $\tau_{ug}=Uniform(0,1), \tau_{ur}=Uniform(1,2), \tau_{ui}=Uniform(2,3), \tau_{uz}=Uniform(3,4)$.
 They reflect the range of expected values of difference between mean time delays of the transfer function  in different bands.
 These should only be considered as suggestions, and can be easily changed.
Priors on expected time lags are non-informative, and their ranges do  not
overlap. Our emphasis is on the fact that multiple models for the AD model and its transfer function could be used in a real-world setting, and that the prior on the time lag should reflect this.
 To determine the similarity of curves after scaling and shifting, we apply PCM which combines  arclength and area, because the choice of parameters can alter the overall length of the curve. {We illustrate the principles of PCM, while details are given in \citep{Jekel2019}}. First, the  arc length of the one  curve is
superimposed onto a section of the other  curve. After that, trapezoids are created between the curves, and the areas of the trapezoids are added together. The {best fit} PCM value is the one {that minimizes the combined area of the trapezoids}.

 The PCM permits the use of non-resampled light curves; nevertheless, we followed the \texttt{PyCS} guideline and resampled them {with homogeneous cadence of $\sim 0.5-1$ day}. The time lag {and its} uncertainty is calculated by fitting the resampled, shifted light curves to the \textit{u}-band light curve using {a  random value drawn from the prior}. The PCM space for each light curve is quite irregular, with multiple local minima. To discern the different minima objectively and without expectation bias, we repeat the procedure  300 times for each of the resampled light curves, using a sample of 100 random-uniform initial time lags. The most common time lag (corresponding to 'dominant PCM minimum') is regarded as the best-fitting value for the specific light curve pair.
Finally, from the distribution of fitted lag values, we compute  PDFs, a mean time lag, and {uncertainties as} 16 and 84 quantile confidence intervals.



\section{Results - cadence strategies for AD time lag measurement}\label{sec:results}

We use the \texttt{OpSim\_v2.0} and \texttt{OpSim\_v1.7}\footnote{\url{https://github.com/lsst/sims_featureScheduler}} cadence simulations.
The annotation adopted here follows the LSST nomenclature {for} the simulations. For example, the label \texttt{baseline\_v2.0\_10yrs\_5 days\_r} indicates that the metric is run on the observing strategy \texttt{baseline},  {from} the \texttt{OpSim}  version  2.0, {with a survey duration of } 10 years, {and that} the time lag to be probed by the metric is 5 days, {in the} LSST \textit{r} {band}.

The v2.0 simulations were developed in response to the Survey Cadence Optimization Committee’s Phase 1 recommendations \citep{Ivezic21}. One of the most significant differences between previous survey simulations is that the baseline survey footprint has been slightly modified: the Wide Fast Deep area (WFD), which typically obtains on the order of 825 visits per pointing split across 6 filters, now explicitly includes a low-dust-extinction area as well as a Galactic Plane extension\footnote{\url{https://github.com/lsst-pst/survey_strategy/blob/main/fbs_2.0/SummaryInfo_v2.0.ipynb}}.

First, we provide the overall analysis in Section \ref{sec:mafsim}, which makes use of our metric \texttt{AGN\_TimeLagMetric} implemented in \texttt{MAF}.
The DDFs \citep[DDFs,][]{2018arXiv181106542B, 2018arXiv181200516S} are the most prominent locations for LSST RM, according to  {our initial analysis} \citep{2021arXiv210512420K}.
The existing baseline survey strategy contains
five DDFs \citep{Jon20}.
 Four of the DDF sites had been established for some time; the fifth field position is at present anticipated to overlap with the Euclid Deep Field South, which needs two LSST pointings to be covered entirely.
The cadence of visits, as well as coadded depths of these DDFs, still needs to be finalized \citep{Jon20}.
The {identifiers and equatorial coordinates} of the DDFs are \citep[see][]{Jon20}: 
 \texttt{ELAISS1} $(9.45^{\circ}, -44.0^{\circ})$,
 \texttt{XMM\_LSS} 
$(35.71^{\circ}, -4.75^{\circ})$, \texttt{ECDFS} $(53.13^{\circ}, -28.1^{\circ})$, \texttt{COSMOS} $(150.1^{\circ} , 2.18^{\circ})$, \texttt{EDFS\_a} $(58.9 ^{\circ}, -49.32^{\circ})$, \texttt{EDFS\_b} $(63.6^{\circ}, -47.6)^{\circ}$. {Note that the Euclid Deep Field South (EDFS) } is divided into two sections \citep[see details in][]{2019arXiv190410439C,EDFS21} {but for} the sake of brevity, we only refer to  the five {unrelated} DDFs when a broad discussion is required.
Then, armed with the best strategies, we aim to simulate photometric time lag recovery as precisely as {possible} in Section \ref{sec:lagrec}.
In our  simulation, we chose a quasar
redshift of $z = 0.5$ {as for higher redshifts a careful choice of bands covering the continuum is  needed }\citep[see details in][]{Jankov21}.

\subsection{Assessment of cadence strategies based on \texttt{MAF}  simulations}\label{sec:mafsim}

Here, we use our metric to investigate the impact of different cadence strategies on the study of AD time lag measurement.
We performed a simulation {using the} \texttt{AGN\_TimeLagMetric} in \texttt{MAF}  as described in Section \ref{sec:metric}.
{We {probed  fiducial} time lags of 1, 2, 3, and 5 days because they are close to the expected mean time delay for our transfer function model.}
Aitoff projection of our metric realization in \texttt{MAF}  simulation for LSST cadences are shown in Appendix (see Figures  \ref{fig:fig1}-\ref{fig:fig9}).  

{In our analysis, we searched for the shortest possible time lag that could be measured reliably by RM in all bands.}
In this and the following subsections, we  show  that the best cadences for simultaneous\footnote{The term simultaneity refers to the fact that our method determines all delays for the light curves quintet at the same time.} time lag measurements in \textit{ugriz} bands are  5-day lags for the DDFs.  {Shorter time lags 
can be observed in 
\textit{i} and \textit{r} filters in DDFs, as will be seen in zoomed-in plots of metric values later on in this section. 
This should not come as a surprise given that these bands may contain cadences that are somewhat more dense \citep{2021arXiv210512420K}. Longer time lags, $\gtrsim$ 7 days, are detectable in each of the filters, as we shall show in the following  that mean cadence samplings are favorable across bands.}
As it could be an issue that the optimal solution matches the upper limit of the probed time-lags, we also tested these cadences  by simulating RM measurements with shorter time delays in all bands.  

\begin{deluxetable*}{llllllll}
	\tablenum{1}
	\tablecaption{Overview of the LSST cadence strategies. A ‘visit’  is currently defined to consists of two back-to-back 15 sec exposures, for a
total of 30 sec of on-sky exposure time. These back-to-back exposures are always in the same filter,
separated only by the 2 second readout time \citep{Jon20}. From first to last column: observing strategy, \texttt{OpSim} version, the amount of area of WFD that receives at least 825 visits per pointing, the amount of area of WFD with low extinction,  total number of visits per point on the sky,  the median number of visits that the most frequently visited
18,000 square degrees receives, the median visits distributed per \textit{u} and \textit{g} band.   \label{tab:overview}}
	\tablewidth{0pt}
	\tablehead{
		&\colhead{\texttt{OpSim}  } & \colhead{Area with} & \colhead{Area with} & \colhead{Nvisit} &
		\colhead{Median} & \multicolumn{2}{c}{Median Nvisit}\\	
		& &$>$825 visits& \colhead{low extinction } &
		\colhead{total} &Nvisit & \textit{u} band  &    \textit{g} band \\
		& & \colhead{[~$\rm{deg}^2$~]}& \colhead{[~$\rm{deg}^2$~]} & & & &   \\
	}
	\decimalcolnumbers
	\startdata
	\texttt{baseline}	&2.0	&12434.14	&16842.94	&2081749.0&	838.0	&54.0	&68.0	\\
		\texttt{carina}&	2.0&	12519.74&	16385.53	&2087222.0	&838.0	&54.0	&70.0	\\
			\texttt{multi\_short}& 2.0	& 151.07&	16639.0&	3584186.0&	696.0	&100.0	&125.0	\\
		\texttt{rolling\_all\_sky\_ns2\_rw0.9} &2.0	&12900.78 &	16599.55	&2088105.0	&839.0	&54.0&	69.0	\\
			\texttt{short\_exp} & 2.0&	10664.07&	16581.09&	2180235.0&	831.0&	56.0&	72.0\\
		\texttt{six\_rolling\_ns6\_rw0.5}	&2.0 &	12342.65&	16665.01&	2084148.0	&837.0&	54.0&	70.0	\\
			\texttt{long\_gaps\_nightsof\_delayed-1} & 2.0&	10288.90	&15954.13&	2067452.0&	830.0	&50.0	&70.0\\
		\texttt{euclid\_dither1} &1.7	&17982.71	&15174.43&	2045493.0&	888.0	&55.0	&79.0	\\
		\texttt{euclid\_dither3} & 1.7&	17967.60&	15161.84&	2045144.0	&888.0	&55.0	&78.0\\	
		\texttt{footprint} &1.7	&15390.97	&16492.96&	2046219.0	&853.0	&52.0	&72.0	\\
		\texttt{u\_long\_ms\_30} &1.7&	18091.81	&15186.18&	2080376.0	&904.0&	58.0	&80.0	\\
	\enddata
	\tablecomments{More details are given in a  collection of information relevant for survey strategy
		  \url{https://github.com/lsst-pst/survey_strategy}. The cadence of visits, as well as coadded depths of  DDFs, still needs to be finalized.}
\end{deluxetable*} 

{Table \ref{tab:overview} outlines some of the most important aspects of the LSST observing strategies used in this work. }
The baseline strategy (\texttt{baseline\_v2.0})  features a modified\footnote{Filter balance is modified in different areas of the sky.} survey footprint with an expanded dust-free area and WFD-level visits in the Galactic Bulge and Magellanic Clouds. Coverage of the Northern Ecliptic Spur, South Celestial Pole, and remainder of the Galactic Plane is maintained, at lower levels. The existing baseline used 2$\times$15 sec exposures per visit, and most visits were in the same  filter \citep[although this was not enforced,][]{Jon20}\footnote{As clarified by \citet{Jon20}: "A ‘visit’ here is an LSST default visit, which consists of two back-to-back 15 sec exposures, for a  total of 30 sec of on-sky exposure time. These back-to-back exposures are always in the same filter, separated only by the 2 second readout time."}.
There are specific programs that, if given specialized observing time via "microsurveys," can produce additional science outside or in support of the basic Rubin LSST science drivers. {One such program is that included in the}  strategy \texttt{carina\_2.0} and is intended to observe {the} Carina nebula \footnote{\url{https://github.com/lsst-pst/survey_strategy/blob/main/fbs_2.0/SummaryInfo_v2.0.ipynb}}. 
Both strategies   show three prominent DDF fields-{\texttt{XMM\_LS} close to equator, \texttt{ECDFS} below previous and \texttt{ELAISS1} close to polar axis} (see Figure \ref{fig:fig1}). For measuring time-lags smaller than 5 days \texttt{carina} cadences in \textit{r} and \textit{z} bands show a very small ares  {with Nyquist sampling or better} at the sky location $\delta=30^{\circ}, \alpha=18-20$h of  the LSST equatorial sky plane.

\texttt{OpSim}s are using a custom dither pattern for the Euclid deep drilling field to better match the Euclid field of view (\texttt{EDFS\_{a}, EDFS\_{b}}).
Also,  \texttt{euclid\_dither1} and \texttt{euclid\_dither3} exhibit  three notable DDFs  with moderate metric value in  a tiny field at the position $\delta=0^{\circ}, \alpha=22$h (Figure \ref{fig:fig2}).
Panel (d) in Figure \ref{fig:fig2} has an additional field of moderate metric value on the right side of main vertical axis of sky symmetry at the position $\delta>0^{\circ}, \alpha=8$h. 
The \texttt{footprint} cadence strategy allows various survey footprints based on increasing the low-extinction area of the WFD footprint.
In all filters except the \textit{u}-band filter, the cadence in \texttt{footprint} strategy exhibits a limited region which is shifting the location for time lags between 1 and 3 days (Figure \ref{fig:fig3}).
However, for the measurement of 5-day time lags in the \textit{i} and \textit{r} bands, the three DDFs (as in prior cadences) occur in panels (e) and (f) of Figure \ref{fig:fig4}.

As the baseline survey strategies involves pairs of visits every few evenings, separated by around 33 min,  \texttt{long\_gaps} simulations include a third visit after a 
time interval of 2 to 7 hours.
In the case of \texttt{long\_gaps\_nightsoff}, long gaps appear every seven nights starting after year five.
DDFs are dominant in both the \textit{i} and \textit{r} bands for a measurement of 5-day time lags, which is comparable to prior occurrences in other metric realizations (see Figure \ref{fig:fig5}).

Microsurveys \texttt{multi\_short\_v2.0}  cadences, emphasizes multiple (four) short (5 sec) exposures in each filter.
For them metric simulations show a distinct topology so that suitable fields are clustered in a wavy band that is almost parallel to the sky equator but with  varying densities (Figures \ref{fig:fig6}-\ref{fig:fig7}).
Among the metric realizations, the \textit{u}-band has the highest density of sky locations appropriate for RM, see Figure \ref{fig:fig6}(d) and Figure \ref{fig:fig6}(i), Figure \ref{fig:fig7}(b) and  Figure \ref{fig:fig7}(g).
The same three DDFs that were appearing in the prior cases are depicted in Figure \ref{fig:fig6}(i) and Figure \ref{fig:fig7}(h). Notably, four DDFs are found in the   Figure \ref{fig:fig7}(e) and Figure \ref{fig:fig7}(f).

The \texttt{rolling\_all\_sky\_ns2} are two-band rolling all sky  cadences with weights 0.5 and 0.9, where the metric realization is comparable to \texttt{long\_gaps\_nightsoff}, \texttt{footprint}, \texttt{carina} (see Figure \ref{fig:fig8}) and \texttt{euclid\_dither} (see Figure \ref{fig:fig9}).

The microsurveys strategy  \texttt{short\_exp} takes up to three short (5 sec) visits in year one.
As illustrated in Figures \ref{fig:fig10}-\ref{fig:fig11}, the topology of the metric realization  is comparable to that of the metric realization on \texttt{rolling\_all\_sky\_ns2}, with the exception that the \textit{u}-band is more dominant.

As the updated baseline contains a two-band rolling cadence with strength $\sim90$\% in the dust-free WFD, \texttt{six\_rolling\_ns6\_rw0.5} simulations modify this rolling cadence in the dust-free WFD to use six regions instead of two, with variable weight ("strength") to the level of rolling.
{The} \texttt{u\_long\_ms\_30} cadences take \textit{u}-band observations as single snaps, and test increased \textit{u}-band exposure times, in this case 30 secs. 
Interestingly, metric realizations on \texttt{six\_rolling\_ns6\_rw0.5} and \texttt{u\_long\_ms\_30\_v1.7} cadences (Figure \ref{fig:fig12}) are as prevalent in DDFs as they are in the case of metric realizations on \texttt{euclid\_dither} cadences.

\begin{figure*}
\gridline{\fig{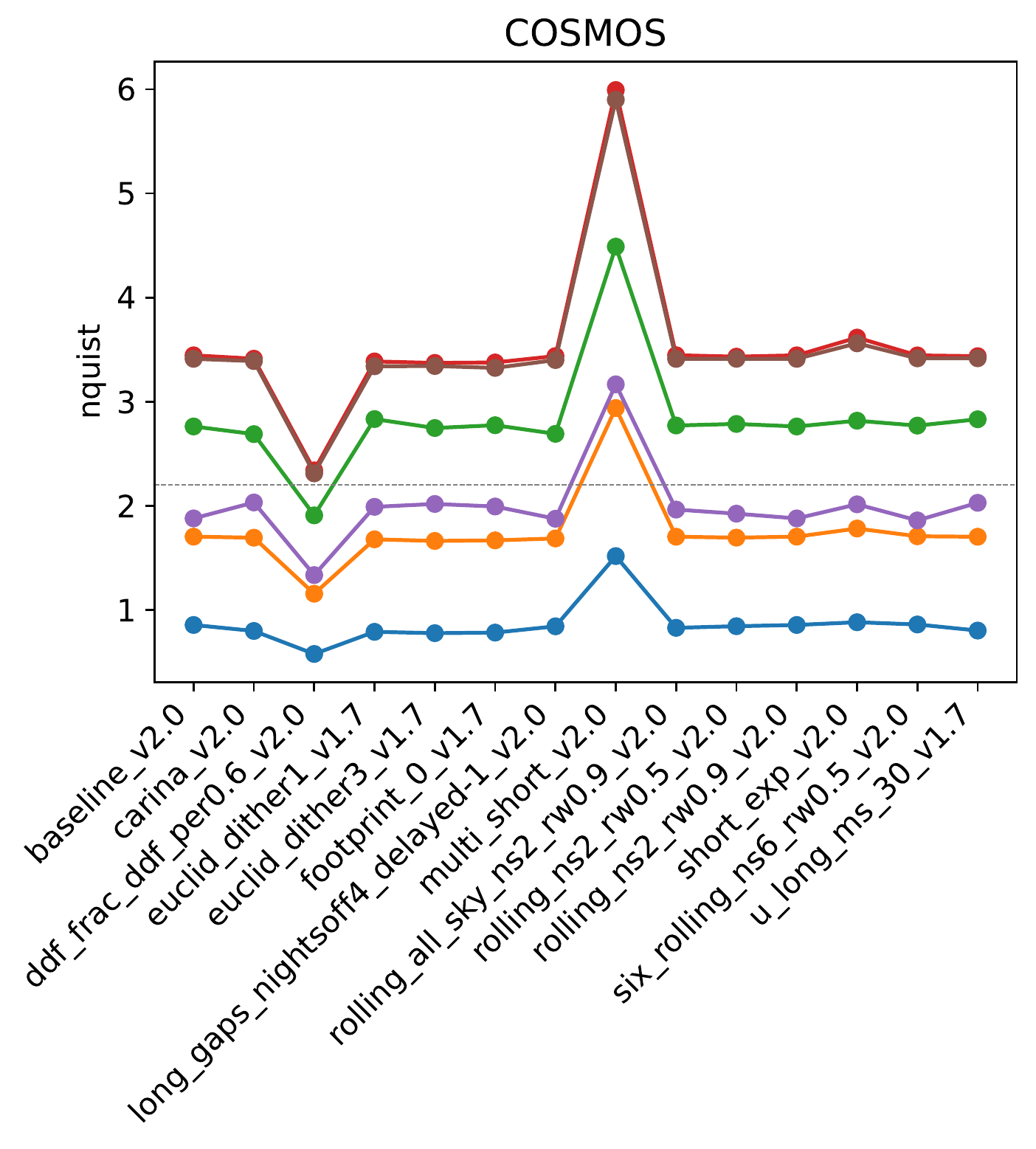}{0.33\textwidth}{(a)}
          \fig{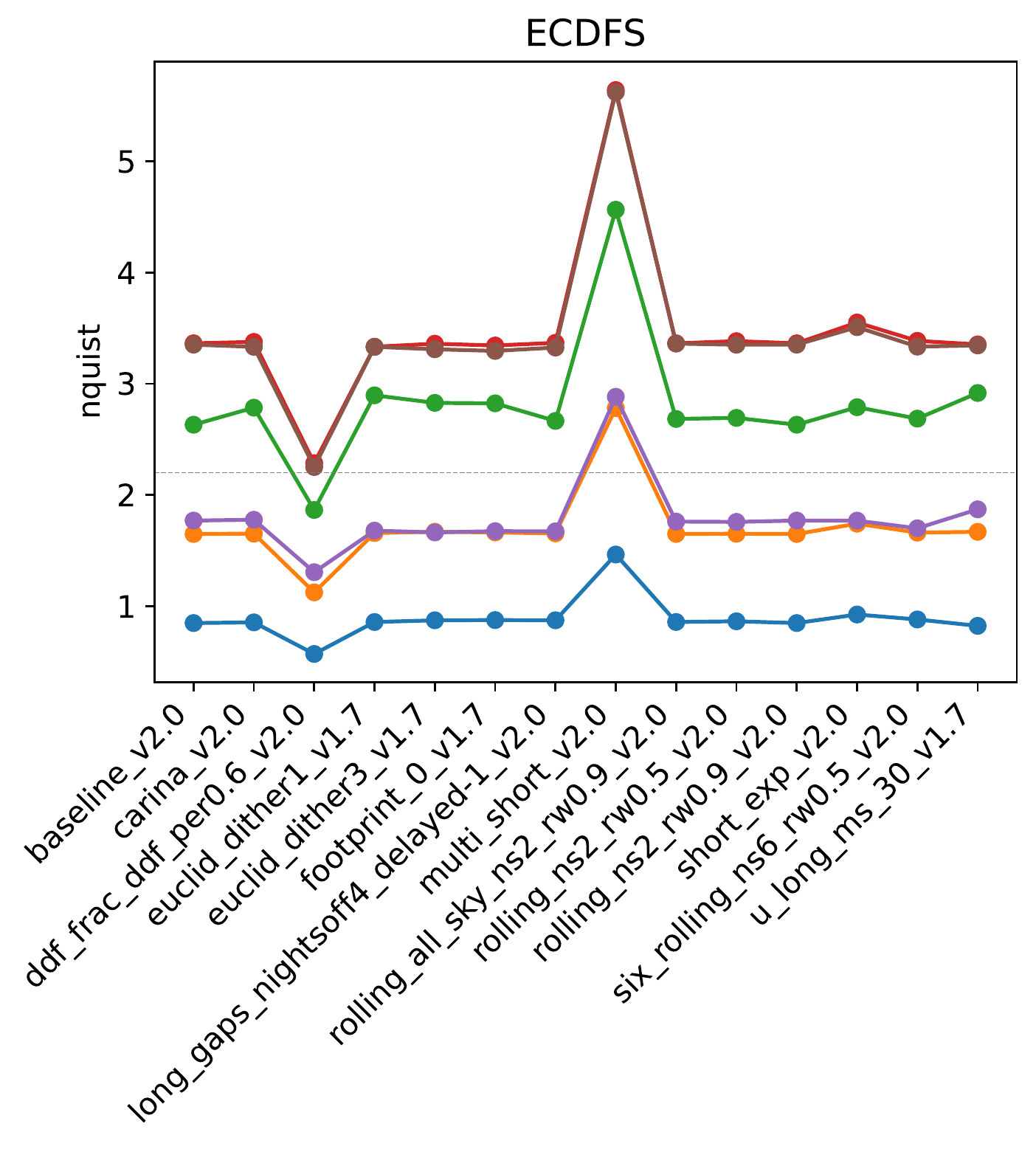}{0.33\textwidth}{(b)}
          \fig{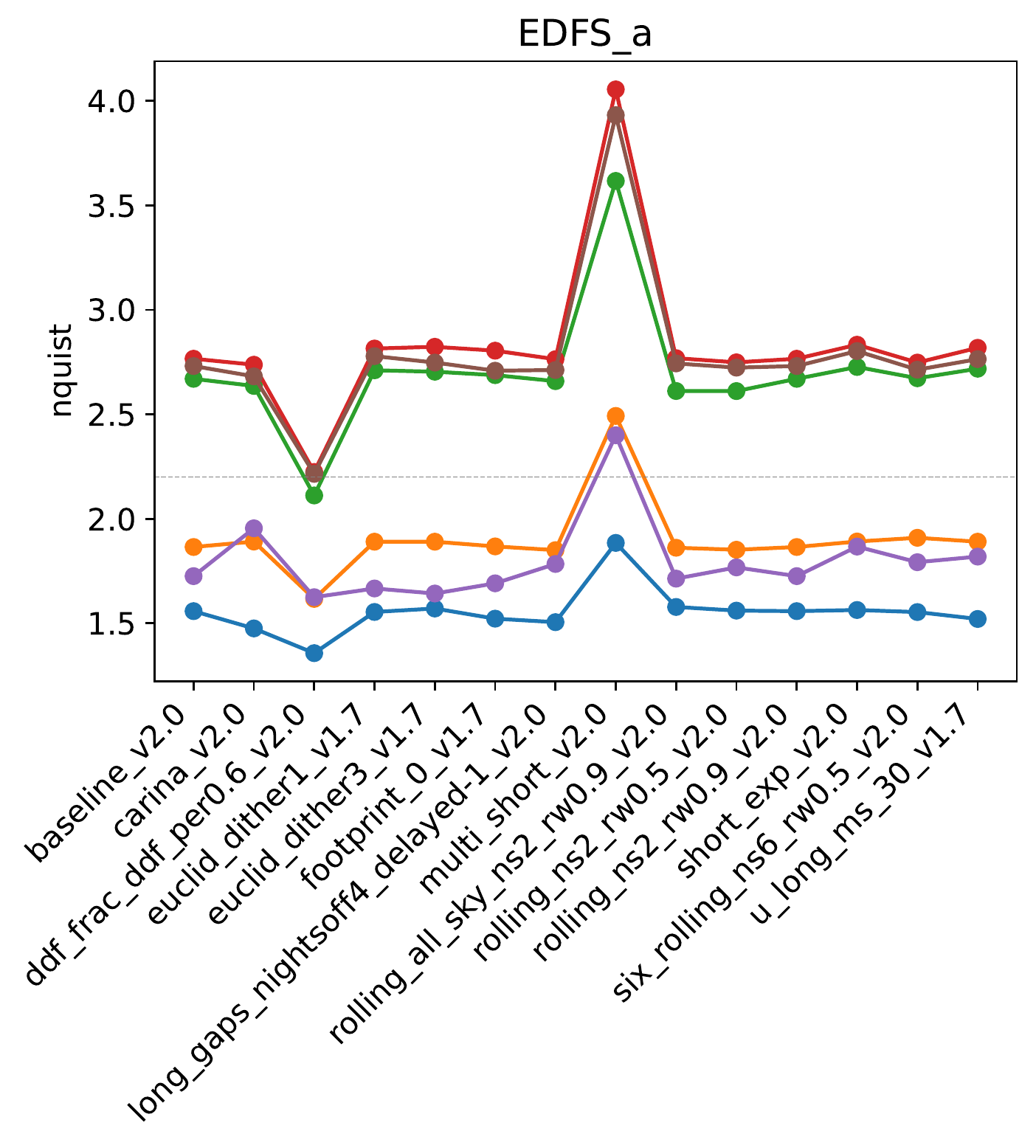}{0.33\textwidth}{(c)}
          }
\gridline{
\fig{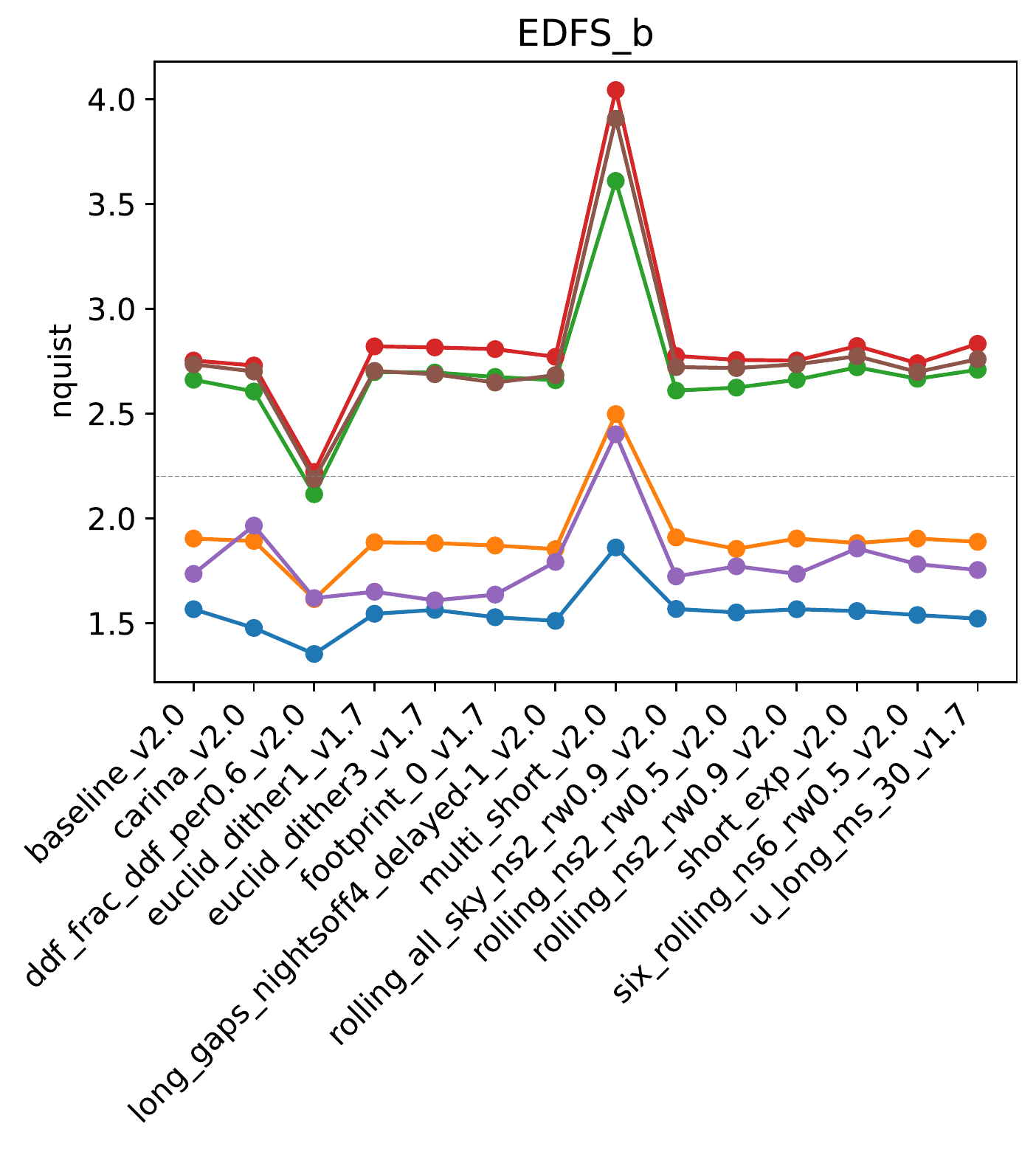}{0.33\textwidth}{(d)}
  \fig{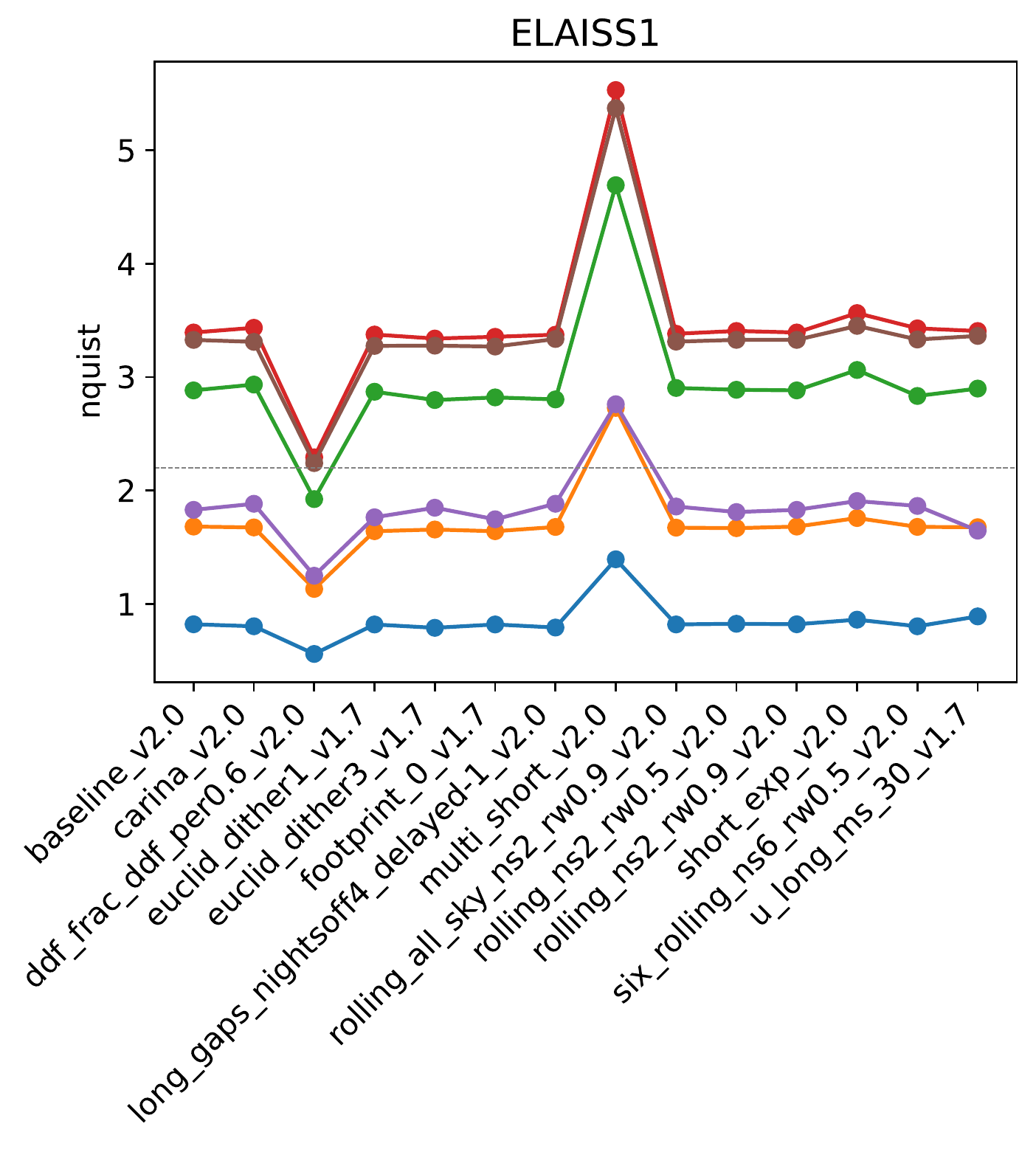}{0.33\textwidth}{(e)}        
          \fig{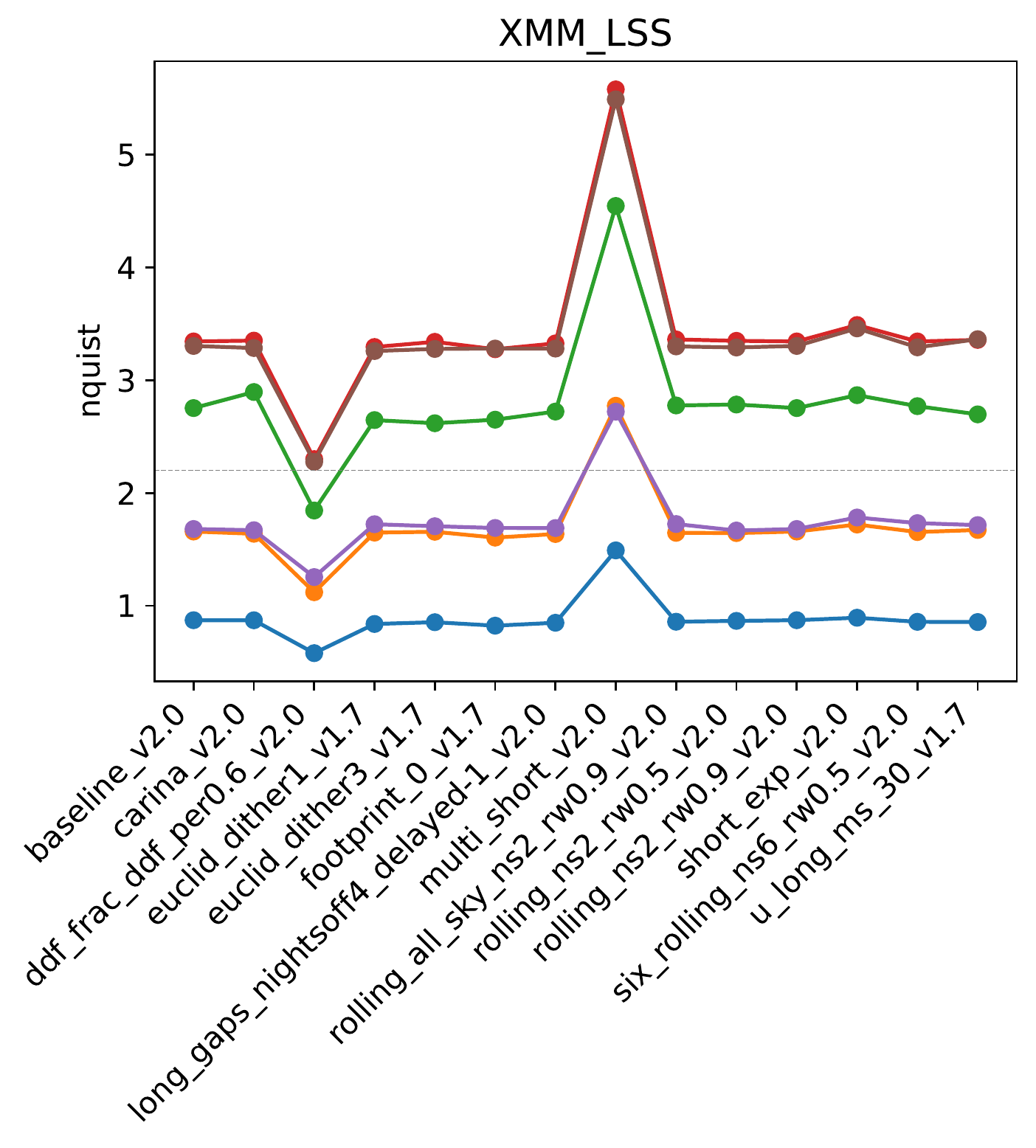}{0.33\textwidth}{(f)}
           }
\caption{The comparison of metric {values above the Nyquist} threshold (y-axis)  for the measurement of 5-day time lags in six DDFs, for various \texttt{OpSim}  strategies indicated on the x-axis and six of the LSST filters stated in the legend. {The solid lines  serve as guides for the eye. The horizontal black line indicates the metric threshold}. DDFs  are indicated in the titles of panels. The passbands \textit{u}, \textit{g}, \textit{r}, \textit{i}, \textit{z}, \textit{y} are denoted with blue, orange, green, red, violet and brown color, respectively.
\label{fig:figcomparison}}
\end{figure*}

 The LSST Science Advisory Council (SAC) report raised a series of questions and identified suggested simulation
experiments to run. Among them are also listed Experiments with DDFs cadences \citep{Jon20}. The observational cadence, which is currently being debated for the LSST DDFs \citep{Jon20}, is indicated as one of  the most critical elements that determine disk size measurements.
The detailed investigation of the distribution of metric realizations {(values above Nyquist criterion)} on  DDFs  is
shown in Figure \ref{fig:figcomparison}.
The highest quality time lag measurement is predicted to be obtained using data from the \texttt{COSMOS}, \texttt{XMM\_LSS} and \texttt{ECDFS} whilst the lowest quality time lag measurement is expected to be obtained using data from the \texttt{EDFS\_a} and \texttt{EDFS\_b} fields.
{Notably, the metric value for the \texttt{multi\_short} cadence in the \textit{u} band is 20-30\% lower than the threshold. As we will see in the following sections on examples of \texttt{multi\_short} cadence, cadences with metric values nearing the threshold could  be adequate to calculate time delays.}
{The co-added depth varies across different filters and observing strategies in each DDF as illustrated in Figs.\ref{fig:fig13}-\ref{fig:fig14}. Currently, the deepest fields are \texttt{ECDFS}, \texttt{XMM\_LSS} and \texttt{ELAISS1}. Interestingly,
\texttt{multishort} strategy consistently produces the highest quality time delays and it also has the largest depth for all DDFs.}

To demonstrate in greater detail the comparison shown in Figure \ref{fig:figcomparison}, we perform \texttt{MAF}  simulation in DDFs with significantly higher resolution (see Figures \ref{fig:zoom121} and \ref{fig:zoom122}) than the original images in the Appendix.
The \texttt{HEALpygnomview} is used to create the high resolution plots.
 The center of each plot is comprised of the point on the DDF itself (as mentioned above) plus a radius of 5 degrees  (this defines the slice utilized to run the metric). The image itself displays the resolution per pixel. The x-axis represents right ascension, and the y-axis represents declination.
The metric realizations are given in two $3\times3$ grids of panels.
The metric outperforms in all DDFs with the exception of \texttt{EDFS\_a} and \texttt{EDFS\_b}.

Based on the similarity of the metric value distributions,  there are three types of DDFs: homogeneous DDFs with the highest metric values (first panel in the top row and  middle panel in the middle row  in the  Figure \ref{fig:zoom121}), inhomogeneous DDFs with dispersed high metric values (second panel in the top row in  panel in Figure \ref{fig:zoom121}; second and third panels in the top and middle row  in Figure \ref{fig:zoom122}), and elliptically distributed DDFs (the last panel in the middle row in Figure \ref{fig:zoom121};  bottom rows in the Figures \ref{fig:zoom121} and \ref{fig:zoom122}).
This example, on the other hand, clearly indicates that  observations in elliptical regions can be coupled with equivalent partitions in homogeneous DDFs in a straightforward manner.

\begin{figure*}
\includegraphics[width=0.9\textwidth]{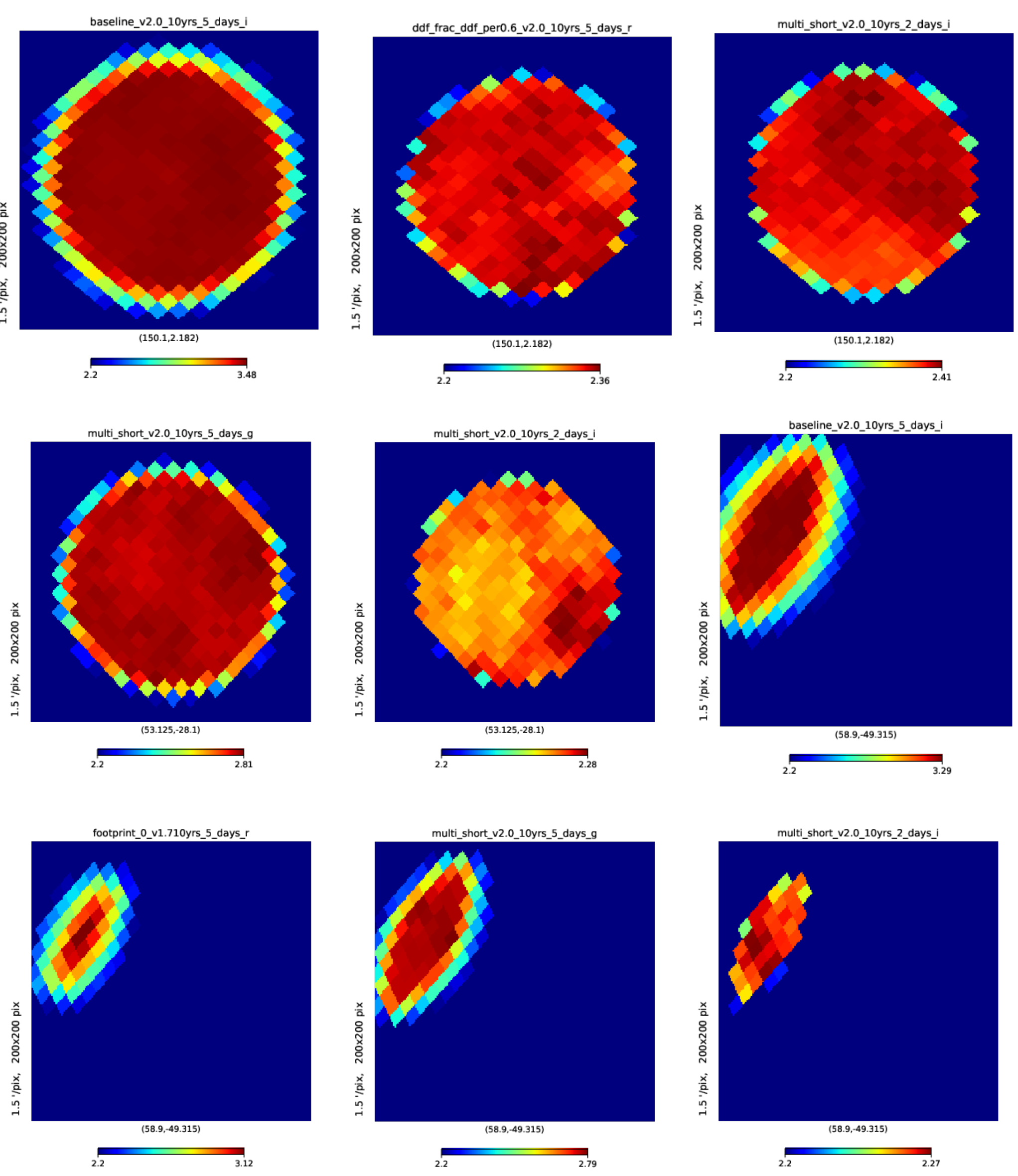}
\caption{Zoom-in of {metric values (indicated in color bars) in DDFs  for different cadence strategies (given in panel titles)} for \textit{r,g,i} bands.  Upper row: \texttt{COSMOS}; Middle row: first and second panel  \texttt{ECDFS}, third panel \texttt{EDFS\_a}; Bottom row: \texttt{EDFS\_a}. 
\label{fig:zoom121}}
\end{figure*}

\begin{figure*}
\includegraphics[width=0.9\textwidth]{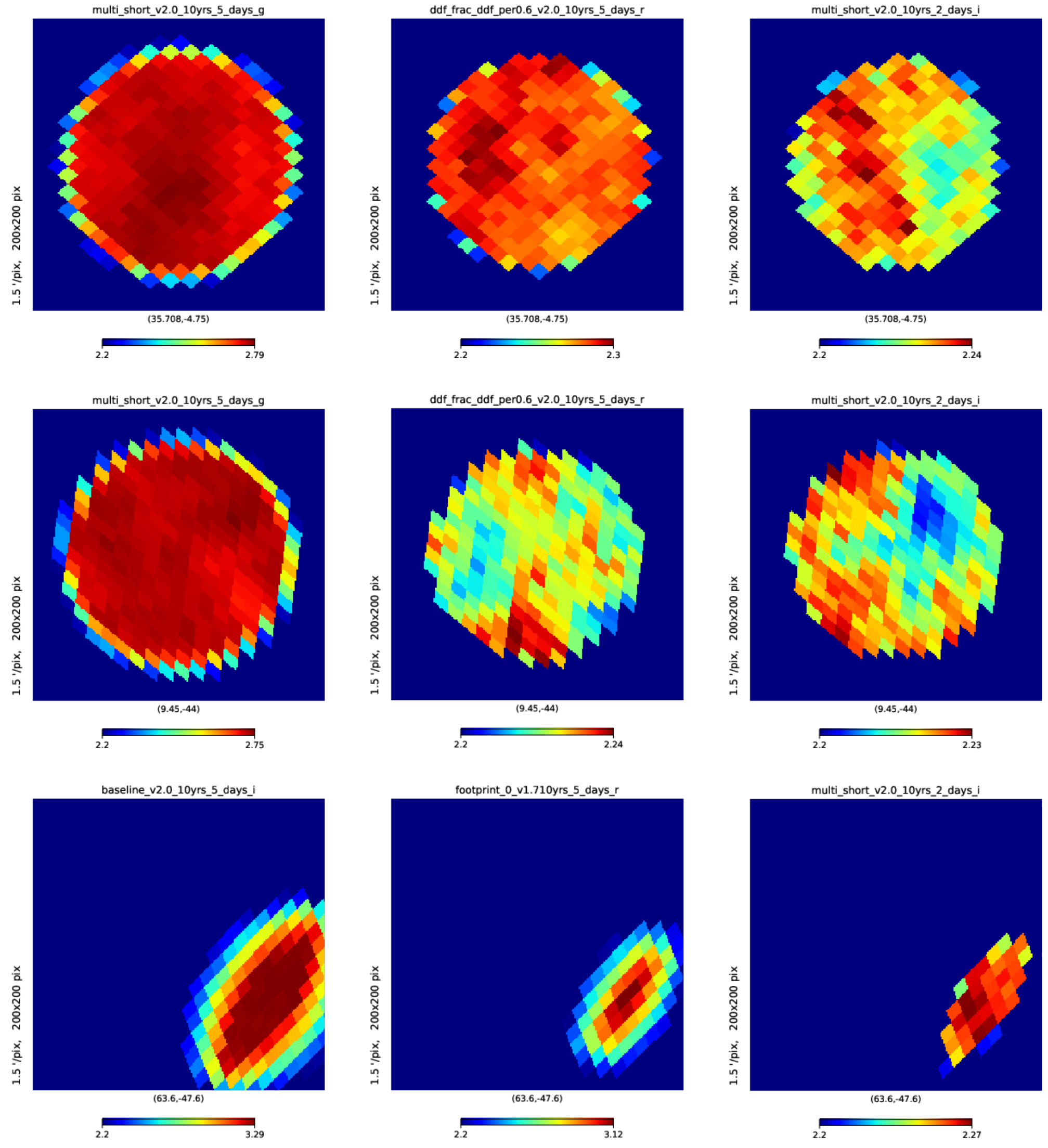}
\caption{The same as Figure \ref{fig:zoom121}, but for   \texttt{XMM} in the top row; \texttt{ELAISS1} in the middle row; \texttt{EDFS\_b} in the bottom row.
\label{fig:zoom122}}
\end{figure*}

{In summary, we found that when the overall cumulative observational season length is  long ($\sim 9 $ yr), the lag measurement likelihood increases because the multiband continuum light curves can track more light curve features. However, survey cadence strategy is also necessary for lag recovery because sampling the accretion disk light curves at the Nyquist frequency is required to resolve the lag accurately. As a result, objects having an estimated full width half-maximum (FWHM) of the AD transfer function less than the minimum sampling  are less likely to have a meaningful lag estimate.
Larger gaps can be tolerated as long as the cadence is small and a long period can be covered, as proven by time lag measurements utilizing the COSMOS and ECDFS fields cadence strategies indicated by our metric.}

\subsection{Effective time lag measurement}\label{sec:lagrec}

Here, we further quantify the potential of measuring time lags using simulated LSST light curves for a  subset of the best observational cadences as suggested by our metric realizations: the two DDFs (\texttt{COSMOS} and \texttt{ECDFS}) observed in \texttt{multi\_short\_v2.0\_10yrs} cadence in the \textit{ugrizy} bands. 

{We use  typical parameters for quasars having median characteristics (SMBH mass $M=2\cdot10^{8}M_{\odot}$, $L/L_{\rm Edd}\sim 0.1$, $\eta=0.1$, disk inclination of  $45^{\circ}$,  azimuth angle of $0^{\circ}$), characteristic DRW timescale $\tau_{\rm DRW}\sim 200$ days and a structure function at infinity
$\sigma_{\rm DRW}=2.5$, because we are interested in quantifying the performance of time lag  measurement  on a \texttt{multishort} cadence {(having multiple (4) short (5s) exposures in each filter)}. }

To create simulated light curves, we first produce the observation schedule and depth of each epoch using the LSST \texttt{OpSim}  \citep{2016SPIE.9910E..13D}.
In each DDF  we created a suite of  quintets of  multiband light curves  by a convolution of the AD transfer function with the DRW driving light curve
(see Section \ref{sec:driving}).

\begin{figure}[ht]
\includegraphics[width=\textwidth]{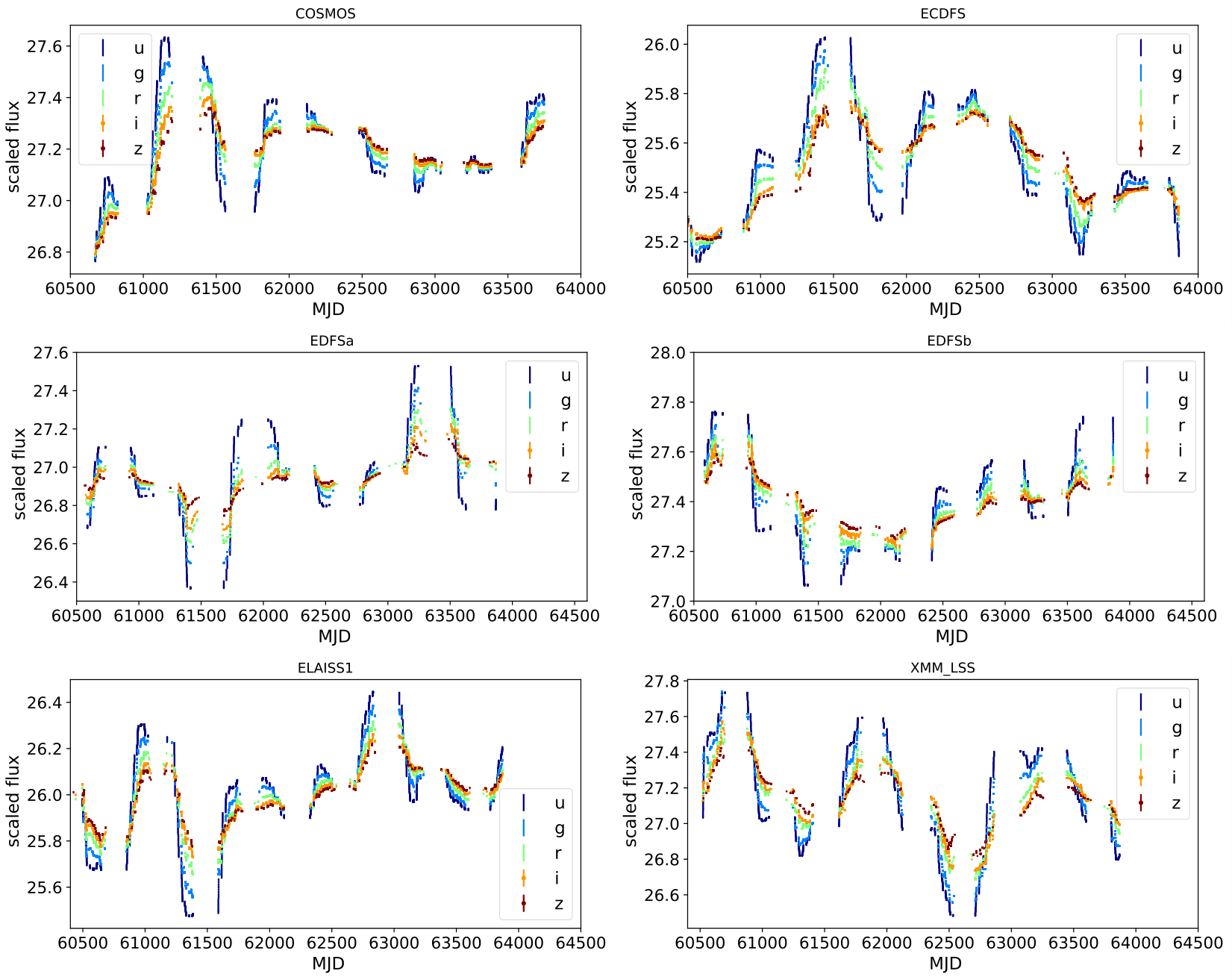}
\caption{Examples of simulated quintets of  LSST light curves for quasars in the LSST DDFs.
We simulate the light curves over a period of 10 years  with  \texttt{multi\_short\_v2.0\_10yrs} cadences. The error bars are set at
$0.005$ mag \textit{rms}.
\label{fig:quintets}}
\end{figure}

Examples of quintets for each DDF are given in Figure \ref{fig:quintets}.  The  warping of light curves in different bands, due to transfer function skewness, is evident. {For example}, warping {can be} seen in \texttt{ELAISS1} DDF around MJD 61000 during one observational season. The photometric noise in all light curves is set to 0.005 mag (i.e. at the level of 0.02\% as the best scenario found in our white paper\footnote{\url{https://github.com/LSST-sersag/white_paper/blob/main/data/table_1.pdf}} thorough simulations \citep{2021MNRAS.505.5012K}.
 The {first row in} Table \ref{tab:idealmeasurement} lists  the true values of AD time lags  $\tau_{uj}=\langle \tau \rangle_{j}- \langle\tau\rangle_{u}, j\in (g,r,i,z)$ as calculated from Equation \ref{eq:meantime delay}. {The remainder show } estimates using our
 procedure to determine time lags  of the light curves quintets (see Section \ref{sec:bayes}  for each {details}).
 
\begin{figure}
\includegraphics[width=0.49\textwidth]{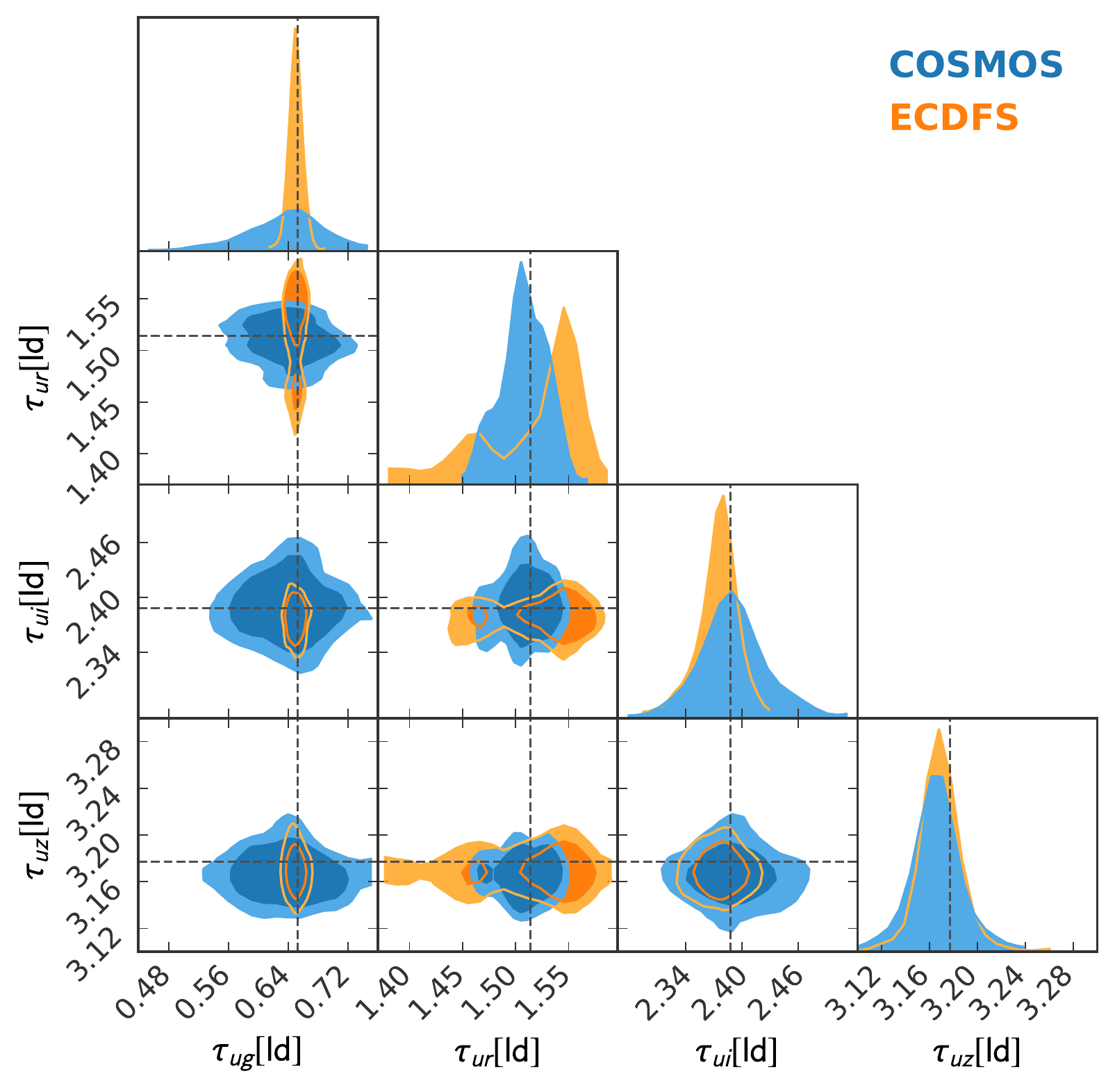}
\includegraphics[width=0.49\textwidth]{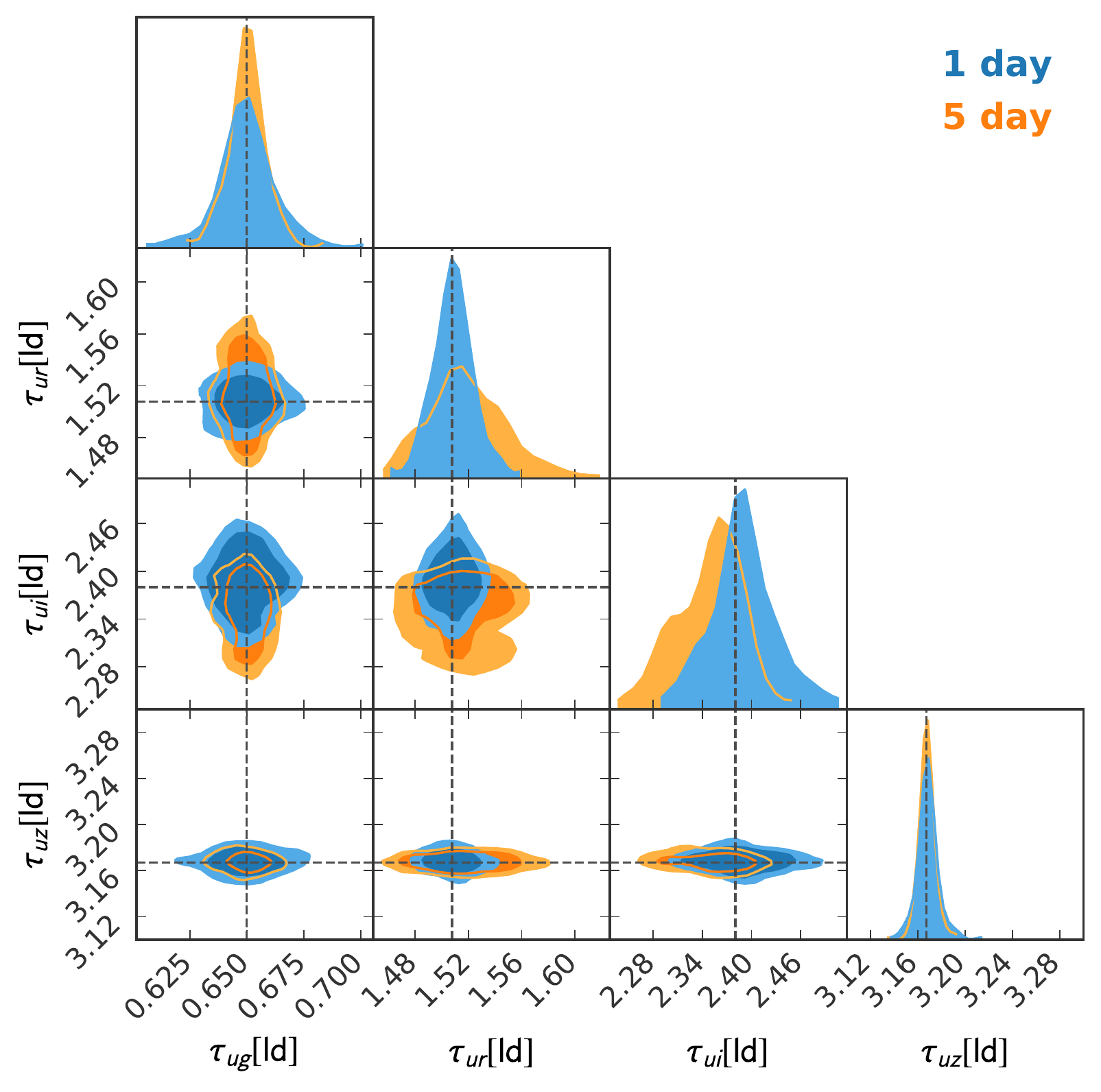}
\caption{Corner plot showing the  PDFs of the time lags of the synthetic light curves with  DDFs \texttt{COSMOS} {(blue)} and \texttt{ECDFS} {(orange)} sampling (left) and homogeneous  1-day (blue) and 5-day (orange) sampling (right).  The distribution for each time lag independently is shown in the histograms along the diagonal, and the 2D distributions as contour plots in the other panels.   The dashed black lines are the true time lag values inferred from the AD model. The contour levels of the 2D joint probability densities are at $1 \sigma$ and $2 \sigma$ regions. }
\label{fig:oneday}
\end{figure}

 The estimated 1D  probability distribution for each time lag and {the} joint probability distribution for each pair of time lags, are
shown in Figure \ref{fig:oneday}. We see that the inferred values of the time lags are both accurate (close to true values) and precise (narrow posterior distribution) for   \texttt{ECDS} light curves in \textit{g}, \textit{r} and \textit{i} bands. {However, the PDF of $\tau_{ug}$ is deformed for} \texttt{COSMOS}.  In general, we see from the 1D histogram of Figure \ref{fig:oneday}  that PDF intervals are
much narrower than the prior ranges (see Subsection \ref{sec:bayes}).
 Table \ref{tab:idealmeasurement} lists the mean of the posterior and uncertainties corresponding to 16th and 84th
quantile of the distribution of measured time lags.
{Because  light curve distortion can make detecting delays in individual segments of light curves difficult, the AD time lag measurement is demonstrated on decadal observational time baseline \citep[see e.g.][]{2020A&A...636A..52C}.}

The light curves in \texttt{OpSim} observation schedules have breaks between a series of continuous observations, with the gaps ranging from 
several days to $\geq 80$ days. To see how these gaps affect lag measurements, we created additional ideally sampled simulated light curves with homogeneous sampling of 1 and 5 days.

By examining the corner plot for recovered time lags of 1 and 5 day homogeneous light curves (right panel in Figure \ref{fig:oneday}), we conclude that the lag distributions are affected by homogeneous sampling.  {For the 1 day sampling}, the PDFs are narrower and more symmetric with respect to true values.
 Table \ref{tab:idealmeasurement} presents summary statistic of distribution of time lag measurement on these ideally sampled quintets of light curves.

\begin{deluxetable*}{cccccc}
\tablenum{2}
\tablecaption{
Summary statistics of recovered time lags from the  distributions for  \texttt{COSMOS}, \texttt{ECDFS}  and homogeneously sampled light curves given in Figure  \ref{fig:oneday}. The columns are: status of time lags,  light curve sampling, the  time lags of \textit{g}-band  vs. reference \textit{u}-band light curves; \textit{r}-band vs \textit{u}-band light curves; \textit{i}-band vs. \textit{u}-band light curves; \textit{z}-band vs \textit{u}-band light curves. \label{tab:idealmeasurement}}
\tablewidth{0pt}
\tablehead{\colhead{Status}&
\colhead{LC sampling} & \colhead{$\tau_{ug}$} & \colhead{$\tau_{ur}$} & \colhead{$\tau_{ui}$} &
\colhead{$\tau_{uz}$}  \\
}
\decimalcolnumbers
\startdata
True &-&0.653&  1.514  &2.388 &3.177	\\
\hline
\hline
Inferred&\texttt{COSMOS} &$0.643^{+0.06}_{-0.04}$ & $1.507_{-0.02}^{+0.027}$ & 	$2.388^{+0.029}_{-0.033}$& $3.166_{-0.017}^{+0.017}$	 \\
Inferred&\texttt{ECDFS} &$0.650^{+0.008}_{-0.008}$ & $1.540_{-0.08}^{+0.012}$ & 	$2.377^{+0.017}_{-0.022}$& $3.168_{-0.009}^{+0.012}$	 \\
Inferred&\texttt{1-day}&$0.649^{+0.01}_{-0.01}$ & $1.508_{-0.01}^{+0.01}$ & 	$2.388^{+0.04}_{-0.04}$& $3.168_{-0.01}^{+0.01}$	 \\
Inferred&\texttt{5-day} &$0.649^{+0.01}_{-0.01}$ & $1.516_{-0.04}^{+0.03}$ & 	$2.369^{+0.02}_{-0.03}$& $3.167_{-0.01}^{+0.01}$	 \\
\enddata
\tablecomments{True values of mean relative time delays $\langle \tau \rangle_{uj}, j\in(g,r,i,z)$ calculated from AD model (see Equation \ref{eq:meantime delay}). The inferred values represent  the 50th quantile with errors corresponding to 16th and 84th quantiles  of the respective probability distributions of calculated time lags.}
\end{deluxetable*}
{
\citet{2011ApJ...732..121B}  have shown that for interpolation and cross correlation methods {applied to} real data, as long as the driving
light-curve is {well} sampled {(in their case almost nightly over the course of 200 days)}, poorer sampling of the output light-curves can
be tolerated.
{We recall that, 
to first order, the centroid of the transfer function is simply
proportional to the luminosity-weighted radius \citep{1991ApJS...75..719K}. However, in the case of LSST multiband continuum transfer functions, their centroids are shifted closer to their FWHM  (see Figure \ref{fig:general}). 
Also, we see that   interband continuum delays which are driven by DRW, have values below or close to transfer function centroids (e.g., Figure \ref{fig:oneday})}. So, it is important to have a sampling that is of the order of the FWHM of the AD transfer function. This condition is most relevant for measuring the time lag between \texttt{u}-band and \texttt{g}-band light curves as their interband time delay is the smallest.}

\section{Discussion}\label{sec:discussion}

Our \texttt{MAF}  {metric} (\texttt{AGN\_TimeLagMetric}) indicates that  the best cadences ({i.e., those with} the largest values above {the} Nyquist threshold)  in the DDFs are those in the \texttt{multi\_short} family,  {taking into account fiducial time lags of} 5 days  in all LSST filters. 
This family of strategies {belong to} the micro-surveys. Micro-surveys aim to support and extend the LSST's primary science drivers, and they were submitted in response to the 2018 Call for White Papers \citep[see][]{2022ApJS..258....1B}. Many of the micro-surveys requiring between $0.3-3\%$ of the overall survey time were simulated as part of \texttt{OpSim}  v2.0 simulations. {The} \texttt{multi\_short} takes four short (5 sec) exposures in each filter, which is about 20 sec in total per filter.  Despite the short exposure time, the high observational cadence allows for  photometric RM analysis as indicated by our simulations as well as  other studies \citep[see ][]{2020ApJS..246...16Y}. However, when considering micro-surveys, it is worthwhile to note that sometimes visits can be extremely short, which can affect RM observables.
The DDFs currently receive about 1 percent  of the total survey time per DDF location. There are five DDFs, so this accounts for 5 \% of the total survey time. Each DDF is a single pointing, with the exception of the Euclid South DDF {which consists of two shallower pointings}.

However,  there are  simulations, where the amount of time allocated for DDFs is varied from 3 percent (\texttt{ddf\_fract\_ddf\_per0.6\_v2.0\_10yrs} ) to 8 percent (\texttt{ddf\_fract\_ddf\_per1.6\_v2.0\_10yrs}). Our \texttt{MAF}  simulations show that there is no {significant} difference between these two {variants} of DDFs when it comes to time lag measurements.

\begin{figure*}
\gridline{\fig{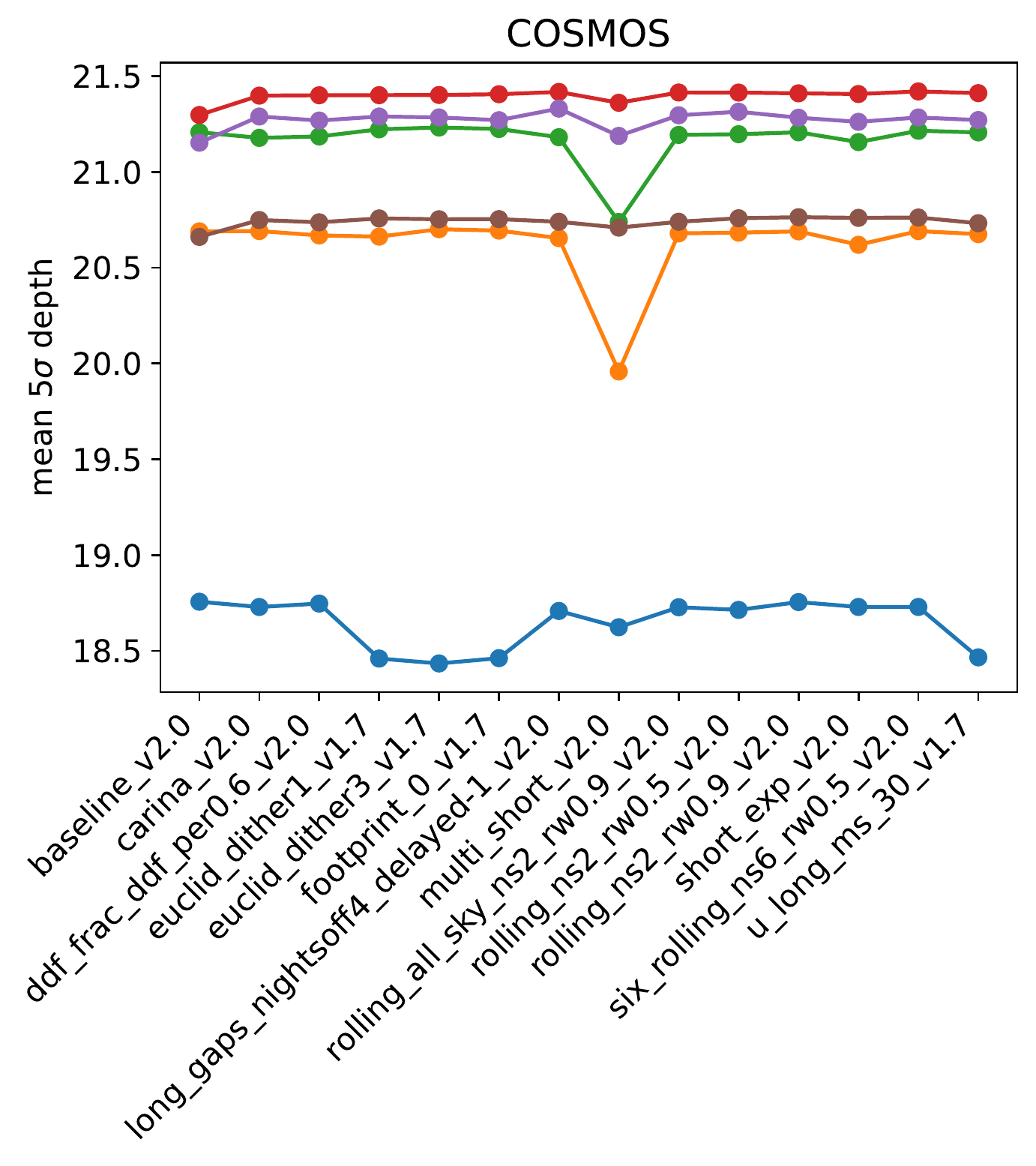}{0.34\textwidth}{}
          \fig{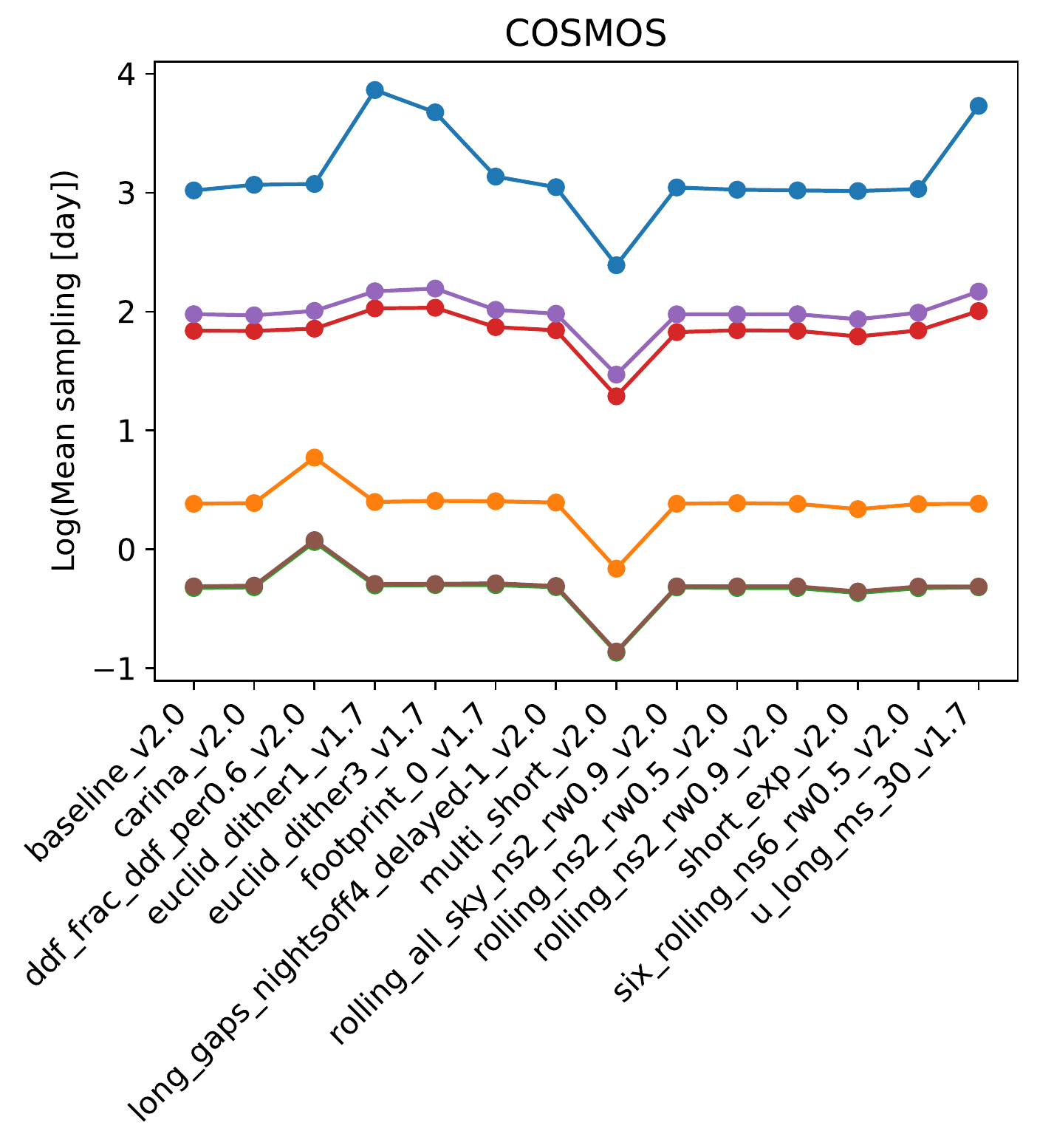}{0.34\textwidth}{}   
       \fig{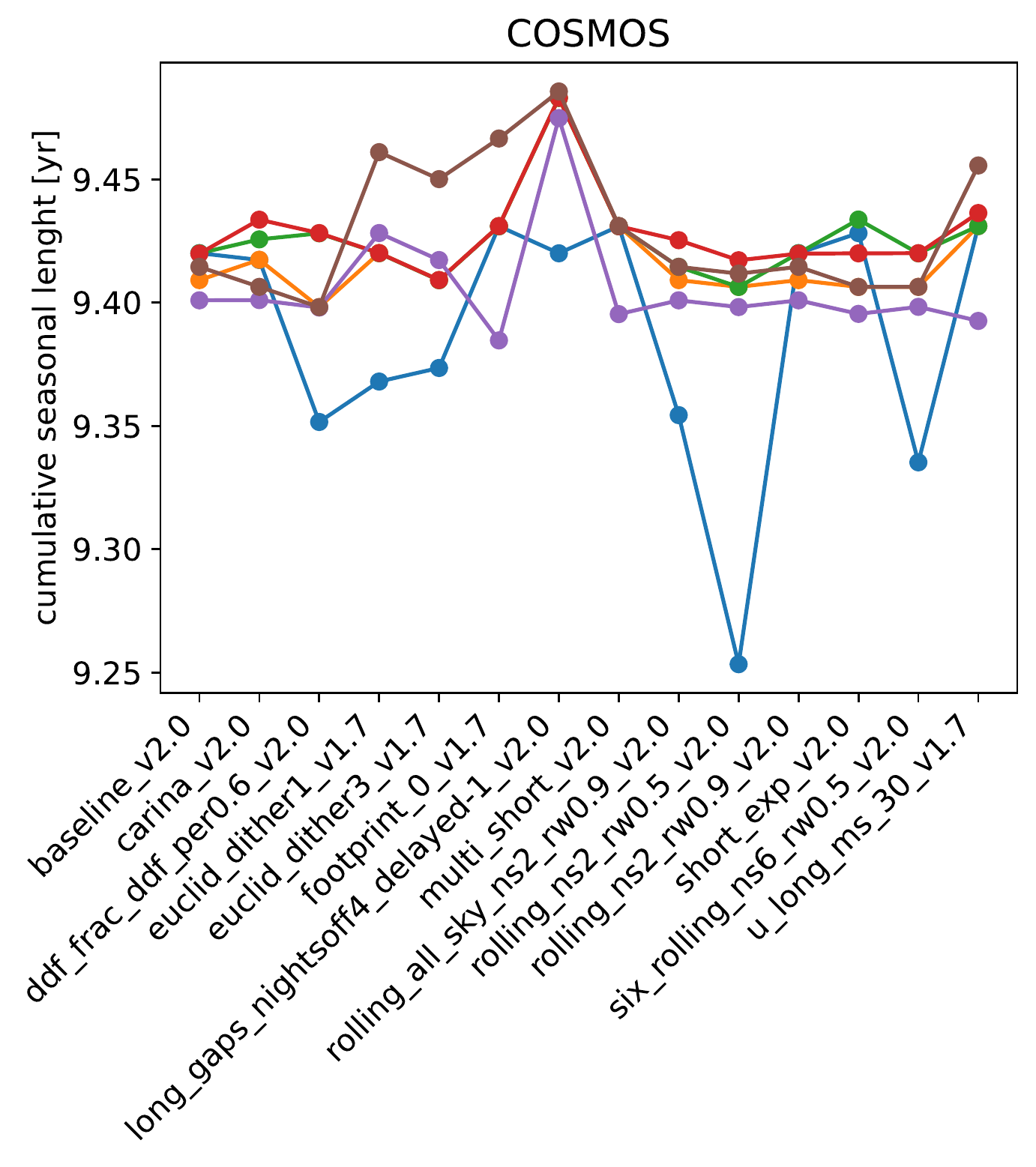}{0.34\textwidth}{}
          }
\gridline{
\fig{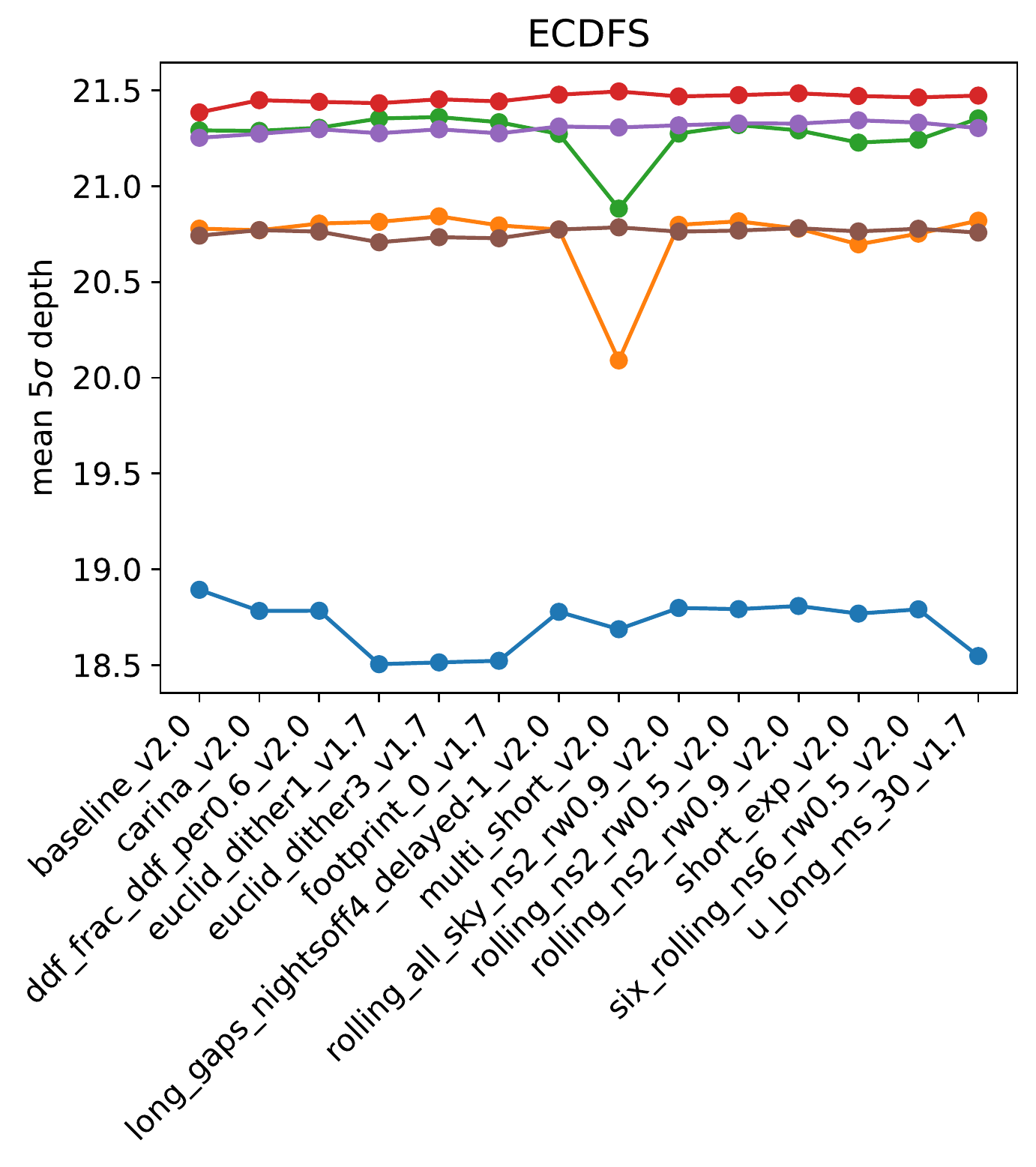}{0.34\textwidth}{}
          \fig{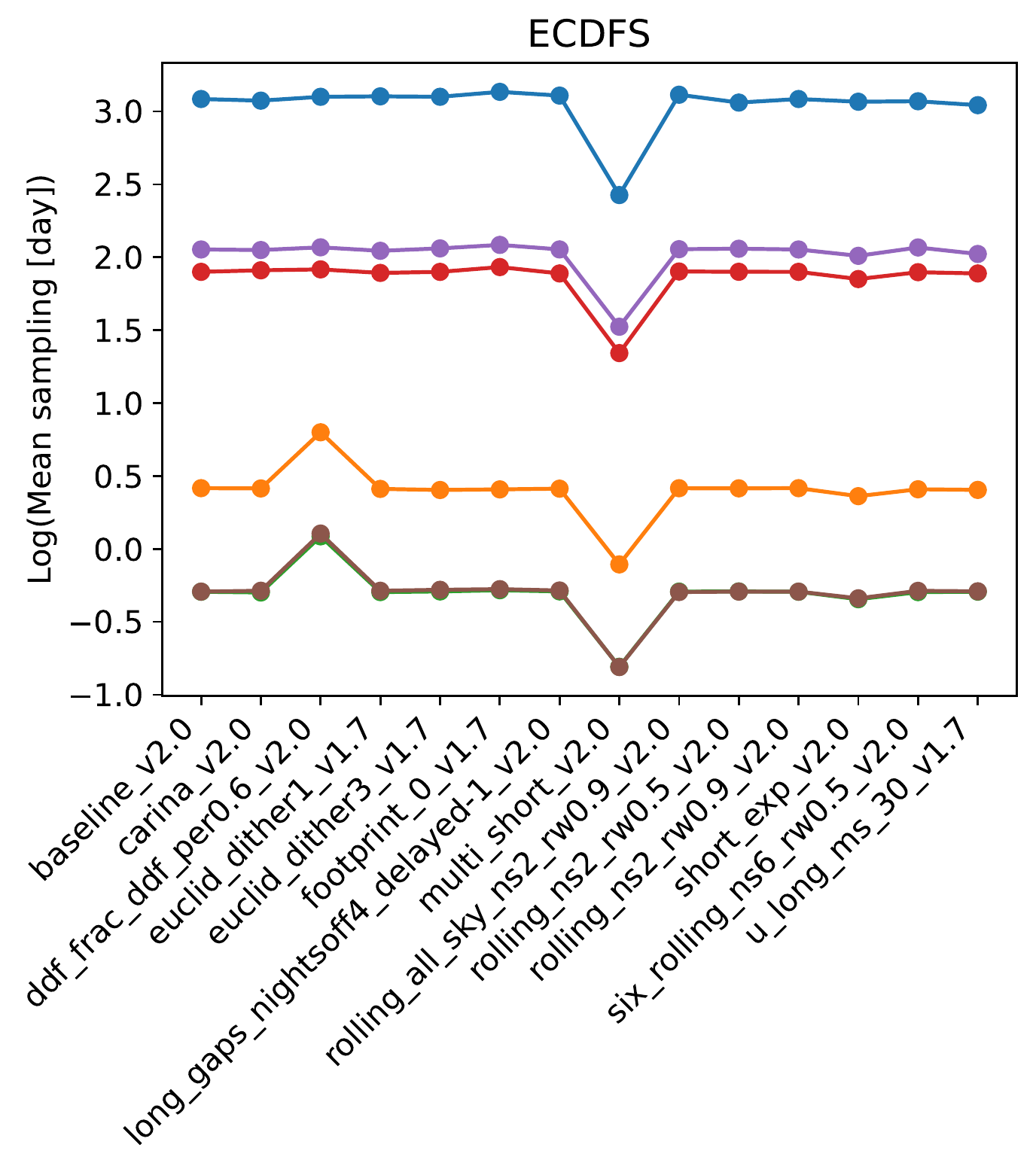}{0.34\textwidth}{}
          \fig{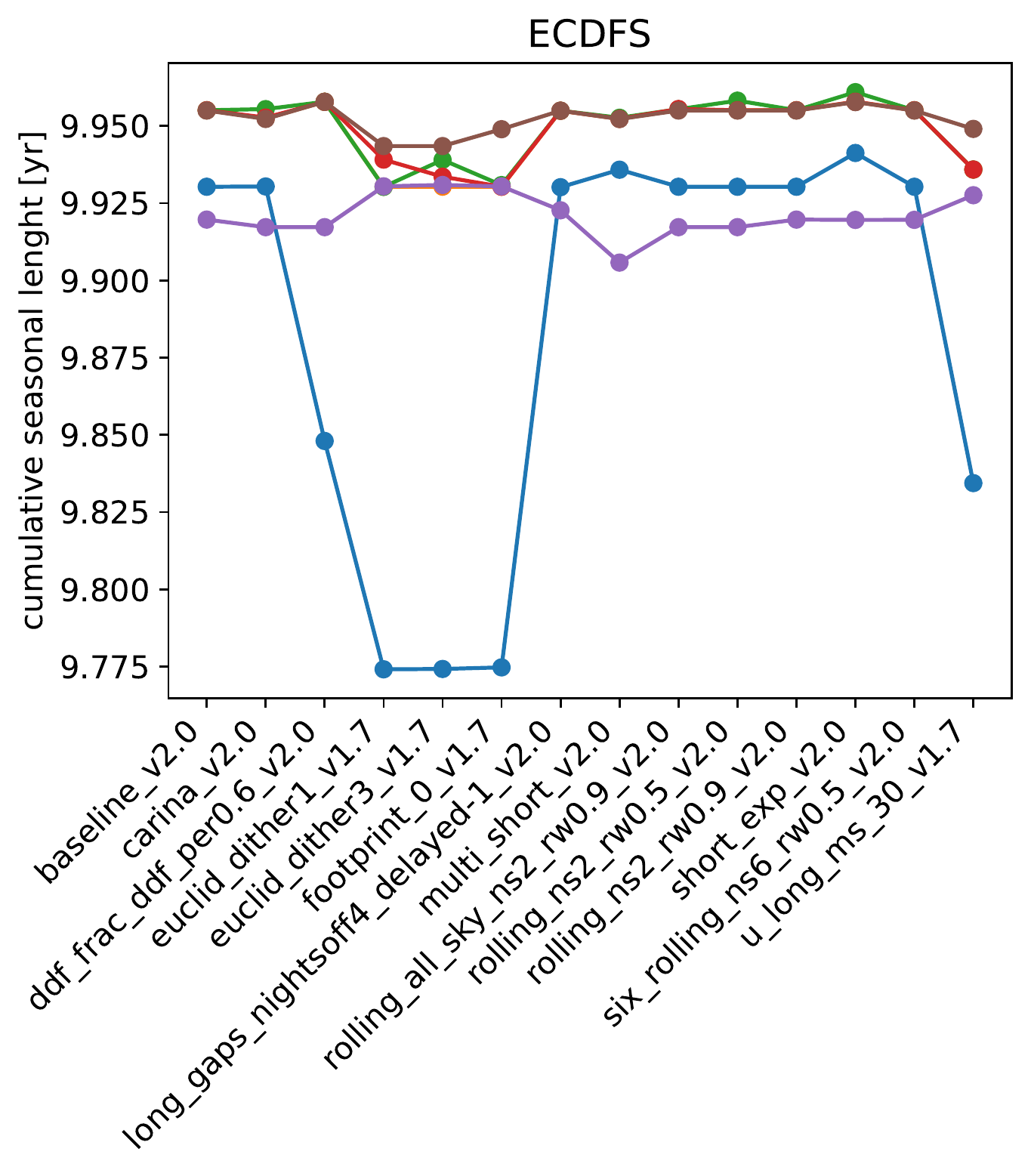}{0.34\textwidth}{}
          }
\caption{Comparison of mean $5\sigma$ depth (left column), mean sampling  (middle column) and mean
cumulative season length (right column) for different
observing strategies (x-axis) in DDFs and \textit{ugrizy} filters. The solid lines guide the eye. The mean  $5\sigma$ depths of the non-\textit{r} and \textit{r} bands are  close. This means that objects as bright as 21 will still be seen in those bands for these overlapping cadences in DDF. Also, fainter objects in non-\textit{r} band will be observable in \textit{r}-band. The passbands \textit{u}, \textit{g}, \textit{r}, \textit{i}, \textit{z}, \textit{y} are denoted with blue, orange, green, red, violet and brown color respectively.
\label{fig:figcomparison1}}
\end{figure*}

\begin{figure*}
\gridline{\fig{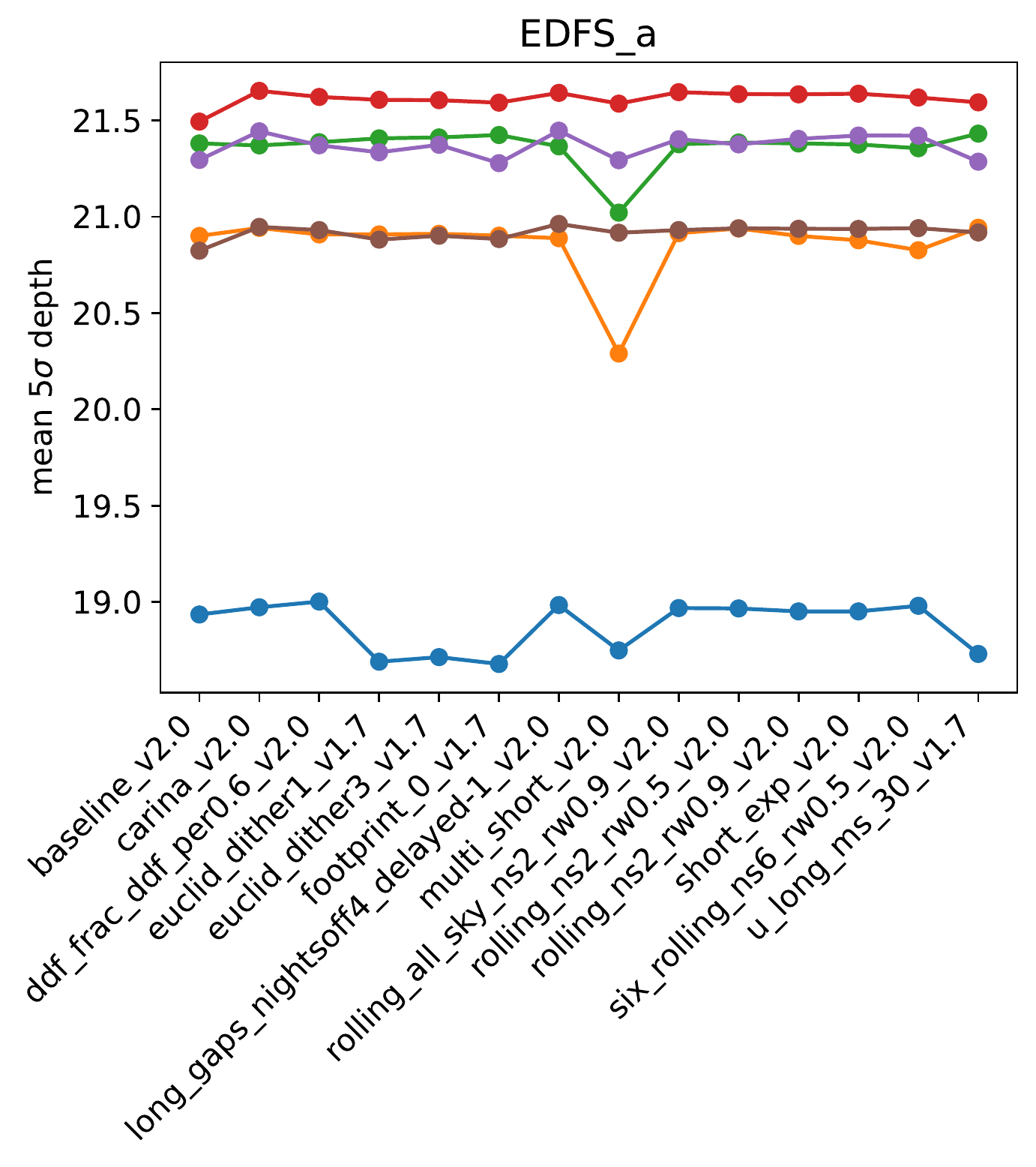}{0.33\textwidth}{}
          \fig{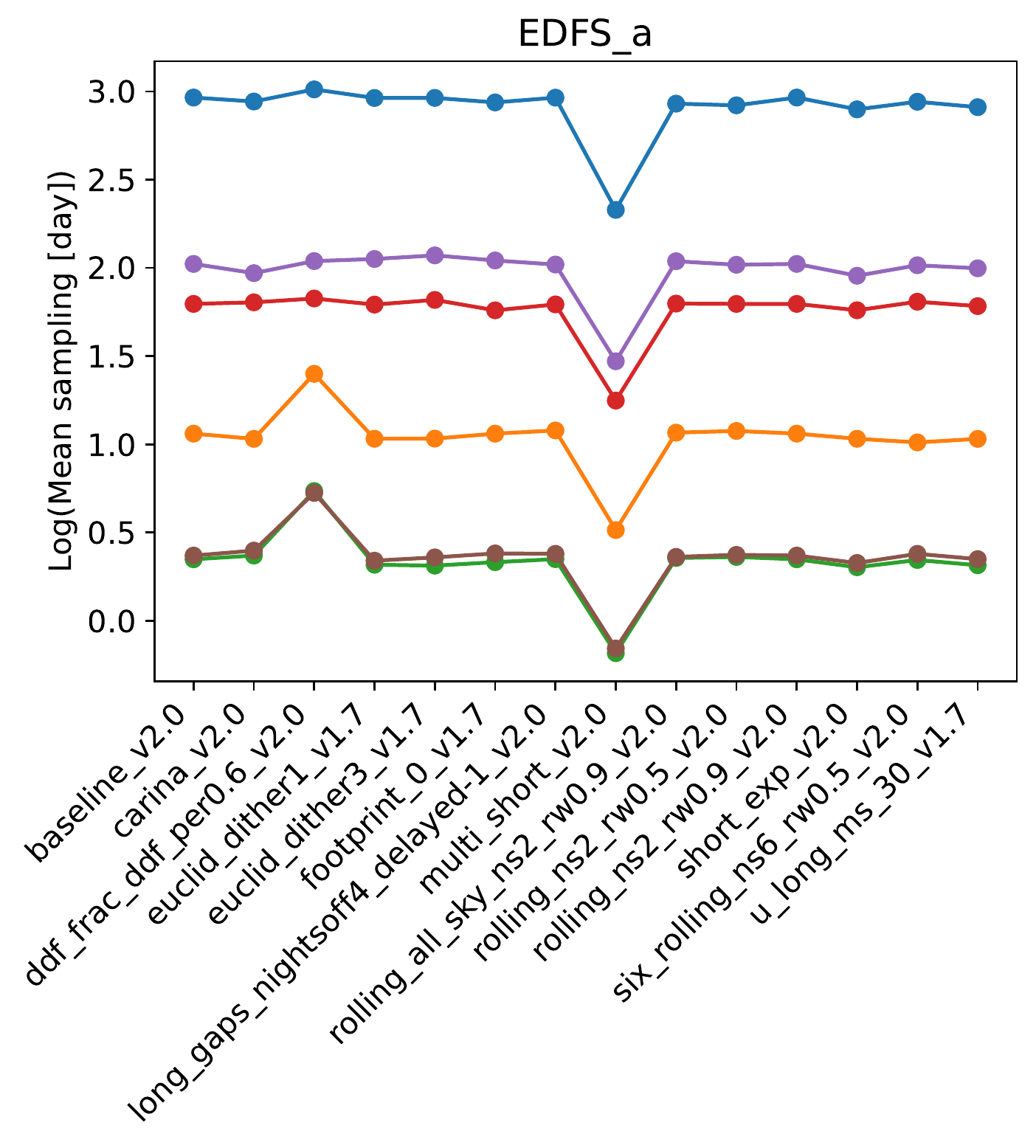}{0.33\textwidth}{}  
        \fig{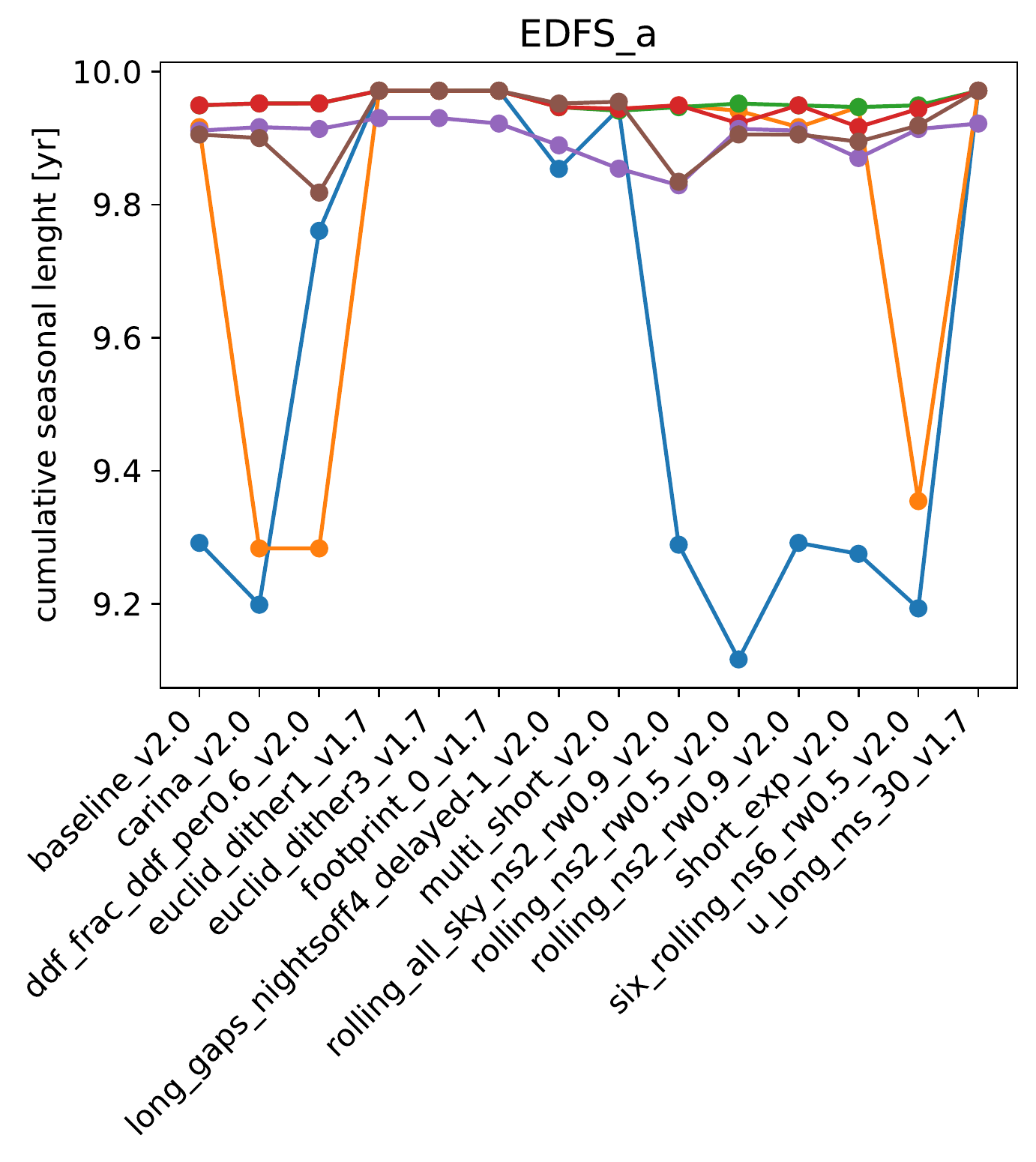}{0.33\textwidth}{}
          }
\gridline{
\fig{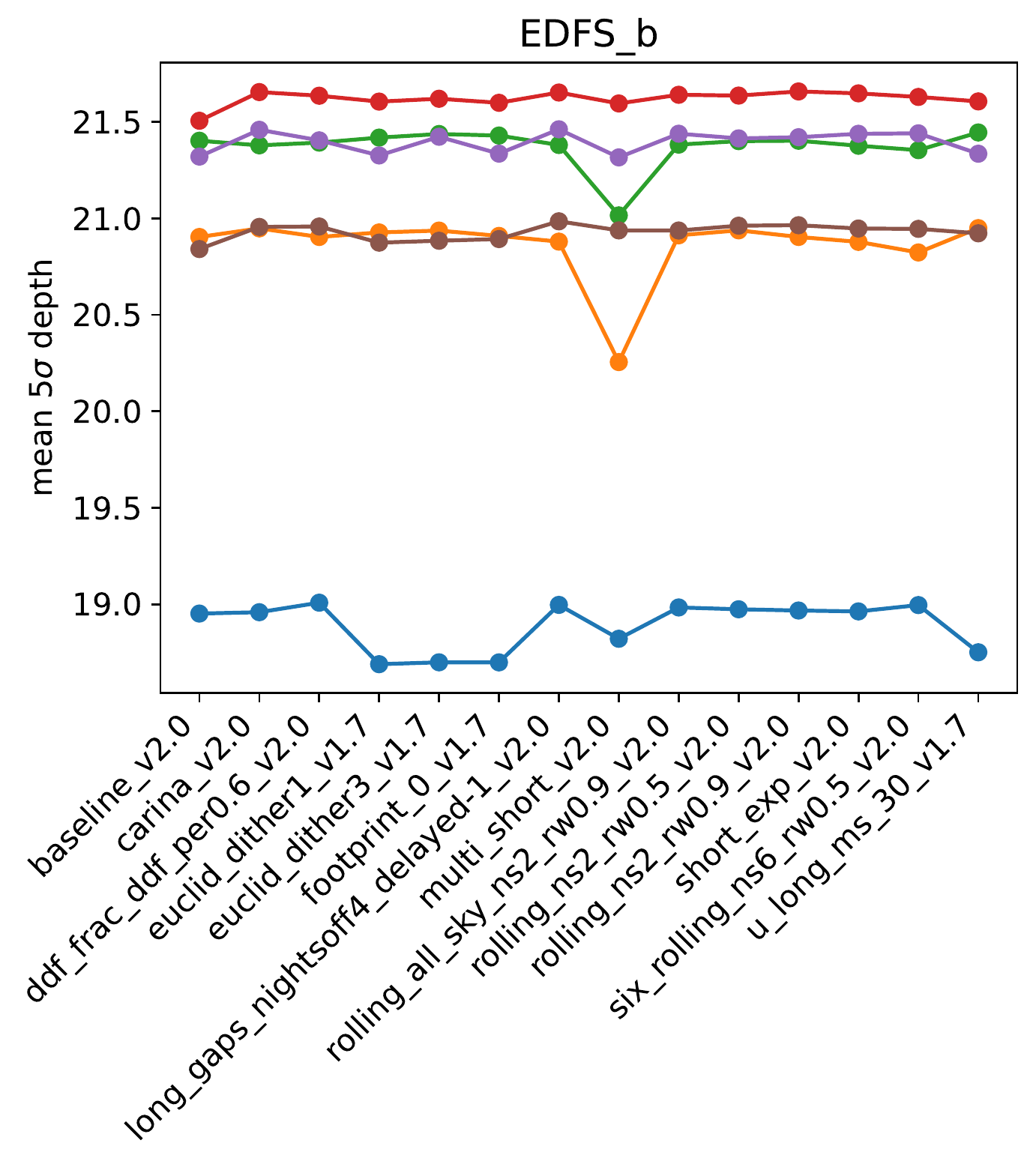}{0.33\textwidth}{}
          \fig{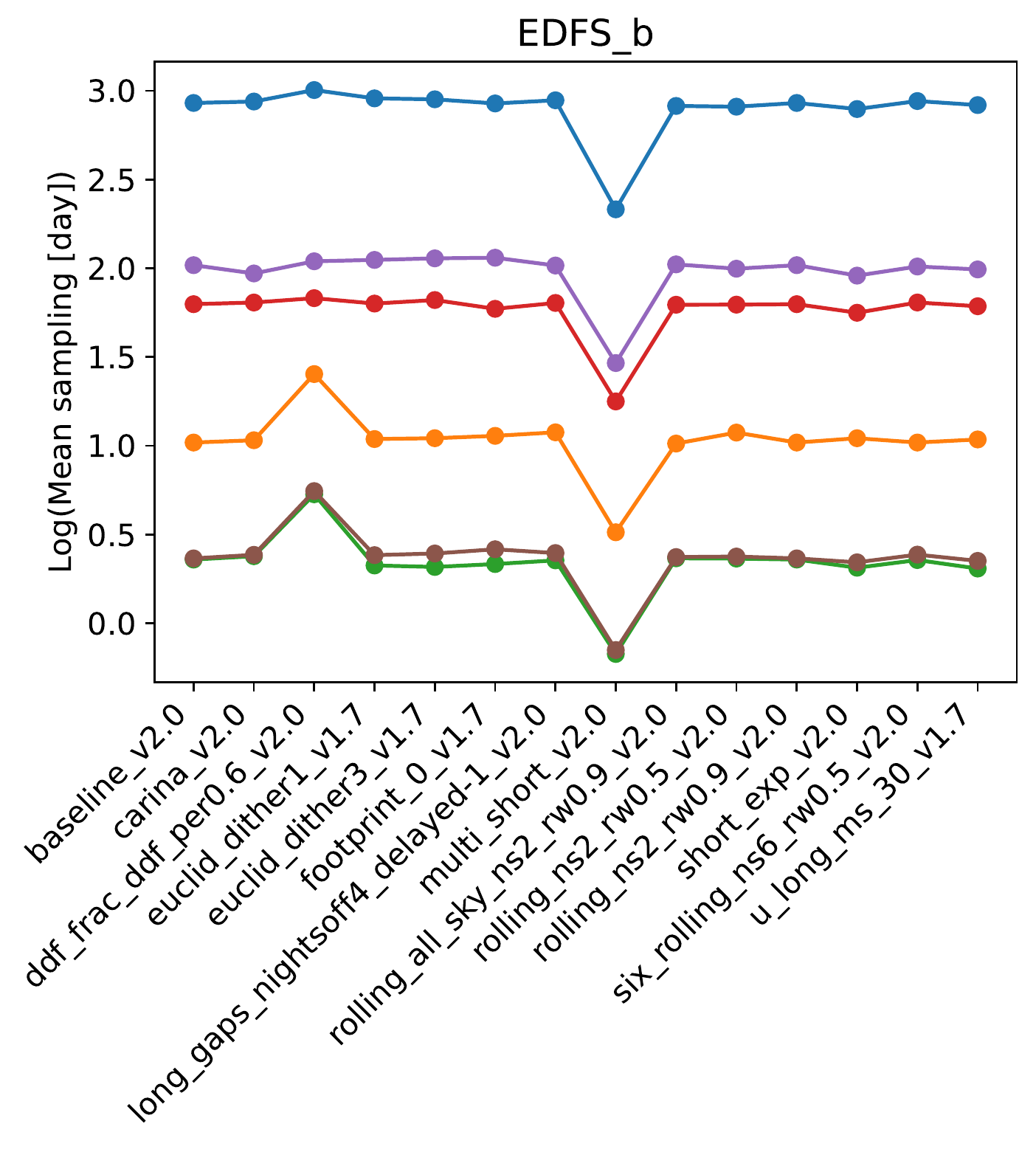}{0.33\textwidth}{}
          \fig{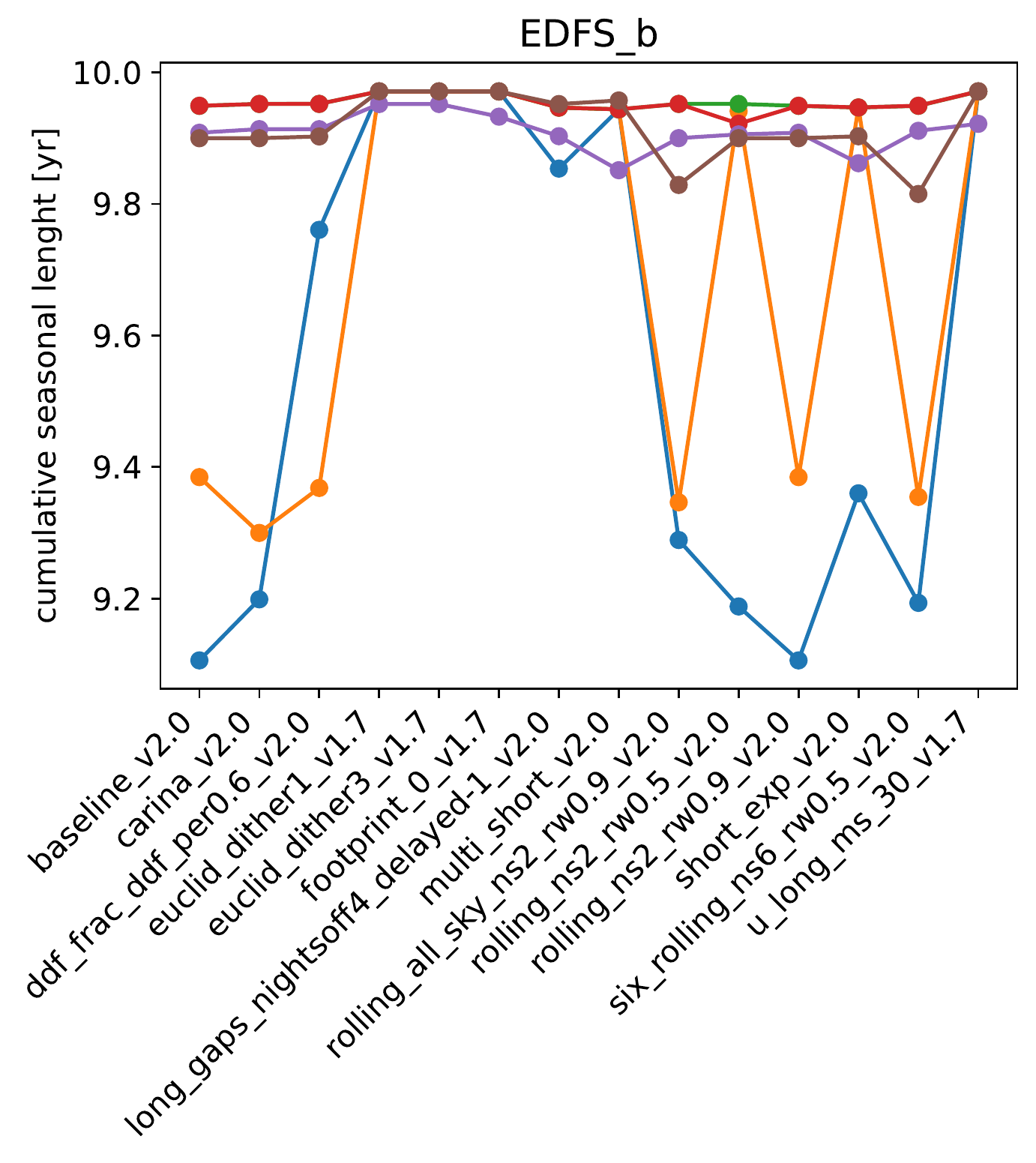}{0.33\textwidth}{}
          }
\caption{The same as Figure \ref{fig:figcomparison1} but for different DDFs.
\label{fig:figcomparison2}}
\end{figure*}

\begin{figure*}
\gridline{\fig{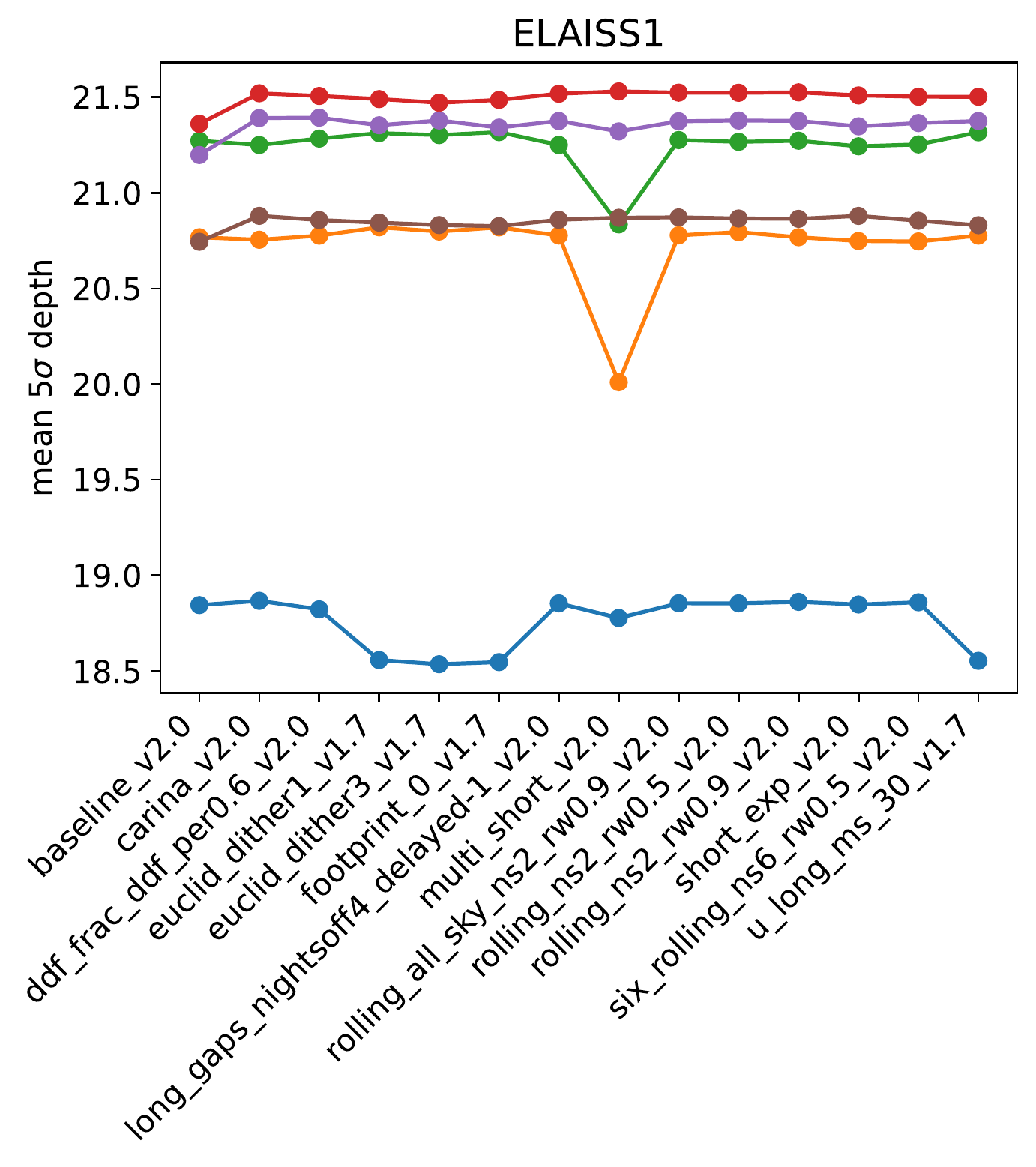}{0.33\textwidth}{}
          \fig{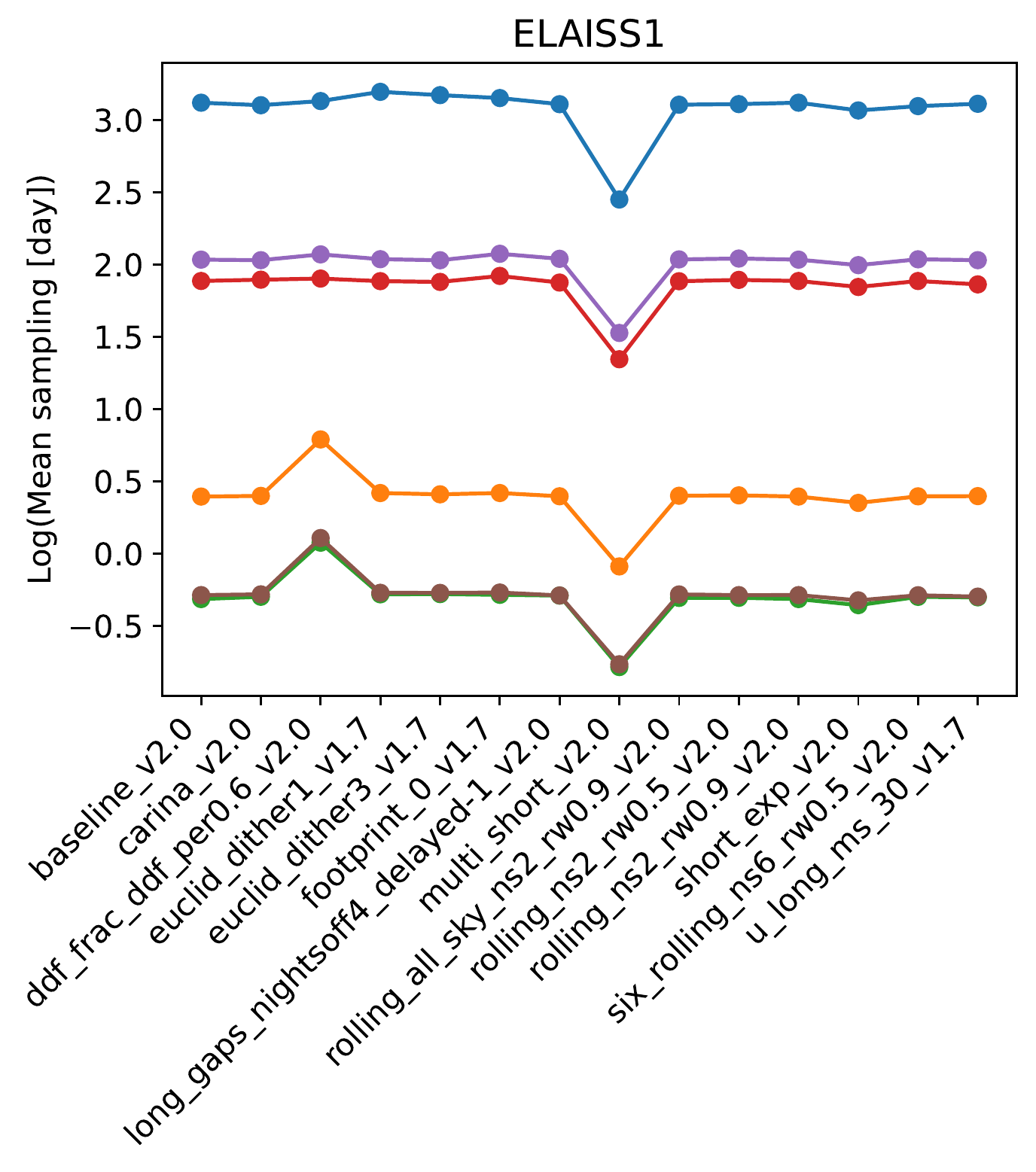}{0.33\textwidth}{}          \fig{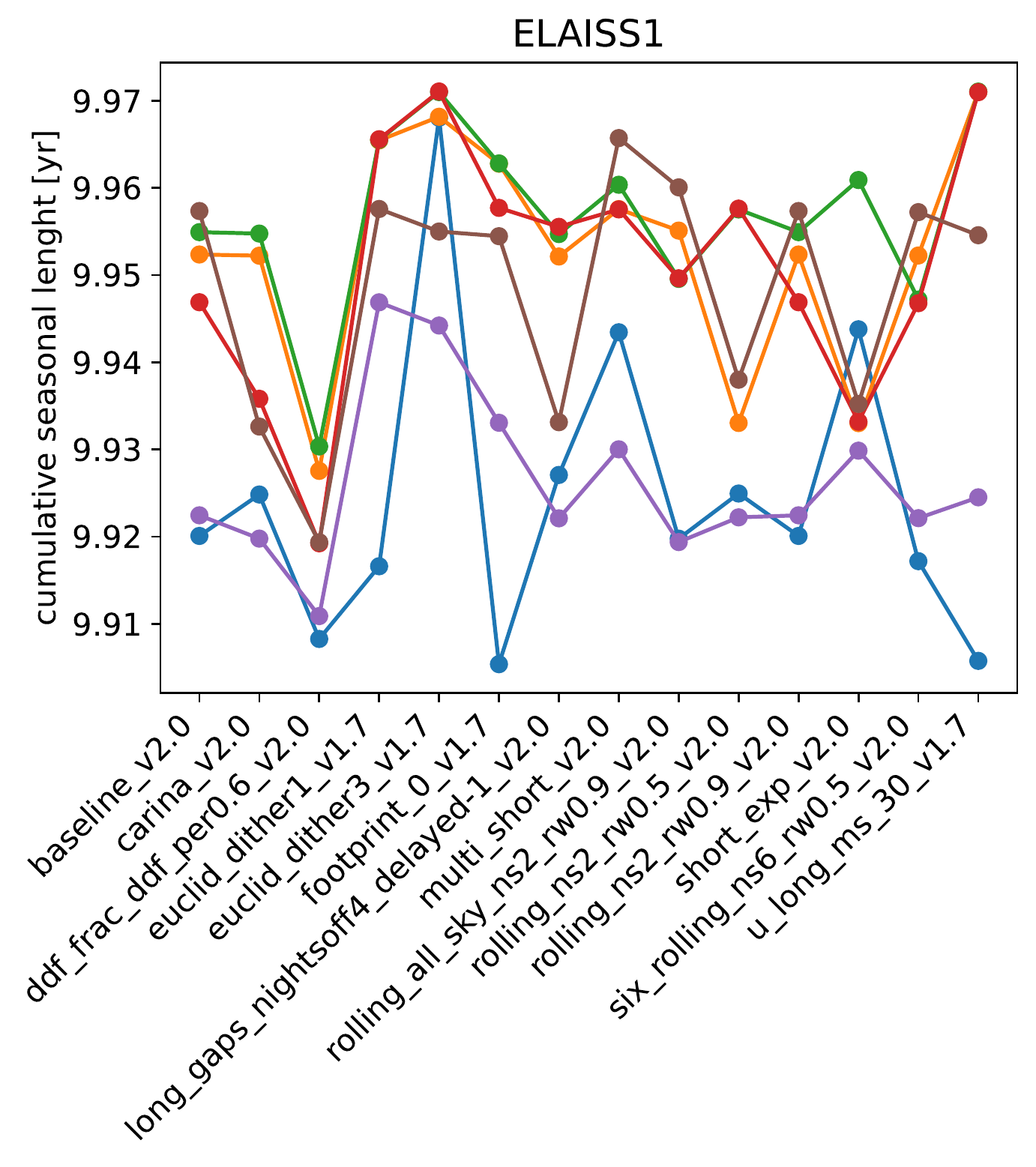}{0.33\textwidth}{}
          }
\gridline{\fig{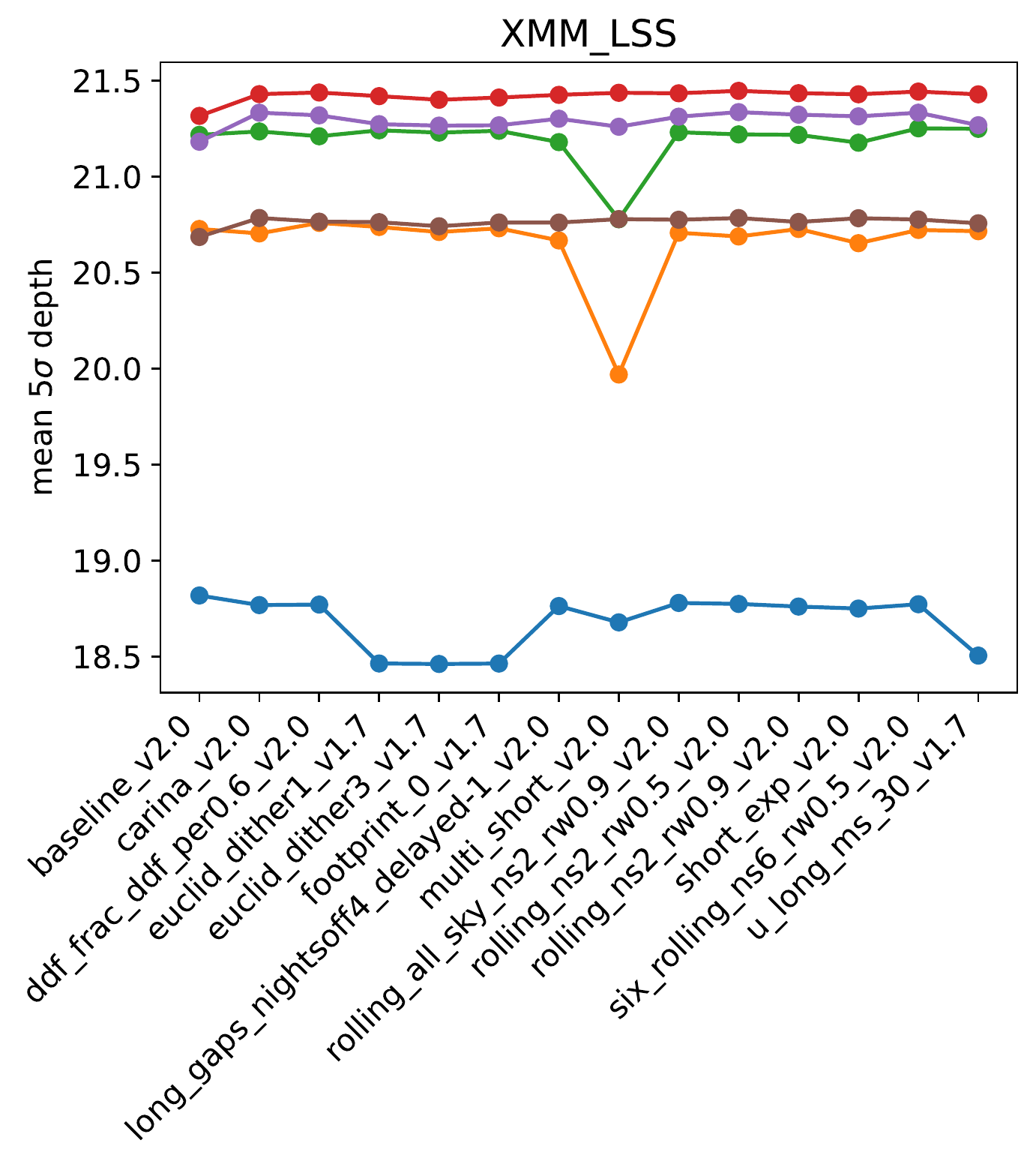}{0.33\textwidth}{}
          \fig{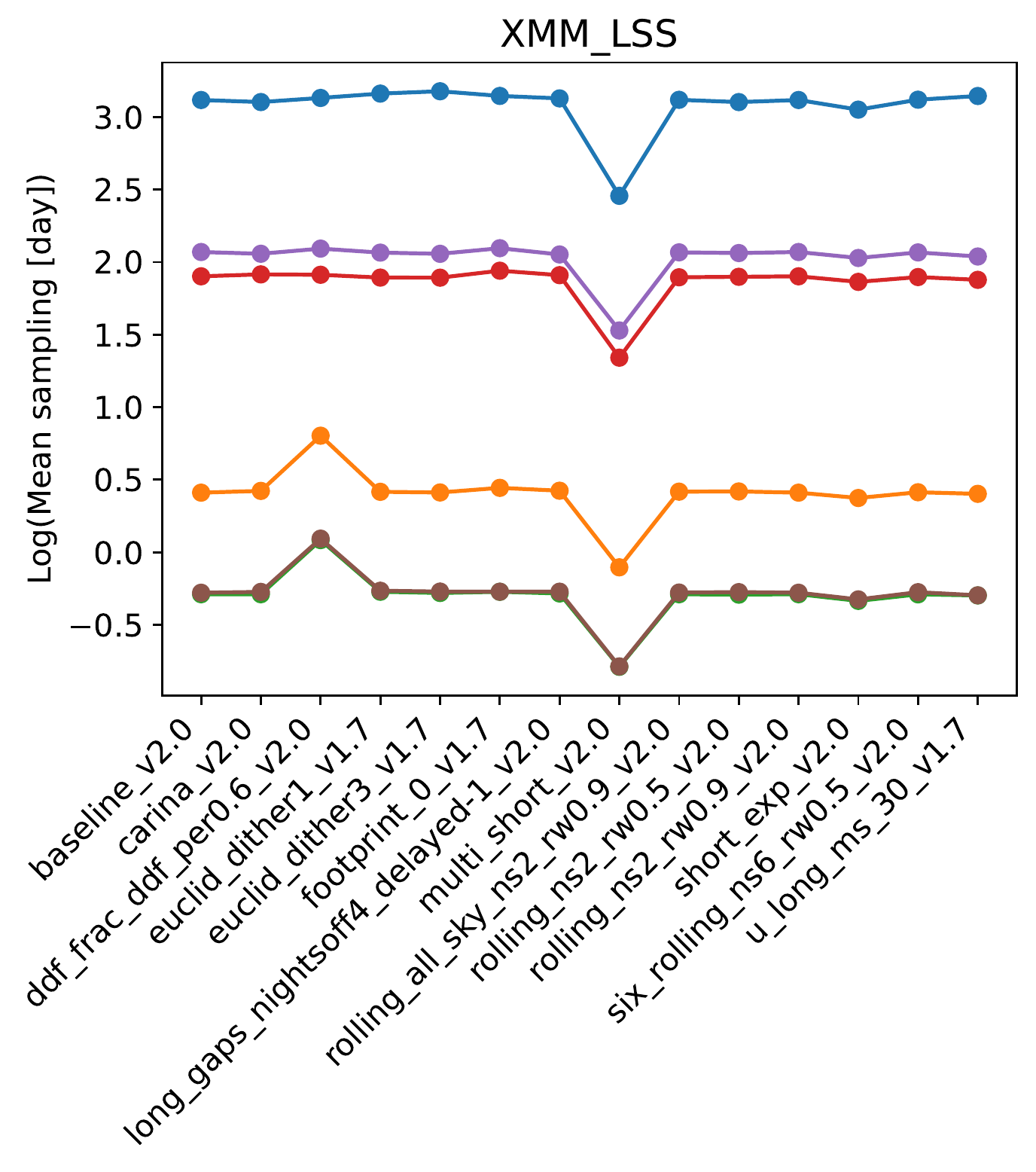}{0.33\textwidth}{}
          \fig{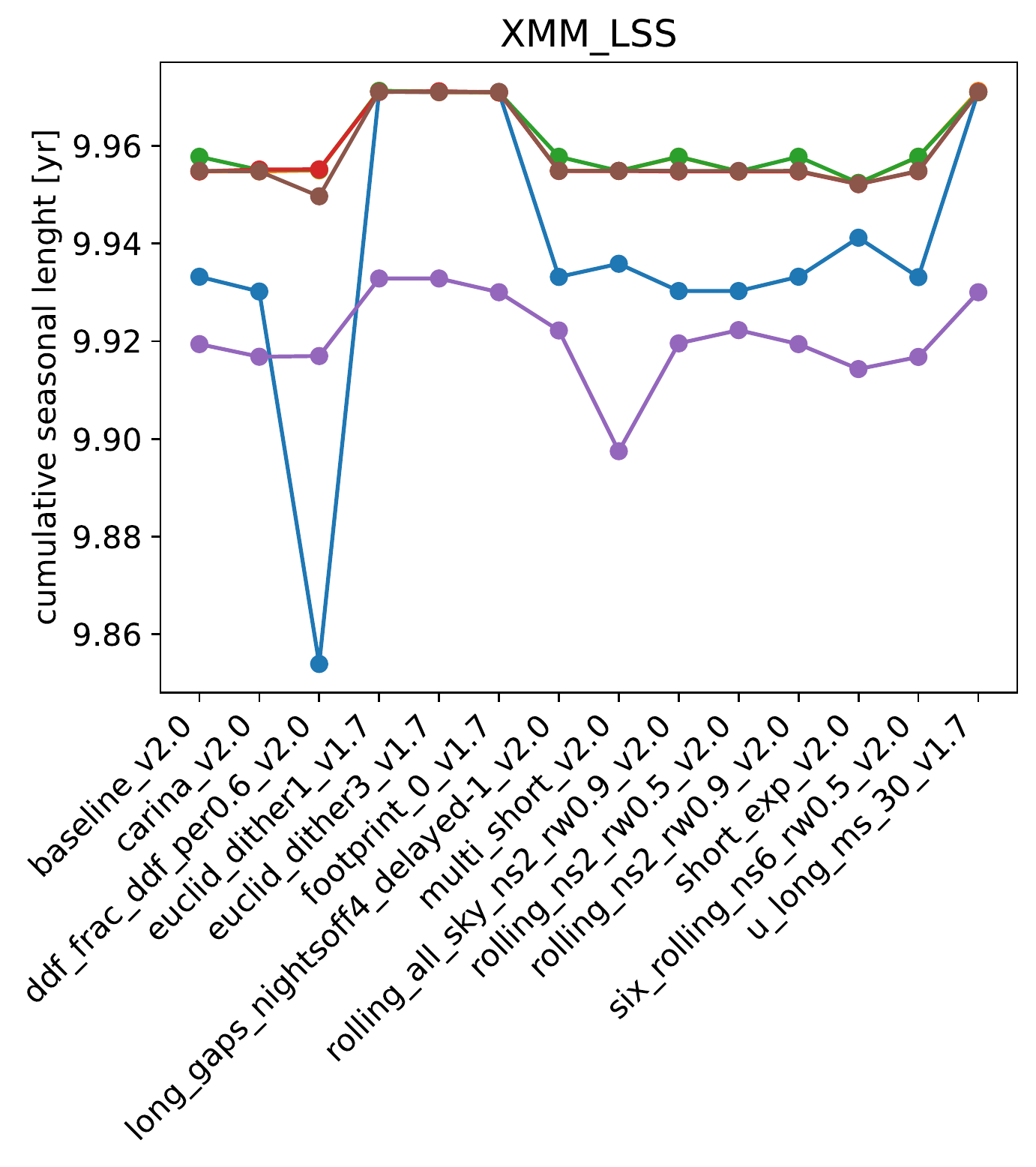}{0.33\textwidth}{}
          }
\caption{The same as Figure \ref{fig:figcomparison1} but for different DDFs.
\label{fig:figcomparison3}}
\end{figure*}

From our  simulations \citep[see also][]{2021MNRAS.505.5012K}, we find that {the most important parameters that affect} the quality of time delay measurements in AGN ADs are  the
cumulative season length {and} the mean sampling (also referred to
as the sampling frequency). A categorization of 14 observing strategies in all DDFs and LSST filters, based on mean $5\sigma$ depth, mean sampling,  and mean cumulative season length  are shown in Figures \ref{fig:figcomparison1}-\ref{fig:figcomparison3}).

The mean cumulative season length, mean sampling and mean $5\sigma$ depth from a simulation of a given observing strategy are calculated by
taking the mean of {these parameters} for  the DDF under consideration  (Figures \ref{fig:figcomparison1}-\ref{fig:figcomparison3}).
The cumulative season length is the {sum of the duration of all observing seasons.} A season gap for an LSST field is defined
{as} no observations in any filter {taking place} for 85 days \citep{2019A&A...631A.161H}. Across all DDFs,
the cumulative season length for
the majority of strategies is
 $\geq 9.5$yr. We note that \texttt{multi\_short} strategy  has the {fastest (smallest)} sampling  but the mean $5\sigma$ depth is shallower.
{
The mean sampling  exhibits a similar pattern of behavior per filter throughout all DDF fields, with subtle variations. Figure \ref{fig:figcomparison1} shows \texttt{COSMOS} field as an example.
The mean sampling for some of the observing strategies in the \textit{u}-band in the \texttt{COSMOS}  {differs from other fields} (e.g., compare mean sampling for families between \texttt{carina} and \texttt{footprint} in \texttt{COSMOS} and \texttt{ECDFS}, see  Figure \ref{fig:figcomparison1}), which is related to time spent on other surveys such as the northern hemisphere, the southern Celestial Pole, and the Galactic Center. {As seen in Figures \ref{fig:figcomparison1}-\ref{fig:figcomparison3}, \textit{y}-band has the best mean sampling; yet,
\textit{y}-band light curve distortion  is the greatest due to AD transfer function asymmetry, making
determining the time lag problematic (not impossible but with the largest error). Thus, for simplicity we do not consider the \textit{y}-band lags.}}

{
We underline that our metric  {pointed} to those strategies that would be appropriate for AD time lag measurements of the order of 5 light days, which is a severe {limitation}.
We investigated alternative cuts for 5 $\sigma$ magnitude depth for brighter $(r<21, g<22)$ and fainter $(r>21,g>22)$ sources.
Adding a cut for 5 $\sigma$ depth in the \textit{r} band $<21.0$ implies that the (vast majority) of visits with $r > 21.0$ will not be counted, but objects as bright as 21 in the \texttt{r} band will still be detectable; on the other hand, using a cut of $r<21$ implies that only visits with a very limited depth will be available (Jones, private communication).
However, no significant adjustments have been made in the overall analysis, as strategies with mean 5 $\sigma$ depth $r\sim 20-21, g\sim 20-21$ have been chosen.}

{Next, we obtain the differential number distribution of quasars with possible  time lag measurements with respect to the \textit{u} band by
integrating the time lag measurement distribution  over the source population:}

$$\frac{dN}{dz}=\int^{\infty}_{L} dL \frac{d^{2}\mathcal{N}}{d L dV}\Omega\frac{dV}{dz} p_{\tau}$$

{\noindent where $\frac{d^{2}\mathcal{N}}{d L dV}$ is a QLF }, $\Omega$ is the fiducial survey area, and $p_{\tau}$ {is the probability of a time lag measurement as defined in Section \ref{sec:expectation}. 
We generated time lags using our AD model for 200 uniformly distributed SMBH masses $(10^{6}, 10^{10}) M_{\odot}$ over 200 equal bins ($d z\sim 0.035$) of redshifts ($z\sim0-7$)}. 
The total number of quasars for which time lags could be measured is
$$N=\int \frac{dN}{dz} dz ~.$$

\begin{figure}
\includegraphics[scale=0.7]{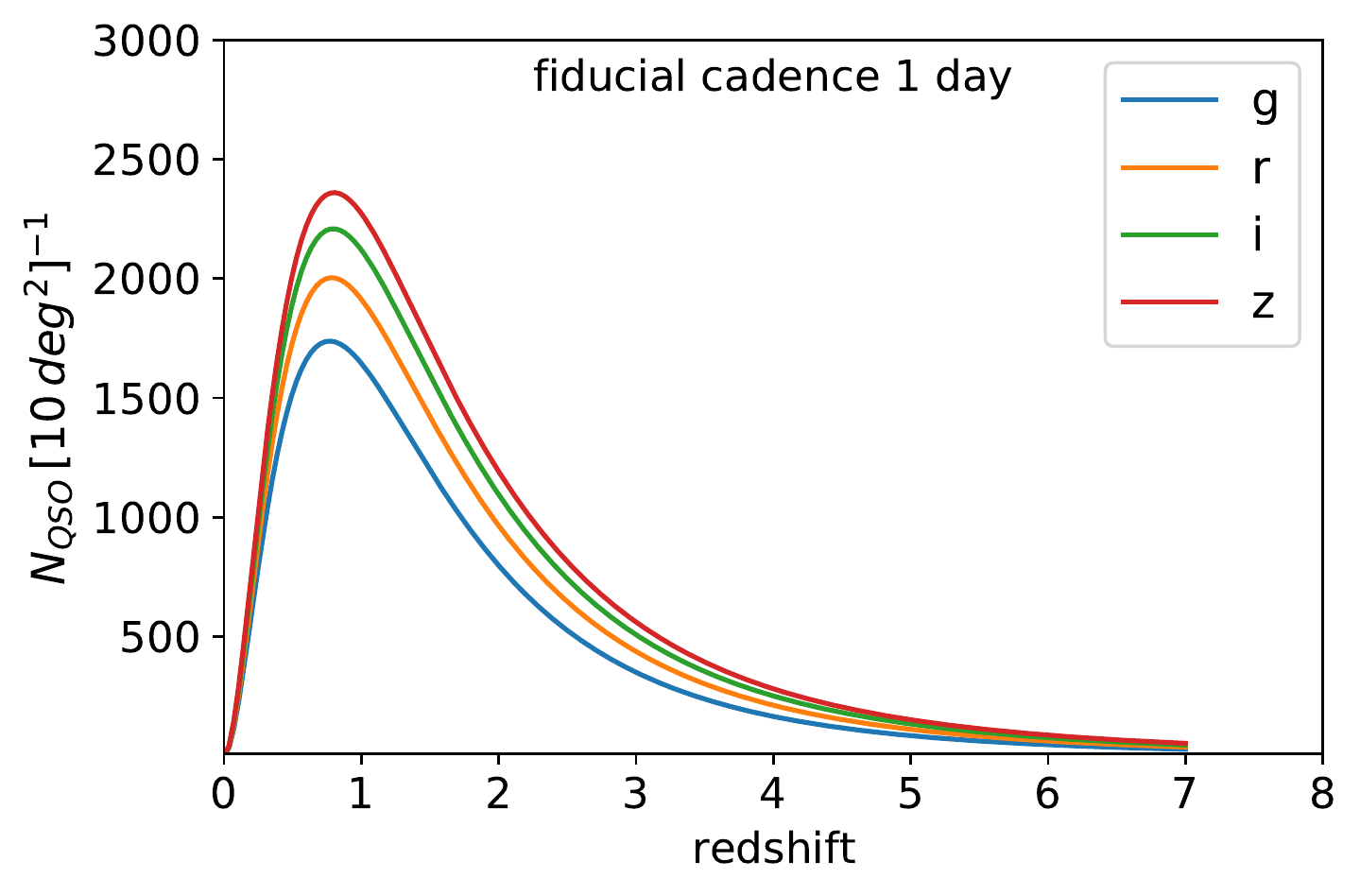}
\caption{The expected number of quasars for which time lag could be measured with respect to \textit{u}-band, in  main LSST filters as a function of the
redshift,  for fiducial $10 \mathrm{deg}^{2}$ survey area and fiducial cadence of 1 day. \label{fig:qsonum}}
\end{figure}

{
In Figure \ref{fig:qsonum}, we plot the expected number of quasars (in  $10 \mathrm{deg}^{2}$ survey area) for which it will be possible to measure time lags in the main LSST bands simultaneously assuming a fiducial cadence $\sim 1$day as a function of redshift. The steepness of these number counts is shallower for \textit{u}-band than for other bands, implying that for time lag measurements using \textit{u}-band, the survey area is far more relevant than the redshift.
Total counts will decrease in the same proportion as cadence increases. As a result, the estimated number of quasars for a 5-day cadence will be 5 times lower than for a 1-day cadence. We emphasize that this is  an upper bound (i.e. optimistic) estimate of expected number of quasars with possible time lag measurements. }

{Also, we approximate the total number 
of quasars whose time lag could be detected in a DDF (in any filter) with confidence $CI$ assuming  an accuracy of the time lag estimate of at least $\Delta=10\%$ for blind search (no data  about mass and redshift) as}:

$$N\sim CI N_{rep} \Delta \frac{\Omega_{LSST}}{\Omega_{rep}}
\frac{t^{cum}_{LSST}}{t^{cum}_{rep}} ~,$$
{\noindent where $N_{rep}\sim 344$ is the number of quasars in VST \texttt{COSMOS} estimated by \citet{2021A&A...645A.103D},   $\Omega_{LSST}\sim 10 \mathrm{deg}^{2}$ is a fiducial element of LSST area, $\Omega_{rep}\sim 1 \mathrm{deg}^{2} $ is an area associated with  $N_{rep}$, and ${t^{cum}_{rep}}\sim 3$yr and $t^{cum}_{LSST}\sim 10$yr are
mean cumulative season lengths for VST \texttt{COSMOS} and LSST, respectively.
Then the  number of quasars for which time lags will be determined at least at $CI\sim 0.95$ and with accuracy rate $\Delta\sim 0.1$ is $N\sim 1089 $ per $10 \deg^{2}$ in any filter.}

{Time lags measured with the actual DDF cadences are} less accurate and precise than those {estimated using} homogeneously sampled light curves (see Figure \ref{fig:oneday}). For instance, we see that the \texttt{COSMOS} and \texttt{ECDFS} PDFs are more skewed than for homogeneously sampled light curves.
Furthermore, the PDF of measured time lags {can be} bimodal, as found for \texttt{r}-band light curves in \texttt{ECDFS}. This is not surprising given that our metric shows that \textit{r} band sampling in \texttt{ECDFS} is of lower quality than in \texttt{COSMOS}, whereas it is the same for other bands (see Figure \ref{fig:figcomparison}). Whilst the secondary peak on the left is much closer to the true value, it is also much less probable.
We showed that using a shifting algorithm with a light curve similarity measure based on area and arc-length might yield temporal lag uncertainties in DDFs of the order of 10\% or less (see Table \ref{tab:idealmeasurement}). 
RM measurement errors are anticipated to be more tolerable in the DDFs, where sampling is expected to be an order of magnitude denser than in the main survey. 
By modelling many different observation tactics for the LSST DDFs, \citet{2020ApJS..246...16Y} discovered that changing the cadence from 3 days to 2-1 days should greatly improve the yield of accretion disk size measurements.
This is in accordance {with} other studies on reverberation mapping. For example, typical time lag errors {of} $\sim 30\%$ {are typically reported by} spectroscopic reverberation {experiments} \citep{2011A&A...535A..73H,2015A&A...576A..73P}. 
{Note as well that} estimated uncertainties of $30\%$ for undersampled data {may be} optimistic \citep{2011A&A...535A..73H}. Enhanced sampling for a spectroscopic reverberation campaign on NGC 4051 revealed that its BLR size is actually a factor of three lower than previously assumed from undersampled data \citep{2009ApJ...702.1353D}. Thus, with well-sampled LSST reverberation data, the {time lag estimations can be improved and tensions with theoretical estimates may be resolved}.

It has been reported that curve-shifting algorithms sensitive to sharp features and operating on the same time scales as the transfer function underestimate multi-band time delays by up to 20\% \citep{2020A&A...636A..52C}.
However, using machine learning methods for a similarity measure of light curves that do not require the same length and number of points, such as PCM, could be advantageous for LSST RM.
Our preliminary tests  show that using curve similarity measures has the potential to improve time lag determination so that {a typical errors might be within } $1\%-10\%$ (see Table \ref{tab:idealmeasurement}).

Because the disk variability for type I AGNs {is almost}  unobscured, measuring intercontinuum delays for these objects will be easier than for type II (due to longer variability time scales and smaller variability amplitudes)\footnote{If type I Seyferts are host dominated, this issue may be applied to them as well.}. The LSST will also increase the number of type I AGNs, and low-luminosity AGNs with smaller black hole masses, so that shorter continuum delays will become more common.
As the magnitude of optical variability on order of $\sim$days  could be anticorrelated with SMBH mass and AGN luminosity \citep{2013ApJ...779..187K}, perhaps AGNs with smaller black hole masses will be better candidates for lag measurements \citep[see also][]{2020ApJS..246...16Y}.

{An alternative  to intensive high-cadence monitoring of a small number of objects is to follow up a larger sample of objects at a cost of a lower cadence.  While individual objects are unlikely to provide high-fidelity lag-wavelength relationships, a large population of objects can be analyzed as a whole \citep[see e.g.,][]{2017ApJ...836..186J, 2018ApJ...862..123M, 2020ApJS..246...16Y, 2019ApJ...880..126H}.
Both \citet{2019ApJ...880..126H} and \citet{2020ApJS..246...16Y} report disk sizes that  are {consistent} with {the} standard  disk model, whereas \citet{2019ApJ...880..126H} demonstrate that if only the best lag measurements are used, disk sizes are skewed to be higher than when the entire sample is taken into account.
However, the multiband AD light curves are warped due to transfer function skewness. We demonstrate that when light curves are denser, as in the case of DDF, where sampling approaches the width of their transfer functions, the results are better. The \textit{u}-band light curve has the shortest wavelength and the smallest width of the transfer function, implying that its higher cadence is very desirable. However, in order to examine X-ray to UV delays, the sample density of light curves in all bands should be greater than the width of their transfer function \citep[see][]{2021MNRAS.503.4163K}.
}

Accretion disk investigations have been challenged by the fact that observations of their sizes are consistently larger than expectations from the thin-disk model by factors ranging from two to three \citep{2020A&A...636A..52C}.
For an accurate measurement of disk size, it is necessary to take into consideration the appropriate transfer functions
\citep{2020A&A...636A..52C,2016MNRAS.456.1960S}.
DDFs \citep{2018arXiv181106542B, 2018arXiv181200516S} where substantial multiwavelength observations have been collected from earlier surveys \citep{2021A&A...645A.103D} {will be of greatest interest}  for detecting time lags by {using these previous data to constrain the transfer functions}.
There are ongoing efforts of creating prime sample of quasars to  further test the applicability of the DRW
model and the stationarity of the stochastic variability process. \citet{2022arXiv220102762S} investigated the optical continuum variability  of   190 quasars within the SDSS Stripe 82 over  20 years long baseline.
This sample contains hundreds of epochs from  SDSS, PanSTARRS-1, the Dark Energy Survey,  and dedicated follow-up photometric monitoring with DECam on the CTIO-4m Blanco telescope, making it one of the best-quality light curve data sets for studying quasar variability. Monitoring of this sample is a  continuous endeavor to photometrically monitor deep extragalactic fields and collect a large amount of multi-wavelength and time-domain data \citep{2022arXiv220102762S}.
Extension of the baseline by another 5 years will allow  these light curves to be smoothly combined with LSST data, so that this quasar sample will be premier for studying optical continuum variability.

{We also briefly discuss the relation of our analysis to an important methodological study  by \citet{2020AJ....160..265H} on multiband GP, the motivation of which is consistent with this work.
Assuming correlations of  the characteristic time-scale ($\tau$) and an asymptotic
rms variability on long time-scales (SF$_{\infty}$) among bands, they proposed a multivariate DRW model for modelling AGN light curves in five bands and with irregular cadences. 
Similar to our study, \citet{2020AJ....160..265H} firstly generated simulated light curves with measurement errors using the DRW model.
Then \citet{2020AJ....160..265H} applied the univariate and multivariate DRW model on simulated light curves and a SDSS spectroscopically-confirmed quasar, and found that the multivariate model can help reveal the possible correlations of true time-scales across five bands  and  time delays between doubly lensed multiband light curves. Although our analysis does not include the more rigorous multiband GP time lag method, the time lag method in our study is sufficient for demonstrating RM within various OpSim runs in terms of optimizing observing cadence and strategy.
}

\section{Final remarks}\label{sec:final}

In this paper, we introduced a metric incorporated in \texttt{MAF}   that {quantifies the accuracy} of LSST  observational strategies for AGN continuum   time lag measurements. 
Our metric {relies on the} Nyquist sampling criterion, which is advantageous when discerning information on small scales structures via electromagnetic signals.
This enables us to  identify specific LSST sky regions (compiled in an atlas) where observing strategies of high quality (satisfying the Nyquist criterion) could be effective for probing continuum time lags {on} scales {of the} AD and {the} transfer function.

A prominent example that will necessitate such spatial scales  and will be explored in the future is the possibility that most observed continuum lags are driven by variable diffuse BLR emission rather than the irradiation disk \citep[see][]{2022MNRAS.509.2637N}.
We anticipate that light curves obtained from fields indicated by our metric will be useful for resolving what happens on long time-scales and {how it diverges from the } lamppost model \citep{2022arXiv220110565N}.

Unlike the spectroscopic approach, photometric RM  does not {typically} allow the BLR emission line and continuum light curves to be {separated}. Under certain conditions,  separation of these processes is possible so the lag can be measured \citep[see details in][]{2012ApJ...747...62C, 2012ApJ...750L..43C}.  LSST sky regions selected by our metric will be ideal for testing {the}  photometric RM methodology on  LSST data \citep{Jankov21}.

It is also interesting to consider what deep learning algorithms \citep[see e.g.,][]{2020ApJ...903...54T, Iva21} might reveal when  trained on multicolor higher cadence data in optimal LSST sky regions (as defined by our metric). 
{Another  feature of our metric is that it can be simply converted to probability and hence combined with other metrics such as quasar counts \citep{Assef21}.}
 
The following is a brief summary of our work:

\begin{itemize}
    \item We compiled  a comprehensive atlas of estimates of the performance of continuum - continuum time lag measurements  as a result of our metric simulation conducted on \texttt{MAF} . The metric points out  as the best strategies those found in  DDFs. The \texttt{multi\_short} strategy has the best sampling ($\lesssim5 $ days) but the mean $5\sigma$ depth is shallower.
\item To illustrate our finding, we simulate light curves using
an AGN model based on a combination of thin-disk, lamppost and damped random walk for the driving function. We
estimate the time lags using shifting method inspired by \texttt{PyCS} but with Partial Curve Mapping (PCM) method to determine the similarity between curves and  identify a time delay measurement through minimization of PCM. Even with inferior sampling quality, like in the \textit{r}-band of \texttt{ECDFS}, the time lags could be determined with accuracy of $\sim 10\%$.

\item {Using our metric together with the artificially generated sample of quasar masses and their QLF, we predicted {an} upper bound of total quasar counts within $0 < z < 7$  for which their time lags might be measured. We estimate that time lag measurements leading to AD size measurements for more than 1000 sources in each DDF per $10 \mathrm{deg}^{2}$ in any filter will be possible.  {We found that the \texttt{baseline} family is suboptimal  in all DDFs, as our metric is below the threshold in \textit{g} and \textit{z} bands.
Similarly, the \texttt{rolling} family  underperforms in the \textit{r} and \textit{g} bands. In contrast, the \texttt{good\_seeing} family of cadences is mainly moderate for determining time lag in all DDFs.}}

\item Based on preliminary numerical results, we propose that, in addition to methods that take into account the transfer function form in detail, LSST continuum reverberation mapping can apply deep learning techniques of similarity measures to potentially improve the time-lag measurement process.

\end{itemize}

\begin{acknowledgments}

We sincerely thank Rachel Street, Federica Bianco, Gordon T. Richards, and William N. Brandt for their essential efforts in coordinating work on cadence evaluation within the Rubin-LSST Science Collaborations. This work was conducted as joint action of  the Rubin-LSST Active Galactic Nuclei (AGN) and  Transients and Variable Stars  (TVS) Science Collaborations.  The authors  express their gratitude to the Vera C. Rubin LSST AGN and TVS Science Collaborations for fostering cooperation and the interchange of ideas and knowledge during their numerous meetings. The authors would like to thank the Vera C. Rubin Observatory \texttt{MAF}  team, specially Lynne Jones, for their significant support in designing and integrating our metric into \texttt{MAF} . The authors warmly acknowledge the LSST Corporation's efforts to train and educate metric teams through many cadence tuning workshops and hackathons.

ABK, DI, IJ, I{\v C}H, L{\v C}P and  VR acknowledge funding provided by University of Belgrade-Faculty of Mathematics  (the contract 451-03-68/2022-14/200104), through the grants by the Ministry of Education, Science, and Technological Development of the Republic of Serbia.
ABK and L{\v C}P thank the support by  Chinese Academy of Sciences President's International Fellowship Initiative (PIFI) for visiting scientist.
BC and SP acknowledge the financial support from the Polish Funding Agency National Science Centre, project 2017/26/A/ST9/-00756 (MAESTRO 9) and MNiSW grant DIR/WK/2018/12. 
CR acknowledges support from the Fondecyt Iniciacion grant 11190831 and ANID BASAL
project FB210003. 
DI acknowledges the support of the Alexander von Humboldt Foundation.
HL acknowledges a Daphne Jackson Fellowship sponsored by the Science and Technology Facilities Council (STFC), UK.
JKD, {L{\v C}P}, ML, OV, SMM acknowledge funding provided by Astronomical Observatory (the contract 451-03-68/2022-14/ 200002), through the grants by the Ministry of Education, Science, and Technological Development of the Republic of Serbia.
RJA was supported by FONDECYT grant number 1191124 and by ANID BASAL project FB210003.
PSS acknowledges funds by ANID grant FONDECYT Postdoctorado Nº 3200250.
SP acknowledges financial support from the Conselho Nacional de Desenvolvimento Científico e Tecnológico (CNPq) Fellowship (164753/2020-6).
SS acknowledges funding provided by the University of Kragujevac - Faculty of Sciences (the contract 451-03-68/2022-14/200122) through the grants by the Ministry of Education, Science, and Technological Development of the Republic of Serbia. YRL acknowledges financial support from the National Science Foundation of China through grant No. 11922304 and from the Youth Innovation Promotion Association CAS.

\end{acknowledgments}




\bibliography{sample631}{}
\bibliographystyle{aasjournal}

\appendix
\section{Atlas of metric \texttt{AGN\_TimeLagMetric} simulation}

\restartappendixnumbering 
\begin{figure*}[ht]
\gridline{\fig{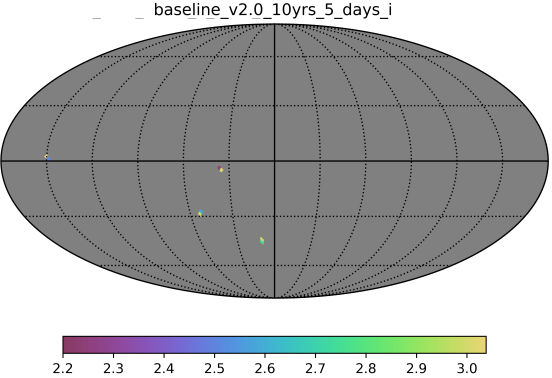}{0.44\textwidth}{(a)}
\fig{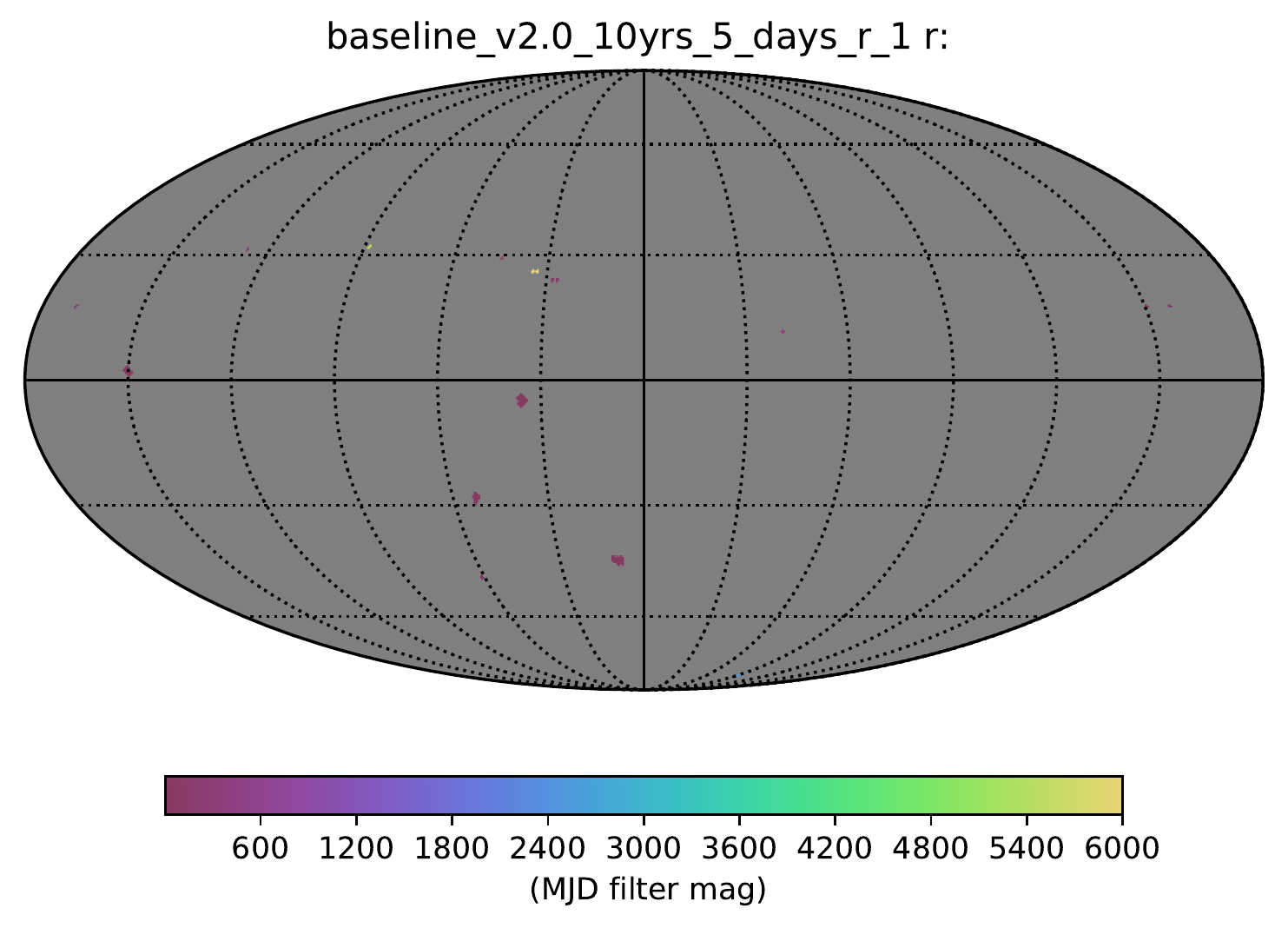}{0.44\textwidth}{(b)}
          }
\gridline{\fig{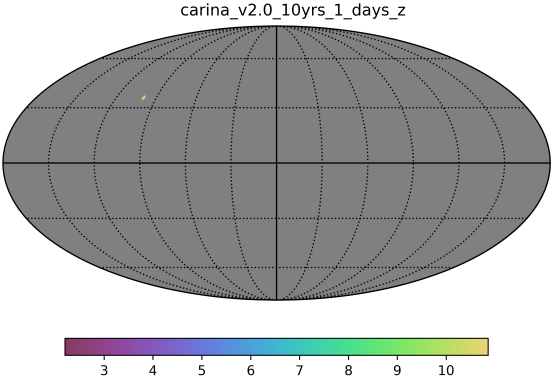}{0.44\textwidth}{(d)}
          \fig{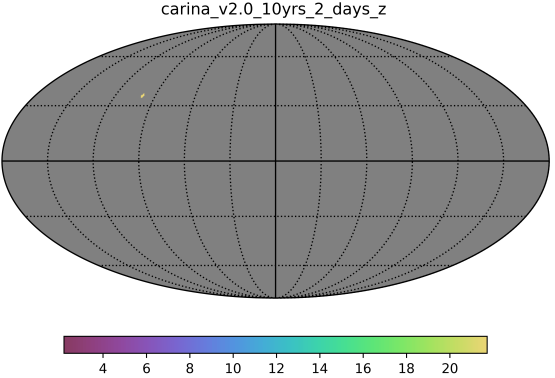}{0.44\textwidth}{(f)}
          }
   \gridline{\fig{ 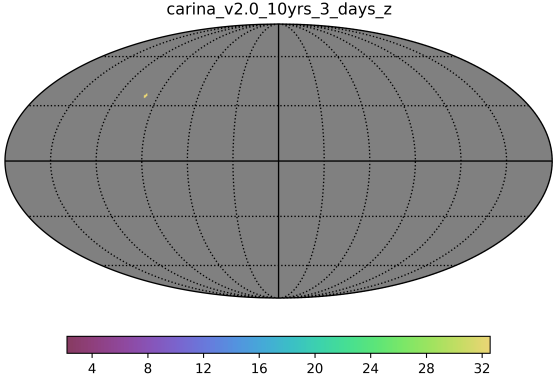}{0.44\textwidth}{(j)}
         \fig{ 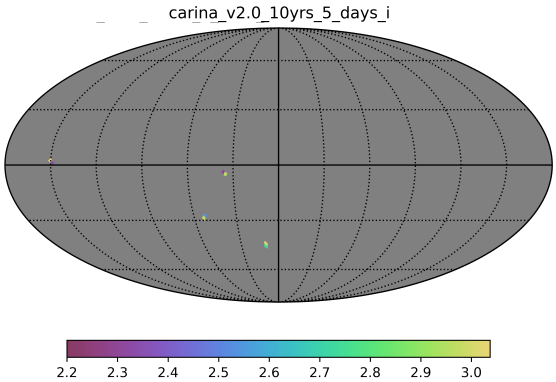}{0.44\textwidth}{(k)}
         }
\caption{Metric realization for cadences from  \texttt{baseline} and \texttt{carina} observing strategies for observed time lags of 1,2,3, and 5 days. The colorbar displays metric values above the Nyquist threshold, which have been scaled for improved visual presentation in the Aitoff projection of celestial coordinates.
\label{fig:fig1}}
\end{figure*}


\begin{figure*}
\gridline{ \fig{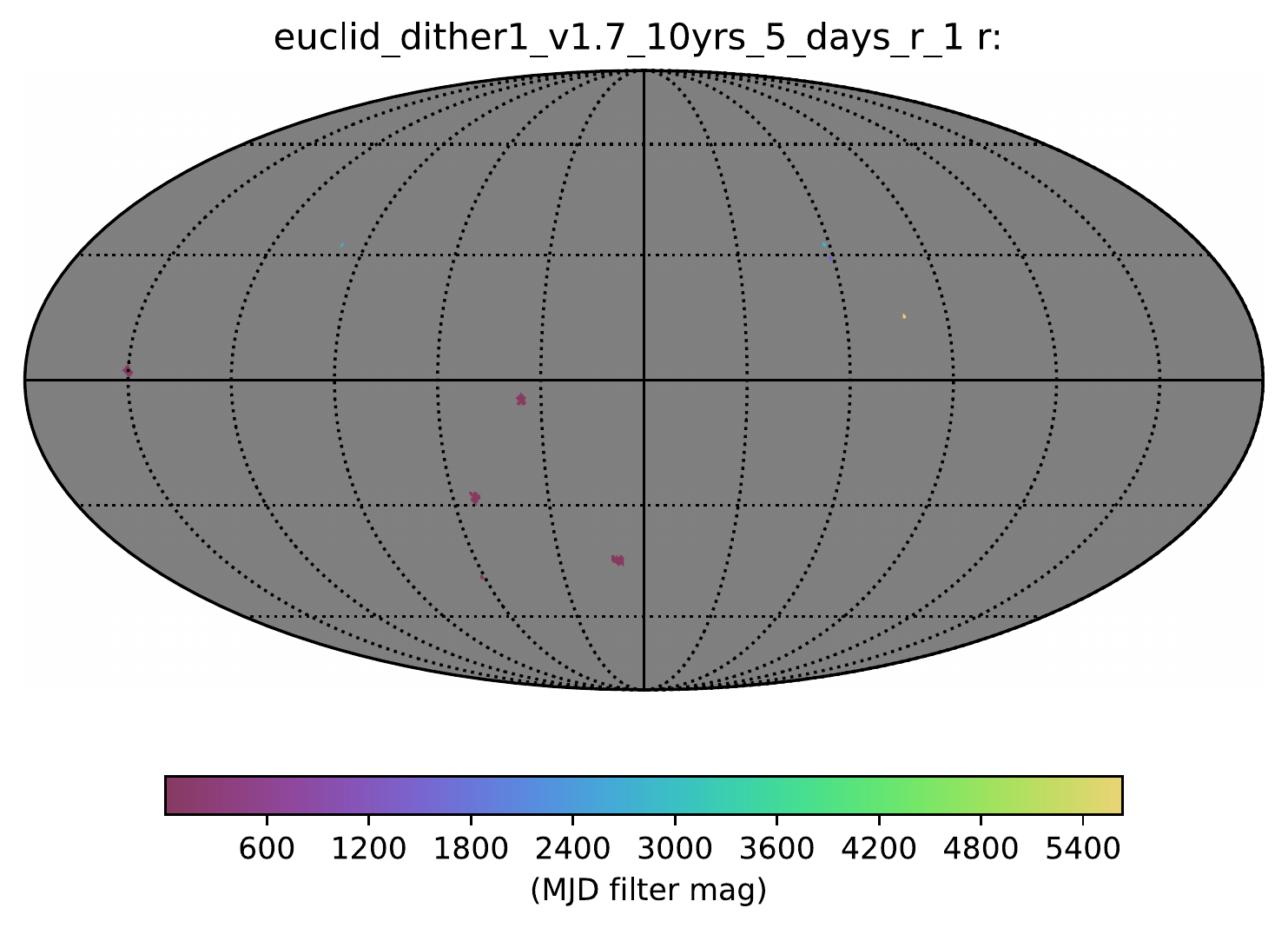}{0.43\textwidth}{(a)}
        \fig{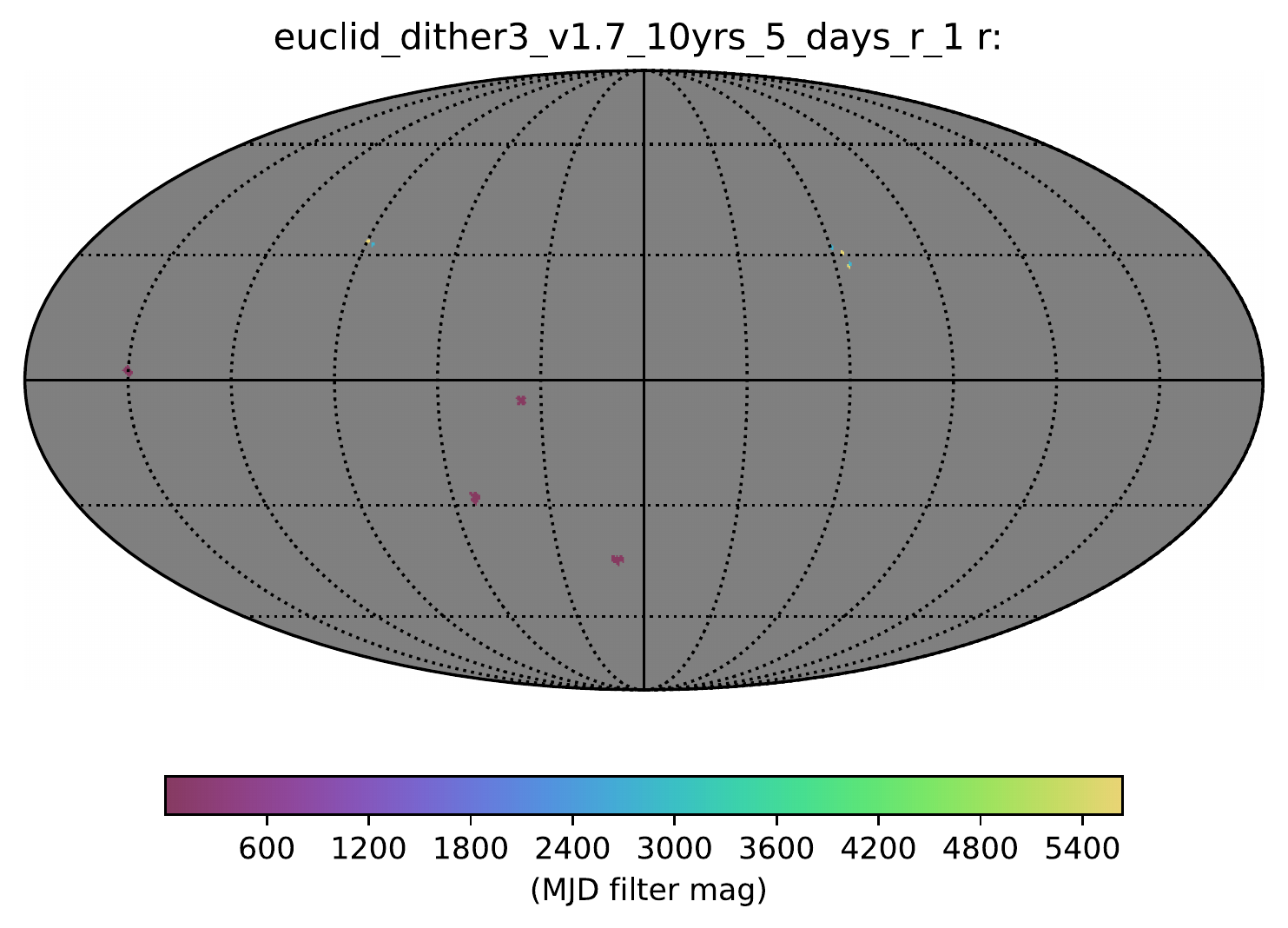}{0.43\textwidth}{(b)}
  }
\caption{The same as Figure \ref{fig:fig1} but for \texttt{euclid\_dither1} and \texttt{euclid\_dither3} observing strategies. In all metric simulations, three DDFs stand out. For better color visibility in the right column, the metric threshold is multiplied by XXX.
\label{fig:fig2}}
\end{figure*}


\begin{figure*}[ht]
\gridline{\fig{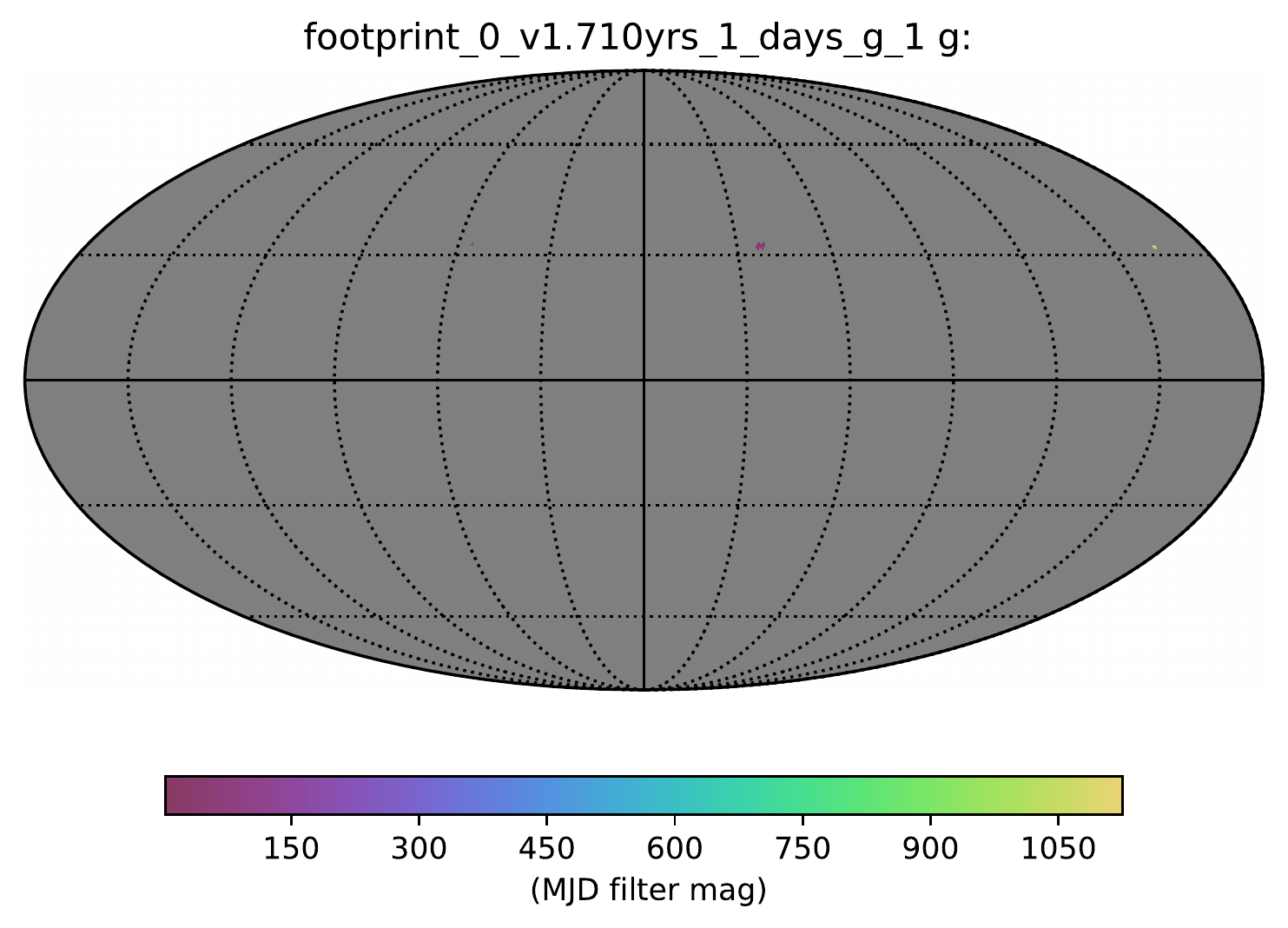}{0.34\textwidth}{(a)}
          \fig{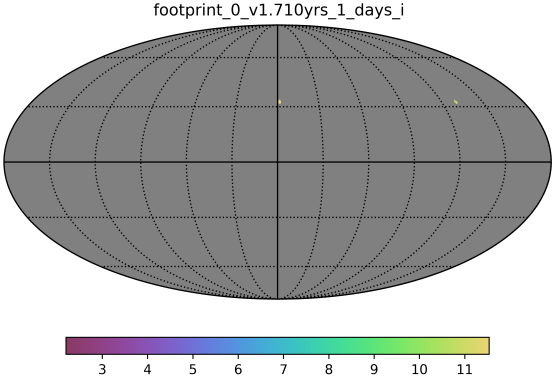}{0.34\textwidth}{(b)}
 \fig{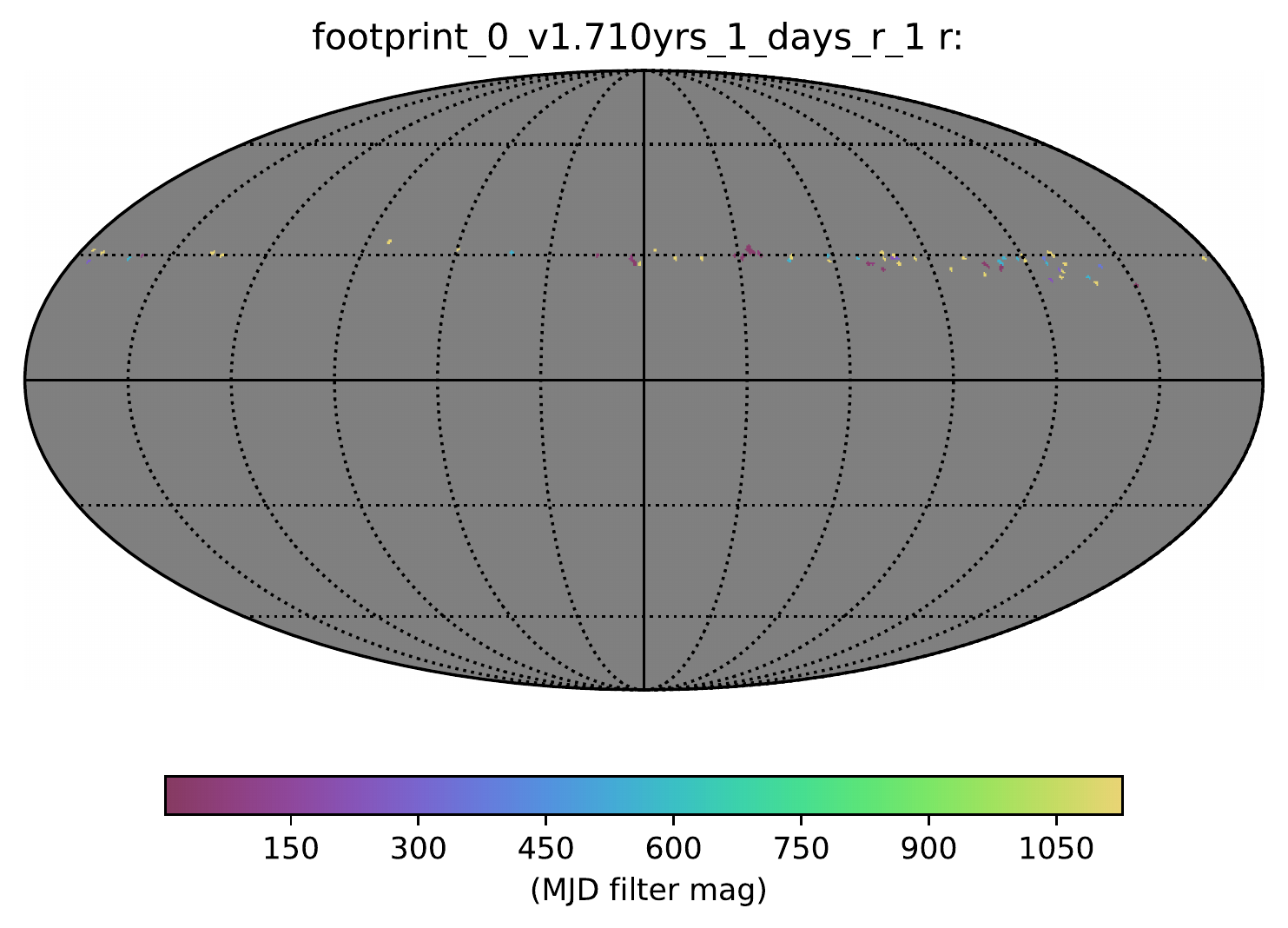}{0.34\textwidth}{(c)}
          }
\gridline{\fig{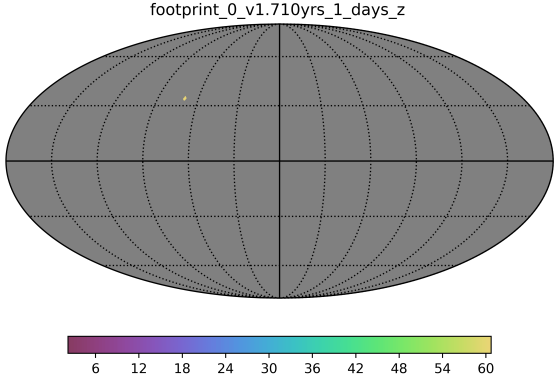}{0.34\textwidth}{(d)}
          \fig{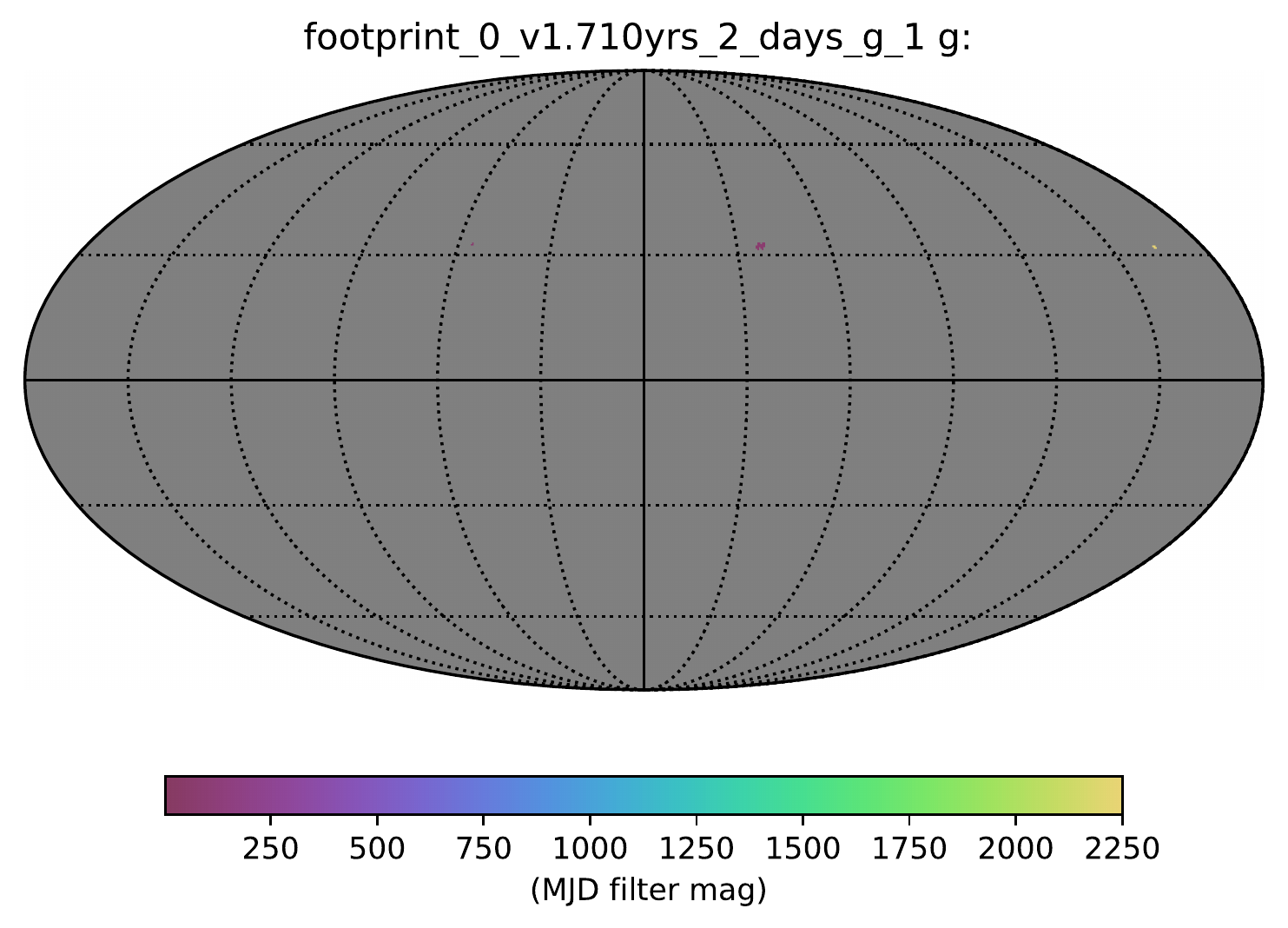}{0.34\textwidth}{(e)}
          \fig{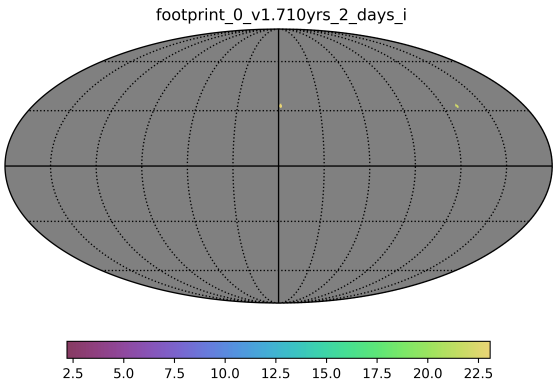}{0.34\textwidth}{(f)}
          }
  \gridline{\fig{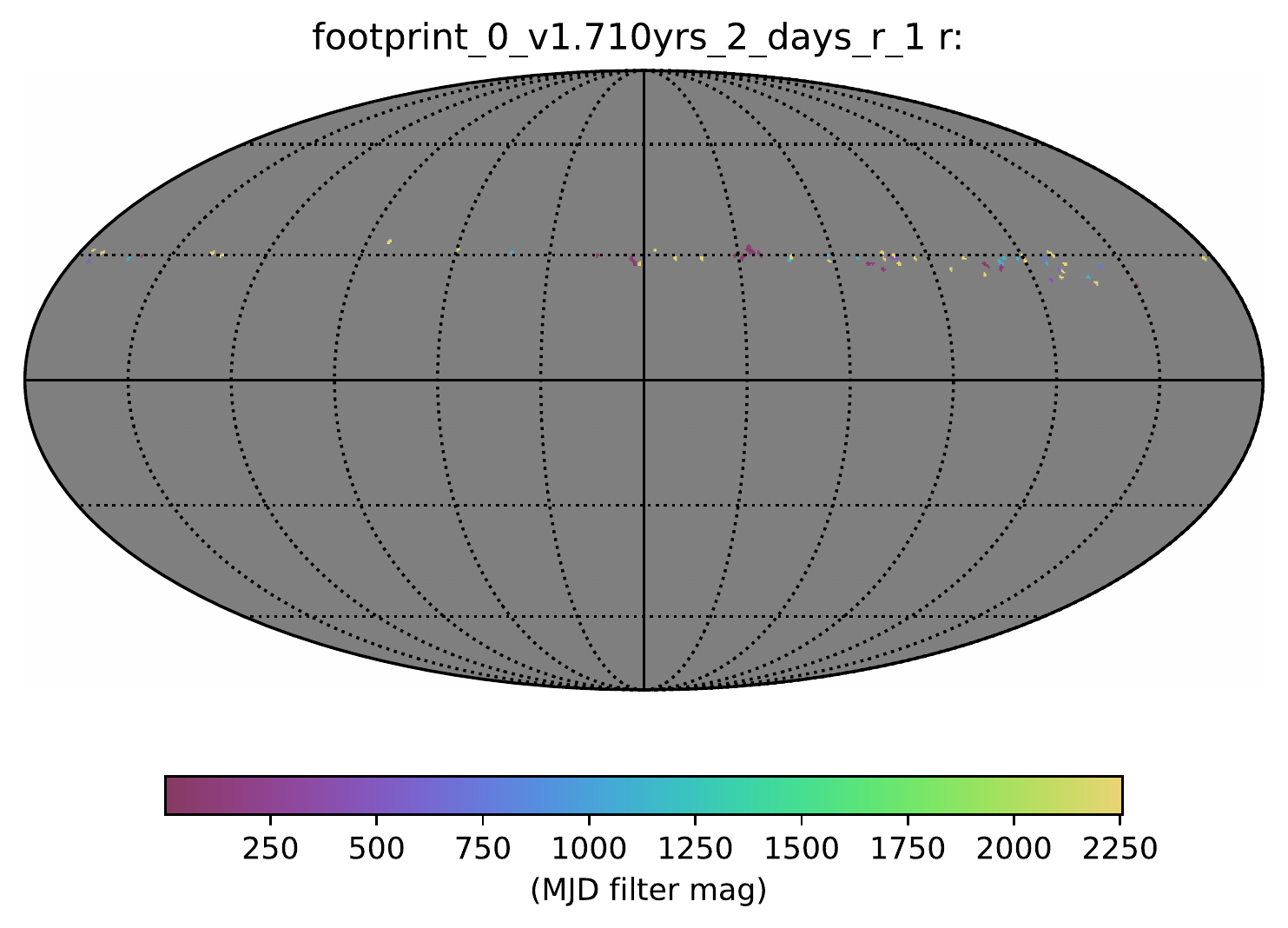}{0.34\textwidth}{(g)} 
   \fig{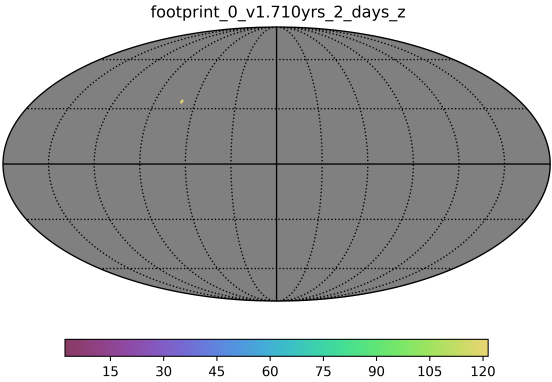}{0.34\textwidth}{(h)}
          \fig{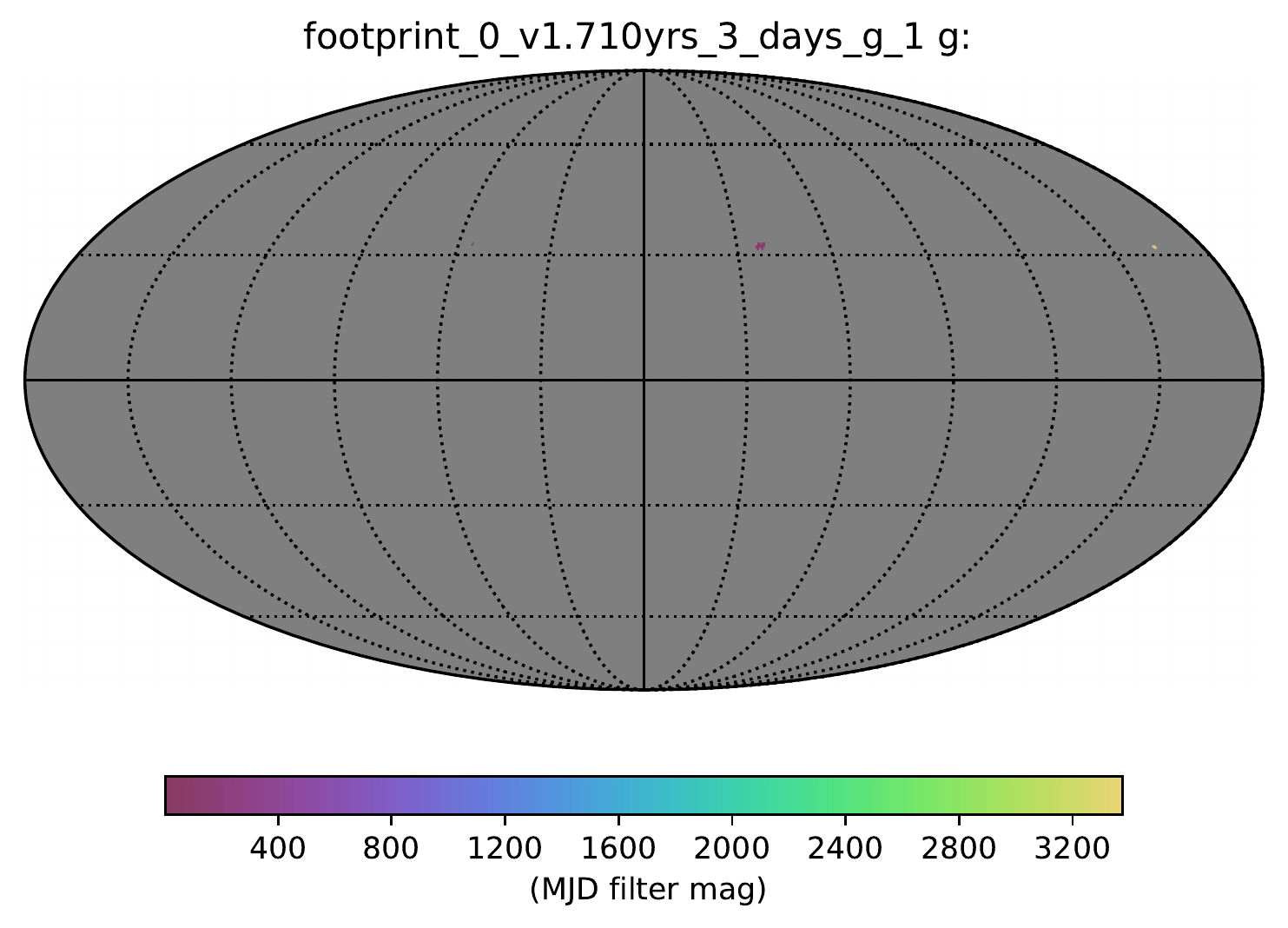}{0.34\textwidth}{(i)}
       }
\caption{Metric realization from \texttt{footprint} observing strategies. There is only one prominent field in each panel (except panel (g)) slightly varying in position on the sky.
\label{fig:fig3}}
\end{figure*}

\begin{figure*}[ht]
\gridline{\fig{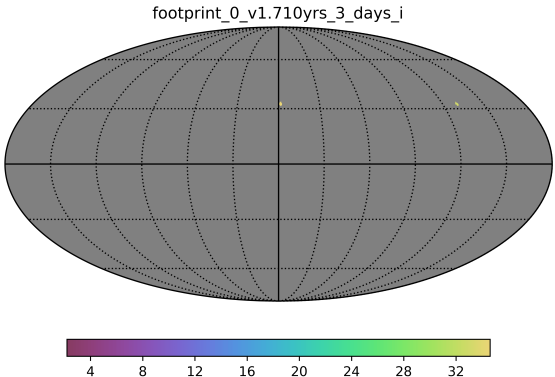}{0.34\textwidth}{(a)}
          \fig{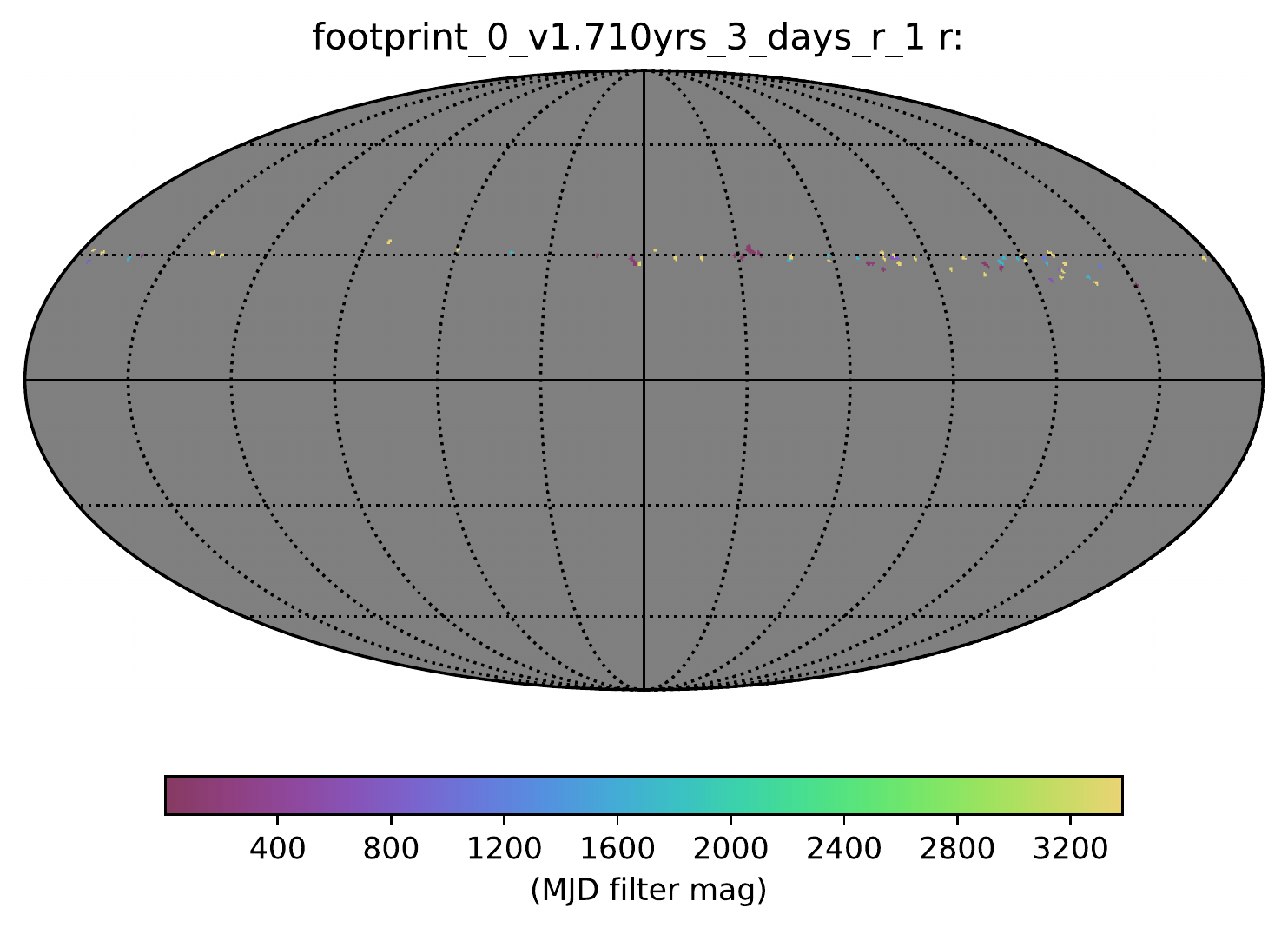}{0.34\textwidth}{(b)}
          \fig{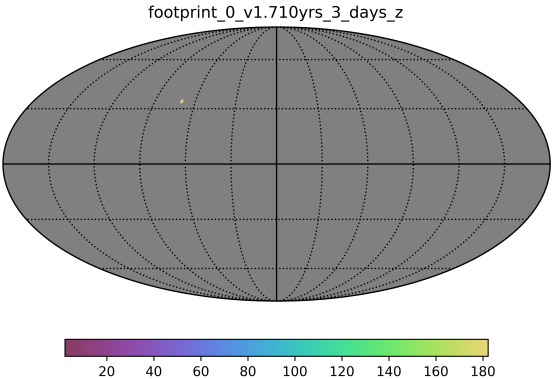}{0.34\textwidth}{(c)}
          }
\gridline{\fig{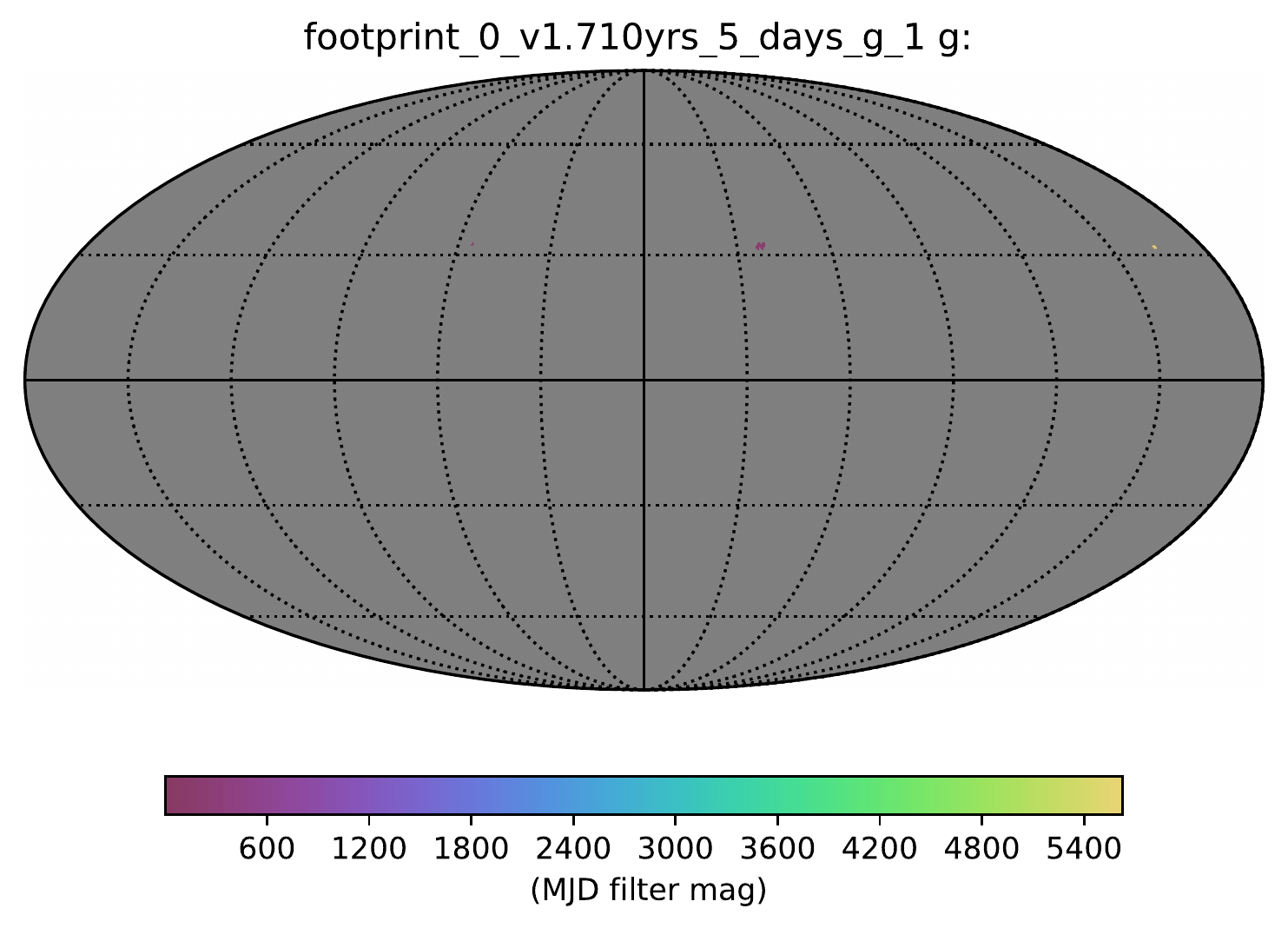}{0.34\textwidth}{(d)}
         \fig{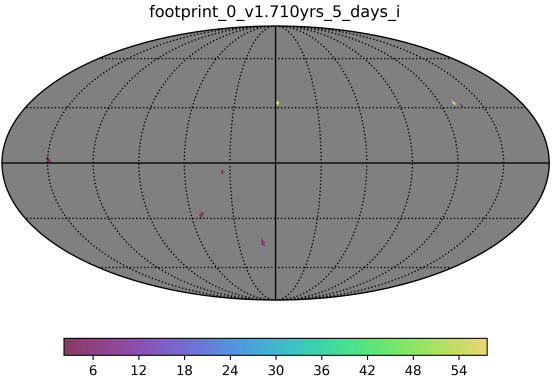}{0.34\textwidth}{(e)}
          \fig{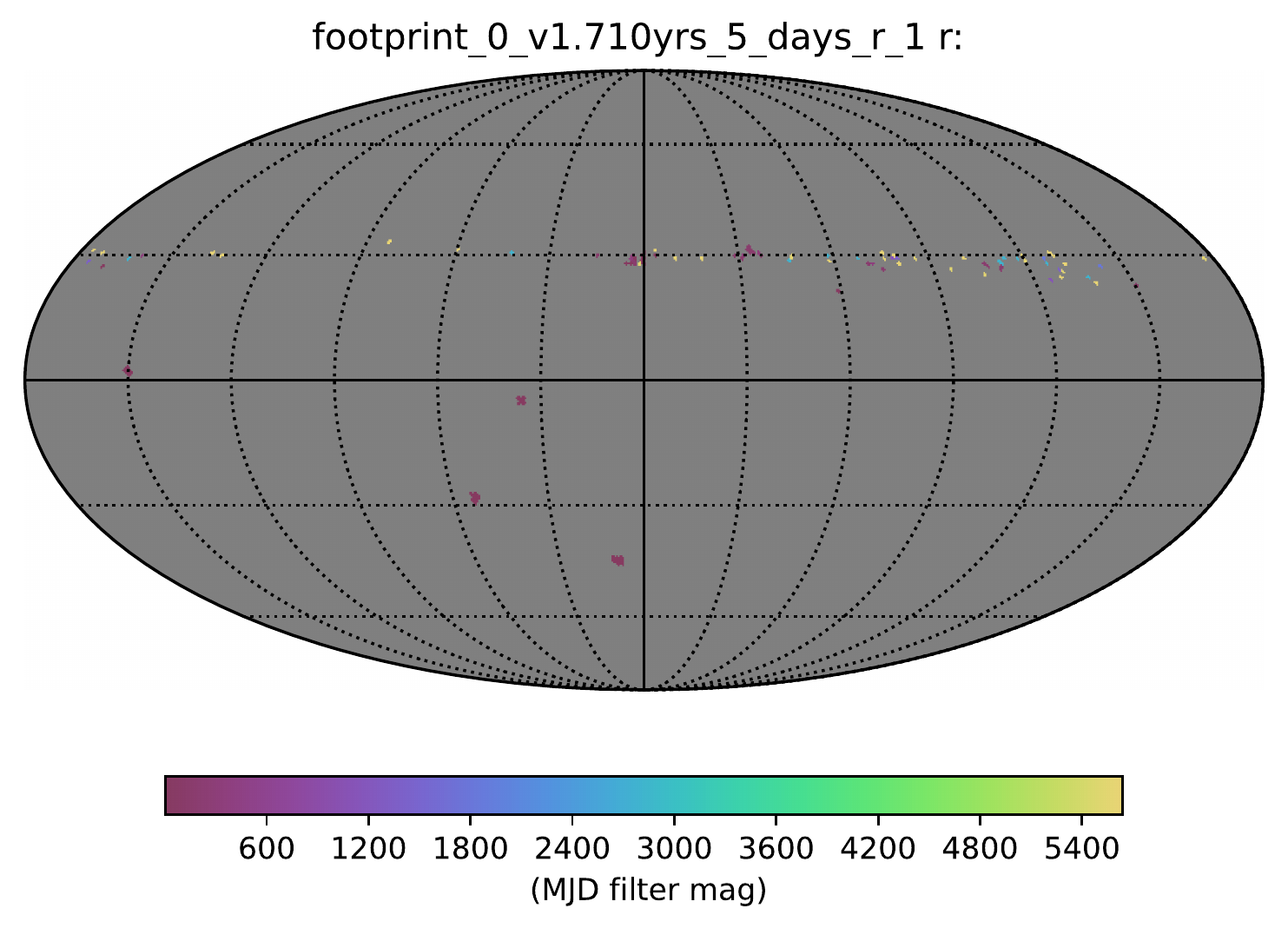}{0.34\textwidth}{(f)}}
          \gridline{
         \fig{  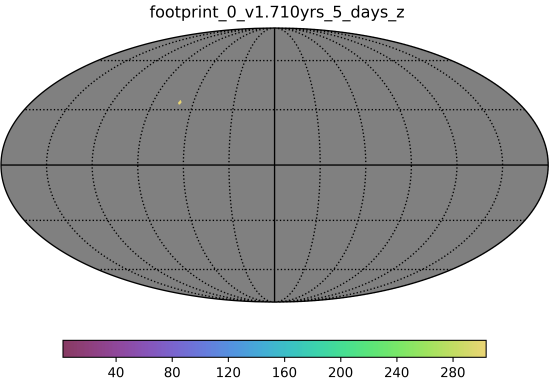}{0.34\textwidth}{(g)}
         }
\caption{Metric realization for cadences from \texttt{footprint} observing strategies, continued. In contrast to Figure \ref{fig:fig3}, DDFs appear in panels (e) and (f).\label{fig:fig4}}
\end{figure*}

\begin{figure*}
\gridline{\fig{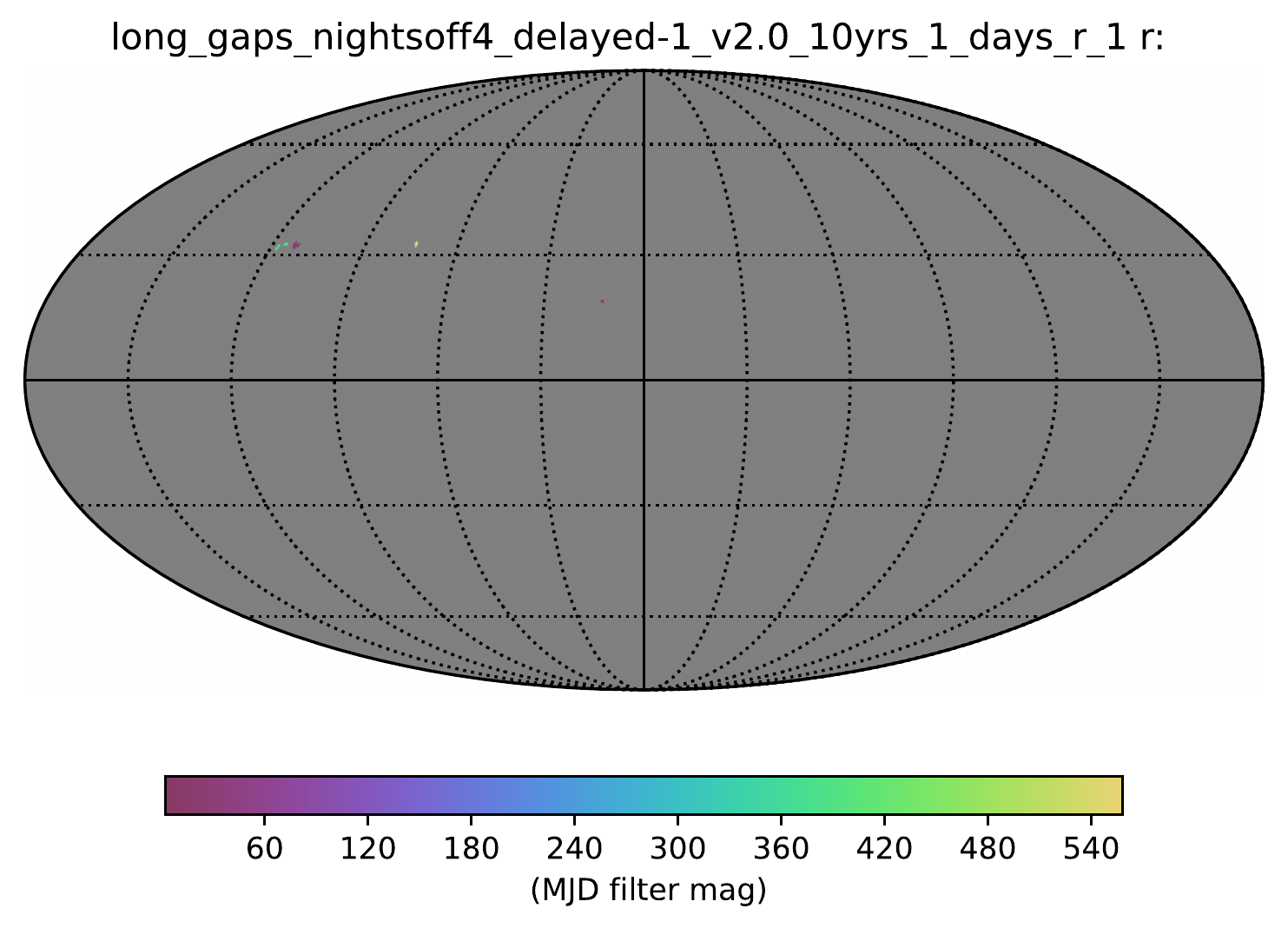}{0.34\textwidth}{(a)}
          \fig{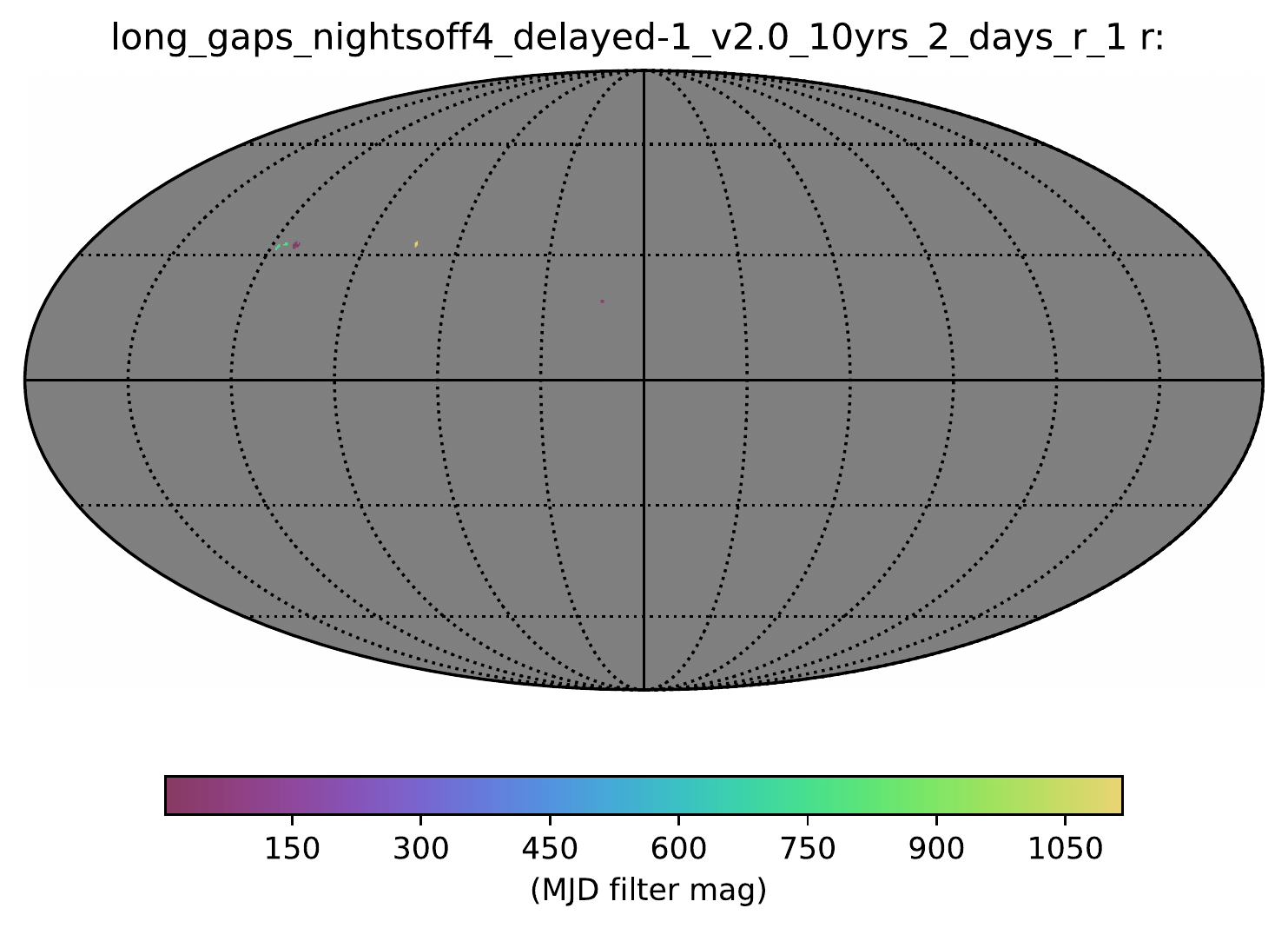}{0.34\textwidth}{(b)}
   \fig{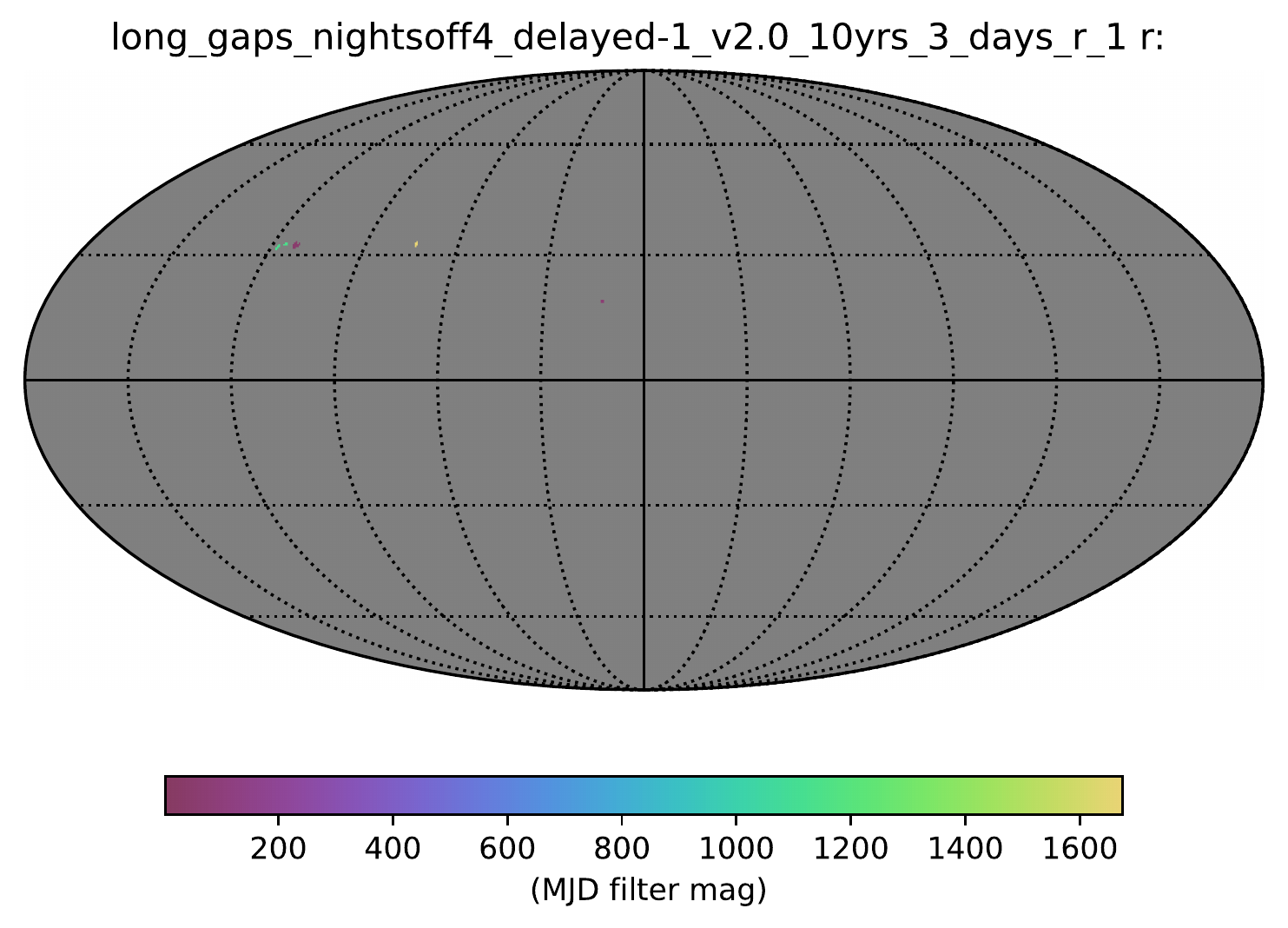}{0.34\textwidth}{(c)}}
     \gridline{    \fig{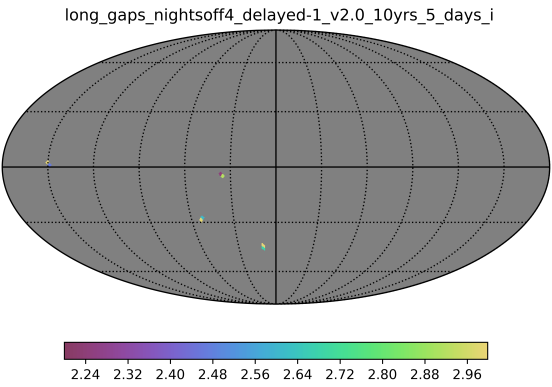}{0.34\textwidth}{(d)}
     \fig{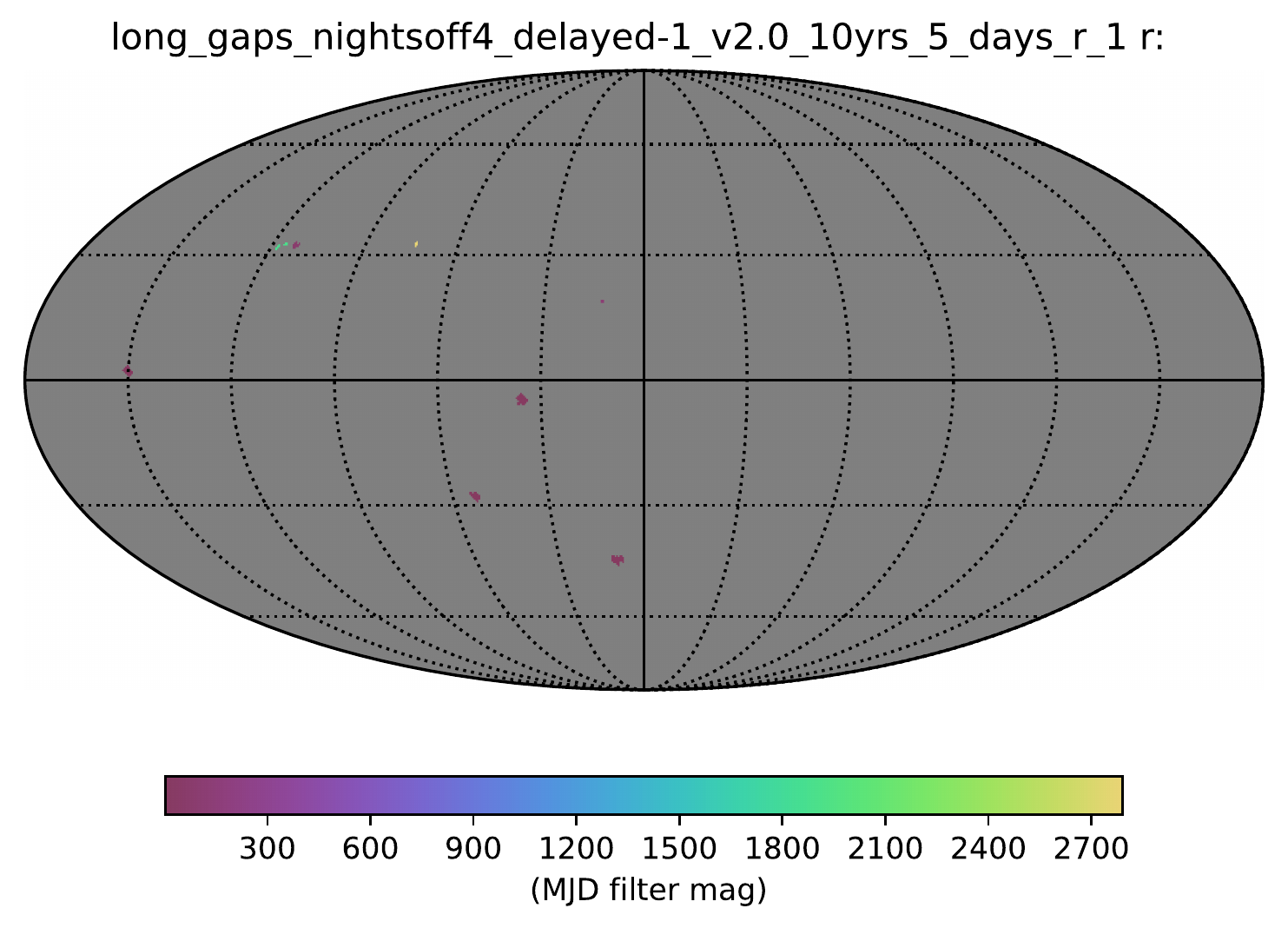}{0.34\textwidth}{(c)}
         }
\caption{Metric realizations on cadences from \texttt{long\_gaps\_nightsoff} observing strategy. The first row of panels shows  the  fields (in the  third and fourth area from the left of the main meridian)  suitable for calculating 1- 3 day delays.  DDFs appear in the bottom row with better metric realization in \textit{i}-band then in \textit{r}-band.
\label{fig:fig5}}
\end{figure*}

\begin{figure*}[ht]
\gridline{\fig{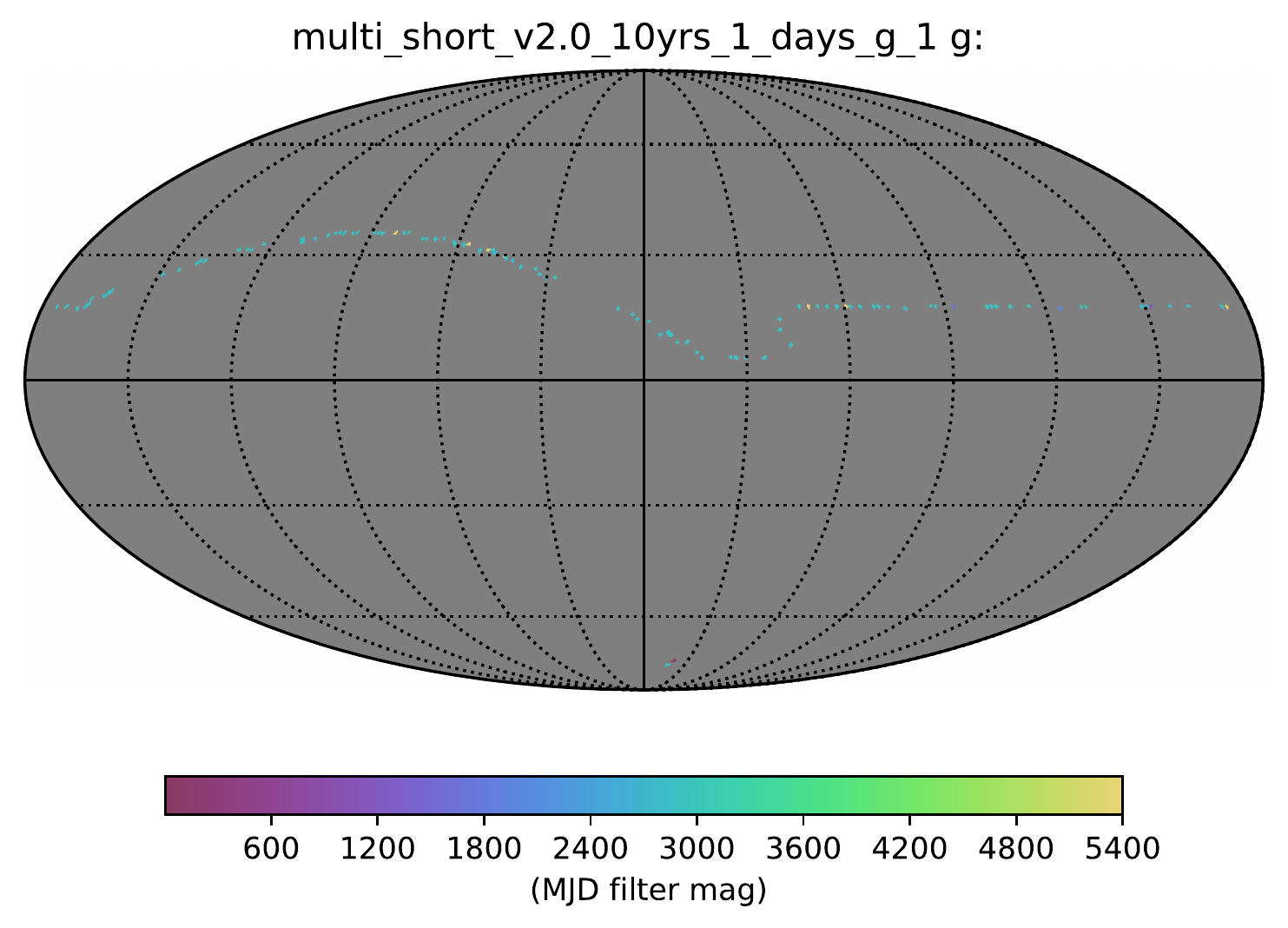}{0.34\textwidth}{(a)}
          \fig{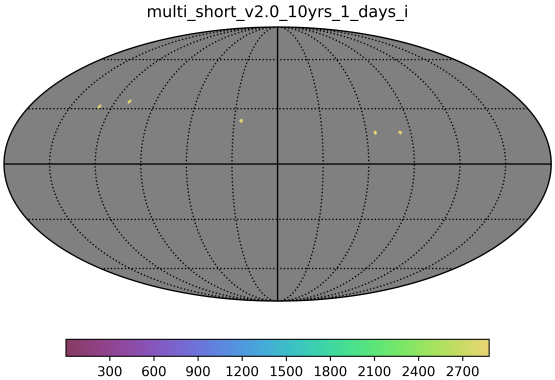}{0.34\textwidth}{(b)}
          \fig{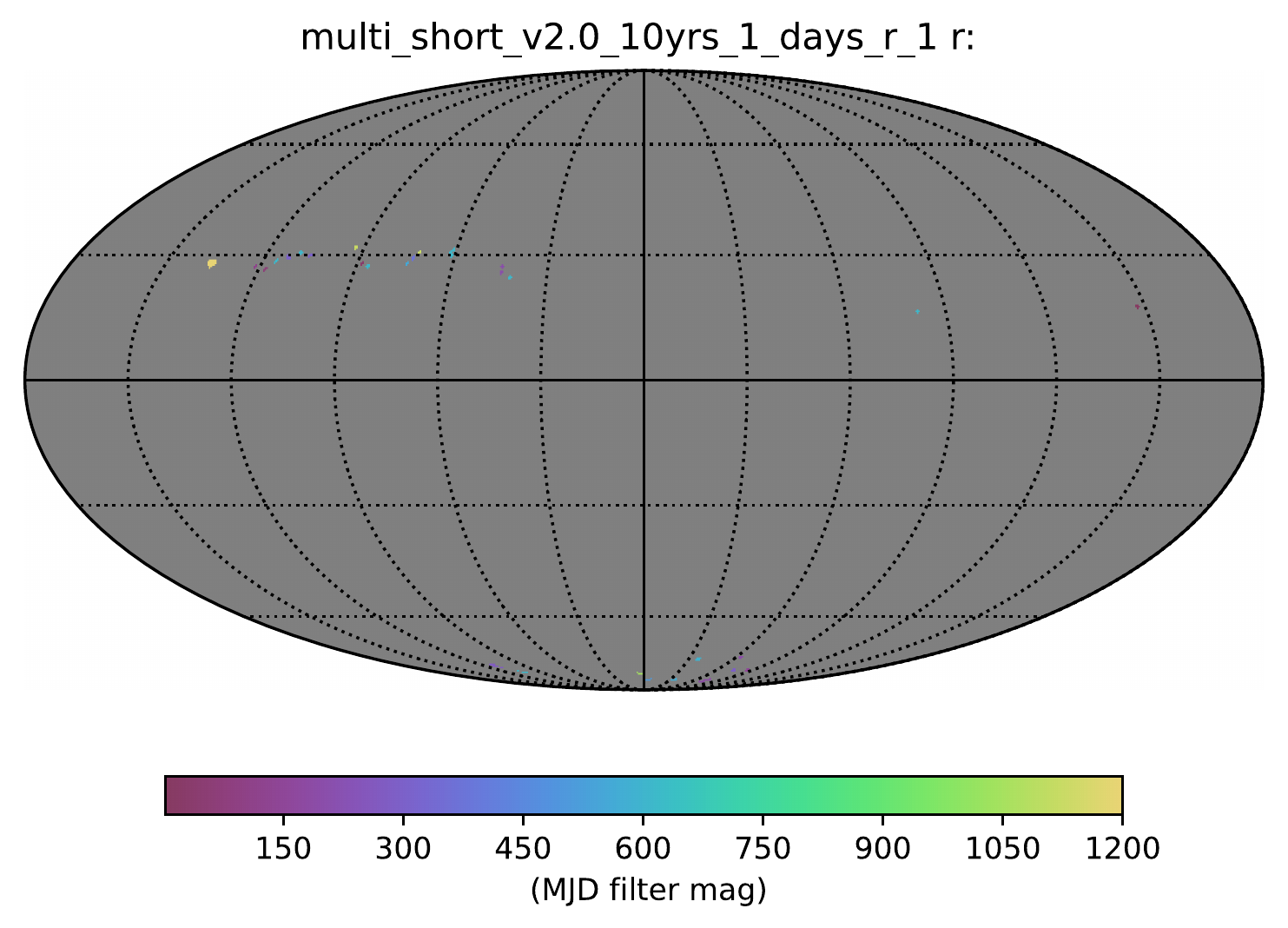}{0.34\textwidth}{(c)}
          }
\gridline{\fig{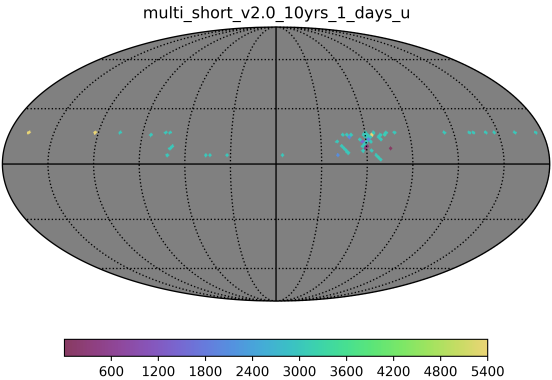}{0.34\textwidth}{(d)}
          \fig{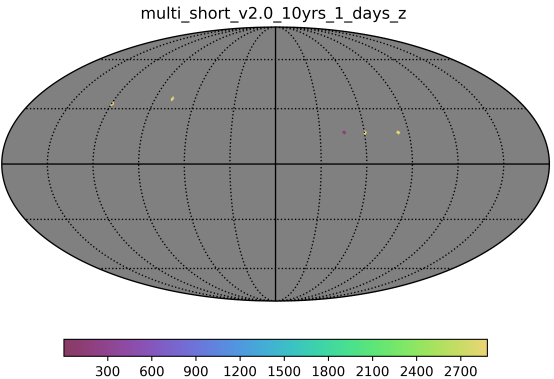}{0.34\textwidth}{(e)}
          \fig{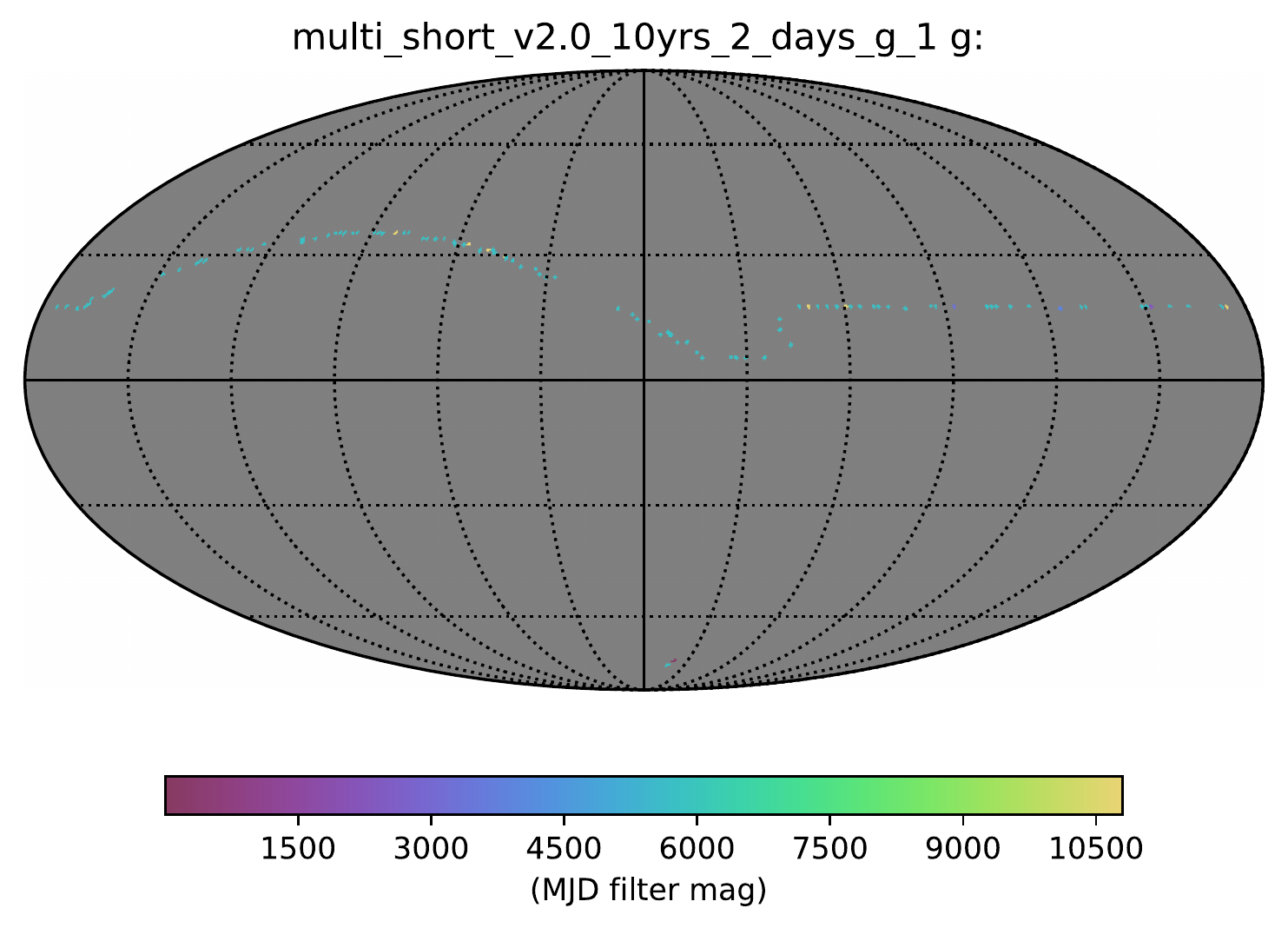}{0.34\textwidth}{(f)}
          }
  \gridline{\fig{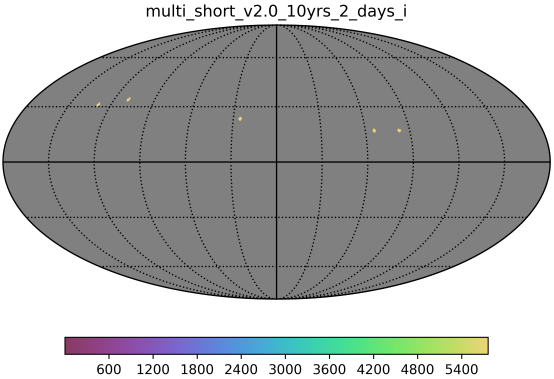}{0.34\textwidth}{(g)}        
          \fig{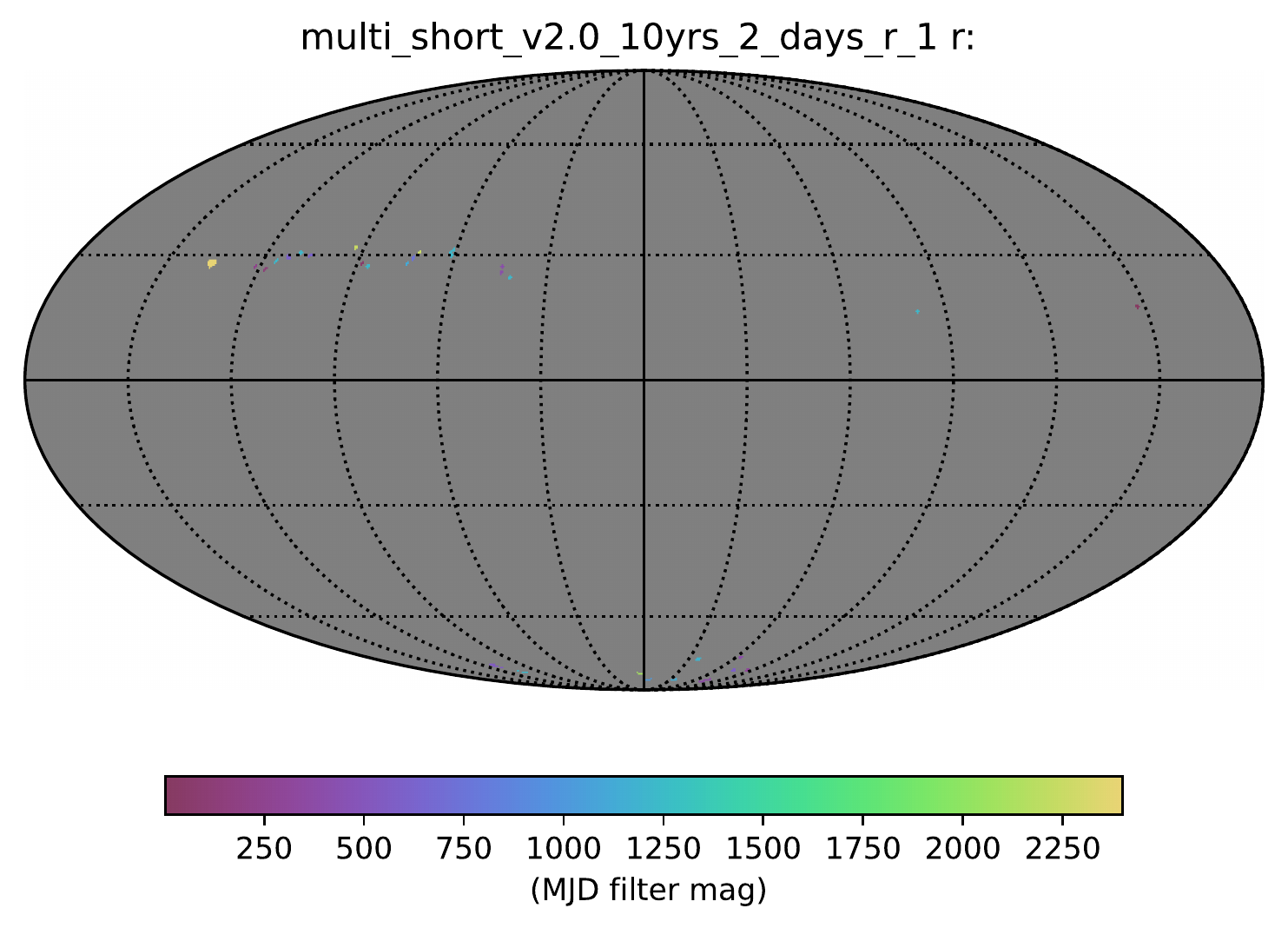}{0.34\textwidth}{(h)}
          \fig{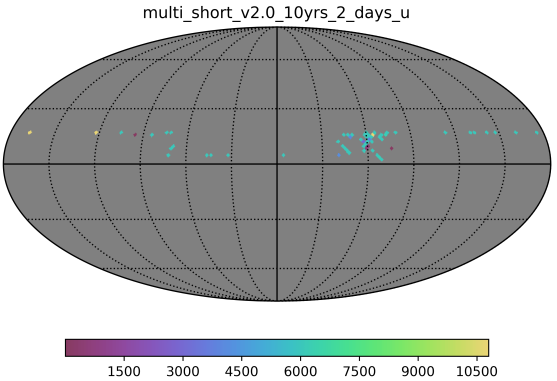}{0.34\textwidth}{(i)}
          }
  
   \gridline{\fig{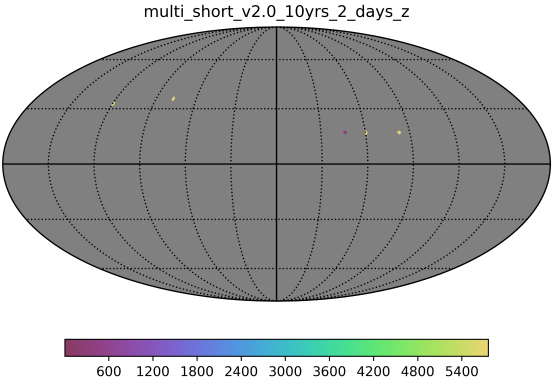}{0.34\textwidth}{(j)}        
          \fig{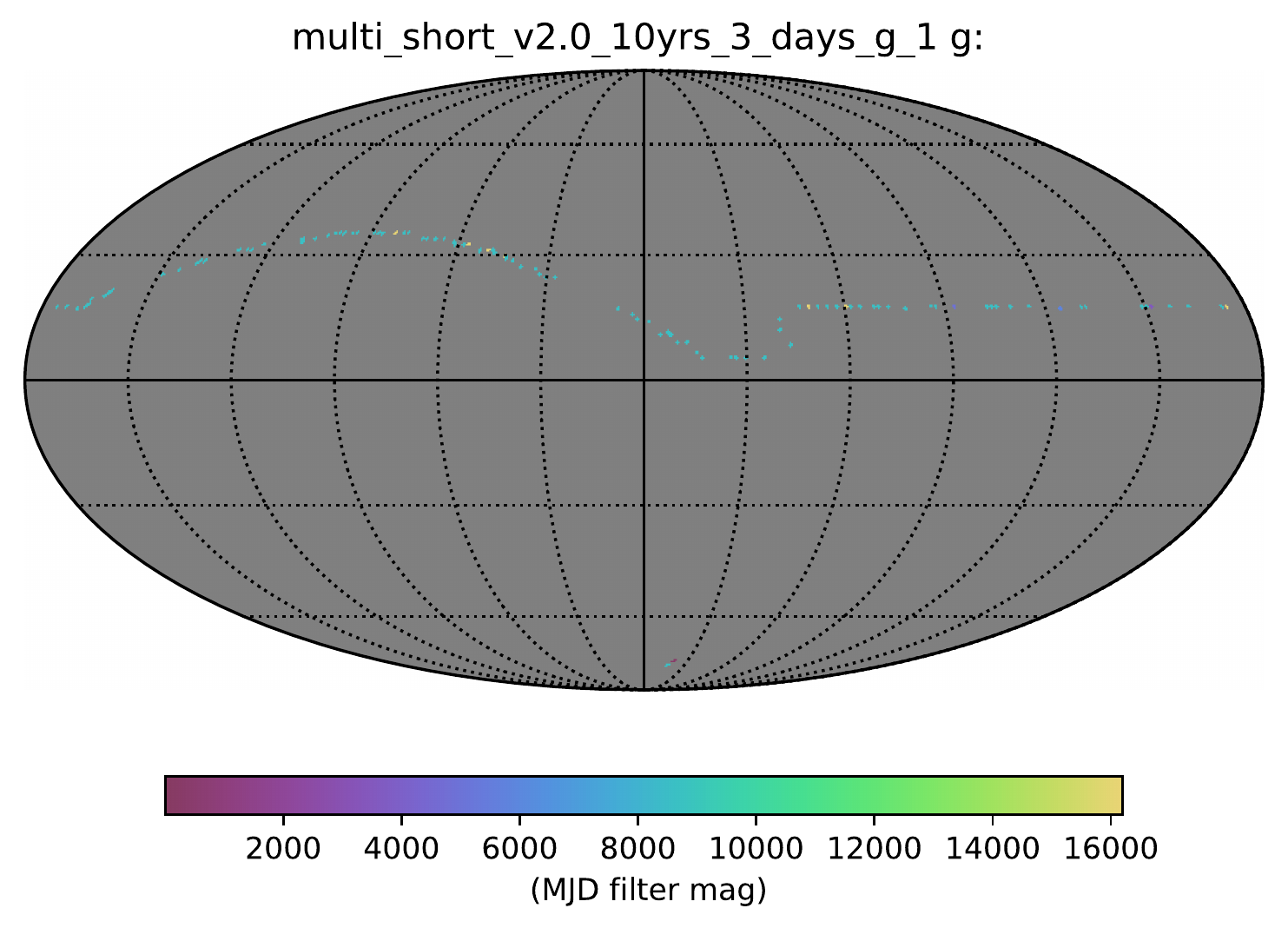}{0.34\textwidth}{(k)}
          \fig{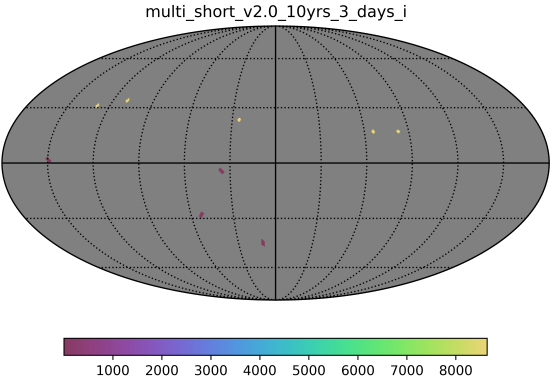}{0.34\textwidth}{(l)}
          }
\caption{Metric realization on \texttt{multishort} cadences. DDFs appear in panel (l).
\label{fig:fig6}}
\end{figure*}

\begin{figure*}[ht]
\gridline{\fig{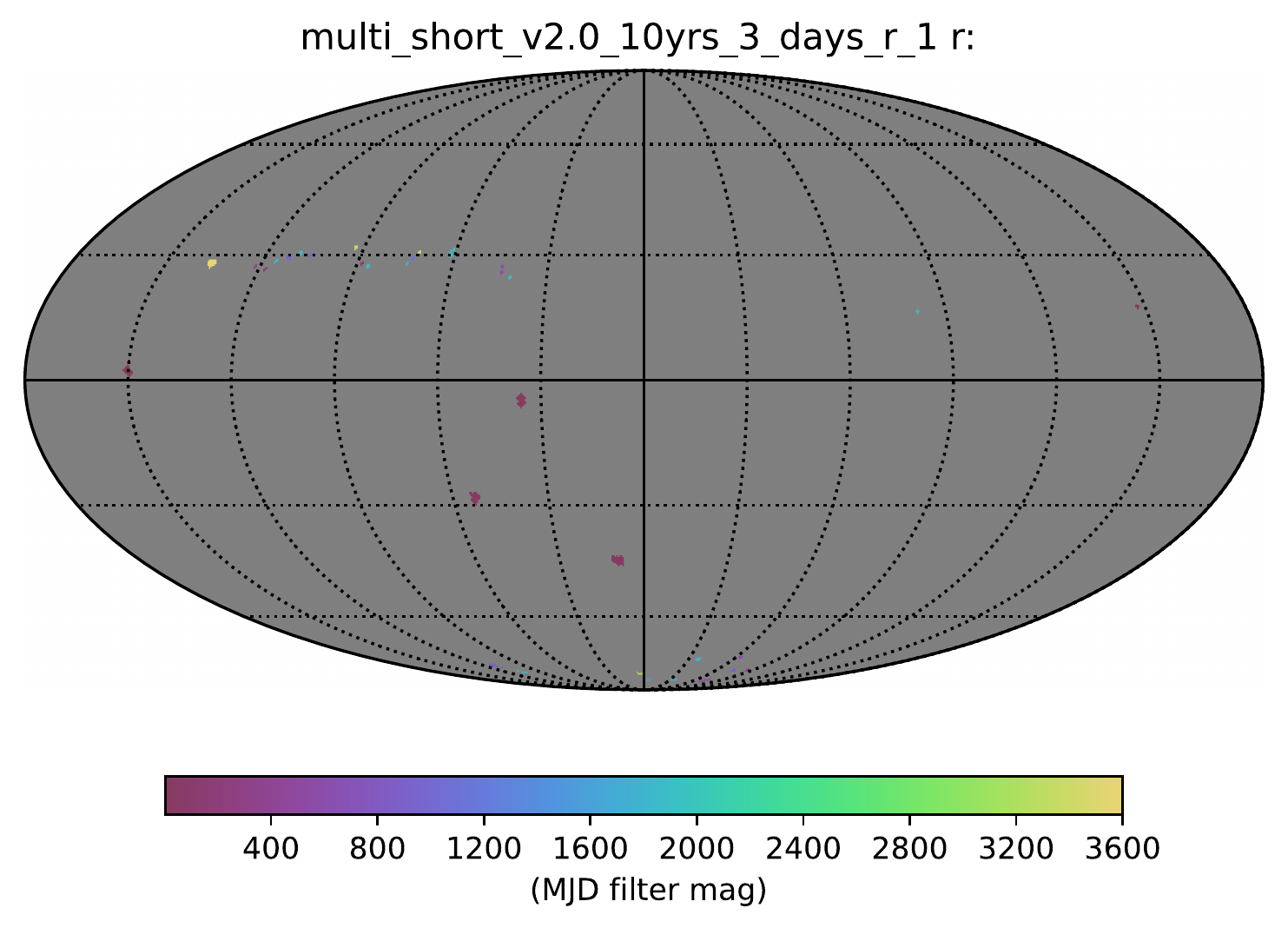}{0.34\textwidth}{(a)}
          \fig{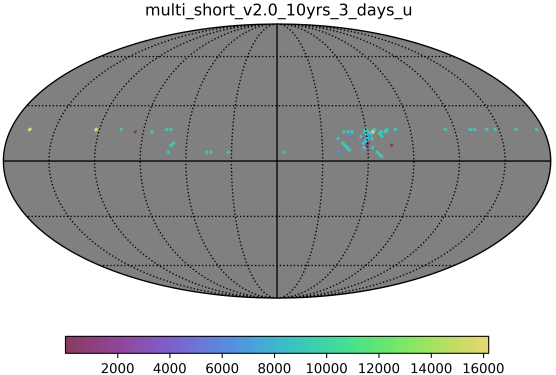}{0.34\textwidth}{(b)}
         \fig{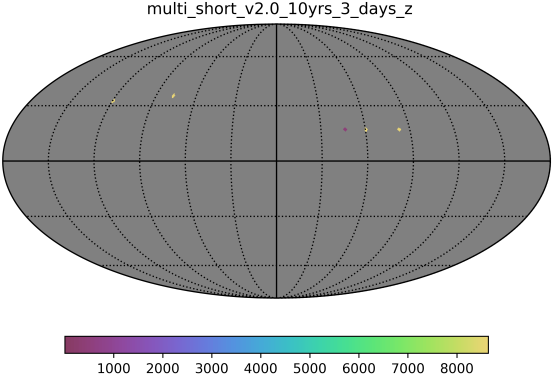}{0.34\textwidth}{(c)}
        }
\gridline{\fig{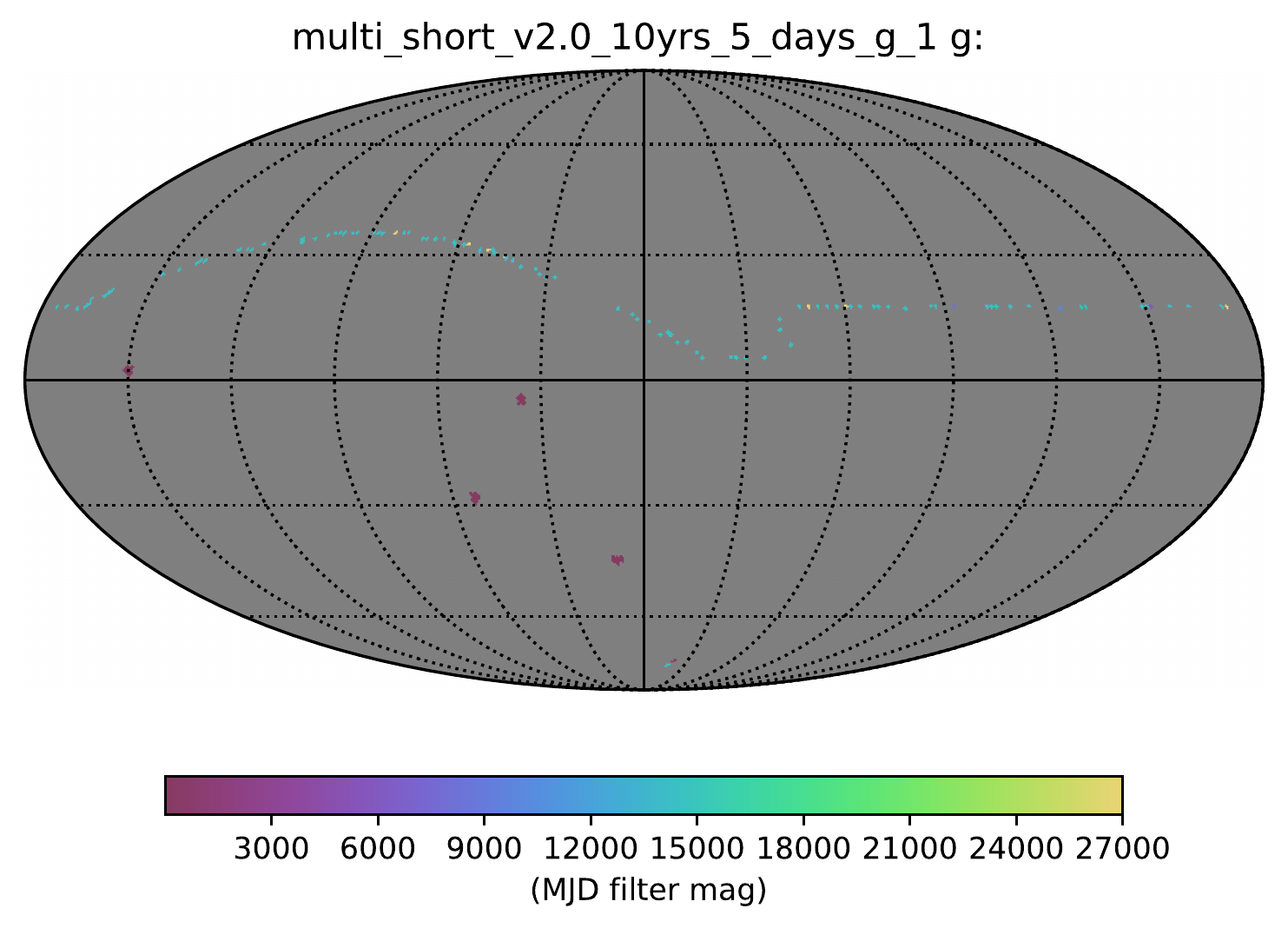}{0.34\textwidth}{(d)}
         \fig{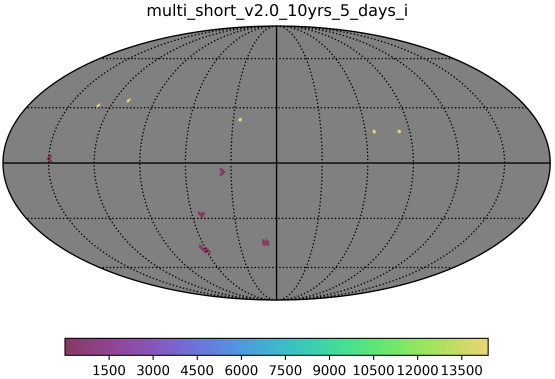}{0.34\textwidth}{(e)}
         \fig{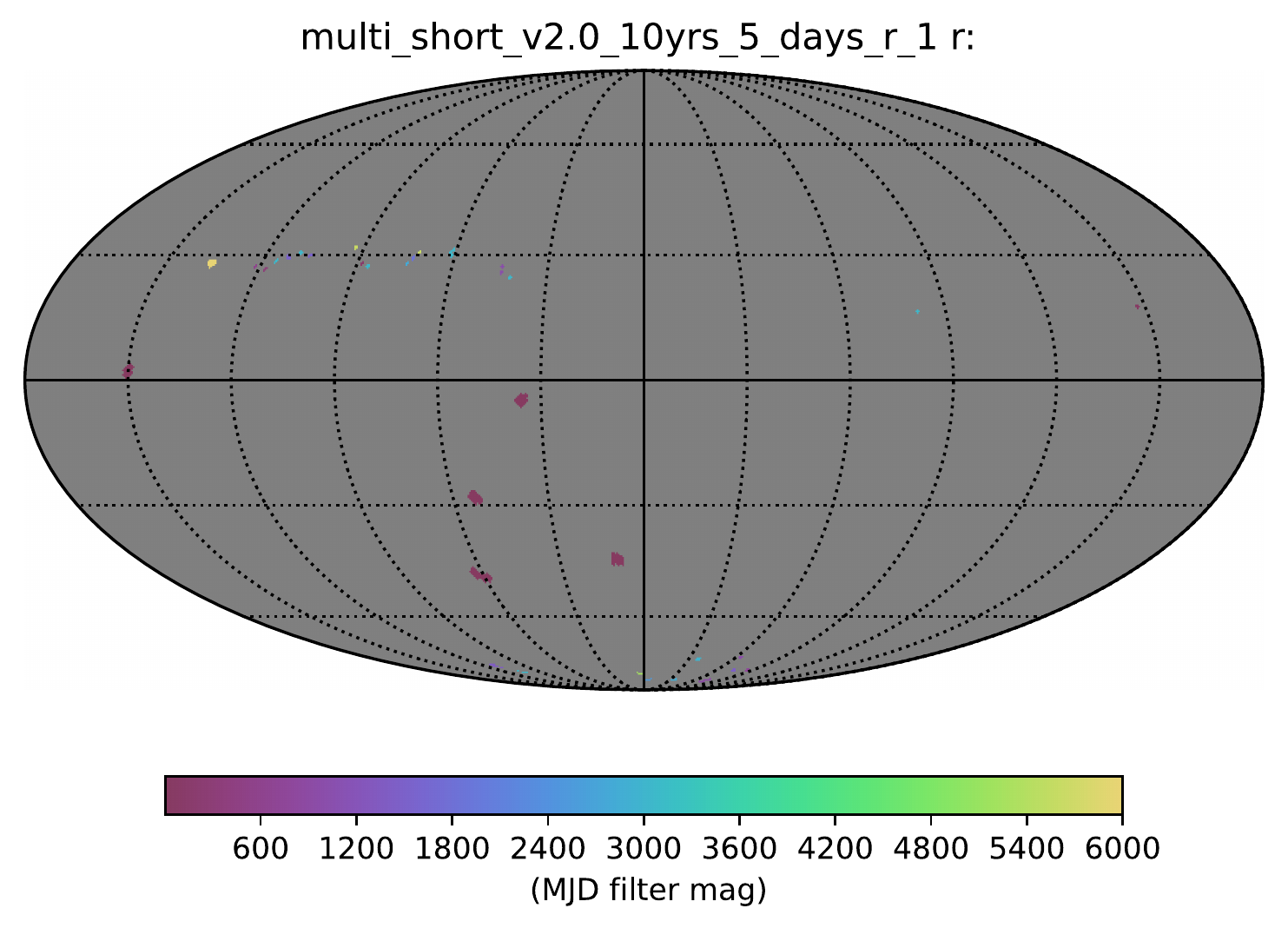}{0.34\textwidth}{(f)}
         }
 \gridline{\fig{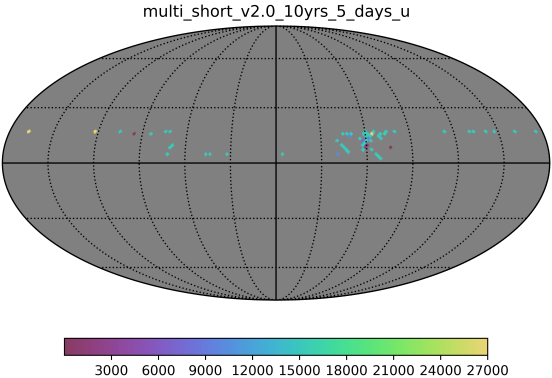}{0.34\textwidth}{(g)}        
          \fig{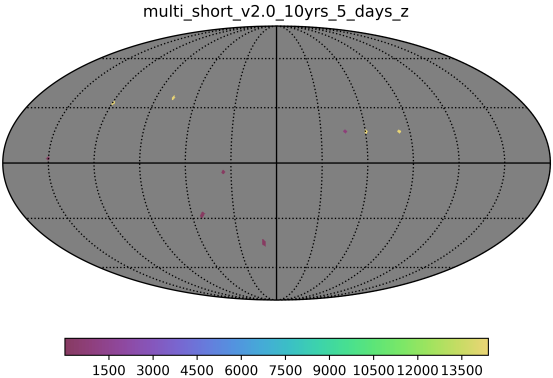}{0.34\textwidth}{(h)}
          }
\caption{Metric realization on \texttt{multishort} cadences continued.
DDFs appear in all panels except (b), (c) and (g).
\label{fig:fig7}}
\end{figure*}

\begin{figure*}[ht]
\gridline{\fig{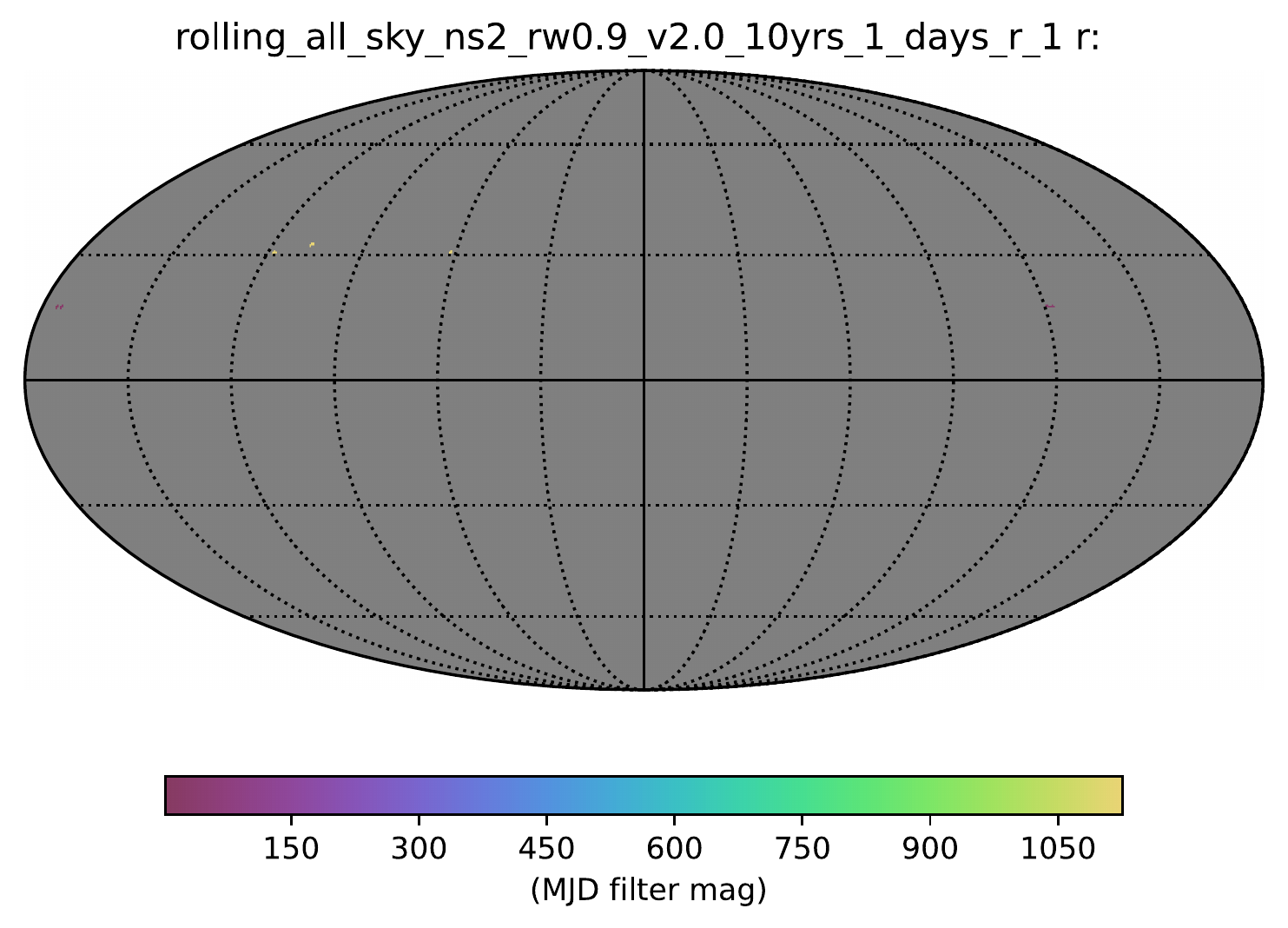}{0.44\textwidth}{(a)}
          \fig{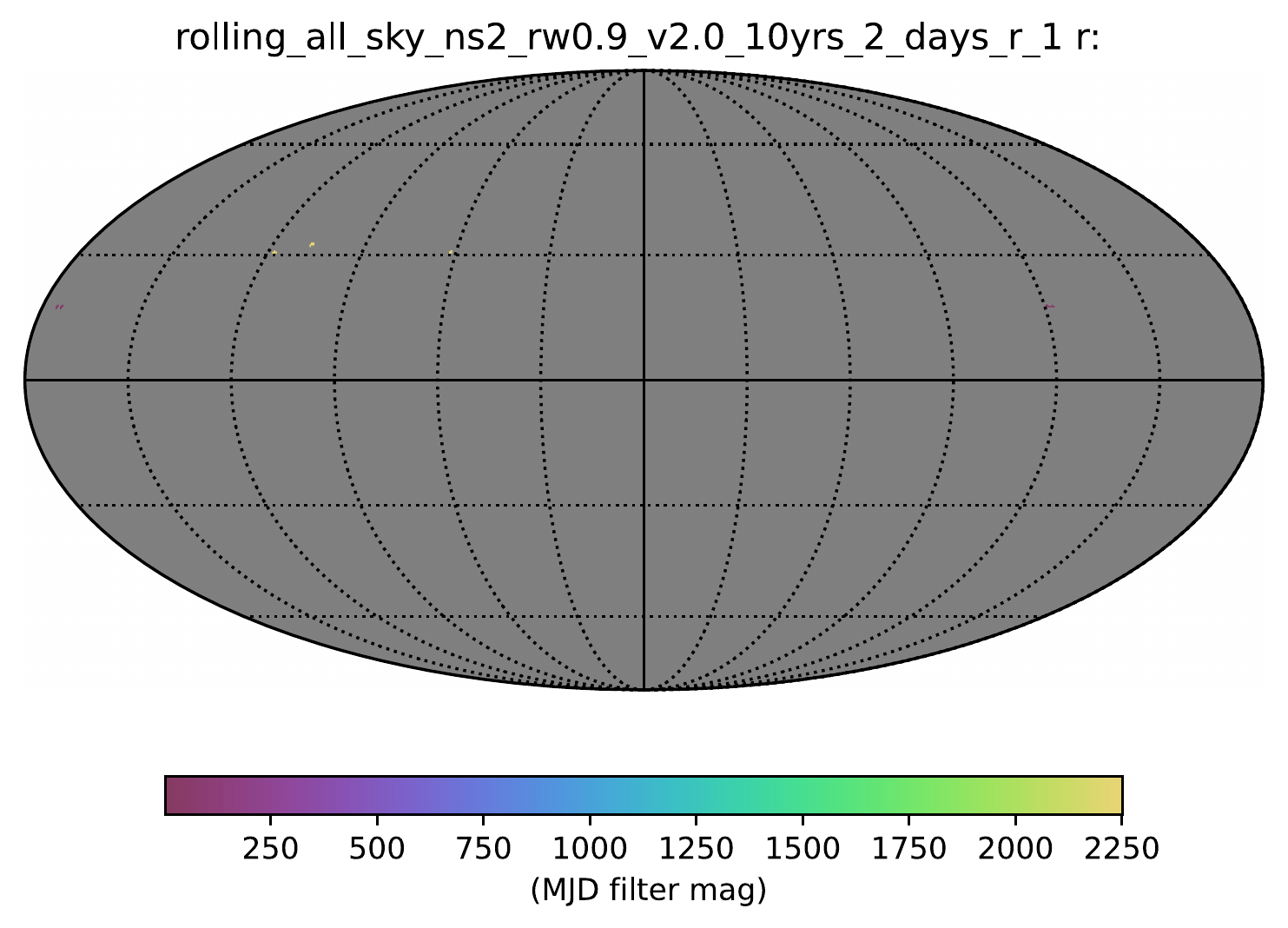}{0.44\textwidth}{(b)}
          }
 \gridline{         
         \fig{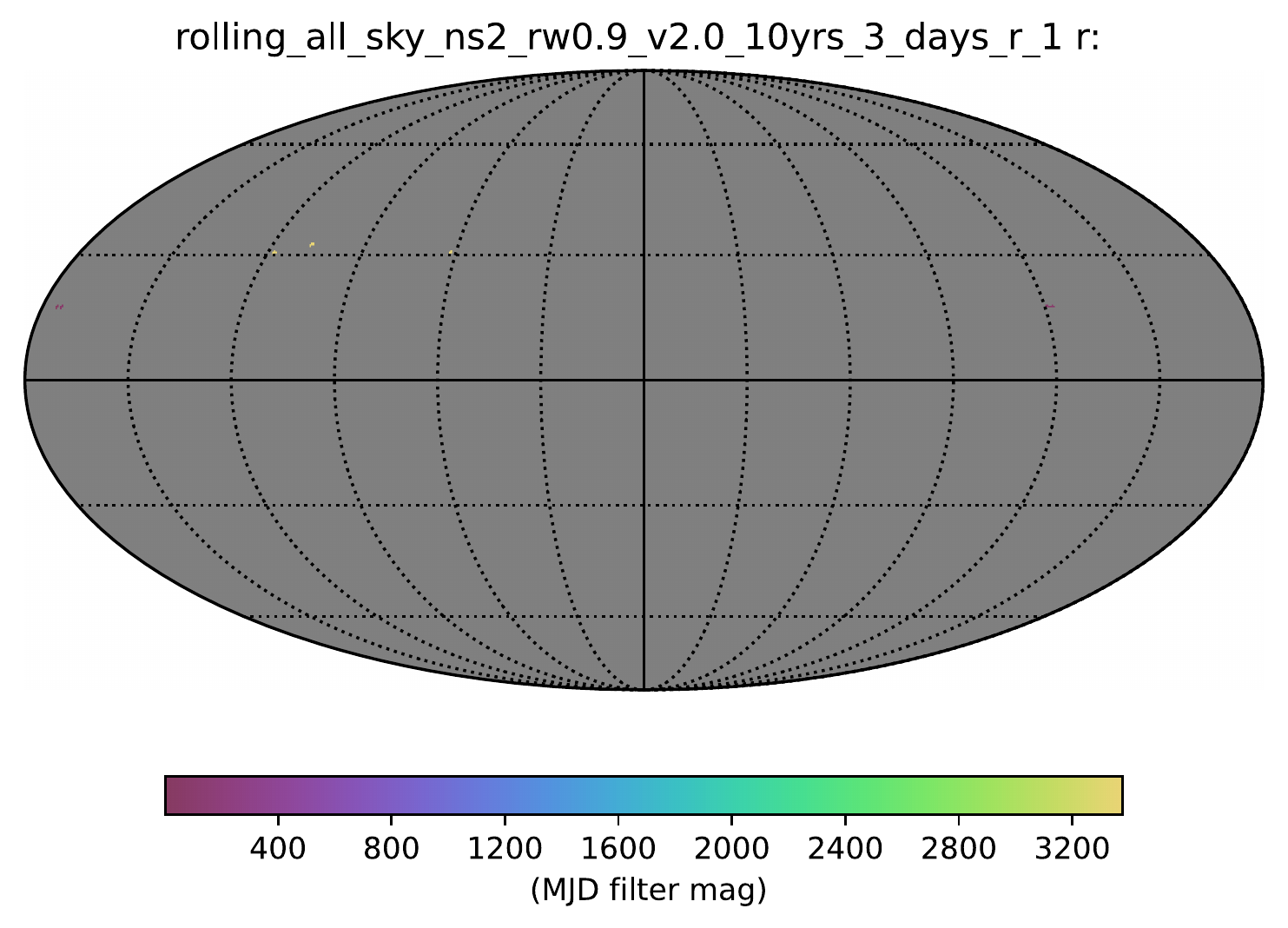}{0.44\textwidth}{(c)}
        \fig{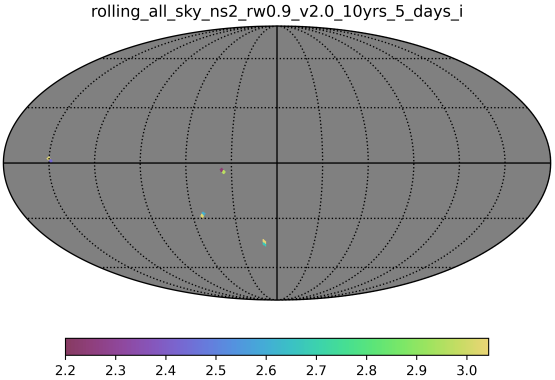}{0.44\textwidth}{(d)}}
        \gridline{
         \fig{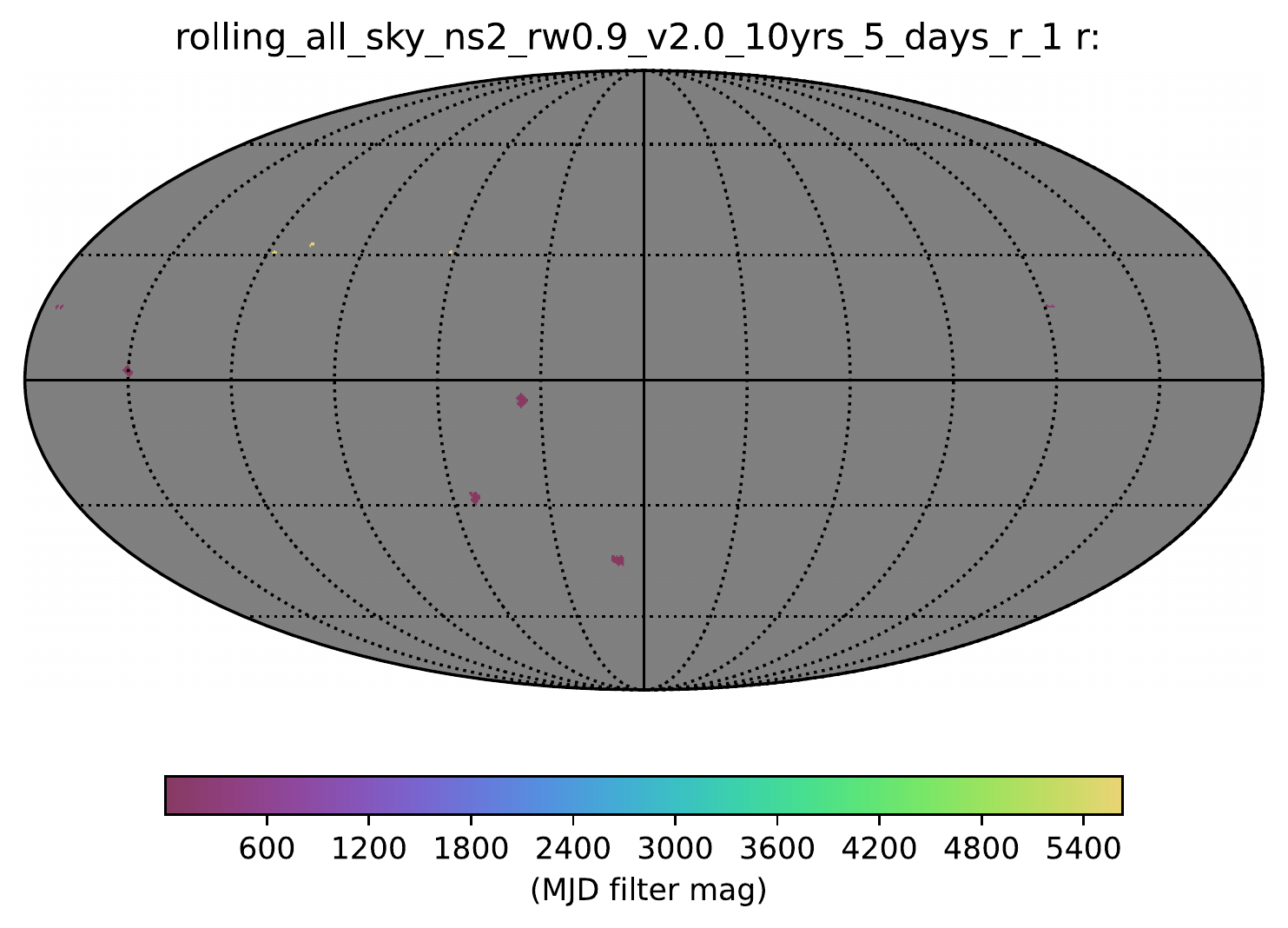}{0.46\textwidth}{(e)}
         \fig{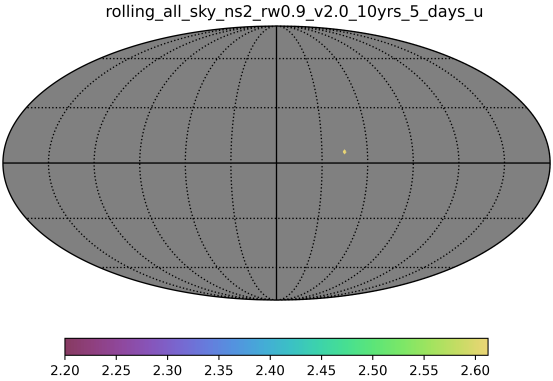}{0.44\textwidth}{(f)}
         }
 
\caption{Metric realization on \texttt{rolling\_all\_sky\_ns2} cadences.  Panels (a)-(c)  show  the  fields (in the fourth area from the left of the main meridian)  suitable for calculating 1- 3 day delays.
DDFs appear in panels (d) and (e).
\label{fig:fig8}}
\end{figure*}

\begin{figure*}[ht]
\gridline{\fig{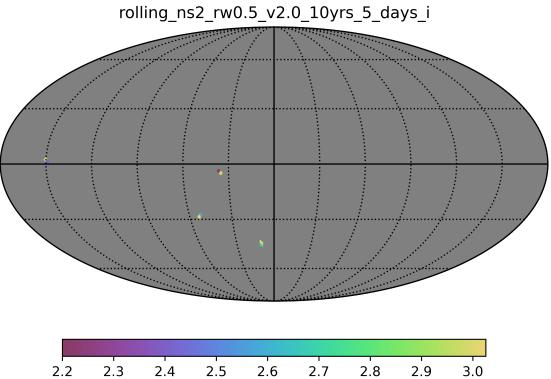}{0.44\textwidth}{(a)}
          \fig{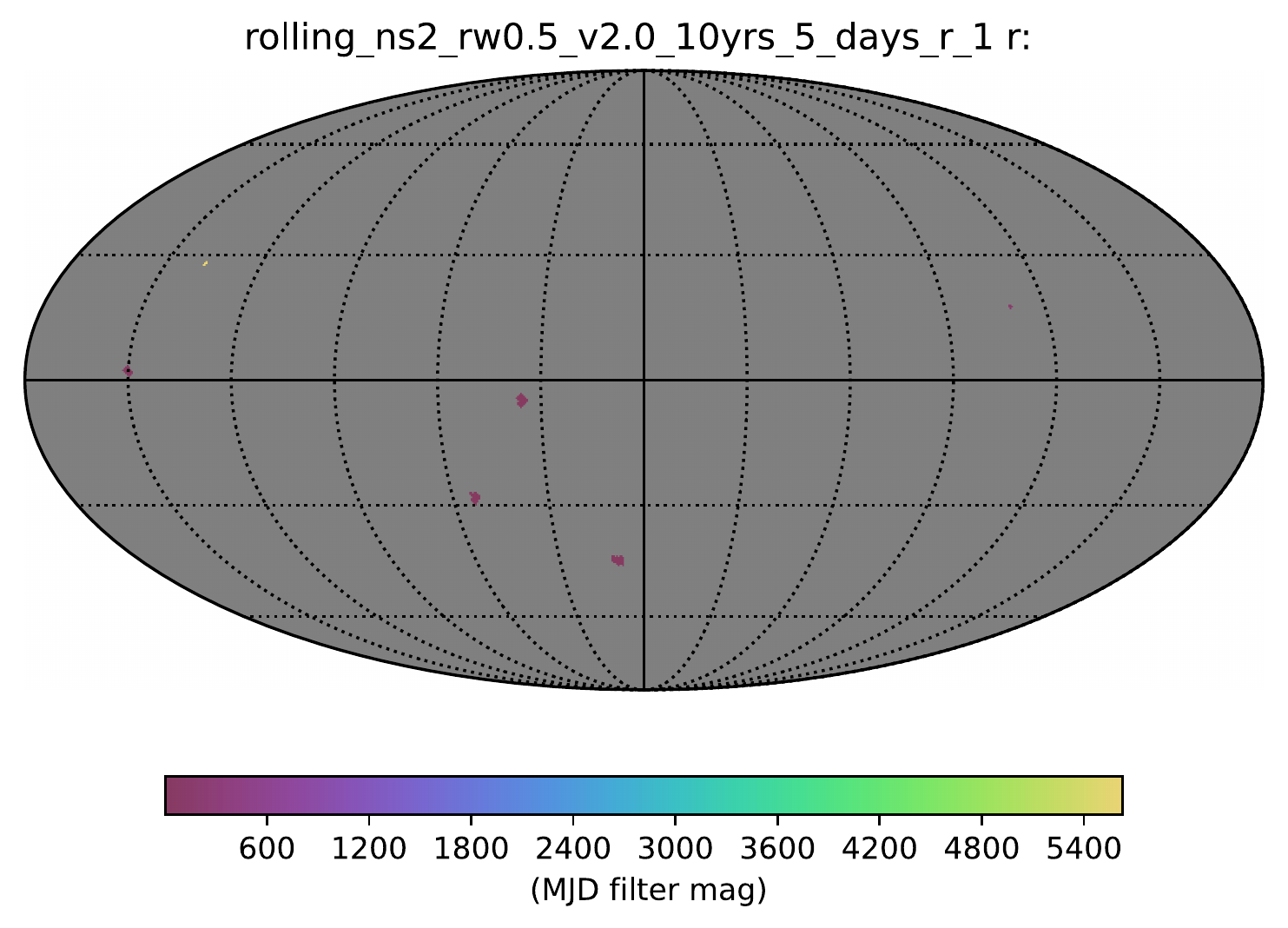}{0.44\textwidth}{(b)}
        }
\gridline{ \fig{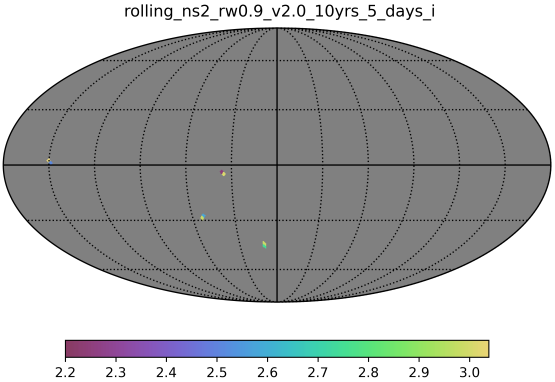}{0.44\textwidth}{(c)}
         \fig{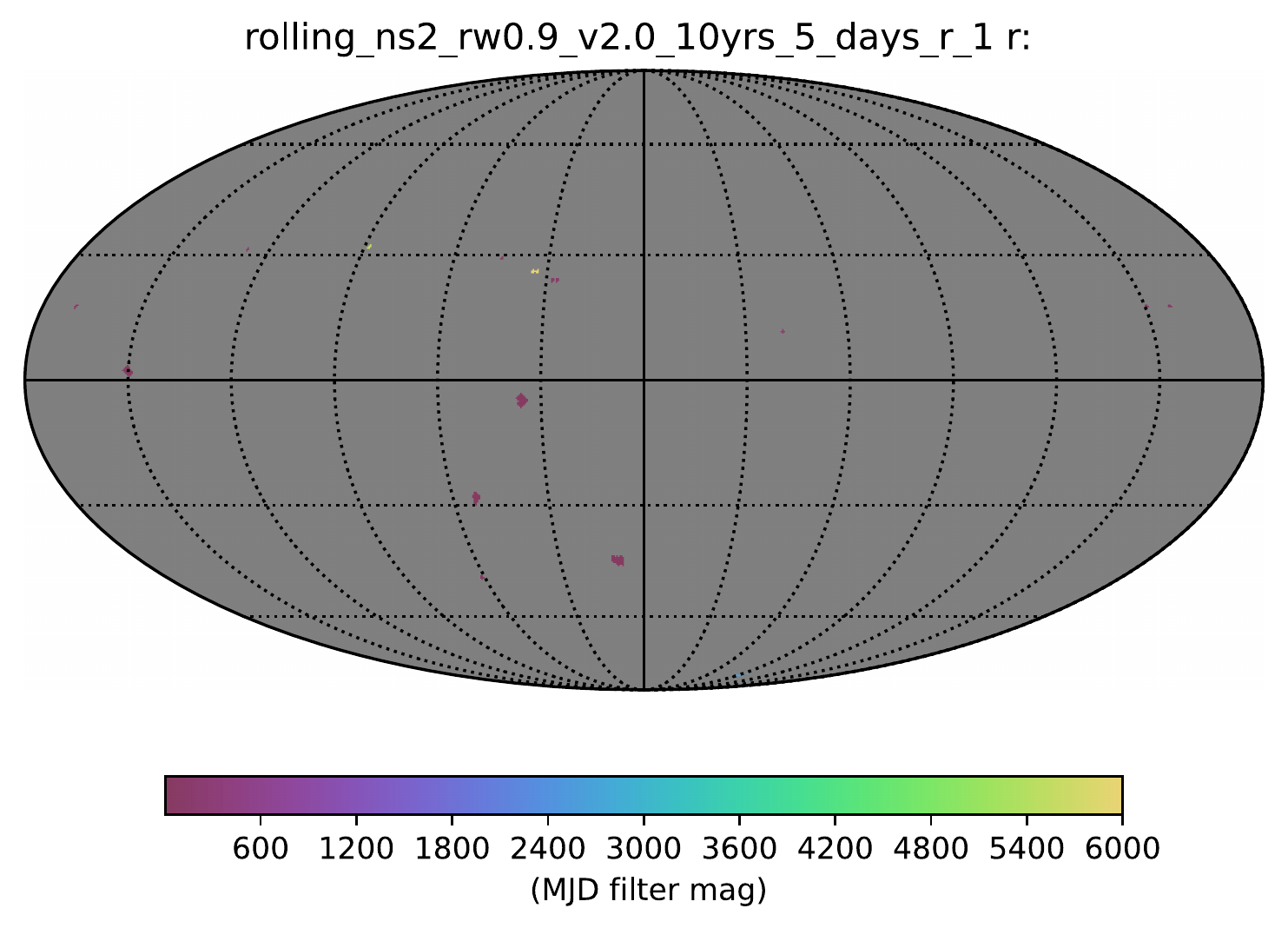}{0.44\textwidth}{(d)}
         }
 
\caption{Metric realization on \texttt{rolling\_all\_sky\_ns2} cadences continued.
DDFs appear in all panels.
\label{fig:fig9}}
\end{figure*}

\begin{figure*}[ht]
\gridline{\fig{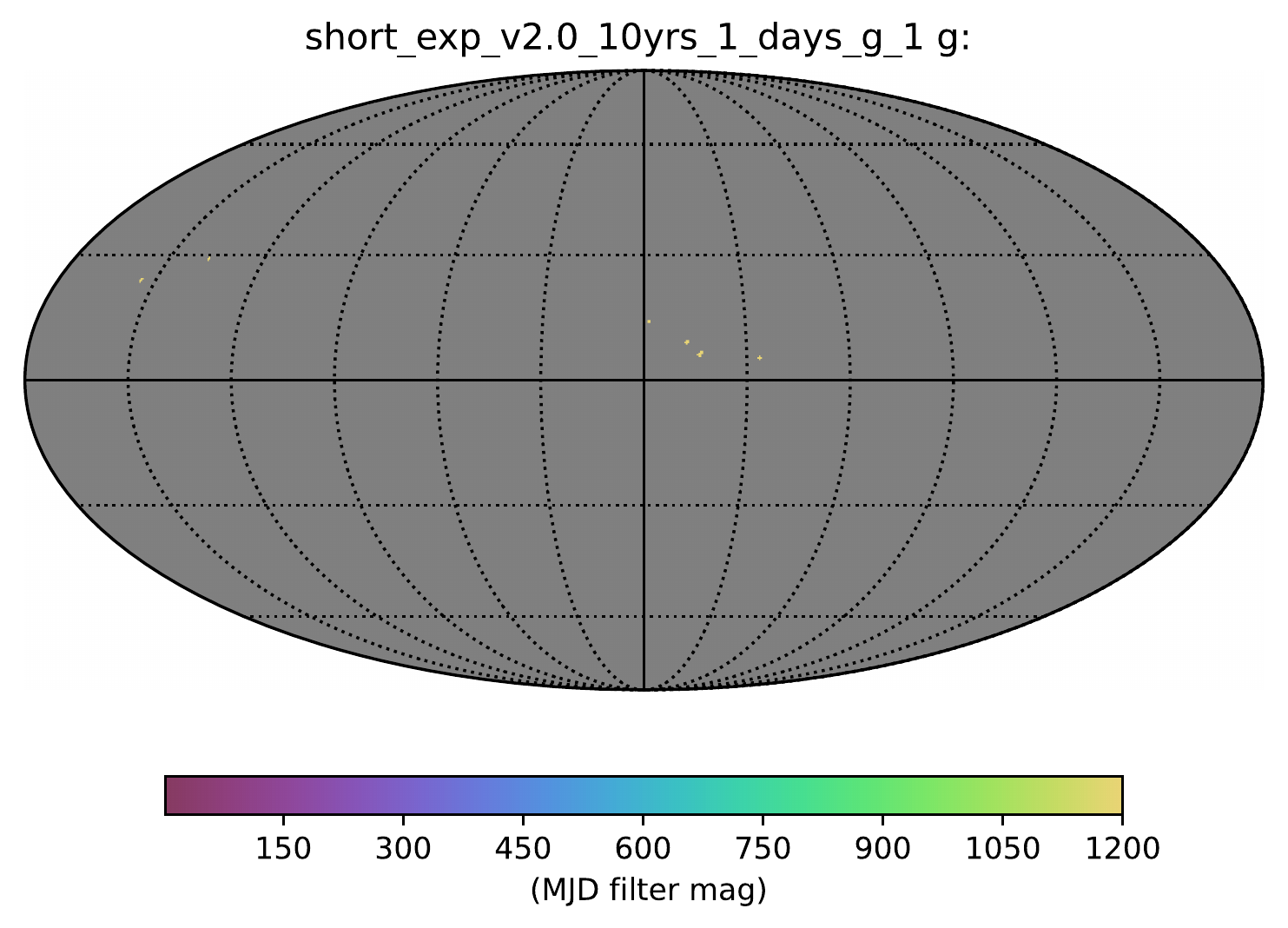}{0.34\textwidth}{(a)}
          \fig{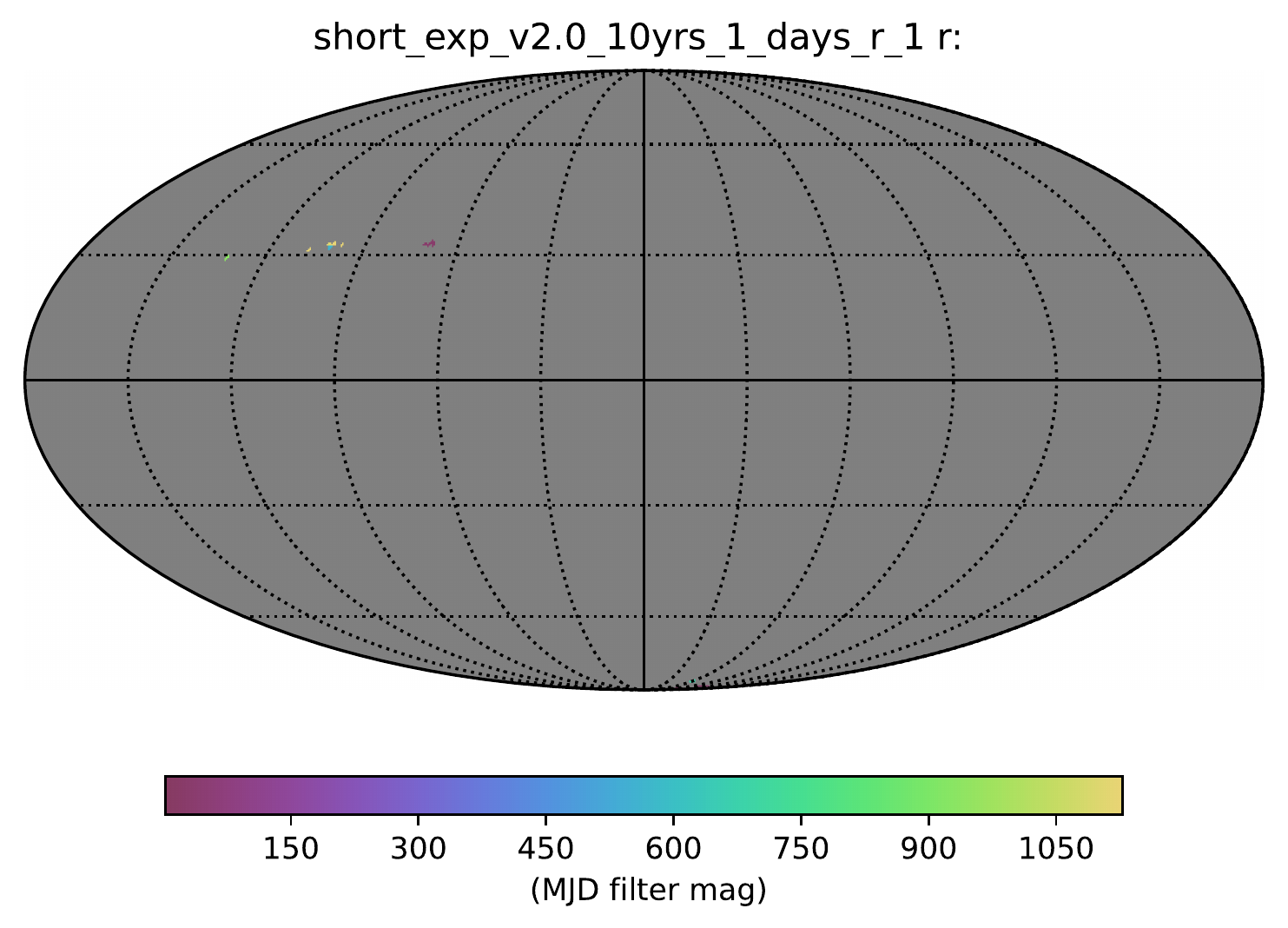}{0.34\textwidth}{(b)}
         \fig{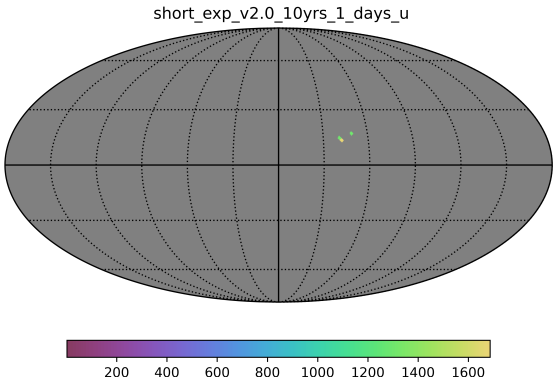}{0.34\textwidth}{(c)}
        }
\gridline{\fig{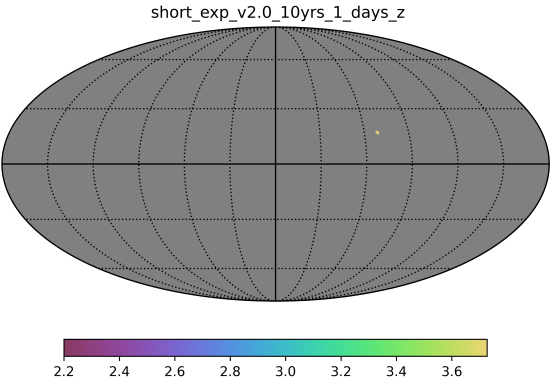}{0.34\textwidth}{(d)}
         \fig{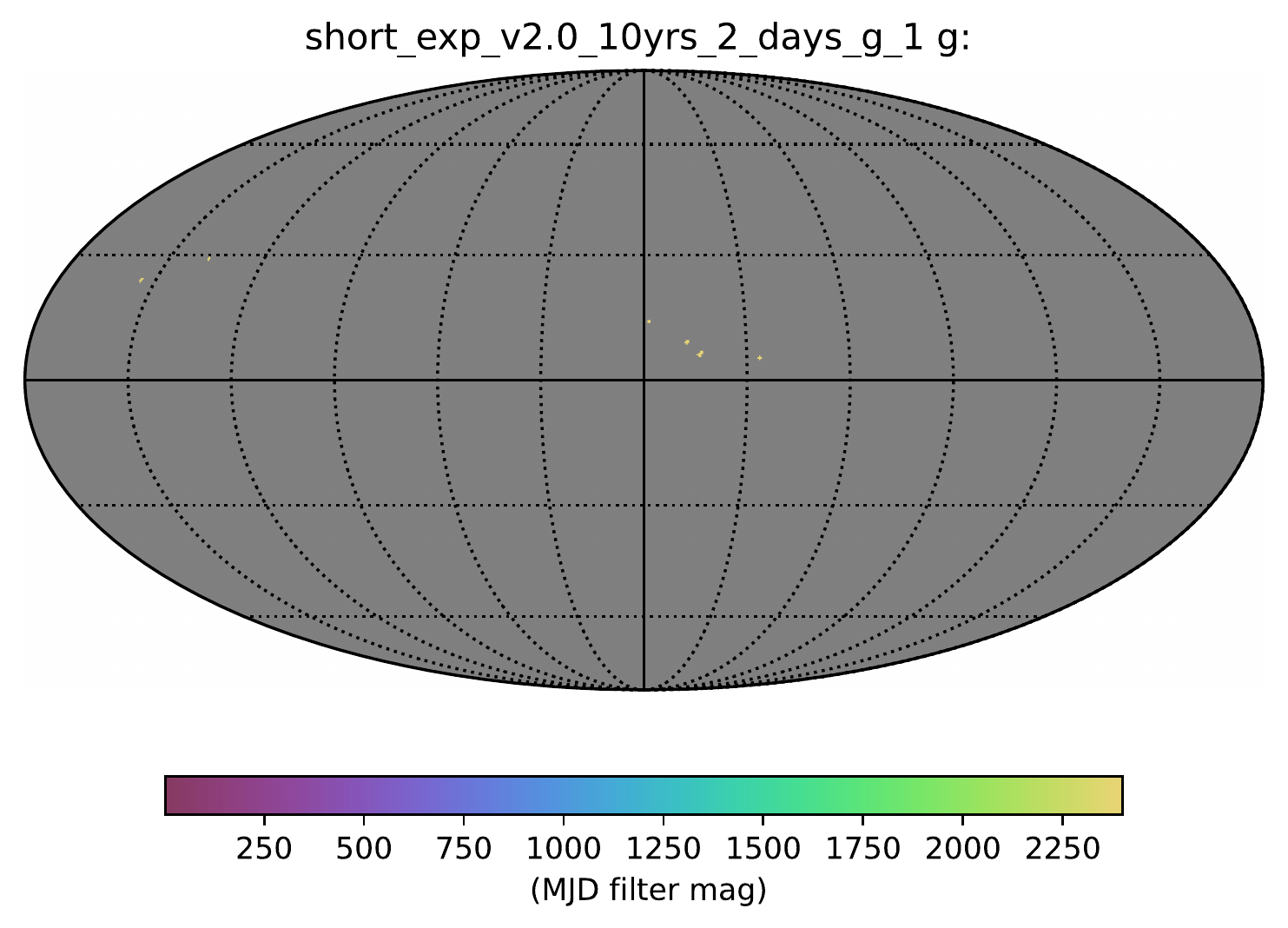}{0.34\textwidth}{(e)}
         \fig{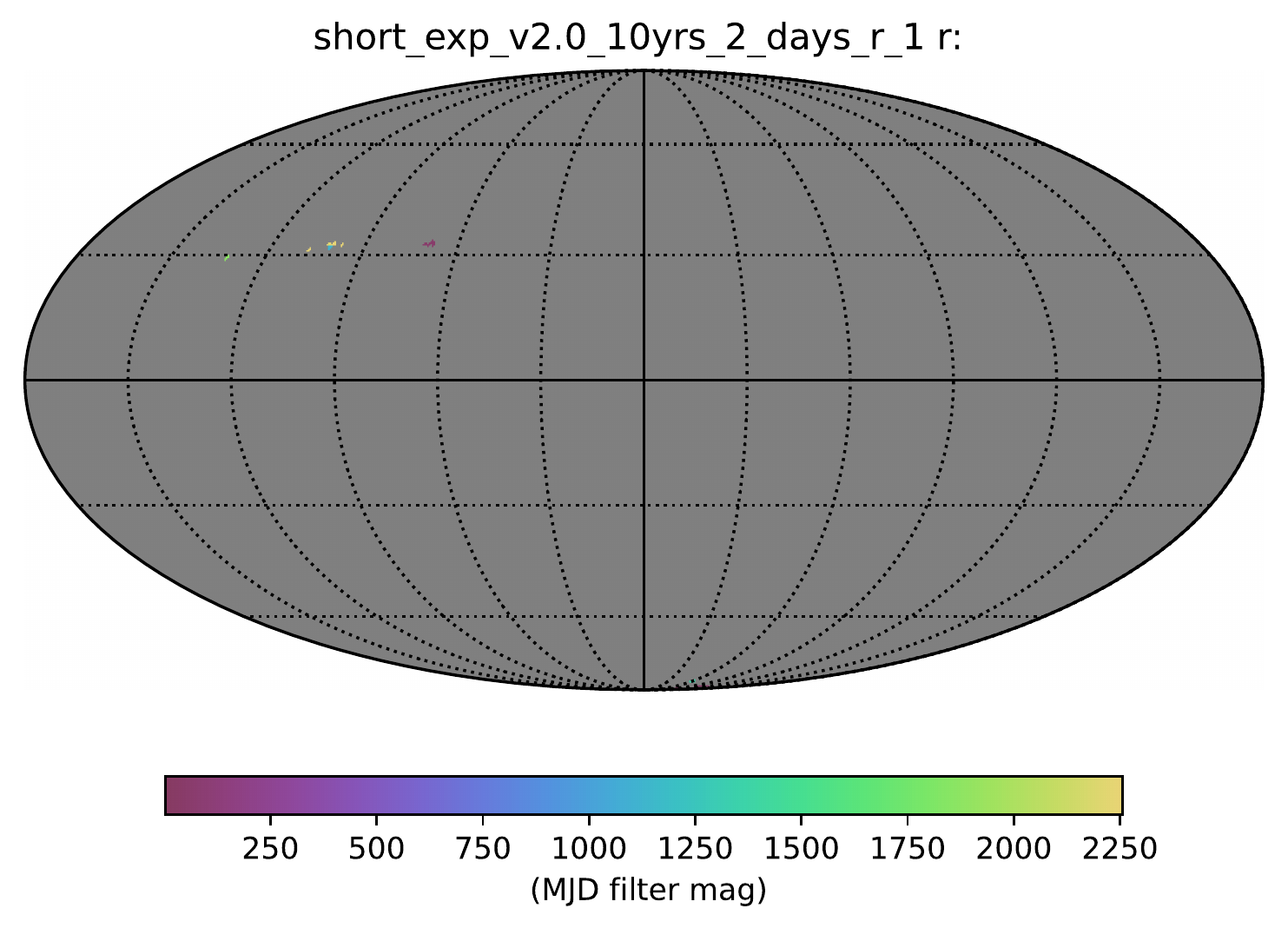}{0.34\textwidth}{(f)}
         }
 \gridline{\fig{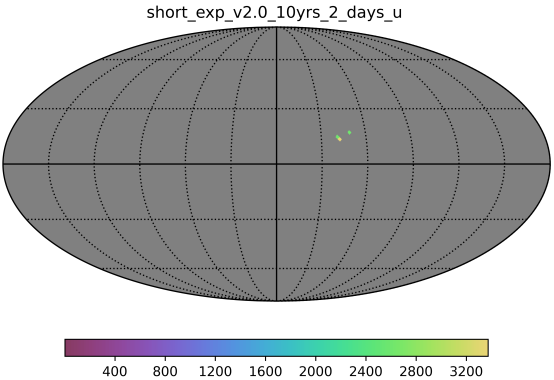}{0.34\textwidth}{(g)}        
          \fig{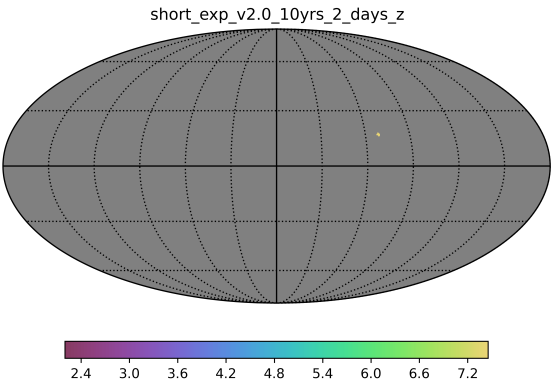}{0.34\textwidth}{(h)}
         \fig{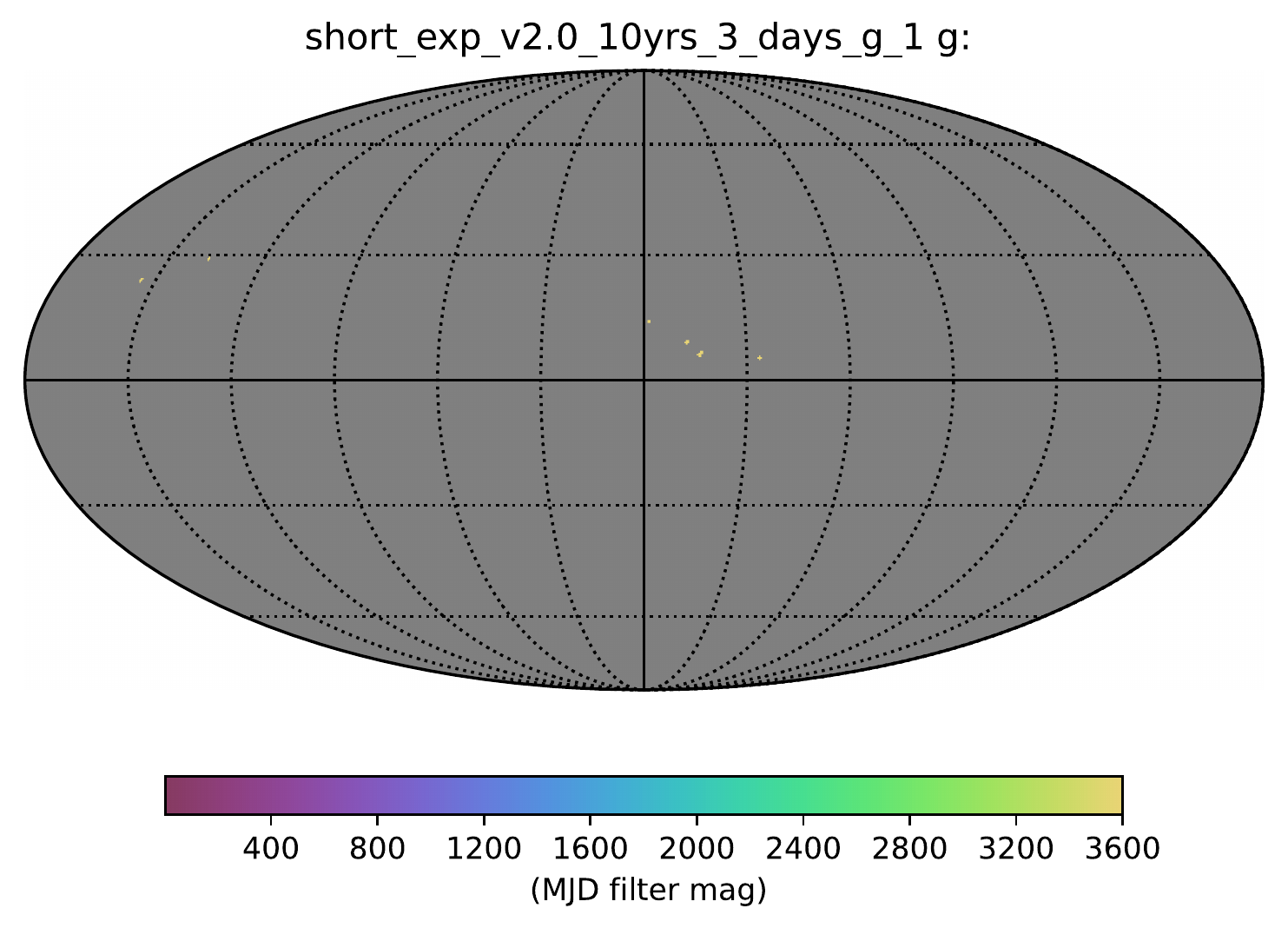}{0.34\textwidth}{(i)}
          }
  \gridline{\fig{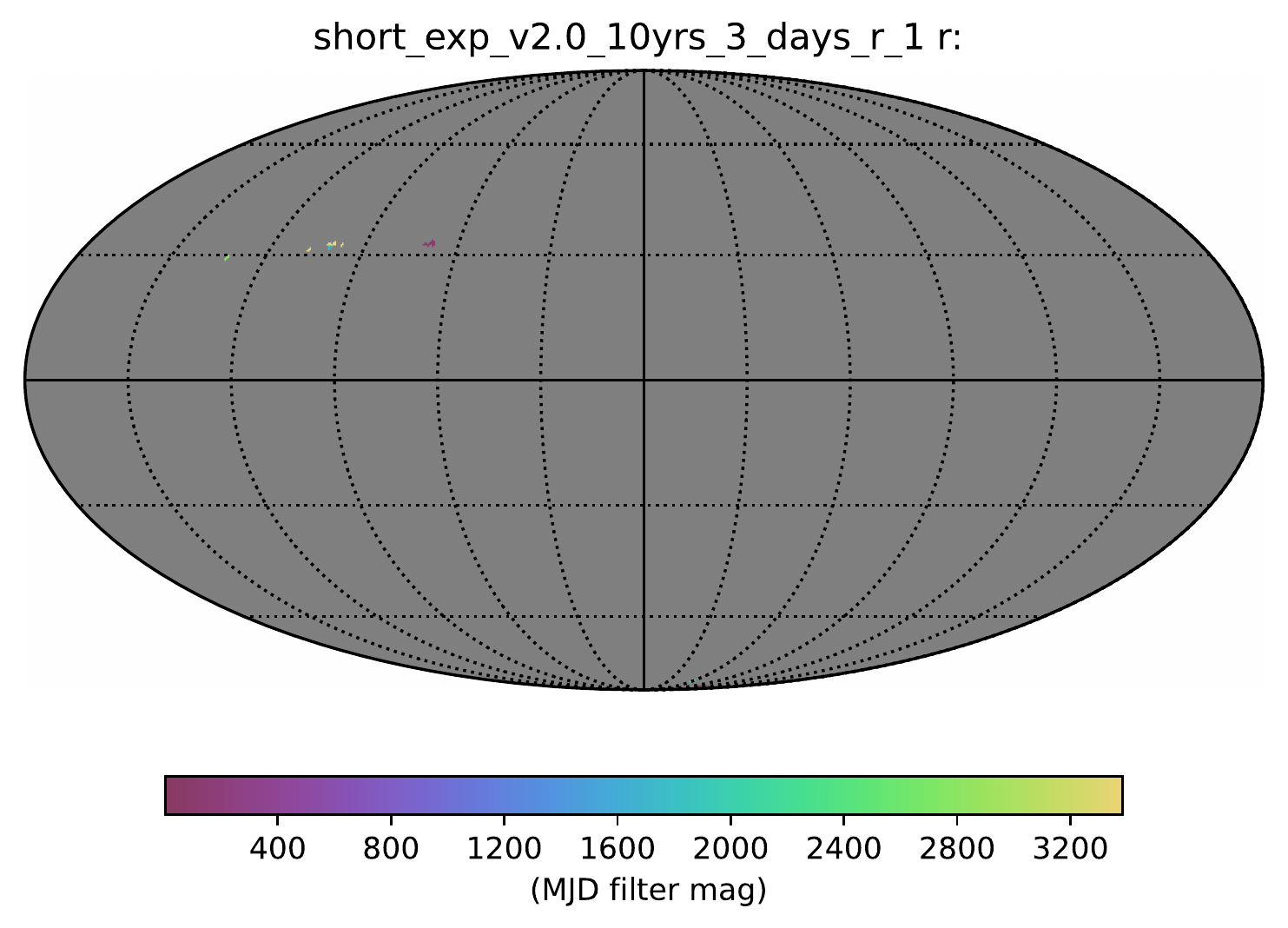}{0.34\textwidth}{(j)}
\fig{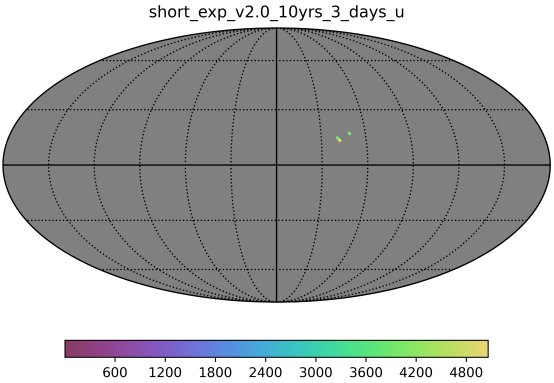}{0.34\textwidth}{(k)}
\fig{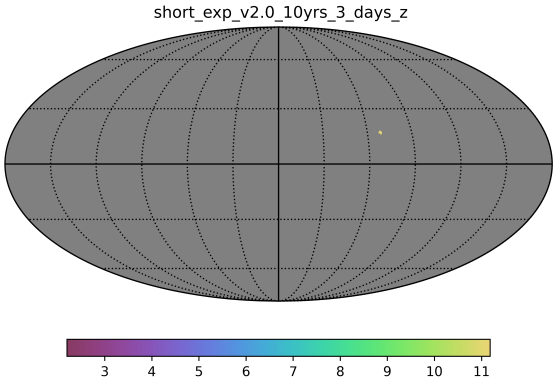}{0.34\textwidth}{(l)}
}
\caption{Metric realization on \texttt{short\_exp} cadences. Metric run emphasizes two fields in \textit{u}-band (panels (c), (g) and (k)).
\label{fig:fig10}}
\end{figure*}

\begin{figure*}[ht]
\gridline{\fig{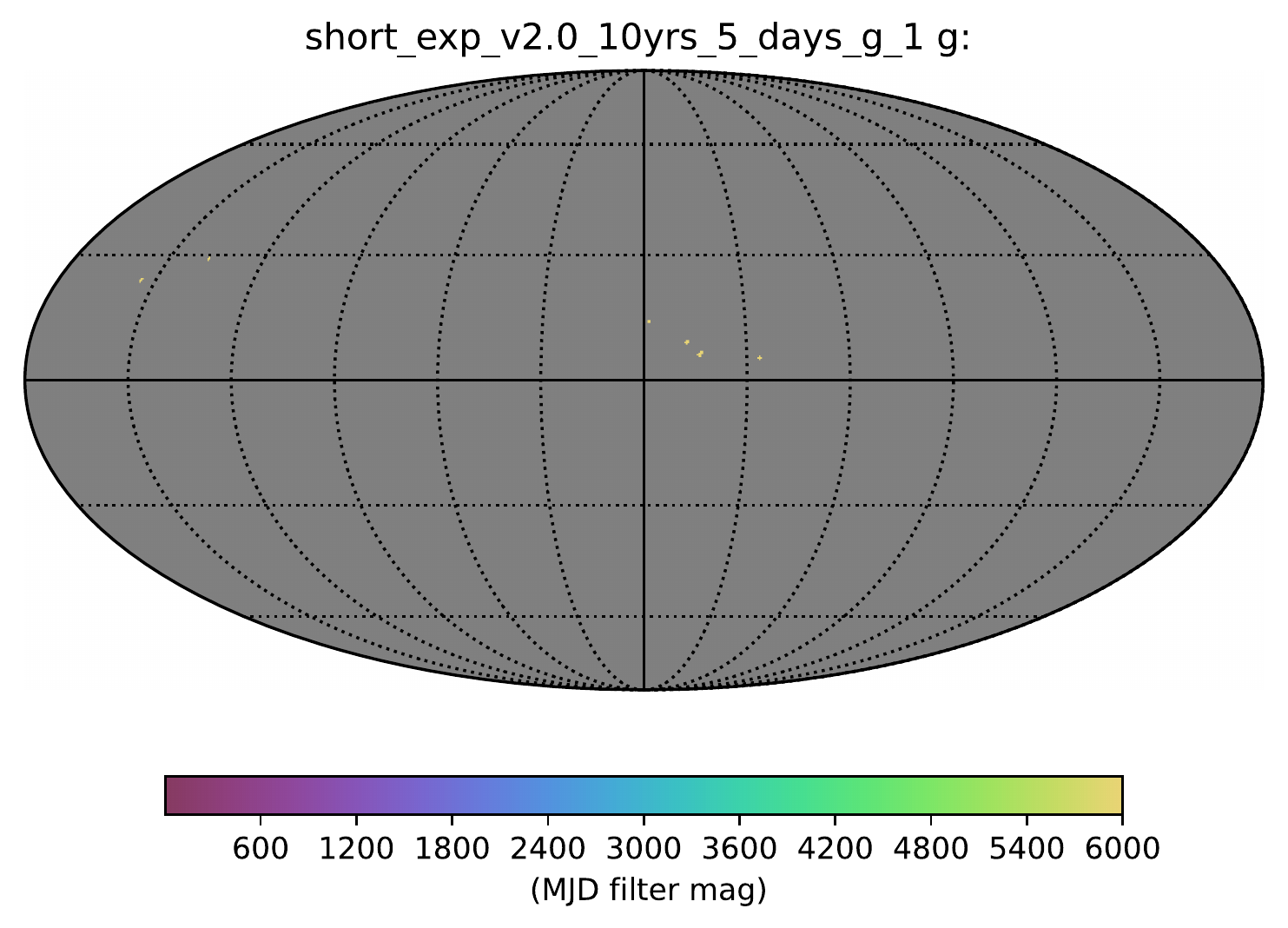}{0.34\textwidth}{(a)}
          \fig{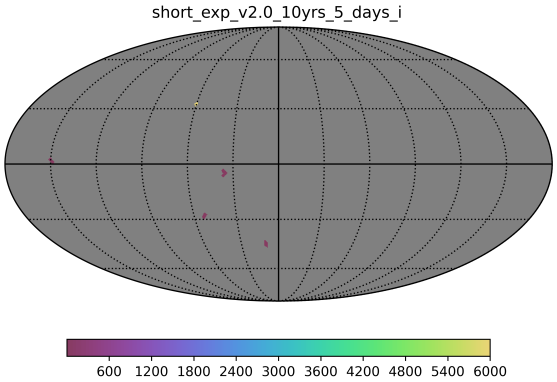}{0.34\textwidth}{(b)}
         \fig{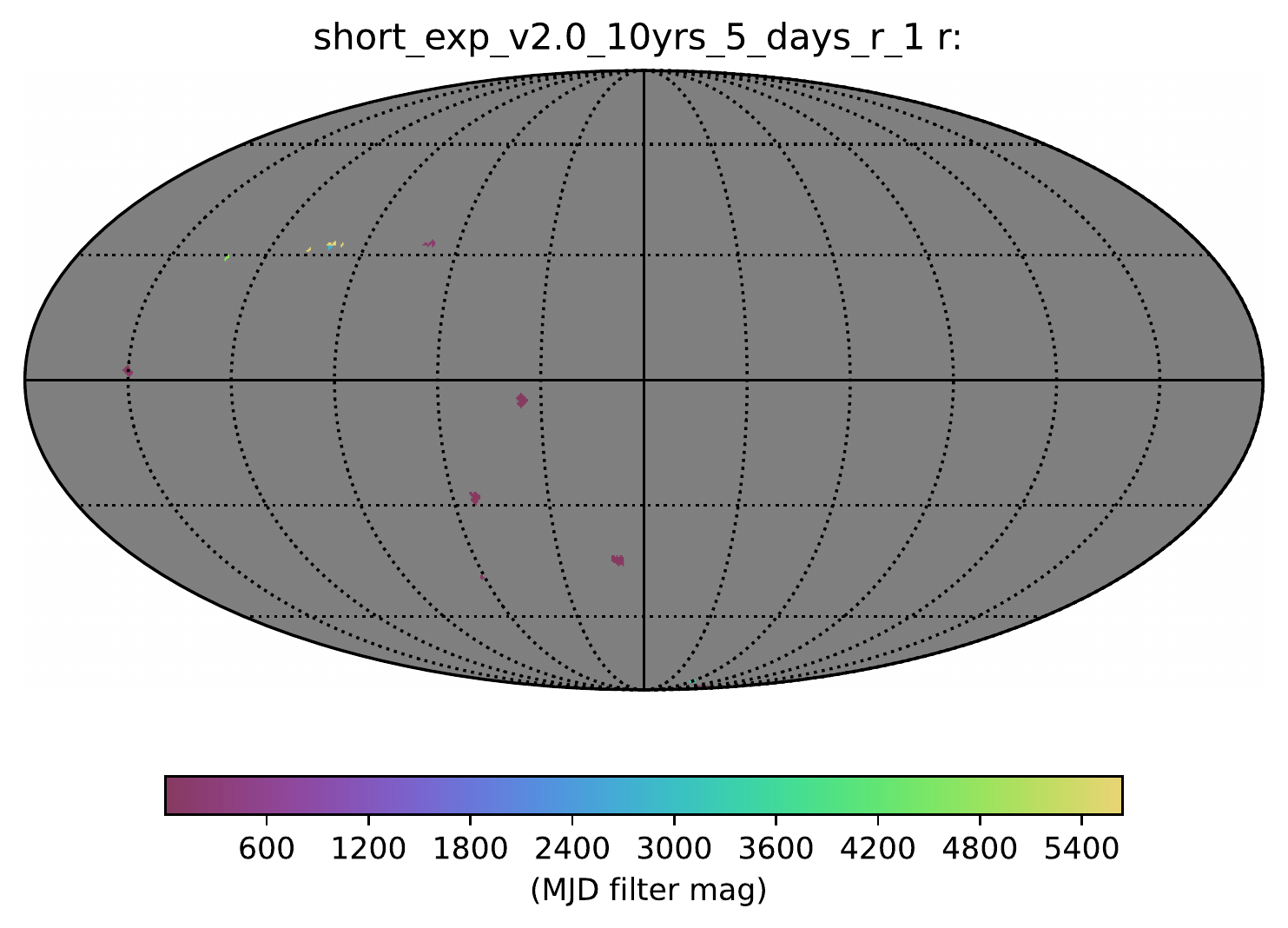}{0.34\textwidth}{(c)}
        }
\gridline{\fig{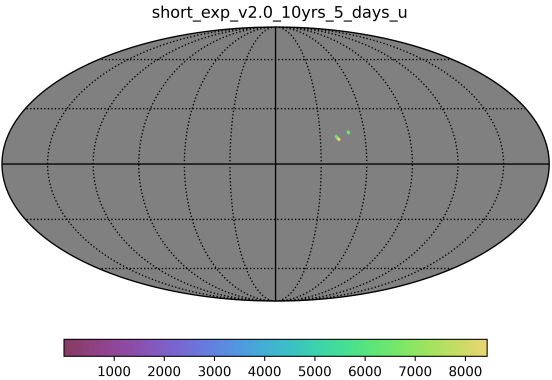}{0.34\textwidth}{(d)}
         \fig{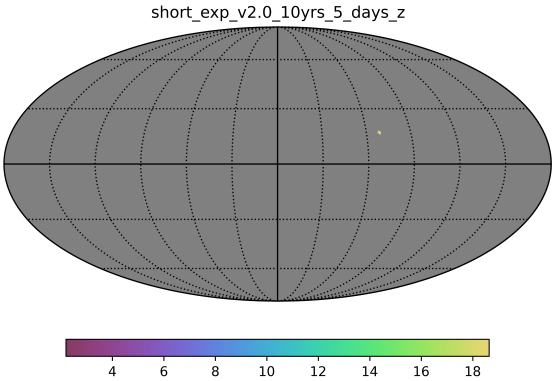}{0.34\textwidth}{(e)}
}
          
\caption{Metric realization on \texttt{short\_exp} cadences continued. Three DDFs are prominent in panels (b) and (c).
\label{fig:fig11}}
\end{figure*}

\begin{figure*}[ht]
\gridline{\fig{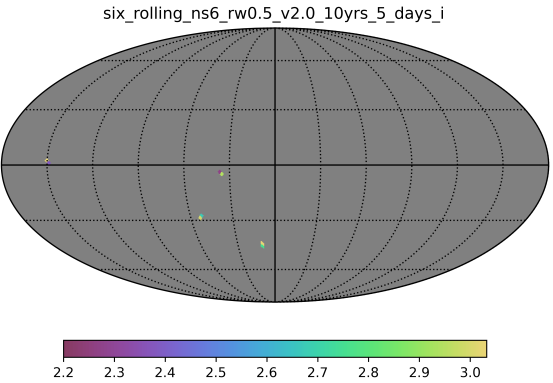}{0.44\textwidth}{(a)}
          \fig{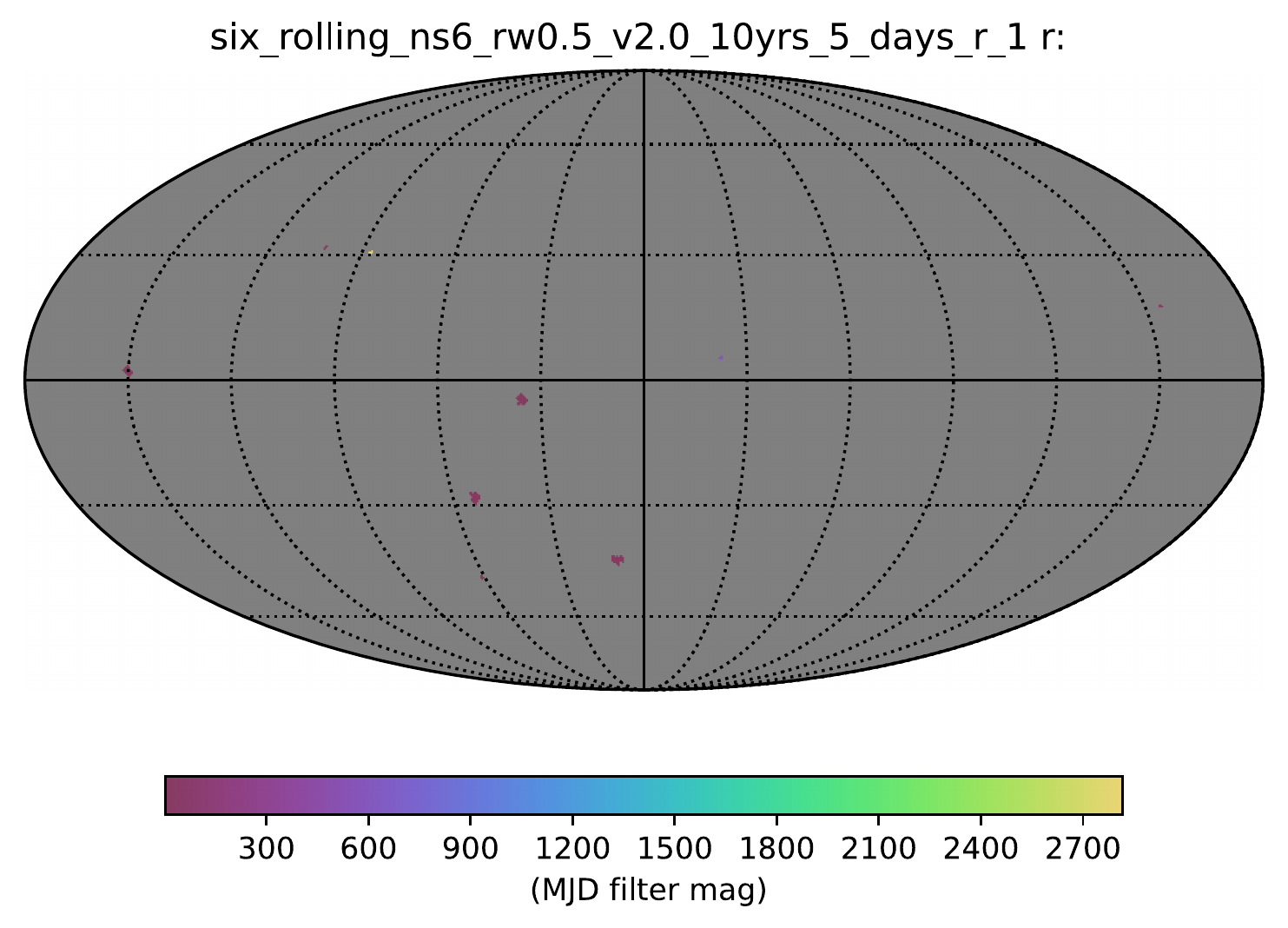}{0.44\textwidth}{(b)}
        }
\gridline{\fig{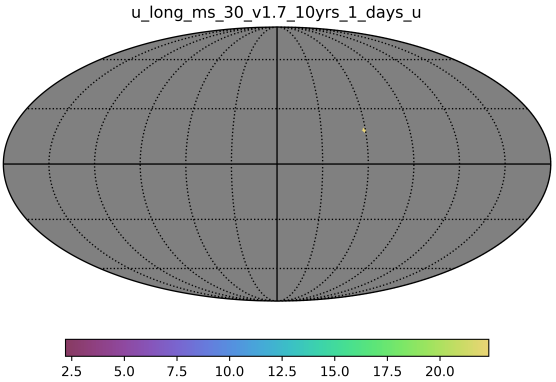}{0.44\textwidth}{(c)}
         \fig{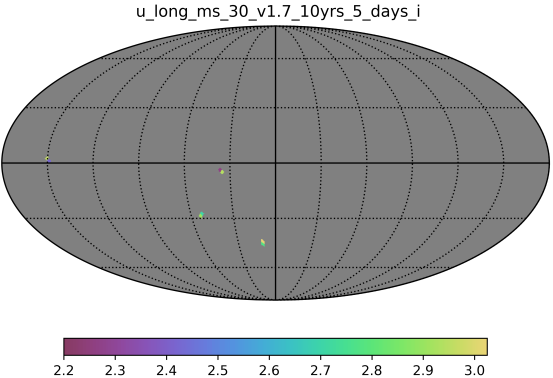}{0.44\textwidth}{(d)}
         }
 \gridline{\fig{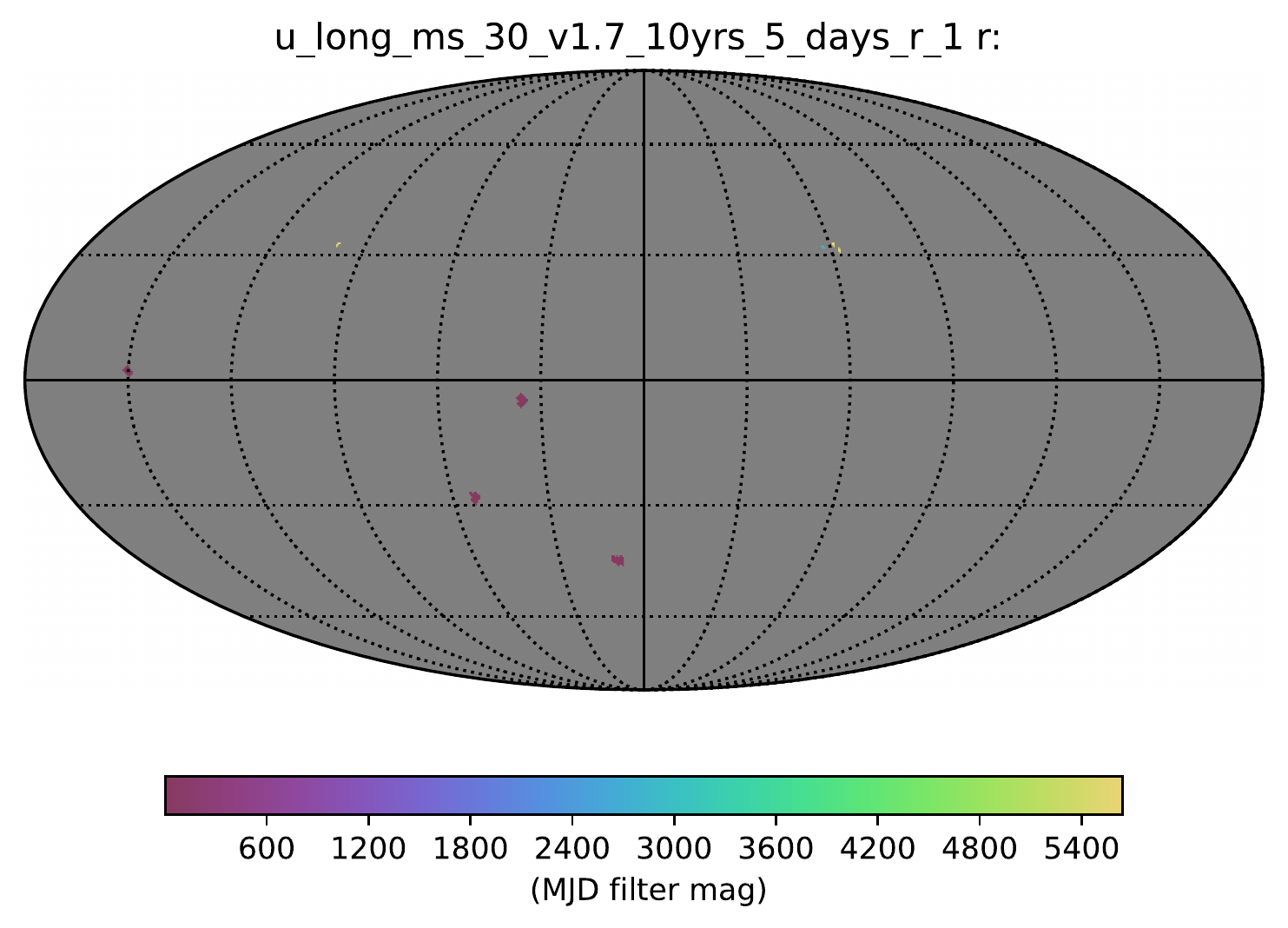}{0.44\textwidth}{(g)}        
          }
\caption{Metric realization on \texttt{six\_rolling} and \texttt{u\_long\_ms\_30} cadences. DDFs are prominent in \textit{i} band (panels (a) and (e))  and \textit{r} band (panels (b) and (g)).
\label{fig:fig12}}
\end{figure*}

\begin{figure}[hbt!]
\gridline{\fig{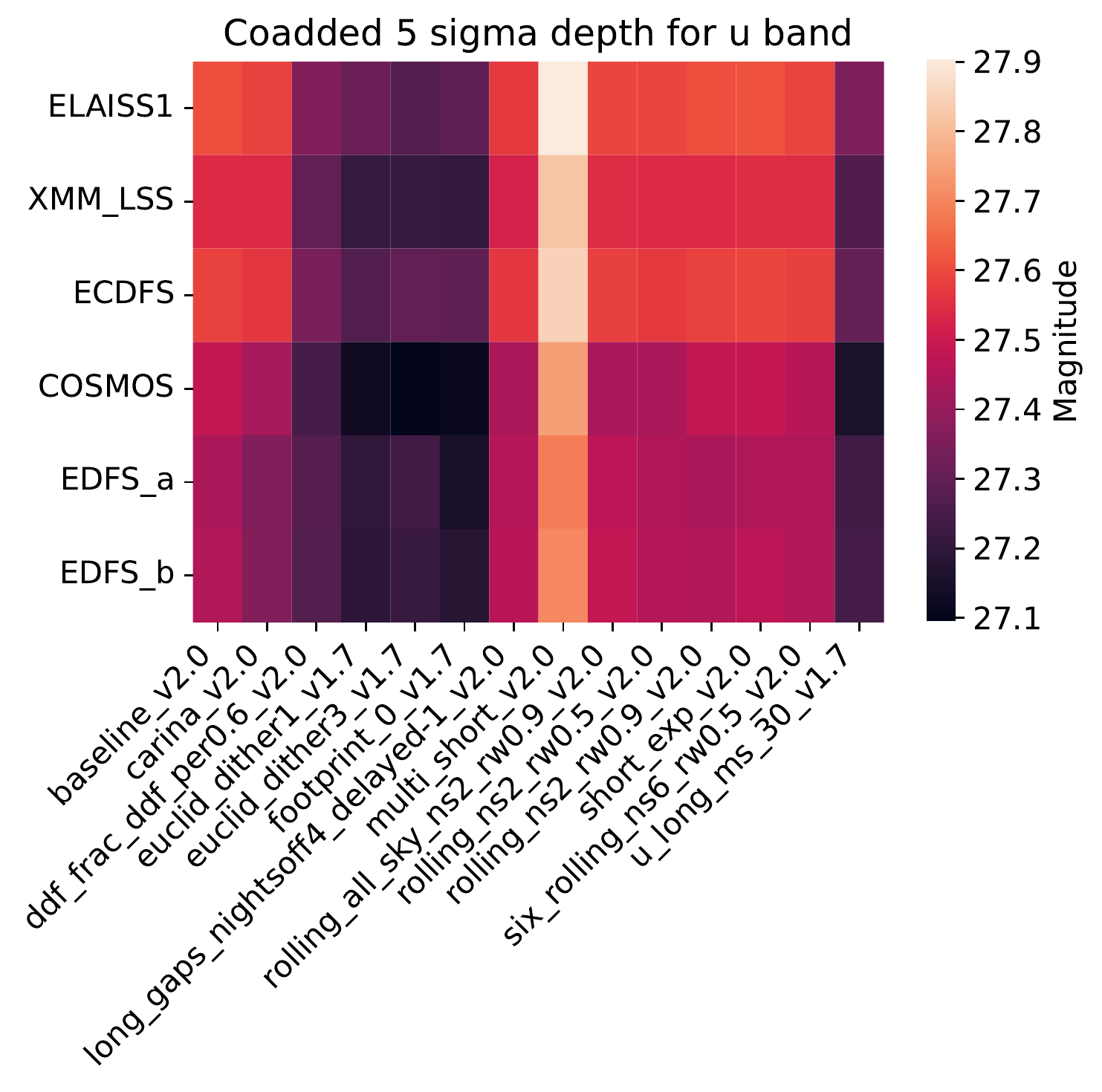}{0.44\textwidth}{(a)}
          \fig{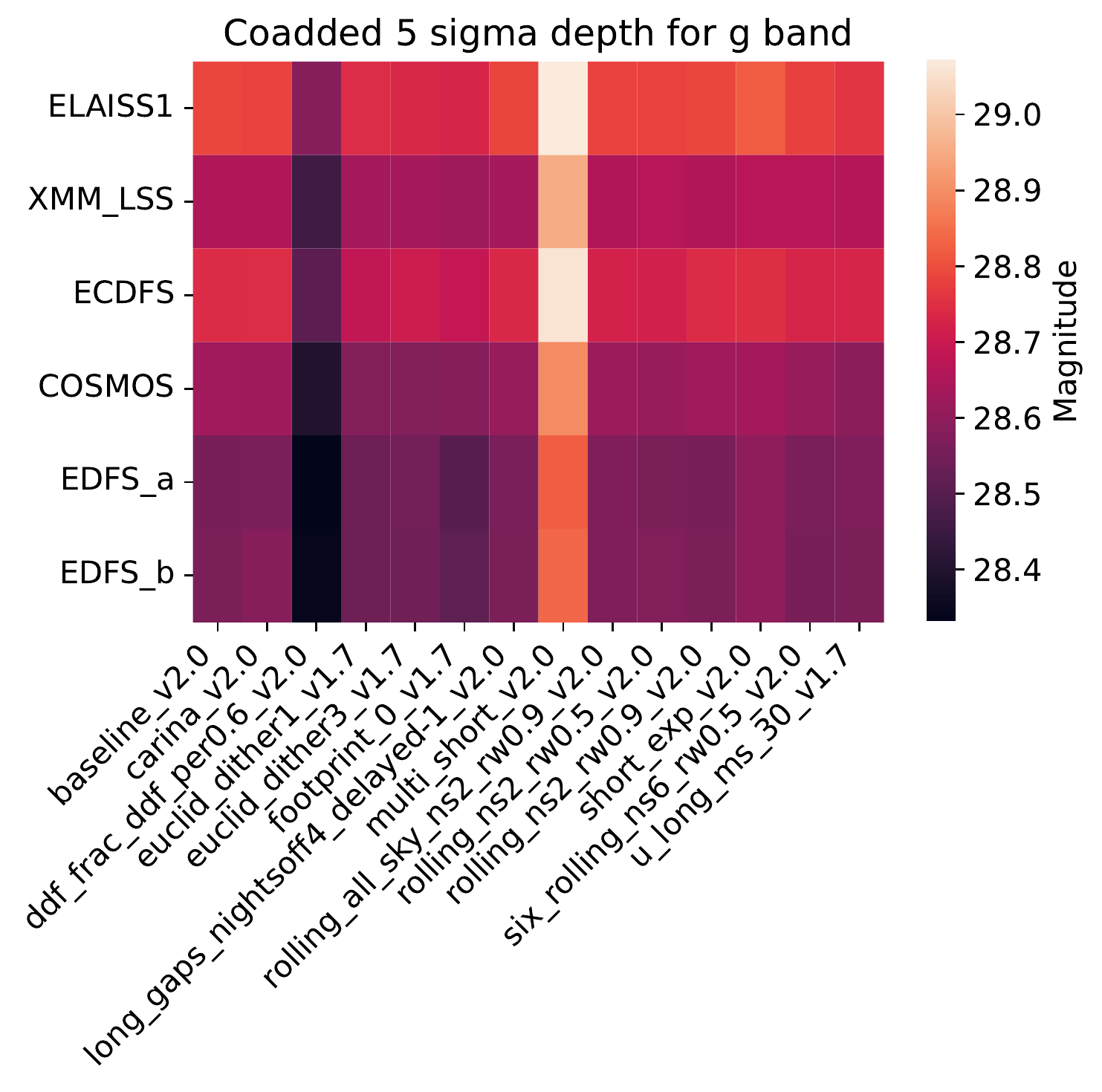}{0.44\textwidth}{(b)}
        }
\gridline{\fig{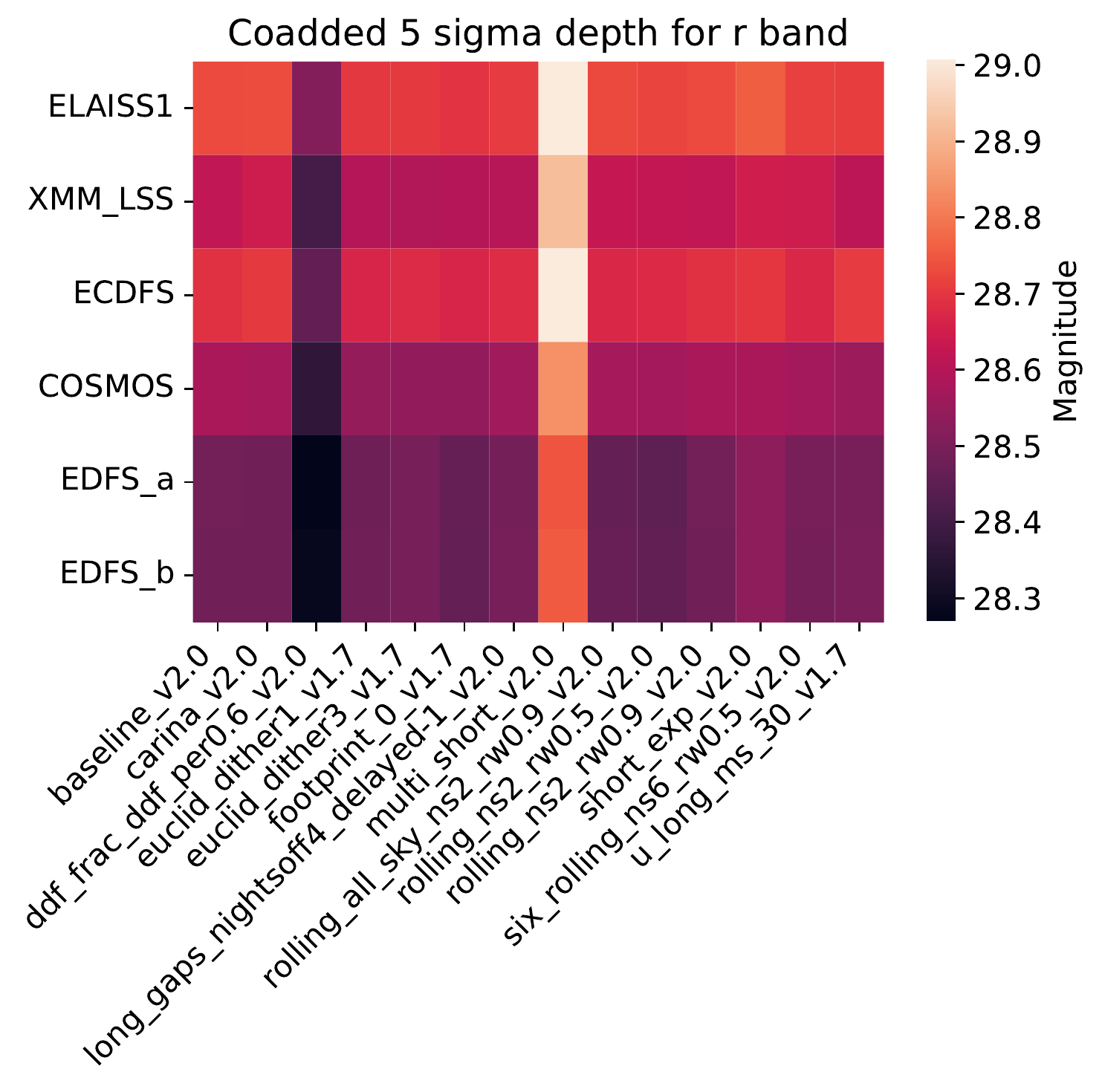}{0.44\textwidth}{(c)}
         \fig{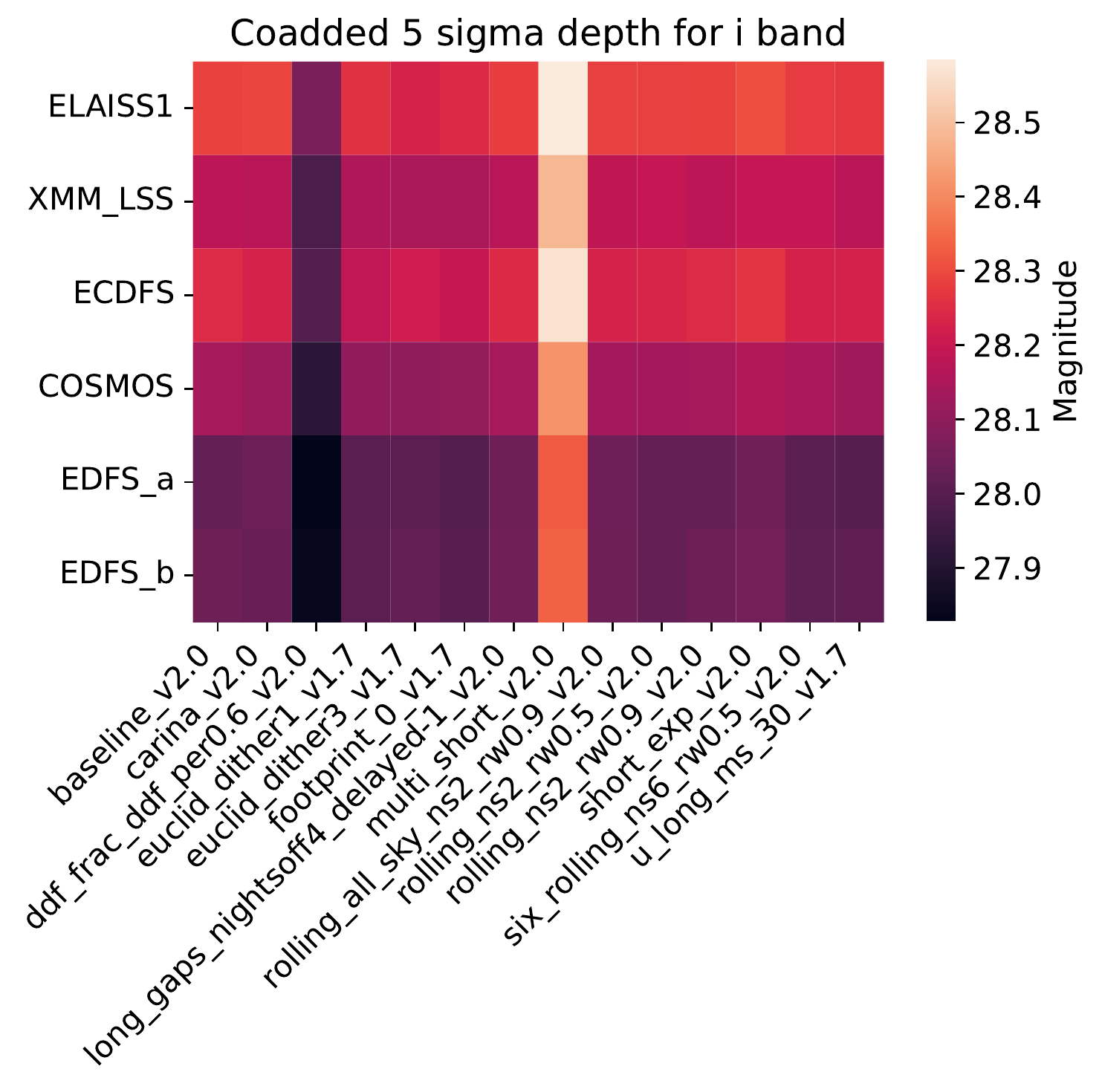}{0.44\textwidth}{(d)}
         }
\caption{Co-added depth in deep drilling fields for different observing strategies in \textit{u,g,r,i} filters. The best strategy for AD reverberation mapping is also the deepest in all bands.
\label{fig:fig13}}
\end{figure}

\begin{figure}[!ht]

 \gridline{\fig{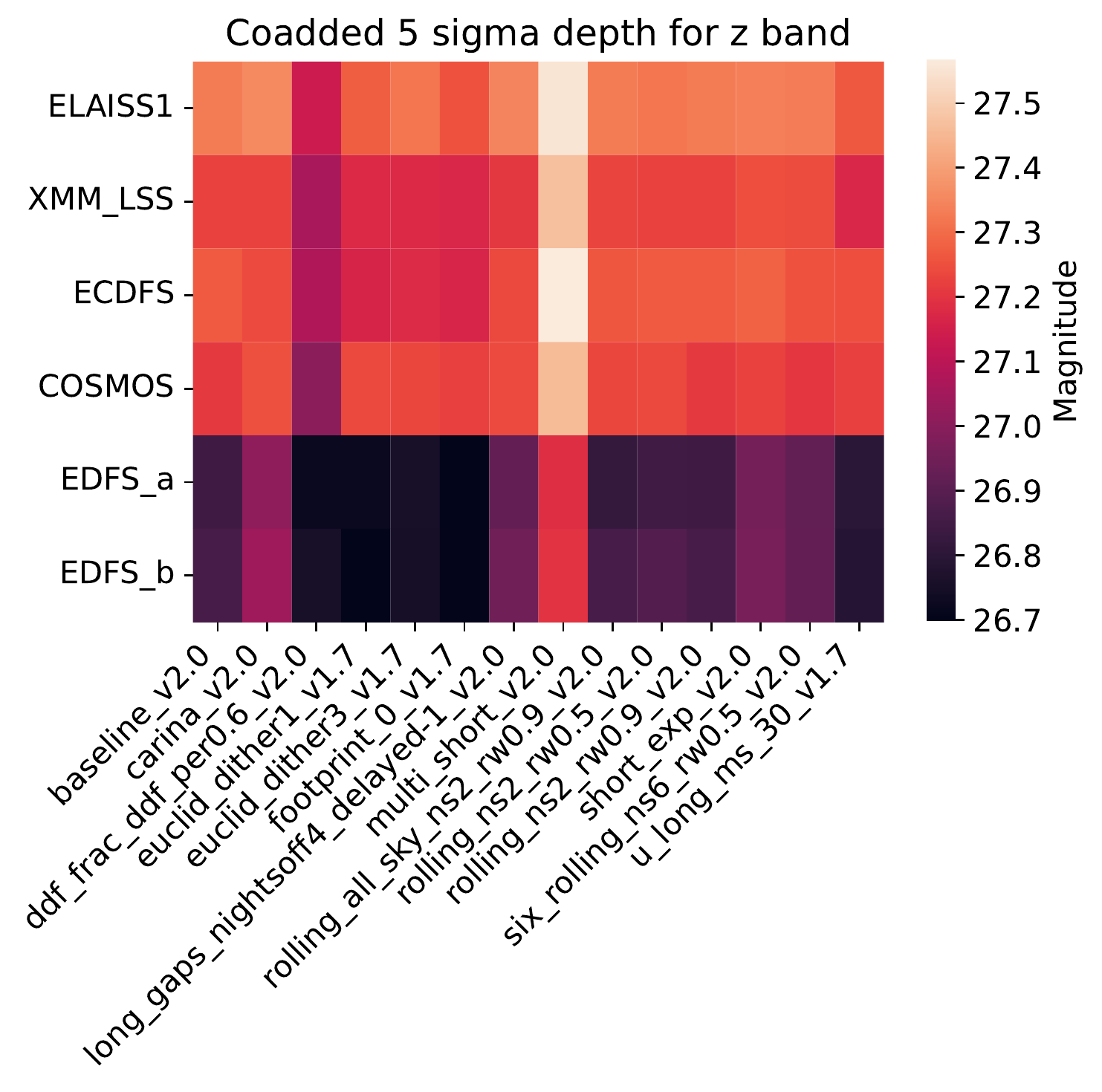}{0.44\textwidth}{(e)}        
     \fig{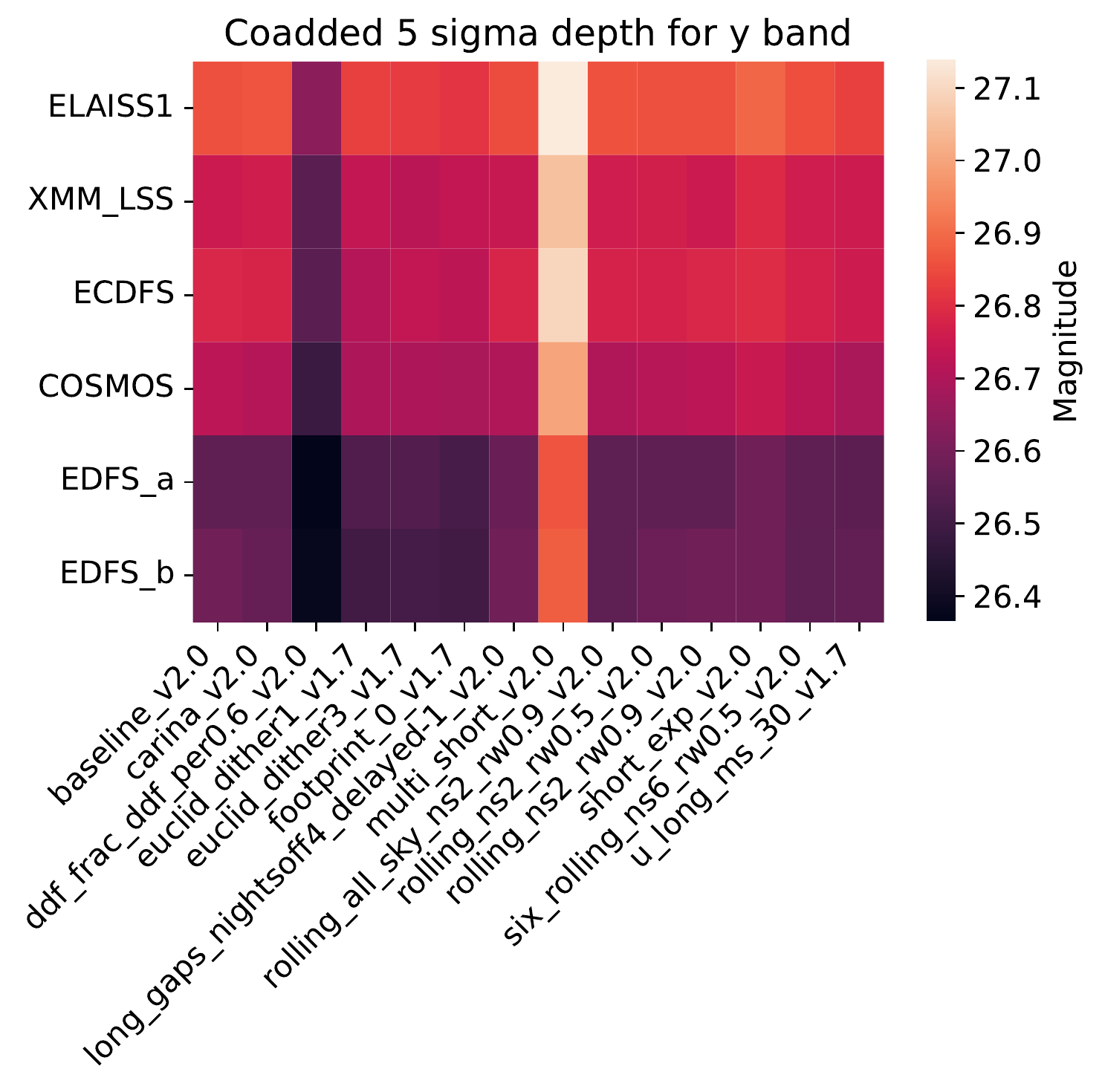}{0.44\textwidth}{(f)}
          }
\caption{The same as Fig \ref{fig:fig13}, but for \textit{z,y} filters.
\label{fig:fig14}}
\end{figure}

\end{document}